\documentclass[
dsingle,
]{Dissertate}

\usepackage{amsmath, amsfonts, amssymb}
\usepackage{mathrsfs}
\usepackage{dsfont}
\usepackage[capitalise]{cleveref}
\usepackage{physics}
\usepackage{datetime2}
\usepackage{soul}

\usepackage{xcolor}

\usepackage{tikz}
\usetikzlibrary{shapes}

\newcommand{\ident}{\mathds{1}}

\begin{document}


\title{
Entanglement with neutral atoms in the simulation of nonequilibrium dynamics of one-dimensional spin models
}
\author{Anupam Mitra}
\previousdegrees{
B.E., Computer Science, Birla Institute of Technology and Science Pilani, Goa, India, 2011 \\
M.Sc., Physics, Birla Institute of Technology and Science Pilani, Goa, India, 2011 \\
M.S., Physics, The University of New Mexico, 2019
}

\advisor{Ivan H. Deutsch}

\committeeInternalOne{Akimasa Miyake}
\committeeInternalTwo{Tameem Albash}

\coadvisorOne{Delightful Researcher}
\coadvisorTwo{Equally D. Researcher}
\committeeInternal{Person Inside}

\committeeExternal{Grant W. Biedermann}

\degree{Doctor of Philosophy}
\degreeabbrv{Ph.D.}
\field{Physics}
\degreeyear{2024}
\degreeterm{Spring}
\degreemonth{May}
\department{Physics and Astronomy}
\defensedate{27\textsuperscript{th} of March, 2024}

%

\frontmatter


\setcounter{chapter}{0}

\chapter{Introduction}
\label{chap:Introduction}

Quantum mechanics was developed to address the inability to explain phenomena related to the nature of the electromagnetic field and its interaction with matter, like the photoelectric effect, the absence of an ultraviolet catastrophe in blackbody radiation, the structure of atoms, and their interactions with electromagnetic radiation. In trying to account for these phenomena, quantum mechanics introduced a dual wave and particle nature of both matter and energy, which has since been referred to as `wave-particle duality'. The wave nature is associated with phenomena like interference for both matter and energy, as seen in experiments with light and electrons. When the superposition principle is applied to quantum systems with multiple components, a direct consequence is the phenomenon of quantum entanglement. Entanglement is the property of a quantum system where the state of the entire system can be described with certainty, but the state of its constituents, if described, can only be described with uncertainty \cite{nielsen2010quantum, bengtsson2017geometry}. Another consequence of the wave nature is the inability to describe the statistical aspects of quantum degrees of freedom with probability distributions. Wavefunctions, density operators, quasi-probability distributions, and other representations have been introduced to attempt to recover the statistical aspects of quantum mechanics. The Born rule prescribes a way to estimate mean or expectation values of observables from these quantities.

Quantum properties like superposition, entanglement, and interference, when introduced to information processing systems, allow a completely new paradigm for solving problems. For example, using interference between computational possibilities \cite{bernstein1993quantum, nielsen2010quantum} integers can be factored using Shor's algorithm \cite{shor1999polynomial, nielsen2010quantum}. Moreover, using amplitude amplification, searching of unstructured databases can be done quadratically faster than a serial assessment of every entry with the search condition \cite{grover1997quantum, grover2005fixed, roget2020grover, kwon2021quantum, nielsen2010quantum}. 
Therefore, quantum mechanics promises a new way of solving computational problems, albeit with some caveats. 

In the domain of many-body physics of interacting quantum systems, the exponentially growing quantum state space has made microscopic, ab initio descriptions of phenomena like quantum chemistry, high-temperature superconductivity, magnetism, coherent energy transport, and many others intractable using classical computers, due to the exponential overhead \cite{bloch2012quantum, blatt2012quantum, houck2012chip, buluta2009quantum, georgescu2014quantum, browaeys2020many, monroe2021programmable}. Fortunately, quantum computation also offers the possibility to simulate quantum phenomena more easily than classical computation, through a mapping between the exponentially large number of variables in the quantum many-body system of interest and a laboratory accessible quantum many-body system \cite{feynman1982simulating, buluta2009quantum, georgescu2014quantum, preskill2012quantum, preskill2018quantum, trivedi2022quantum, daley2022practical}. Furthermore, addressing many-body physics problems seems within reach of current quantum devices \cite{preskill2018quantum, buluta2009quantum, georgescu2014quantum, monroe2021programmable, houck2012chip, browaeys2020many, deutsch2020harnessing, preskill2022physics}. We will return to quantum simulation in Sec.~\ref{sec:IntroQuantumSimulation}.

Meanwhile, over the past few decades, there has been much development in the preparation and manipulating of quantum systems, with the ability to observe and control these systems at individual and multiple particle levels. For atomic and molecular platforms like trapped ions \cite{blatt2008entangled, bruzewicz2019trappedion}, neutral atoms \cite{bloch2008quantum, weiss2017quantum, henriet2020quantum}, molecules \cite{shuman2010laser, hornberger2012colloquium, gershenfeld1998quantum, bohn2017cold}, this has been facilitated by remarkable developments in laser cooling and trapping of individual atoms and molecules \cite{wineland1979laser, stenholm1986semiclassical, shuman2010laser}. Moreover, solid-state platforms like superconducting circuits \cite{clarke2008superconducting, dicarlo2009demonstration, devoret2013superconducting, kjaergaard2020superconducting}, defects like nitrogen-vacancy centers \cite{childress2013diamond, gurudev2007quantum, pezzagna2021quantum} have seen rapid advances in fabricating, characterizing and controlling the quantum states of mesoscopic assemblies of atoms in substrates. This has benefited from the advances in semiconductor manufacturing that have revolutionized classical computing \cite{kjaergaard2020superconducting}.

At the confluence of these advances along with other advances in quantum optics, quantum information theory, and coherent control of quantum process, has arisen the interdisciplinary discipline of quantum information science \cite{benioff1980computer, benioff1982quantum1, benioff1982quantum2, deutsch1985quantum, deutsch1998quantum, deutsch2000machines, deutsch2020harnessing}. Modern endeavors in quantum information science address multiple challenges related to understanding, controlling and developing quantum systems, developing algorithms and procedures for solving computational problems using quantum computers, and ways of mitigating unwanted effects in quantum systems through quantum error correction, quantum error mitigation, dynamical decoupling, etc. \cite{nielsen2010quantum, lidar2013quantum, wilde2013quantum, akulin2014dynamics, bengtsson2017geometry, bruss2019quantum, preskill2022physics}. This has been sometime referred to as the second or third quantum revolution \cite{dowling2003quantum, celi2017quantum, deutsch2020harnessing}.

Any non-trivial quantum information processing (QIP) system requires two building blocks -- operations that transform the state of individual components of the systems, and operations that transform the state of more than one component of the system through interactions, thereby generating quantum entanglement. The latter, which can introduce correlations between components of the system, requires interactions between the components. In fact, these correlations in quantum states contribute to the potential exponential advantage of QIP over classical information processing \cite{preskill2012quantum, akulin2014dynamics, bruss2019quantum, bengtsson2017geometry, preskill2018quantum}. 

Implementation of these operations on individual components and multiple components is essential for the development of QIP. In this dissertation, we study quantum entanglement with neutral atoms encoding quantum bits, in the context of simulation of nonequilibrium dynamics of one-dimensional (1D) spin models. In the first part of the dissertation, we discuss entanglement generation in QIP systems with neutral atoms. In the second part of the dissertation, we discuss the role of entanglement in quantum simulation of critical phenomena. We next review some concepts from information theory and introduce a few concepts related to quantum entanglement.

%
\section{Classical and quantum information}
%
In classical information theory, a commonly chosen unit of information is a binary digit, typically contracted to `bit', which can have two values. These two values can have arbitrary labels, as long as they are distinct. The choice of labels is typically determined by how bits are processed by an information processing system. Common choices include $0$ and $1$ when considering operations like addition, conjunction, and disjunction and $\pm 1$ when considering multiplication. Independent of the choice of labels, the two values are represented using two physically distinguishable states of an information processing system. The most well-known representation of a bit is one using two distinct voltages in an electronic circuit, which is one of the fundamental building blocks of digital computers.

The proverbial cement holding the building block of bits together involves the derivation of one or more output bits from one more input bits. This is typically implemented using logic gates, which implement a boolean function from input bits to output bits. A remarkable fact is that logic gates that take only one input bit, that is one-bit gates, and logic gates that take only two input bits, or two-bit gates, can be used to construct any function of an arbitrary number of bits. This fact has enabled the revolution in digital classical computers which in turn have revolutionized the way we create, process, and store information\footnote{Information creation, processing, and storage activities benefiting from the digital classical computer revolution include research discussed in this dissertation and writing of this dissertation.}.

In the absence of complete knowledge of the states of bits, we use probability distributions over the possible states of bits. For example, a single bit may be in the state $0$ with probability $p_0$ and in the state $1$ with probability $p_1$. The normalization convention for probability requires $p_0 + p_1 = 1$. For $n$ bits, we would have probabilities $p_{\vec{b}}$ where $\vec{b}$ is an ordered string of bits from the set $\{0, 1\}^{n}$. As before, normalization would require that the $p_{\vec{b}}$'s sum to $1$. Now, consider describing the state of a subset of $m$ bits of the total $n$ bits. If the $n$-bit, state is known with complete certainty, that is only of the $p_{\vec{b}}$'s is $1$, all other $2^{n}-1$ are $0$, the state of any $m<n$ bits is also known with complete certainty. While this may seem obvious and holds for classical information, it does not hold when we consider quantum entanglement.

The uncertainty in the state of the bits is quantified by assessing how far each probability, $p_{\vec{b}}$, is from the landmarks $0$ corresponding to impossibility and $1$ corresponding to certainty. This is often measured using Réyni entropies, which take the form
\begin{equation}
  \mathcal{S}_{k} = - \frac{1}{k-1}\log \left(\sum_{j} p_{j}^{k} \right)
  \label{eq:IntroReyniEntropies},
\end{equation}
where $k$ is the order of the Réyni entropy. For the special case of $k=1$, we get the von Neumann entropy
\begin{equation}
  \mathcal{S}_{1} = - \sum_{j} p_{j} \log p_{j} 
  \label{eq:IntroVonNeumannEntropy},
\end{equation}
which is related to thermodynamic entropy, for the special case of $k=0$, $\mathcal{S}_0$, counts the number of possibilities with non-zero probability, and for the special case of $k \to \infty$, $\mathcal{S}_{\infty}$ measures how far the largest probability is from $1$. The base of the logarithm determines the units of entropy, with base $2$ being used to measure entropy in bits\footnote{A bit represents both a binary-digit and a unit of entropy.} and base $\mathrm{e}$ being used to measure entropy in nats.

\begin{figure}
    \centering
    \includegraphics[width=0.4\textwidth]{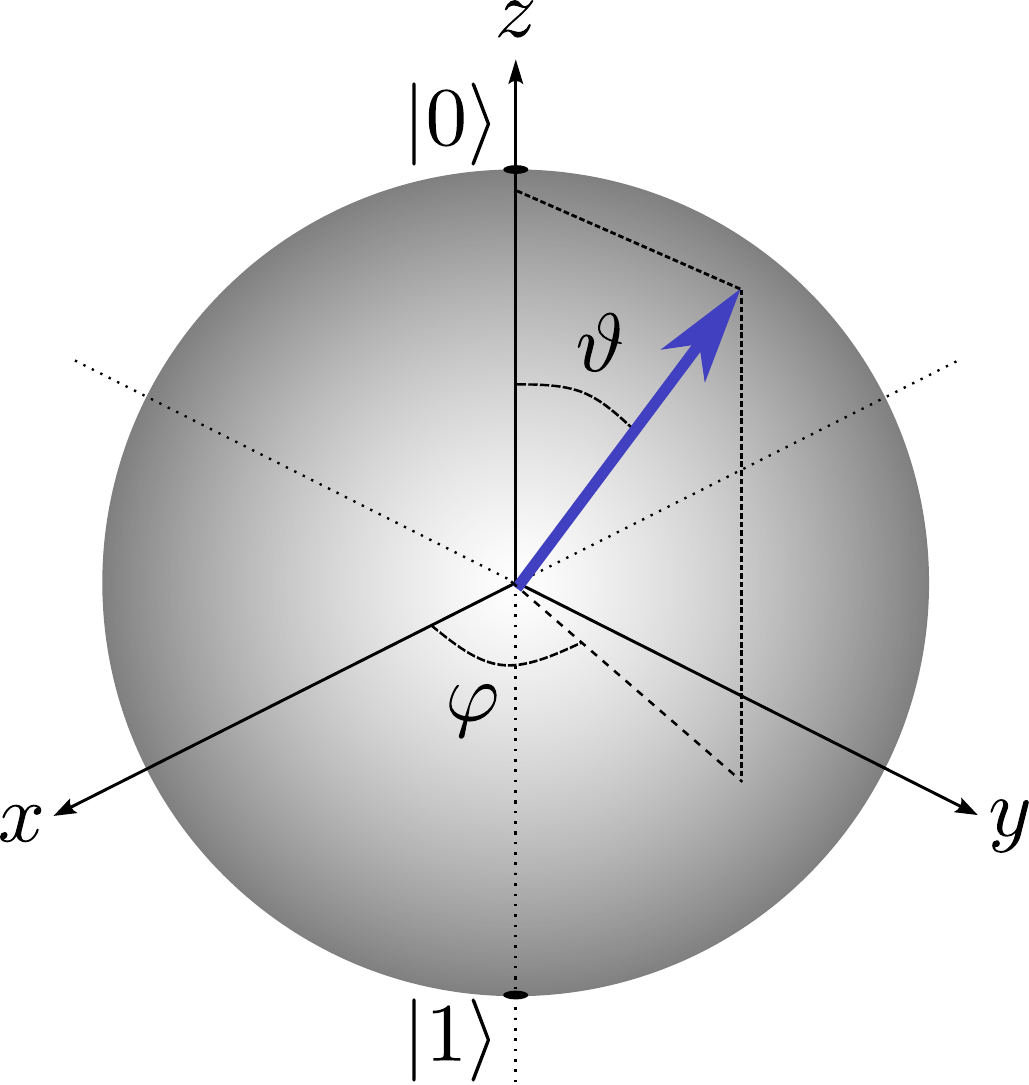}
    \caption{Bloch sphere representation of a qubit. The arrow corresponds to a pure state with polar angle $\vartheta$ and azimuthal angle $\varphi$.}
    \label{fig:IntroBlochSphere}
\end{figure}

Analogous to a bit for classical information, for quantum information, we use a quantum bit, or `qubit'. A state of a qubit can be described by basis vectors with two distinct labels, for example, $\{|0\rangle, |1\rangle\}$. The state space of a qubit is isomorphic to that of a spin-1/2 system, where the basis states are labeled $\{| \uparrow \rangle, | \downarrow \rangle\}$. These states correspond to two completely distinguishable, or orthogonal, states of the qubit. The superposition principle of quantum mechanics allows a qubit in a pure state, that is, with complete certainty in its state, to be in any state of the form
\begin{equation}
  |\psi_{\mathrm{1-qubit}} \rangle
  = a_{0}|0\rangle + a_{1}|1\rangle,
\end{equation}
with $a_{0}$ and $a_{1}$ being complex numbers, or complex amplitudes, satisfying $|a_{0}|^2 + |a_{1}|^2 = 1$, as a normalization convention. The space of a qubit can be represented using the Bloch sphere, as shown in Fig.~\ref{fig:IntroBlochSphere}. Pure states, corresponding to complete certainty are on the surface of the the sphere. Two antipodal points, $(x=0, y=0, z=\pm 1)$, by convention are chosen to be the computational basis states $\{| 0 \rangle, | 1 \rangle\}$. Mixed states are within the sphere, in the Bloch ball. A pure state can be described by the polar angle $\vartheta$ and $\varphi$, describing the position on the surface of the sphere. In addition to these angles, a mixed state is specified by the length of the vector, from the origin.

Now, considering two of these qubits, a state can be described by the basis vectors $\{|0,0\rangle, |0,1\rangle, |1,0\rangle, |1,1\rangle\}$, where the two labels specify the state of the two qubits. In particular, a pure state of two qubits can be written as
\begin{equation}
  |\psi_{\mathrm{2-qubit}} \rangle
  = a_{0,0} |0,0\rangle
  + a_{0,1} |0,1\rangle
  + a_{0,0} |0,0\rangle
  + a_{0,1} |0,1\rangle,
\end{equation}
which cannot be factored into products of two one-qubit states as
\begin{equation}
  |\psi_{\mathrm{2-qubit}} \rangle
  \neq
  \left(
    a_{0}|0\rangle + a_{1}|1\rangle
  \right)
    \left(
    b_{0}|0\rangle + b_{1}|1\rangle
  \right).
\end{equation}
If we try to describe the state of one of the qubits, independent of the other, we get a generally mixed state, whose state is not completely certain,
\begin{equation}
  \rho
  = p_{0} | 0 \rangle\langle 0 |
  + p_{1} | 1 \rangle\langle 1 |
  + c_{0,1} |0 \rangle\langle 1|
  + c_{1,0} |1 \rangle\langle 0|,
\end{equation}
where $p_0$ ($p_1$) is the probability of finding the qubit in state $|0\rangle$ ($| 1 \rangle$), and $c_{0,1}$ and $c_{1,0}$, which are complex conjugates of each other, represent coherences between states $|0\rangle$ and $|1\rangle$.

As we increase the number of qubits whose state we are trying to describe, the number of complex amplitudes $a$'s we need grows exponentially with the number of qubits. Indeed, the state of an $n$-qubit system lives in an exponentially large state space, which for pure states is the vector space spanned $n$-fold tensor products of the basis vectors, $\{|0\rangle, |1\rangle \}$, that is $\{|0\rangle, |1\rangle \}^{\otimes n}$
\begin{equation}
  \vert \psi_{n\mathrm{-qubit}} \rangle =
  \sum_{\vec{b} \in \{0, 1\}^n} a_{\vec{b}} \, \vert \vec{b} \rangle,
  \label{eq:IntroManyBodyWavefunction}
\end{equation}
where $a_{\vec{b}}$ represents the probability amplitude for the state corresponding to $|\vec{b}\rangle$, sometimes referred to as a computational basis state. The vector space where pure quantum states of a quantum system live is usually referred to as the Hilbert space for the system. The inability to factor these $2^n$ complex amplitudes is at the heart of potential quantum advantage in solving computational problems.

For some applications, we are interested in the state of a subset of the qubits, or a subsystem of the system. The marginal state of a subsystem, $A$, which is the state of subsystem $A$, obtained after averaging over all possibilities of the rest of the system is represented using a reduced density operator,
\begin{equation}
  \rho^{(A)} = \mathrm{Tr}_{\setminus A} \left( \rho^{(A \cup \setminus A)} \right)
  \label{eq:IntroMarginalState},
\end{equation}
which is generally a mixed state without complete certainty in the state of subsystem $A$. The uncertainty in the state of the marginal state of a subsystem is quantified by entropies of the form Eqs.~\eqref{eq:IntroReyniEntropies}, and \eqref{eq:IntroVonNeumannEntropy}. These are called entanglement entropies, as the entropy measures uncertainty in the state of a subsystem due to its entanglement with the rest of the system.

\begin{figure}[ht]
    \centering
    (a)\\
    \includegraphics[width=0.8\textwidth]{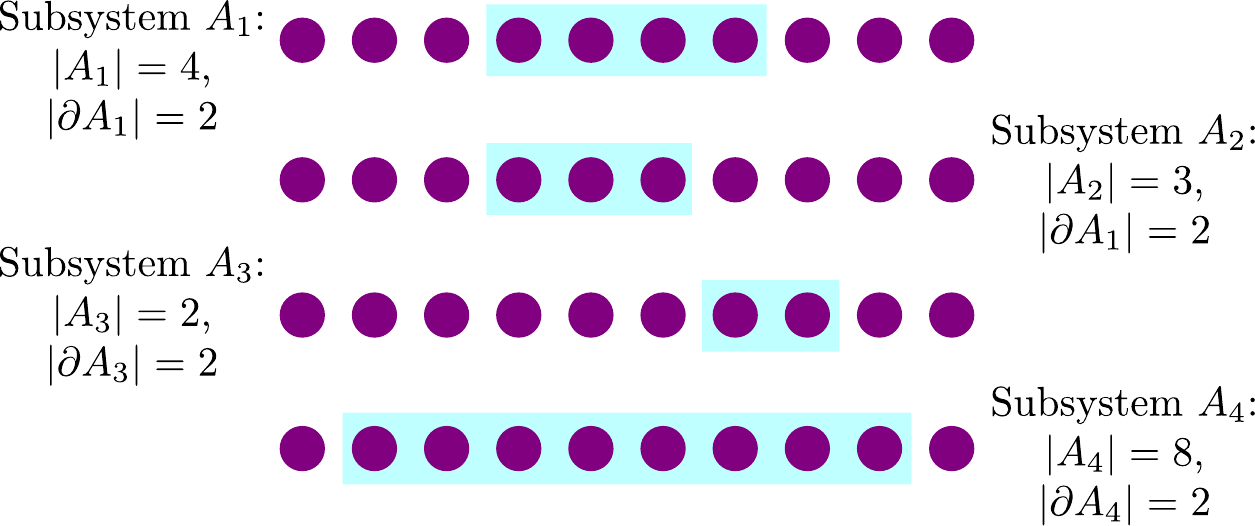}
    \\ \vspace{1cm}
    (b)\\
    \includegraphics[width=0.8\textwidth]{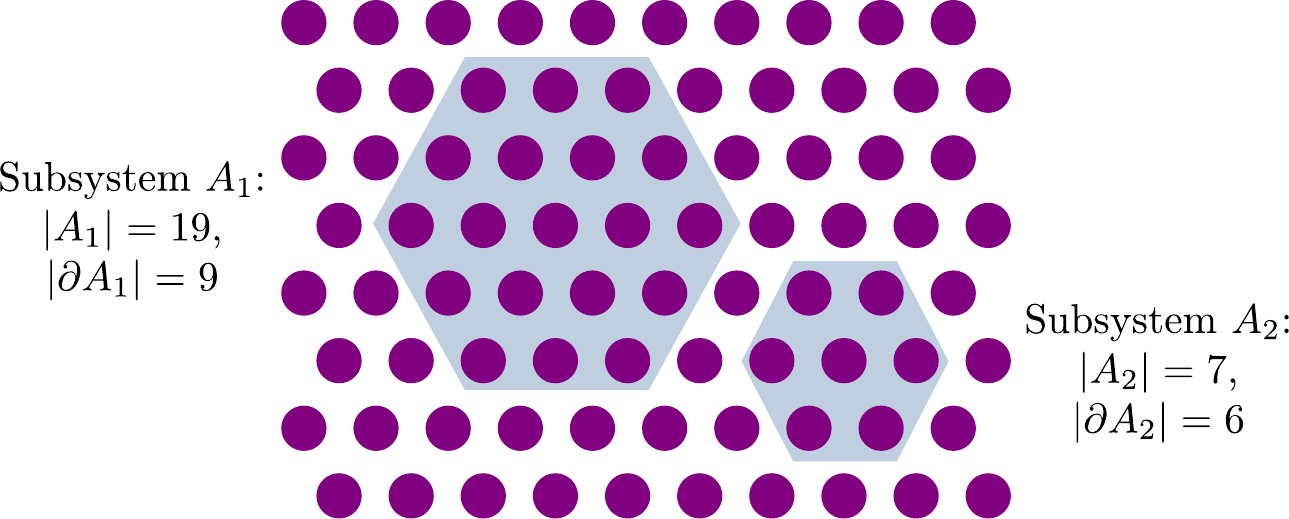}
    \caption{Scaling of entanglement entropy of a subsystem. Area law corresponds to the entanglement entropy scaling with the size of the boundary of the subsystem, $\mathcal{S}_{k} \propto |\partial A|$. Volume law corresponds to entanglement entropy scaling with the size of the bulk of the subsystem, $\mathcal{S}_{k} \propto |A|$.
    (a) In 1D, the size of the boundary is independent of the size of the system, while the size of the bulk grows with the size of the system. (b) In 2D, the size of the boundary and bulk grow with the size of the system.}
    \label{fig:IntroAreaLawVolumeLaw}
\end{figure}

Considering the entanglement entropy of a subsystem $A$, due to its entanglement with other degrees of freedom, there are two regimes of interest. One regime is when the entanglement entropy scales with the size of the boundary of the system, sometimes described as $\mathcal{S}_k \propto |\partial A|$. This regime is called the `area-law' regime as the entanglement entropy scales with the surface area of the boundary between $A$ and other degrees of freedom. The other regime is when the entanglement entropy scales with the size of the bulk of the systems, sometimes represented as $\mathcal{S}_k \propto |A|$. This regime is called the `volume-law' regime as the entanglement entropy scales with the volume of $A$. We show examples of area and volume scaling for one dimension (1D) and two dimensions (2D) in Fig.~\ref{fig:IntroAreaLawVolumeLaw}. Next, we review the generation of entanglement as a dynamical process in composite quantum systems.

%

%
\section{Entanglement in quantum dynamics}
\label{sec:IntroEntanglementDynamics}
%
In any physical system implementing QIP, the evolution of the quantum states is determined by an equation of motion. In the ideal situation of the system being isolated, and ignoring relativistic effects, the transformation of the system is described by a unitary map, $\hat{U}$, which obeys the time-dependent Schrodinger equation
\begin{equation}
  \frac{\mathrm{d}}{\mathrm{d} t} \hat{U} 
  = -\frac{\mathrm{i}}{\hbar} \hat{H} \hat{U},
  \label{eq:IntroTimeDependentSchrodinger}
\end{equation}
where $\hat{H}$ is the Hamiltonian of the system, which is a quantum operator representing the total energy of the system. A pure state $|\psi(t=0)\rangle$, at time $t=0$ evolves to a state $|\psi(t)\rangle = \hat{U}(t, 0) |\psi(t=0)\rangle$, at time $t$. Here $\hat{U}(t, 0)$ is the unitary time evolution operator which is the solution to the differential equation, Eq.~\eqref{eq:IntroTimeDependentSchrodinger}, with the initial condition that at $t=0$, the propagator is the identity operator, $\hat{U}(0, 0) = \mathds{1}$.

The dynamics of a quantum system, interacting with degrees of freedom outside it (sometimes collectively called `the environment'), can be described using a time-dependent Schrödinger equation, where the Hamiltonian $\hat{H}$ and the time evolution operator $\hat{U}$ act jointly on the degrees of freedom within the system and outside it \cite{nielsen2010quantum, lidar2019lecture, manzano2020short}. In general, this leads to entanglement between the constituents of the system and the environment. Lack of knowledge about the state of the environment leads to decoherence \cite{nielsen2010quantum, lidar2019lecture, manzano2020short}.

\begin{figure}
    \centering
    \includegraphics[width=0.6\textwidth]{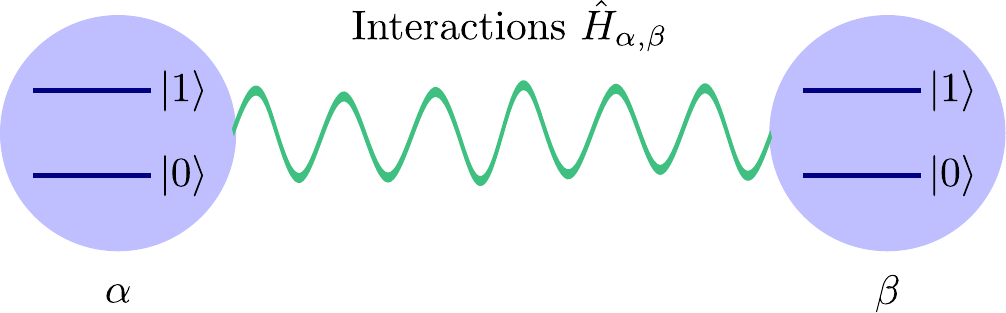}
    \caption{Schematic for a generation of entanglement between subsystems $\alpha$ and $\beta$ of a composite system. Generation of entanglement requires an interaction Hamiltonian, $\hat{H}_{\alpha, \beta}$ as in Eq.~\eqref{eq:IntroInteractEntangler}. In this figure, both $\alpha$ and $\beta$ represent qubits with states $|0\rangle$ and $|1\rangle$.}
    \label{fig:IntroInteractionEntanglement}
\end{figure}

Generating quantum entanglement, from an unentangled quantum state, requires a Hamiltonian that has the components of the system interacting with each other. For example, to generate entanglement between two qubits, $\alpha$ and $\beta$, we would need a Hamiltonian of the form
\begin{equation}
    \hat{H}_{\mathrm{2-qubit}}
    = \hat{H}_{\alpha} \otimes \mathds{1}_{\beta}
    + \mathds{1}_{\alpha} \otimes \hat{H}_{\beta}
    + \hat{H}_{\alpha, \beta},
  \label{eq:IntroInteractEntangler}
\end{equation}
where $\hat{H}_{\alpha}$ ($\hat{H}_{\beta}$) acts only on the physical degree of freedom representing qubit $\alpha$ ($\beta$), and $\hat{H}_{\alpha, \beta}$ acts on both, due to some form of interaction between the degrees of freedom. Therefore, physical systems and models that provide interacting Hamiltonian terms of the form $\hat{H}_{\alpha, \beta}$ are capable of preparing entangled quantum states, from unentangled quantum states. A schematic of such a Hamiltonian for two qubits is shown in Fig.~\ref{fig:IntroInteractionEntanglement}. This has two implications for the subject at hand. Firstly, to generate quantum entanglement, we need Hamiltonians of the form Eq.~\eqref{eq:IntroInteractEntangler}. In the first part of the dissertation, we will focus on the specific platform of neutral atoms and discuss entanglement generation using a Hamiltonian of this form, which is obtained through excitation to Rydberg states \cite{saffman2010quantum, saffman2016quantum, weiss2017quantum, henriet2020quantum, browaeys2020many}. Neutral atoms, due to their weak interactions with each other and the environment in the ground states, serve as robust carriers of quantum information and on-demand excitation to Rydberg states allow the implementation of entanglement generation \cite{saffman2010quantum, saffman2016quantum, henriet2020quantum, browaeys2020many}. Several schemes have been introduced to generate entanglement between atomic qubits by harnessing the strong interactions between Rydberg atoms \cite{jaksch2000fast, beterov2013quantum, keating2015robust, beterov2016two, beterov2018adiabatic, levine2019parallel, mitra2020robust, saffman2020symmetric, beterov2020application}. Moreover, many of these protocols have also been demonstrated in multiple atomic species \cite{zhang2010deterministic, isenhower2010demonstration, wilk2010entanglement, levine2018high, omran2019generation, levine2019parallel, graham2019rydberg, madjarov2020high, jo2020rydberg, martin2021molmer, bluvstein2022quantum, graham2022multi, schine2022long, ma2022universal, evered2023high, bluvstein2024logical}. Yet challenges remain for implementations of faster and robust entangling gates for neutral atoms, with both fundamental and technical questions. This platform is promising for multiple forms of scalable QIP, including simulation of quantum many-body physics \cite{zeiher2016many, zeiher2017coherent, omran2019generation, borish2020transverse, browaeys2020many, henriet2020quantum, bernien2017probing, keesling2019quantum, ebadi2021quantum, scholl2021quantum, bluvstein2021controlling, scholl2022microwave, ohler2022self}, general purpose QIP \cite{brennen1999quantum, isenhower2010demonstration, saffman2010quantum, saffman2016quantum, levine2019parallel, cohen2021quantum, notarnicola2021randomized, burgers2022controlling, wu2022erasure, bluvstein2022quantum, graham2022multi, ebadi2022quantum, jandura2022optimizing, evered2023high, bluvstein2024logical}, metrology and sensing enhanced with quantum properties \cite{kaubruegger2019variational, van2021impacts, schine2022long}, and solving of hard optimization problems \cite{keating2013adiabatic, nguyen2023quantum, andrist2023hardness, ebadi2022quantum, kim2022rydberg, byun2022finding}.

Secondly, any model where the components have interactions with Hamiltonians between two components of the form Eq.~\eqref{eq:IntroInteractEntangler} can generate quantum entanglement. Specifically, in the absence of fine-tuned parameters and special symmetries, such models are capable of generating entanglement during time evolution. At short times, the entanglement entropy of a subsystem with the rest of the system scales with the size of the boundary, in an area-law scaling of entanglement. Eventually, the entanglement scales with the size of the system, in a volume-law scaling of entanglement, making the exact simulation of quantum systems intractable except for small systems \cite{preskill2012quantum, schollwock2011density, paeckel2019timeevolution}. 

In this dissertation, we will consider some dynamical phenomena where volume-law entanglement is generated, but quantities of interest can be approximated well using classical methods without incurring the overhead of volume-law entanglement. The overhead is due to the need to keep track of an exponentially large number of coefficients in a superposition of the form Eq.~\eqref{eq:IntroManyBodyWavefunction}. The approximations involve tensor networks, which factor many-body wavefunctions into a graph of products of higher-order tensors, with entanglement content of the wavefunctions determining the size of the tensors \cite{schollwock2011density, orus2019tensor, biamonte2019lectures, cirac2021matrix}. This enterprise is successful due to many quantum states of interest being describable using a polynomial (in the system size) number of parameters. There are area-law states in which entanglement entropy of reduced states of subsystems scale with the boundary of the subsystems. In 1D systems, area-law systems have entanglement entropy that scales independently of system size, as shown in Fig.~\ref{fig:IntroAreaLawVolumeLaw}. Ground states of gapped 1D systems, away from criticality, are believed to be tractable through 1D tensor network states, called matrix product states (MPS) \cite{perez2006matrix, schollwock2011density}. While there have been many studies of nonequilibrium dynamics using tensor networks, the tractability of these problems is not clear even in 1D.

Dynamics of entanglement also plays a key role in the phenomenology of quantum systems. Some recent work suggests that entanglement is the key to statistical mechanics, both in the description of typical quantum states of multiple degrees of freedom and in the dynamical process of equilibration, that is the approach to equilibrium \cite{pop06ent, Gol2006, rei07typ, sug13can, Leb2007, Dub2012, Mor2018}. The dynamics of quantum entanglement play an important role in the field of quantum simulation, which we review next.



%
\section{Quantum simulation}
\label{sec:IntroQuantumSimulation}
%

A prominent task for a quantum information processor is the simulation of another quantum system \cite{feynman1982simulating, buluta2009quantum, georgescu2014quantum, preskill2012quantum, preskill2018quantum, browaeys2020many, deutsch2020harnessing, monroe2021programmable, preskill2022physics, chinni2022reliability, daley2022practical}. This idea of quantum simulation involves mapping the variables describing a system of interest to the variables describing a system in the laboratory, which we call the `simulator'. These quantum simulators have the potential to solve problems in condensed-matter physics \cite{bernien2017probing, hofstetter2018quantum, keesling2019quantum, browaeys2020many, ebadi2021quantum, scholl2021quantum, monroe2021programmable, yang2020observation}, quantum chemistry \cite{aspuruguzik2005simulated, buluta2009quantum, lanyon2010towards, georgescu2014quantum, mccaskey2019quantum, arguello2019analogue, lee2023evaluating}, and high-energy physics studying elementary particles \cite{martinez2016real, klco2018quantum, bauer2023quantum}. Moreover, quantum simulation of many-body physics is considered a near-term goal for QIP \cite{preskill2012quantum, preskill2018quantum, deutsch2020harnessing, preskill2022physics, daley2022practical, trivedi2022quantum}, due to the common lore that the information needed to solve the problems of interest depends on coarse-grained macro properties. These macro properties are expected to be more robust to imperfections both from experimental noise and inadequate control to achieve universal quantum computation, making them quantities of the interest for noisy intermediate scale quantum (NISQ) experiments \cite{preskill2018quantum, trivedi2022quantum, daley2022practical, trivedi2022quantum}.

In quantum simulators, the primary paradigm is quantum dynamics where a quantum system evolves under a partially or completely known Hamiltonian, which may be time-independent or time-dependent. Equilibrium properties can be accessed from quantum dynamics through steady-state behavior, for example in quantum quenches \cite{mitra2018quantum, zhang2017observationmany, halimeh2017dynamical, halimeh2017prethermalization, Zun2018, piccitto2019dynamical, daley2022practical}. A recent experiment using up to fifty-three trapped ${}^{171} \mathrm{Yb}^{+}$ ions probed the non-equilibrium dynamics of a dynamical quantum phase transition (DQPT) in transverse field Ising models with long-range interactions in one dimension \cite{zhang2017observationmany}. The spin-1/2 degrees of freedom were encoded in the hyperfine clock states of each ion. Using light shifts and spin-dependent optical dipole forces, the effective Hamiltonian of the system was a transverse field Ising model. Using the quantum quench paradigm, the authors explored the phase transition assessed using the steady state of expectation values of local observables. These dynamical phases were also corroborated using domain sizes. For $53$-spins, the Hilbert space dimension is $2^{53}$, making the complete description of the wavefunction impossible with state-of-the-art computers. Yet, for the quantities of interest to study the phase diagram, a less precise description of the state may suffice. Moreover, imperfections in the system and decoherence would change the state of the system with respect to the ideal state expected by the model. The quantities needed to study the dynamical phases were insensitive to the imperfections and decoherence in the system \cite{zhang2017observationmany}. In the second part of the dissertation, we will study the simulation of these quantities using classical methods.

Another paradigm of accessing equilibrium properties, specifically those of zero temperature equilibrium states, is using quasi-adiabatic passages \cite{farhi2000quantum, farhi2001quantum, albash2018adiabatic}. In these quasi-adiabatic passages, an initial state of the ground state of a known Hamiltonian is adiabatically, that is slowly compared to time-scales corresponding to the Hamiltonian, driven to the ground state of a Hamiltonian of interest. The system stays in the instantaneous ground state, as per the adiabatic theorem \cite{kato1950adiabatic, nenciu1980adiabatic}. This was originally implemented in superconducting quantum devices with tunable couplers to implement interacting spin-models \cite{johnson2011quantum}. These devices were used to find solutions to optimization problems \cite{pudenz2015quantum, albash2018demonstration, izquierdo2021testing}, showing evidence of better performance than some classical methods for some problems. In recent experiments with neutral atoms, interacting in their Rydberg states under versions of the Rydberg blockade, ground states of spin-1/2 models with anti-ferromagnetic order were reached, through adiabatic passages from easy-to-prepare product states \cite{bernien2017probing, keesling2019quantum, ebadi2021quantum, scholl2021quantum, semeghini2021probing, ebadi2022quantum}. Moreover, using the flexibility of atom rearrangements in 1D  \cite{bernien2017probing, keesling2019quantum, omran2019generation} and two-dimensional (2D) geometries \cite{de2018analysis, ebadi2021quantum, scholl2021quantum}, ground states encoding solutions to optimization problems have been probed using adiabatic passages \cite{ebadi2022quantum, kim2022rydberg, byun2022finding}. Solving optimization problems through probing ground states is at the interface of quantum simulation and quantum computation, these optimization problems. These problems are often $\mathsf{NP}$-hard, and do not lend themselves to approximate solutions with tractable classical methods \cite{albash2018adiabatic, ebadi2022quantum}. We will use this powerful tool of quasi-adiabatic passages for generating entangling gates for neutral atoms in Chap.~\ref{chap:AdiabaticRydbergDressing}.

%
\section{Outline of the dissertation}
\label{sec:IntroOutline}
%

There are two entanglement-related themes in the dissertation. The first theme is about generating quantum entanglement robustly for QIP with neutral atoms. The second theme is about an application of QIP involving the study of phenomena in quantum many-body physics. 

In the first part of the dissertation, consisting of Chap.~\ref{chap:NeutralAtoms} and Chap.~\ref{chap:AdiabaticRydbergDressing}, we discuss some building blocks of quantum computers using qubits encoded in energy levels of neutral atoms. In Chap.~\ref{chap:AdiabaticRydbergDressing}, which is based on Refs. \cite{mitra2020robust, mitra2023neutral}, we introduce the neutral atom M{\o}lmer-S{\o}rensen gate, using adiabatic Rydberg dressing interleaved in a spin-echo sequence. We show that this paradigm facilitates an entangling gate that is robust to many quasi-static imperfections in the control parameters of the Hamiltonian. We also discuss some implementation details and the fundamental bounds of this approach.

In the second part of the dissertation, consisting of Chap.~\ref{chap:SpinModels}, Chap.~\ref{chap:QuenchDQPT} and Chap.~\ref{chap:QuenchChaos}, we discuss the quantum simulation of phase transitions in quench dynamics and the critical phenomena occurring at these transitions. In Chap.~\ref{chap:SpinModels}, we review the transverse field Ising model and notions of thermalization in closed quantum systems. We also discuss notions of microstates which have a complete description of the state of a many-body system and macrostates which consist of a set of microstates which are consistent with a macroproperty. Next, in Chap.~\ref{chap:QuenchDQPT} and Chap.~\ref{chap:QuenchChaos}, which are based on Ref. \cite{mitra2023macrostates}, we explore quench dynamics in 1D transverse field Ising models to study critical behavior near a $\mathbb{Z}_2$ symmetry breaking phase transition. We consider a dynamical quantum phase transition (DQPT) which is related to an equilibrium phase transition occurring for thermal states in long-range interacting models and only for ground states in long- and short-range interacting models.

Finally, in Chap.~\ref{chap:SummaryOutlook}, we summarize the findings of the dissertation with a unified perspective and discuss possible directions for future work.

%
\section{List of publications}
%

The dissertation is based on the following publications and preprint.

\begin{itemize}

\item Chap.~\ref{chap:AdiabaticRydbergDressing} is based on the following publications.

\begin{itemize}

\item \cite{mitra2020robust} ``Robust M{\o}lmer-S{\o}rensen gate for neutral atoms using rapid adiabatic Rydberg dressing''; \\
\textbf{Anupam Mitra}, Michael J. Martin, Grant W. Biedermann, Alberto M. Marino, Pablo M. Poggi, Ivan H. Deutsch; \\
arXiv:1911.04045; \textit{Phys. Rev. A 101, 030301(R)} (2020).

\item \cite{mitra2023neutral} ``Neutral atom entanglement using adiabatic Rydberg dressing''; \\
\textbf{Anupam Mitra}, Sivaprasad Omanakuttan, Michael J. Martin, Grant W. Biedermann, Ivan H. Deutsch; \\
arXiv:2205.12866; \textit{Phys. Rev. A 107, 062609} (2023)

\end{itemize}

\item Chap.~\ref{chap:QuenchDQPT} and Chap.~\ref{chap:QuenchChaos} are based on the following preprint.

\begin{itemize}

\item \cite{mitra2023macrostates} ``Macrostates vs. Microstates in the Classical Simulation of Critical Phenomena in Quench Dynamics of 1D Ising Models''; \\
\textbf{Anupam Mitra}, Tameem Albash, Philip Daniel Blocher, Jun Takahashi, Akimasa Miyake, Grant W. Biedermann, Ivan H. Deutsch; \\
\textit{arXiv preprint} arXiv:2310.08567.

\end{itemize}
\end{itemize}

Other publications and preprints prepared during the course of the Ph.D. are the following.

\begin{itemize}

\item \cite{martin2021molmer} ``A Mølmer-Sørensen Gate with Rydberg-Dressed Atoms''; \\ Michael J. Martin, Yuan-Yu Jau, Jongmin Lee, \textbf{Anupam Mitra}, Ivan H. Deutsch, Grant W. Biedermann; \\
\textit{arXiv preprint} arXiv:2111.14677.

\item \cite{omanakuttan2021quantum} ``Quantum optimal control of ten-level nuclear spin qudits in \textsuperscript{87}Sr''; \\
Sivaprasad Omanakuttan, \textbf{Anupam Mitra}, Michael J. Martin and Ivan H. Deutsch; \\
arXiv:2106.13705; \textit{Phys. Rev. A 104, L060401} (2021).

\item \cite{omanakuttan2023qudit} ``Qudit entanglers using quantum optimal control''; \\
Sivaprasad Omanakuttan, \textbf{Anupam Mitra}, Michael J. Martin, Ivan H. Deutsch; \\
arXiv:2212.08799; \textit{PRX Quantum 4, 040333} (2023).

\end{itemize}


%
\chapter{Background: Quantum information with Rydberg atoms}
\label{chap:NeutralAtoms}
%

\section{Introduction}

The neutral atom platform, involving individual atoms optically trapped in arrays of optical tweezers or optical lattice is a promising platform for scalable quantum information processing. Applications for which this platform provides a promising platform include quantum simulations \cite{zeiher2016many, zeiher2017coherent, omran2019generation, borish2020transverse, browaeys2020many, henriet2020quantum, bernien2017probing, keesling2019quantum, ebadi2021quantum, scholl2021quantum, bluvstein2021controlling, scholl2022microwave, ohler2022self}, quantum computation  \cite{brennen1999quantum, isenhower2010demonstration, saffman2010quantum, saffman2016quantum, levine2019parallel, cohen2021quantum, notarnicola2021randomized, burgers2022controlling, wu2022erasure, bluvstein2022quantum, graham2022multi, ebadi2022quantum, jandura2022optimizing, evered2023high, bluvstein2024logical} and quantum metrology~\cite{kaubruegger2019variational, van2021impacts, schine2022long} and finding solutions to hard optimization problems \cite{keating2013adiabatic, nguyen2023quantum, andrist2023hardness, ebadi2022quantum, kim2022rydberg, byun2022finding}. In this chapter, we introduce the neutral atom qubits, which are long-lived and at the heart of ultraprecise atomic clocks \cite{ludlow2015optical} and can be arranged in flexible trapping geometries \cite{endres2016atom, barredo2016atom, barredo2018synthetic, kumar2018sorting}. In the longer term, this system is a promising platform for universal fault-tolerant quantum computing \cite{bluvstein2022quantum, bluvstein2024logical}. We also introduce the paradigm of Rydberg excitation to generate entanglement between atoms \cite{saffman2010quantum, saffman2016quantum, weiss2017quantum, henriet2020quantum, browaeys2020many}.

Implementation of universal quantum computation requires the ability to generate any one-qubit gate with arbitrary precision and the ability to generate any one of several possible entangling two-qubit gates \cite{nielsen2010quantum}. The entangling two-qubit gates are considered more challenging as they require interactions between atoms \cite{saffman2010quantum, saffman2016quantum, weiss2017quantum, henriet2020quantum, browaeys2020many}. Moreover, the source of these interactions can potentially lead to interactions with extraneous degrees of freedom and thus decoherence. In this chapter, we discuss the implementation of entanglement between neutral atoms using Rydberg states.

%
\section{Neutral atoms qubits}
%

\begin{figure}
    \centering
    \includegraphics[width=0.96\textwidth]{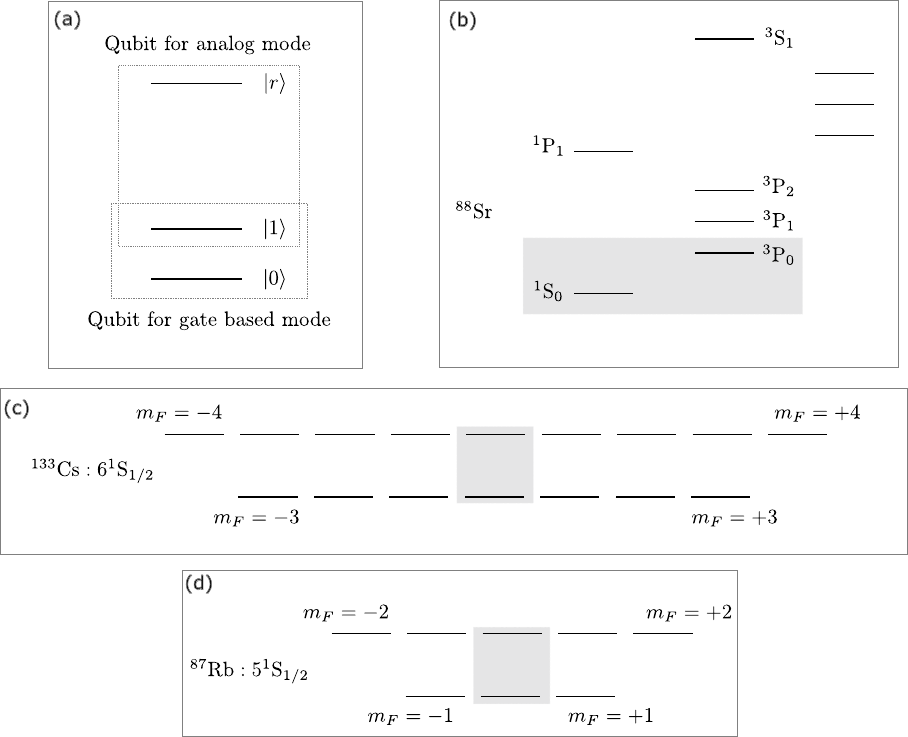}
    \caption{Energy levels for neutral atom qubits. (a) Ground state qubit used for gate-based mode of operation and ground-Rydberg qubit used for analog mode of operation. (b) Ground state qubit for ${}^{88}\mathrm{Sr}$ involving the clock states ${}^{1}\mathrm{S}_0$ and ${}^{3}\mathrm{P}_0$ in the ground subspace with principal quantum number $n = 5$. (c) Ground state qubit for ${}^{133}\mathrm{Cs}$ involving the hyperfine clock states $F=3, m_F=0$ and $F=4, m_F=0$ in the ground subspace with principal quantum number $n=6$ and angular momentum quantum numbers ${}^{2}\mathrm{S}_{1/2}$. (d) Ground state qubit for ${}^{87}\mathrm{Rb}$ involving the hyperfine clock states $F=2, m_F=0$ and $F=1, m_F=0$ in the ground subspace with principal quantum number $n=5$ and angular momentum quantum numbers ${}^{2}\mathrm{S}_{1/2}$.
    }
    \label{fig:NeutralAtomQubits}
\end{figure}

Ground states of neutral atoms are long-lived states with minimal interactions with other degrees of freedom \cite{brennen1999quantum, brennen2000entangling, briegel2000quantum, saffman2010quantum, saffman2016quantum, weiss2017quantum, henriet2020quantum}. This facilitates a carrier of quantum information that is robust to perturbations due to interactions with unwanted degrees of freedom. Two levels with long lifetimes in the ground manifold of neutral atoms are typically chosen for atomic clocks, with the frequency difference between them serving as a frequency reference. Typical choices for qubit states are clock states, which for alkali atoms are separated by microwave frequencies, and for alkaline earth atoms are separated by optical frequency, that is,
\begin{equation}
\begin{aligned}
    \ket{g_0} 
    \equiv \ket{(ns),\, {}^2\mathrm{S}_{1/2}, F, m_{F}=0}
    , &&
    \ket{g_1} 
    \equiv \ket{(ns),\, {}^2\mathrm{S}_{1/2},\, F', m_{F'}=0} 
\end{aligned}
\end{equation}
for alkali atoms and
\begin{equation}
\begin{aligned}
    \ket{g_0} 
    \equiv \ket{(ns)^2,\, {}^1\mathrm{S}_0} 
    , &&
    \ket{g_1} 
    \equiv \ket{(nsnp),\, {}^3\mathrm{P}_0}
\end{aligned}
\end{equation}
for alkaline earth-like atoms. Some examples are shown in Fig.~\ref{fig:NeutralAtomQubits}(b), (c), (d).

%

Neutral atoms have state-dependent interactions between each other \cite{saffman2010quantum}. Neutral atoms in their ground state interact primarily via short-range Van der Waals interactions, scaling as $R^{-6}$ where $R$ is the distance between the atoms. Beyond a distance of a few tens of nanometers, the interactions are primarily due to magnetic dipole-dipole interactions, scaling as $R^{-3}$. Beyond a few micrometers, these interactions are quite small, of the order of a few Hertz in frequency units \cite{saffman2010quantum}. This enables the neutral atoms to operate independently of each other. Exciting the atoms to Rydberg states with large principal quantum numbers endows the atoms with large dipole moments and results in stronger interactions which are about $10^{12}$ stronger than the interactions between ground state atoms \cite{saffman2010quantum}. These interactions may be Van der Waals involving the same Rydberg state for every atom, scaling as $R^{-6}$ or resonant dipole-dipole when atoms are excited to different Rydberg states, that exhibit a Förster resonance, scaling as $R^{-3}$ \cite{saffman2010quantum, browaeys2020many}. In this chapter and the next, we will consider the case of every atom being excited to the same Rydberg state, with electromagnetic field(s) near-resonantly driving the excitation to it from a ground state.

Neutral atoms with Rydberg excitation are typically used in two modes of operation, as illustrated in Fig.~\ref{fig:NeutralAtomQubits} (a). The first mode is the analog mode, which is commonly used for implementing interacting spin models \cite{bernien2017probing, omran2019generation, keesling2019quantum, ebadi2021quantum, scholl2021quantum, semeghini2021probing, bluvstein2021controlling}. In this mode, the interactions between the atoms are always present in the time-independent Hamiltonian. The other mode is gate-based, which is commonly used to implement quantum circuits \cite{levine2019parallel, bluvstein2021controlling, bluvstein2022quantum, graham2022multi, norcia2023mid, graham2023mid, lis2023mid, singh2023mid, ma2023high, bluvstein2024logical}. In this mode, the interactions between the atoms are turned on for implementing entangling gates and turned off for implementing one-qubit gates.

Transitions between two energy levels are driven by an external electromagnetic field that is near-resonant to the frequency difference between them. Due to the highly unequal energy differences between states of atoms and selection rules due to angular momentum conservation, it is possible to selectively address an atomic transition between two energy levels. Therefore the approximation of an atom as a quantum system with two energy levels is quite successful when addressing these two levels. The one-atom Hamiltonian for each atom, in the rotating frame at the frequency used to address the transition, in the rotating wave approximation, reads
\begin{equation}
  \hat{H}_{\text{1-atom}} 
  = \frac{\hbar\Omega}{2} \left(\cos(\phi) \hat{\sigma}_x 
  + \sin(\phi) \hat{\sigma}_y \right)
  - \frac{\hbar\Delta}{2} \hat{\sigma}_z
  ,
  \label{eq:HamiltonianOneAtom}
\end{equation}
where 
\begin{equation}
\begin{aligned}
    \hat{\sigma}_x  
    & = \vert g_1 \rangle \langle g_0 \vert + \vert g_0 \rangle \langle g_1 \vert
    \\
    \hat{\sigma}_y  
    & = \mathrm{i} \vert g_1 \rangle \langle g_0 \vert -\mathrm{i} \vert g_0 \rangle \langle g_1 \vert
    \\
    \hat{\sigma}_z 
    & = \vert g_0 \rangle \langle g_0 \vert - \vert g_1 \rangle \langle g_1 \vert
    ,
\end{aligned}
\end{equation}
for driving the $| g_0 \rangle \leftrightarrow | g_1 \rangle$, transition,
\begin{equation}
\begin{aligned}
    \hat{\sigma}_x  
    & = \vert r \rangle \langle g_1 \vert + \vert g_1 \rangle \langle r \vert
    \\
    \hat{\sigma}_y  
    & = \mathrm{i} \vert r \rangle \langle g_1 \vert -\mathrm{i} \vert g_1 \rangle \langle r \vert
    \\
    \hat{\sigma}_z 
    & = \vert r \rangle \langle r \vert - \vert g_1 \rangle \langle g_1 \vert
    ,
\end{aligned}
\end{equation}
for driving the $| g_1 \rangle \leftrightarrow | r \rangle$, transition and $\Omega$, $\Delta$, and $\phi$ are the Rabi frequency, detuning, and phase of the corresponding drive. Modulation of the Rabi frequency $\Omega$, detuning $\Delta$ and phase $\phi$ allows the implementation of any one-qubit gate in the subspace spanned by $\{ |g_0\rangle, |g_1\rangle\}$ or $\{ |g_1\rangle, |r\rangle\}$, which is a unitary transformation from the $2$-dimensional representation of the group $\mathbb{SU}(2)$.

%
\section{Rydberg states and their interactions}
%

Entanglement between neutral atoms mediated by Rydberg states is facilitated by the large dipole moments of Rydberg atoms, leading to strong electric dipole-dipole interactions \cite{saffman2010quantum, saffman2016quantum, weiss2017quantum, henriet2020quantum, browaeys2020many}. Indeed, the interaction between atoms excited to Rydberg states is the electric dipole-dipole interaction, which in atomic units reads
\begin{equation}
    \hat{V}_{\mathrm{dd}}
    = \frac{e^2}{R^3}
    \left(
    \hat{\vec{R}}_\alpha \cdot \hat{\vec{R}}_\beta
    - 3 (\hat{\vec{R}}_\alpha \cdot \vec{\mathrm{e}}_R )
    (\hat{\vec{R}}_\beta \cdot \vec{\mathrm{e}}_R )
    \right)
    \label{eq:ElectricDipoleDipole}
    ,
\end{equation}
where $\hat{\vec{R}}_\alpha$, $\hat{\vec{R}}_\beta$ are the position operators for atoms $\alpha$ and $\beta$ respectively, $\vec{\mathrm{e}}_{R}$ is the unit vector along the relative vector. $\vec{R}_\alpha - \vec{R}_\beta$, between the two atoms, $R = |\vec{R}_\alpha - \vec{R}_\beta|$ is the separation between the atoms, and $e$ is the magnitude of charge of an electron.

When atoms are excited to the same Rydberg state, $|r, r\rangle$ (which could be a pair such as $|n \mathrm{S}, n\mathrm{S}\rangle$), $\hat{V}_{\mathrm{dd}}$ in Eq.~\eqref{eq:ElectricDipoleDipole} vanishes in first-order perturbation theory as atomic orbitals have zero dipole moment \cite{browaeys2020many}. However, it plays a role in second-order perturbation theory, having a Van der Waals energy shift of the state $|r, r\rangle$ by $V_{\mathrm{VdW}} = C_6 R^{-6}$. For Rydberg excited atoms separated by a few micrometers, the Van der Waals interaction can be of the order of a few mega Hertz to a few giga Hertz, in frequency units.

When atoms are excited to different Rydberg states $|r_1, r_2\rangle$ (which could be a pair such as $|n \mathrm{S}, n\mathrm{P}\rangle$ or $|n \mathrm{P}, n\mathrm{S}\rangle$), the states are directly coupled in Eq.~\eqref{eq:ElectricDipoleDipole}, giving rise to entangled states of the form $|n \mathrm{S}, n \mathrm{P} \rangle \pm | n \mathrm{P}, n \mathrm{S} \rangle$ participating in the interaction \cite{browaeys2020many}. The interaction leads to a `flip-flop' interaction of the form $V_{\mathrm{rd}} \, \left(\hat{\sigma}^+_{\alpha} \hat{\sigma}^-_{\beta} + \hat{\sigma}^-_{\alpha} \hat{\sigma}^+_{\beta}\right)$, with $V_{\mathrm{rd}} = C_3 R^{-3}$ \cite{browaeys2020many}. This term flips spins states between atoms $\alpha$ and $\beta$. Finally, the resonant dipole interaction is anisotropic, making them suitable for configurations in two and higher-dimensional geometries.

In addition to strong interactions (a few tens of mega Hertz to giga Hertz), Rydberg states have long lifetimes (about a hundred microseconds) \cite{saffman2010quantum, saffman2016quantum, weiss2017quantum, browaeys2020many}. The lifetime is limited due to decay processes occurring through interactions with the electromagnetic field. At zero temperature, the lifetime is finite due to interactions with the nonzero field amplitudes in the vacuum, a process called spontaneous emission. At nonzero temperatures, interactions with thermal photons, in stimulated emission processes, of the ambient electromagnetic field can reduce the lifetime further. For alkali atoms, the Rydberg state lifetimes scale as $\tau_r \propto n^{3}$, where $n$ is the principal quantum number.

The fidelity of implementing an entangling gate is limited by the interaction $V$ and the Rydberg state lifetime $\tau_r$ -- the smaller $V \tau_r$, the larger the theoretically achievable fidelity \cite{saffman2010quantum, saffman2016quantum, mitra2023neutral}. We will return to this fundamental scaling in the next chapter.

%
\section{Analog mode: neutral atom arrays with Rydberg excitation}
%

We first consider the analog mode of operation for neutral atom arrays. Here a ground state $|g\rangle$ and a Rydberg state $|r\rangle$ of each atom participate in the dynamics. Each atom can be excited to its Rydberg state $\vert r \rangle$ from its ground state $\vert g \rangle$ with Rabi frequency $\Omega$ and detuning $\Delta$ as in Eq.\eqref{eq:HamiltonianOneAtom}. The Hamiltonian reads
\begin{equation}
  \hat{H}_{\mathrm{Rydberg-array}} 
  = \frac{\hbar\Omega}{2}\sum_{\ell} \hat{\sigma}_\ell^x
  - \frac{\hbar\Delta}{2}\sum_{\ell} \hat{\sigma}_\ell^z
  + \sum_{\ell_1, \ell_2} V_{\ell_1, \ell_2}
  (| r \rangle\langle r |)_{\ell_1} \otimes (| r \rangle\langle r |)_{\ell_2}
  ,
\end{equation}
where $V_{\ell_1, \ell_2}$ is the interaction between atom $\ell_1$ and atom $\ell_2$ when they are excited to their Rydberg states. Each of the projectors onto Rydberg states in the last term can be written as
\begin{equation}
 \left(\vert r \rangle \langle r \vert \right)_{\ell}
  = \frac{1}{2} \left(\mathds{1}_\ell + \hat{\sigma}^z_\ell\right)
  ,
\end{equation}
so that the Hamiltonian reads
\begin{equation}
  \hat{H}_{\mathrm{Rydberg-array}}
  = \frac{\hbar\Omega}{2}\sum_{\ell} \hat{\sigma}_\ell^x
  - \frac{\hbar\Delta}{2}\sum_{\ell} \hat{\sigma}_\ell^z
  + \sum_{\ell_1, \ell_2} V_{\ell_1, \ell_2}
  \left(\frac{\mathds{1}_{\ell_1} + \hat{\sigma}_{\ell_1}^z}{2}\right)
  \left(\frac{\mathds{1}_{\ell_2} + \hat{\sigma}_{\ell_2}^z}{2}\right).
\end{equation}

If the atoms are sufficiently far, the Van der Waals interaction, scaling as $R^{-6}$ can be well approximated using a nearest neighbor interaction. For a regular 1D lattice with spacing $b$ with only nearest neighbor interactions,
$V_{\ell_1, \ell_2} = C_6 b^{-6} \delta_{\ell_1, \ell_1-1} \equiv V \delta_{\ell_1, \ell_1-1}$.

Thus, the 1D nearest neighbor Hamiltonian is
\begin{equation}
  \hat{H}_{\mathrm{1d-nn}} 
  = \frac{\hbar\Omega}{2}\sum_{\ell} \hat{\sigma}_\ell^x
  + \sum_{\ell} \left(\frac{V'_{\ell}}{2}- \frac{\hbar\Delta}{2}\right) \hat{\sigma}_\ell^z
    + \frac{V}{4} \sum_{\ell} 
  \hat{\sigma}_{\ell}^z \hat{\sigma}_{\ell+1}^z,
\end{equation}
where $V'_{\ell} = V$ for bulk atoms and $\tilde{V}_{\ell} = V/2$ for the edge atoms as
\begin{equation}
    V'_{\ell} = \begin{cases}
    V & \ell \in \text{bulk} \\
    \frac{1}{2} V & \ell \in \text{edges} \\
\end{cases}
.
\end{equation}
This Hamiltonian corresponds to that of an Ising model with transverse and longitudinal fields with parameters
\begin{equation}
  B_{\perp} = \frac{\hbar\Omega}{2}
  , \quad
  B_{\parallel}(\ell) = \left(\frac{V'_{\ell}}{2}- \frac{\hbar\Delta}{2}\right)
  , \quad
  J = \frac{V}{4},
\end{equation}
where $J$ is the Ising interaction, $B_{\parallel}$ is the longitudinal field and $B_{\perp}$ is the transverse field. A transverse field Ising model with Hamiltonian can be implemented by setting detunings such that $\hbar\Delta = V$ for bulk atoms and $\hbar\Delta = V/2$ for edge atoms.

%
\subsection{Rydberg blockade}
%

The phenomenon of Rydberg blockade occurs between atoms $\ell_1$ and $\ell_2$ if the excitation of one of these atoms to its Rydberg state $\vert r \rangle$ prevents the excitation of the other atom to its Rydberg state. This occurs when the interaction energy is much larger in magnitude than the Rabi frequency of ground-to-Rydberg excitation, $\vert V_{\ell_1, \ell_2} \vert \gg \hbar \Omega$ and the detuning of the round-to-Rydberg excitation, $\vert V_{\ell_1, \ell_2} \vert \gg \hbar \Delta$.

In this regime the Hamiltonian between atoms $\ell_1$ and $\ell_2$ reads
\begin{equation}
  \hat{H}_{\mathrm{block}}(\ell_1, \ell_2)
  = \frac{\hbar\Omega}{2} \left(
  \hat{P}_{\ell_1} \hat{\sigma}^x_{\ell_2} + \hat{\sigma}^x_{\ell_1} \hat{P}_{\ell_2}
  \right)
  - \frac{\hbar\Delta}{2}
  \left( \hat{\sigma}^z_{\ell_1} + \hat{\sigma}^z_{\ell_2} \right)
  ,
\end{equation}
where
\begin{equation}
  \hat{P}_{\ell}
  = \left(\vert g \rangle \langle g \vert \right)_{\ell}
  = \frac{1}{2} \left(\mathds{1}_\ell - \hat{\sigma}^z_\ell\right)
\end{equation}
is a projector onto the ground state for atom $\ell$. This will play a role in the discussion of the gate-based mode of operation in Sec.~\ref{sec:GateBasedRydberg}.

Consider now the case of a one-dimensional array with atoms label $\ell \in \{1, \cdots, n\}$ with nearest-neighbor Rydberg blockade and open boundary conditions. In this case, the Hamiltonian reads
\begin{equation}
\begin{aligned}
  \hat{H}_{\mathrm{1d-nn-block}}
   &
  = \underbrace{
  \frac{\hbar\Omega}{2}
  \sum_{\ell \in \{2, \cdots n-1 \}}
  \hat{P}_{\ell-1} \hat{\sigma}^x_{\ell} \hat{P}_{\ell+1}
  }_{\text{bulk}}
  \\ &
  +
  \underbrace{
  \frac{\hbar\Omega}{2} \hat{\sigma}^x_{1} \hat{P}_{\ell + 1}
  + \frac{\hbar\Omega}{2} \hat{P}_{\ell - 1} \hat{\sigma}^x_{n}
  }_{\text{edges}}
  \\ &
  - \frac{\hbar\Delta}{2} \sum_{\ell}
  \hat{\sigma}^z_{\ell}
  ,
  \end{aligned}
\end{equation}
which has terms of the form $\hat{P}_{\ell-1} \hat{\sigma}^x_{\ell} \hat{P}_{\ell+1}$ in the bulk, that is excluding the left and right edges. Using an alternative notation for the Pauli operators $\hat{X}_{\ell} \equiv \hat{\sigma}^x_{\ell}$, this term reads $\hat{P}_{\ell-1} \hat{X}_{\ell} \hat{P}_{\ell+1}$. This form leads to the name ``PXP Hamiltonian'' \cite{turner2018weak, serbyn2021quantum, moudgalya2022quantum}. The PXP model with this Hamiltonian has been paradigmatic, leading to the discovery of quantum many-body scars \cite{turner2018weak, serbyn2021quantum}, which provide a mechanism of hitherto unknown weak ergodicity breaking \cite{turner2018weak, serbyn2021quantum, moudgalya2022quantum}.

Note that the effective term $\hat{P}_{\ell-1} \hat{\sigma}^x_{\ell} \hat{P}_{\ell+1}$ has three-spin terms in it can be written as
\begin{equation}
  \begin{aligned}
  \hat{P}_{\ell-1} \hat{\sigma}^x_{\ell} \hat{P}_{\ell+1}
  &
  = \frac{1}{2}\left(\mathds{1}_{\ell-1} - \hat{\sigma}^z_{\ell-1}\right)
  \hat{\sigma}^x_{\ell}
  \frac{1}{2}\left(\mathds{1}_{\ell+1} - \hat{\sigma}^z_{\ell+1}\right)
  \\ &
  = \frac{1}{4}
  \left( \hat{\sigma}^x_{\ell}
  + \hat{\sigma}^z_{\ell-1} \hat{\sigma}^x_{\ell}
  + \hat{\sigma}^x_{\ell} \hat{\sigma}^z_{\ell+1}
  + \underbrace{
    \hat{\sigma}^z_{\ell-1} \hat{\sigma}^x_{\ell} \hat{\sigma}^z_{\ell+1}
  }_{\text{three-spin term}}
  \right)
  ,
  \end{aligned}
\end{equation}
which has effective three-spin terms of the form $\hat{\sigma}^z_{\ell-1} \hat{\sigma}^x_{\ell} \hat{\sigma}^z_{\ell+1}$.

Constrained dynamics of the Rydberg blockade have been of potential to generate entanglement for quantum information processing \cite{jaksch2000fast, lukin2001dipole}. This was empirically observed in pairs of ${}^{87}\mathrm{Rb}$ atoms separated by $4\,\mathrm{\mu m}$ in Ref.~\cite{gaetan2009observation} and $10\,\mathrm{\mu m}$ in Ref.~\cite{urban2009observation}. It also plays a key role in the gate-based mode, as we see in the next section.

%
\section{Gate-based mode: two-qubit operations}
\label{sec:GateBasedRydberg}
%

The other mode of operating neutral atoms with Rydberg excitation is the gate-based mode. The qubit states are typically ground states, $|0\rangle \equiv |g_0\rangle$ and $|1\rangle \equiv |g_1\rangle$. One of these qubit states say $|1\rangle$, can be excited to Rydberg states. Gates which are single-qubit without generating entanglement, are implemented by addressing the qubit directly without involving any Rydberg state. Two-qubit entangling gates are implemented using the interactions between two atoms in the Rydberg states.

\begin{figure}
    \centering
    \includegraphics[width=0.8\textwidth]{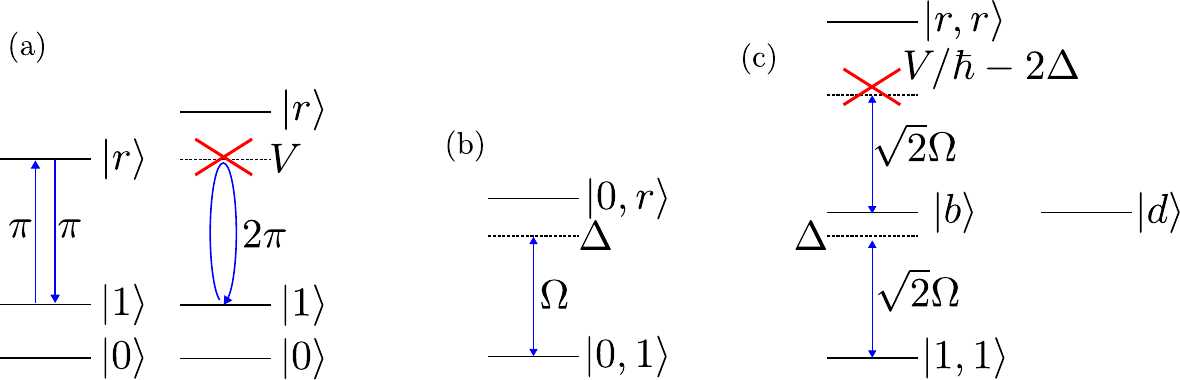}
    \caption{Energy levels for independent and symmetric addressing schemes for entangling gates using the Rydberg blockade. (a) Independent addressing scheme for the $\pi_{\mathrm{c}}$, $2\pi_{\mathrm{t}}$, $\pi_{\mathrm{c}}$ pulse scheme proposed by Jaksch \textit{et al} \cite{jaksch2000fast}. (b) Symmetric addressing for two atoms with only one Rydberg excitation for the states $|0,1\rangle$ and $|0,r\rangle$ with Rabi frequency $\Omega$ and detuning $\Delta$. A similar scheme exists for the states $|1,0\rangle$ and $|r,0\rangle.$ (c) Symmetric addressing for two atoms with the possibility of more than one Rydberg excitation for the states $|1,1\rangle$, bright state $|b\rangle$, and $|r,r\rangle$. The dark state $|d\rangle$ does not participate due to destructive interference between excitation paths. In (a) and (c), excitation to $|r,r\rangle$ is prevented by the Rydberg blockade}
    \label{fig:RydbergAddressingSchemes}
\end{figure}
%

%
\subsection{Independent addressing of two atoms}
%

A well-known scheme of implementing an entangling using Rydberg excitation of neutral atoms is the $\pi_{\mathrm{c}}$, $2\pi_{\mathrm{t}}$, $\pi_{\mathrm{c}}$ pulse scheme proposed in the pioneering work of Jaksch \textit{et al} \cite{jaksch2000fast}. In this scheme, the $| 1 \rangle \leftrightarrow | r \rangle$ transition of each atom is addressed independently and the gate is implemented through the conditional dynamics of a target atom, determined by the state of the control atom. The Hamiltonian reads
\begin{equation}
\begin{aligned}
    \hat{H}
    &
    = 
    \left(
    -\Delta_{\mathrm{c}}(t) \left( | r \rangle \langle r | \right)_{\mathrm{c}}
    + \frac{\Omega_{\mathrm{c}}(t)}{2}
    \left(
    | r \rangle \langle 1 | + | 1 \rangle \langle 1 |
    \right)_{\mathrm{c}}
    \right) \otimes \mathds{1}_{\mathrm{t}}
    \\ &
    + \mathds{1}_{\mathrm{c}} \otimes
    \left(
    -\Delta_{\mathrm{t}}(t) \left( | r \rangle \langle r | \right)_{\mathrm{t}}
    + \frac{\Omega_{\mathrm{t}}(t)}{2}
    \left(
    | r \rangle \langle 1 | + | 1 \rangle \langle 1 |
    \right))_{\mathrm{t}}
    \right)
    \\ &
    + V | r \rangle \langle r | \otimes | r \rangle \langle r |
    ,
\end{aligned}
\end{equation}
where explicit time dependence of $\Omega_{\mathrm{c}}$, $\Delta_{\mathrm{c}}$, $\Omega_{\mathrm{t}}$, and $\Delta_{\mathrm{t}}$ is shown to emphasize the pulsed nature of this scheme. As before, Rydberg blockade occurs when $|V| \gg \hbar \Omega_{\mathrm{c}}, \hbar \Omega_{\mathrm{t}}$. While the blockade condition is strictly not necessary for the control atom, in practice similar Rabi frequncies are used for both control and target atoms. The sequence involves the following three steps as shown in Fig.~\ref{fig:RydbergAddressingSchemes} (a).

\begin{enumerate}
    \item $\pi$-pulse on the $|1\rangle \leftrightarrow |r\rangle$ transition on the control atom.
    \item $2 \pi$-pulse on the $|1\rangle \leftrightarrow |r\rangle$ transition on the target atom.
    \item $\pi$-pulse on the $|1\rangle \leftrightarrow |r\rangle$ transition on the control atom.
\end{enumerate}

Using this pulse sequence under a perfect Rydberg blockade, ignoring population in $| r, r \rangle$, the computational basis states evolve as
\begin{equation}
\begin{aligned}
  \ket{0, 0} 
  & \xrightarrow[]{\pi_\mathrm{c}} & \ket{0, 0}
  & \xrightarrow[]{2\pi_{\mathrm{t}}} & \ket{0, 0}
  & \xrightarrow[]{\pi_\mathrm{c}} & \ket{0, 0},
  \\
  \ket{0, 1}   
  & \xrightarrow[]{\pi_\mathrm{c}} & \ket{0, 1}
  & \xrightarrow[]{2\pi_{\mathrm{t}}} & - \ket{0, 1}
  & \xrightarrow[]{\pi_\mathrm{c}} & - \ket{0, 1},
  \\
  \ket{1, 0}
  & \xrightarrow[]{\pi_\mathrm{c}} & \mathrm{i} \ket{r, 0}
  & \xrightarrow[]{2\pi_{\mathrm{t}}} & \mathrm{i} \ket{r, 0}
  & \xrightarrow[]{\pi_\mathrm{c}} & - \ket{1, 0},
  \\
  \ket{1, 1} 
  & \xrightarrow[]{\pi_\mathrm{c}} & \mathrm{i} \ket{r, 1}
  & \xrightarrow[]{2\pi_{\mathrm{t}}} & \mathrm{i} \ket{r, 1}
  & \xrightarrow[]{\pi_\mathrm{c}} & - \ket{1, 1},
\end{aligned}
\end{equation}
where the effect of the Rydberg blockade is seen for the initial state $\ket{1, 1}$ when the $2\pi_{\mathrm{t}}$ pulse has no effect due to the control atom being in the Rydberg state.

This implemented a gate that is equivalent to a Controlled-Z (CZ) gate up to local unitary transformations on each qubit \cite{jaksch2000fast, saffman2010quantum}. A variant of this gate was demonstrated with a pair of ${}^{87}\mathrm{Rb}$ atoms in Ref.~\cite{isenhower2010demonstration} and further demonstrated with ${}^{133}\mathrm{Cs}$ atoms in a two-dimensional array in Ref.\cite{maller2015rydberg}.
%
\subsection{Simultaneous addressing of two atoms}
%

A more convenient scheme for using the Rydberg blockade to implement entangling gates between neutral atoms is to use simultaneous addressing. Each atom is addressed by the same laser field(s)\footnote{We allow for the possibility of a two-photon ground-to-Rydberg transition, which we consider in the next chapter.}, making the excitation of one atom to its Rydberg state indistinguishable from the excitation of the other atom to its Rydberg state. In this case, it is convenient to write the Hamiltonian in the basis of permutation symmetric and anti-symmetric states \cite{jaksch2000fast, keating2015robust, levine2018high, levine2019parallel, mitra2020robust, schine2022long, mitra2023neutral}. The two-atom Hamiltonian reads
\begin{equation}
    \hat{H}_{\mathrm{2-atom}} =
    -\hbar\Delta(\ketbra{b}{b} + \ketbra{d}{d})
    + \frac{\sqrt{2}\hbar \Omega}{2}
    \left( | b \rangle\langle d | + |1,1\rangle \langle b \right)
    + \left(V - 2\hbar \Delta\right) |r,r\rangle \langle r,r|
    ,
\end{equation}
where $|b\rangle$, $|d\rangle$ are the bright (symmetric) and dark (antisymmetric) superpositions or one atom in $|1\rangle$ and one atom in $|r\rangle$ respectively $|b\rangle \equiv (|1, r\rangle + |r, 1\rangle)/\sqrt{2}$ and $|d\rangle \equiv (|1, r\rangle - |r, 1\rangle)/\sqrt{2}$. The Rabi frequency of excitation to the bright state is enhanced by a factor of $\sqrt{2}$ due to constructive interference between two excitation paths. The Rabi frequency of excitation to the dark state is zero due to destructive interference between two excitation paths.

As before, under Rydberg blockade $|V| \gg \hbar \Omega$, excitation to the two-atom Rydberg state $| r, r \rangle$ is suppressed. Therefore, a $\pi$ pulse driving the transition from $| 1, 1 \rangle$ to $|b\rangle$ produces a maximally entangled ground-Rydberg Bell state, $| b \rangle$, as recently demonstrated in Ref.~\cite{levine2018high}.

Generating entanglement in the qubit subspace, however, requires another independent field driving a different transition. In a first demonstration of the generation of Rydberg blockade based entanglement using symmetric addressing, using ${}^{87}\mathrm{Rb}$ atoms, an additional ground-to-Rydberg drive on the $|0\rangle \leftrightarrow |r\rangle$ transition was used to prepare the Bell state $(|0, 1\rangle + |1, 0\rangle)/\sqrt{2}$ \cite{wilk2010entanglement}. During Rydberg excitation and de-excitation, a phase shift, $\varphi = \vec{k}\cdot \vec{v}$ is imparted to each atom through a Doppler shift in the laser frequency seen by the atoms \cite{wilk2010entanglement, keating2015robust, robicheaux2021photon}. This leads to two contributions, one from the center of mass velocity $\varphi_{\mathrm{cm}} = \vec{k}\cdot (\vec{v}_{\alpha}+\vec{v}_{\beta})/2$, and one from the relative velocity, $\varphi_{\mathrm{rel}} = \vec{k}\cdot (\vec{v}_{\alpha}-\vec{v}_{\beta})/2$ \cite{keating2015robust, mitra2020robust, robicheaux2021photon}. The adiabatic Rydberg dressing scheme mitigates these effects through the robustness of the dressed states \cite{keating2015robust} and the ability to cancel dominant sources of error through a spin echo sequence \cite{mitra2020robust}. Moreover, the Rydberg dressing paradigm \cite{johnson2010interactions} was used to create a spin-flip blockade in the dressed ground subspace, by importing aspects of the Rydberg blockade to it \cite{jau2016entangling}.

Recently an implementation of the Controlled-Z gate was demonstrated using two pulses on the $| 1 \rangle \leftrightarrow | r \rangle$, such that the state $| 1, 1 \rangle$ accumulated a phase $\pi$, relative to the phase accumulated by the states $| 0, 1 \rangle$ and $| 1, 0 \rangle$ \cite{levine2019parallel}. This was achieved through two detuned pulses with $\Delta/\Omega = 0.377$, of duration $1.3664\pi/\Omega$ with a relative phase of $\phi = 1.24212 \pi$ between them. The parameters were fixed so that $\varphi_{j,k}$, the phases accumulated by the computational basis state $|j, k\rangle$ satisfied
\begin{equation}
    \varphi_{1,1} - (\varphi_{0,1} + \varphi_{1,0})
    = \pi
    ,
    \label{eq:GateEntanglingPhaseCondition}
\end{equation}
yielding a gate that is equivalent to a Controlled-Z gate up to local unitaries. This gate with a duration $2.733 \pi/\Omega$ is faster than the $\pi_{\mathrm{c}}, 2\pi_{\mathrm{t}}, \pi_{\mathrm{c}}$ pulse sequence of Jaksch \textit{et. al.} \cite{jaksch2000fast}, using independent addressing. This accumulated phase is purely due to the different trajectory of the initial state $| 1, 1 \rangle$ on the Bloch sphere due to the Rydberg blockade and makes the gate the Controlled-Z gate.  Adiabatic gates, using dynamical phases accumulated during adiabatic passages satisfying Eq.~\eqref{eq:GateEntanglingPhaseCondition}, through Rydberg dressing, have also been proposed \cite{keating2015robust, mitra2020robust, mitra2023neutral}.

Simultaneous addressing has been used to design entangling gates using quantum optimal control \cite{jandura2022optimizing, ma2023high, buchemmavari2024entangling, bluvstein2022quantum, evered2023high, bluvstein2024logical} and also for the adiabatic Rydberg dressing paradigm \cite{keating2015robust, mitra2020robust, martin2021molmer, schine2022long, mitra2023neutral}, the latter of which will be the focus of the next chapter. These approaches have the potential to implement robust entangling gates between neutral atoms. Recently, entangling gates with fidelity over 0.99 have been demonstrated with ${}^{87} \mathrm{Rb}$ atoms \cite{evered2023high} and used to implement quantum computation with logical qubits with error correction \cite{bluvstein2022quantum, bluvstein2024logical}.

%
\section{Conclusion}
%

In this chapter, we reviewed the use of neutral atoms as platforms for quantum information processing, both as carriers of qubits and as representing spin-1/2 degrees of freedom for quantum simulation of interacting spin-1/2 models. We discussed the encoding of qubit or spin-1/2 energy levels in the ground-state manifold of these atoms. Furthermore, we discussed interactions between neutral atoms, which can be strong when these atoms are excited to Rydberg states with large principal quantum numbers. The interactions between Rydberg-excited neutral atoms, especially the phenomenon of Rydberg blockade, and the ability to excited atoms to Rydberg states facilitates the implementation of both interacting spin models and two-qubit entangling gates from the Lie-group $\mathbb{SU}(4)$ as operations needed for a general purpose quantum computers. 

This background material sets the stage for the next few chapters discussing the implementation of two-qubit entangling gates and the role of entanglement in critical phenomena in interacting spin models. 

\chapter{Neutral atom entanglement using adiabatic Rydberg dressing}
\label{chap:AdiabaticRydbergDressing}

%
\section{Introduction}
\label{sec:RydbergDresssingIntro}
%
A variety of protocols have been introduced to generate entanglement between atomic qubits by harnessing the strong interactions between Rydberg atoms \cite{jaksch2000fast, beterov2013quantum, keating2015robust, beterov2016two, beterov2018adiabatic, levine2019parallel, mitra2020robust, saffman2020symmetric, beterov2020application}. Moreover, many of these protocols have also been demonstrated in alkali atoms including cesium and rubidium \cite{zhang2010deterministic, isenhower2010demonstration, wilk2010entanglement, levine2018high, omran2019generation, levine2019parallel, graham2019rydberg, jo2020rydberg, martin2021molmer, bluvstein2022quantum, graham2022multi, evered2023high, bluvstein2024logical} and in alkaline earth atoms including strontium and ytterbium  \cite{madjarov2019strontium, schine2022long, ma2022universal}. In this chapter, we introduce the adiabatic Rydberg dressing paradigm and discuss the implementation of a M{\o}lmer-S{\o}rensen (MS) gate \cite{molmer1999multiparticle, sorensen1999quantum} for neutral atoms \cite{mitra2020robust, martin2021molmer, mitra2023neutral}. The MS gate is intrinsically robust to quasi-static inhomogeneities in intensity and in detuning, such as those arising from inhomogeneous laser amplitudes seen by atoms, displacement of atoms during Rydberg excitation, Doppler shifts in the Rydberg-exciting laser at finite temperature, Stark shifts from stray electric fields and other effects \cite{mitra2020robust}. Moreover, the MS gate can theoretically get close to the fundamental bound in entangling gate fidelity, which is achievable using Rydberg excitation, which is set by the interaction energy between Rydberg states and the radiative lifetime of the Rydberg states \cite{wesenberg2007scalable, mitra2023neutral}.

The MS gate is based on rapid adiabatic Rydberg dressing \cite{johnson2010interactions, keating2015robust, mitra2020robust, martin2021molmer, schine2022long, mitra2023neutral}, a powerful tool for robustly creating entanglement in atomic-clock qubits. In this approach, the Rydberg character is adiabatically admixed into one of the qubit states through a chirp of the laser frequency and/or intensity ramp \cite{mitra2020robust, martin2021molmer, schine2022long, mitra2023neutral}. The resulting light shift of the dressed state is affected by the strong interaction (a few tens of $\mathrm{MHz}$ to a few $\mathrm{GHz}$) between Rydberg states, leading to entanglement \cite{mitra2020robust, martin2021molmer, schine2022long, mitra2023neutral}. This tool has been implemented to create Bell states of clock qubits in the microwave \cite{jau2016entangling} and optical regimes \cite{schine2022long} and for studies of many-body physics \cite{zeiher2016many, zeiher2017coherent, borish2020transverse}. Schemes for implementing two-qubit entangling quantum logic gates based on adiabatic Rydberg dressing have been studied theoretically \cite{keating2015robust, mitra2020robust, beterov2018adiabatic, beterov2020application, robicheaux2021photon, mitra2023neutral} and recently demonstrated \cite{martin2021molmer, schine2022long}.

While the basic entangling interaction is due to the interactions between Rydberg states with strength $|V|$, in protocols that employ the Rydberg blockade, the speed of the gate is limited by the effective Rabi frequency of the coupling laser $\Omega_{\mathrm{eff}}$, as in the seminal work of \cite{jaksch2000fast}. Rydberg dressing under a strong blockade, where the admixture of the doubly excited Rydberg states is small and often negligible requires $\hbar \Omega_{\mathrm{eff}} \ll |V|$. As such, one cannot achieve the fundamental scaling in the gate error rate set by the ratio $2\pi\hbar \Gamma/|V|$ for a characteristic decoherence rate $\Gamma$ \cite{saffman2010quantum, saffman2016quantum}. Adiabatic Rydberg dressing can also be used outside the strong blockade regime, thereby strongly increasing the entangling energy, or may be used to maintain atoms separated beyond the blockade radius where they can be more easily individually addressed, yet still achieve fast gates. Rydberg-mediated entanglement beyond the strong blockade regime has been demonstrated using finely tuned two-atom Rabi oscillations \cite{jo2020rydberg}. In addition, some quantum simulation schemes implementing interacting spin models do not assume strong blockade in a multi-atom array, allowing implementation of elaborate interaction graphs between atoms in one-dimensional \cite{bernien2017probing, keesling2019quantum, omran2019generation} and two-dimensional geometries \cite{de2018analysis, ebadi2021quantum, scholl2021quantum, ebadi2022quantum, bluvstein2022quantum, graham2022multi}.

We show here that by going beyond the perfect blockade regime one can use adiabatic Rydberg dressing to reach the fundamental scaling of entangling gate fidelity \cite{wesenberg2007scalable}. Such an approach may become more feasible, for example, using bound states of doubly excited Rydberg macrodimers \cite{sassmannshausen2016observation} that have been well resolved \cite{sassmannshausen2016observation, hollerith2021realizing}, and can be employed for such coherent control of entanglement \cite{hollerith2021realizing}. In addition, we find that one can implement entangling gates in the weak blockade regime using an adiabatic Rydberg dressing scheme that requires only a limited population in the doubly-excited Rydberg state, similar to \cite{jo2020rydberg} and unlike some other protocols for entangling gates \cite{jaksch2000fast, wilk2010entanglement, isenhower2010demonstration, levine2018high, levine2019parallel, saffman2010quantum, saffman2016quantum}. Thus, protocols that extend beyond the perfect blockade regime may enable even more powerful schemes for neutral atom quantum information processing.

%
\section{Adiabatic Rydberg dressing}
%
We consider an atom with two long-lived clock states to serve as the qubit states $\ket{0}$ and $\ket{1}$. The state $\ket{1}$ is optically coupled to a Rydberg state $\ket{r}$, using symmetrically addressing uniform laser fields. The Hamiltonian takes the form
\begin{equation}
    \label{eq:Hamiltonian}
    \hat{H} = \hat{H}_1 \otimes \ketbra{0}{0} +  \ketbra{0}{0} \otimes \hat{H}_1 + \hat{H}_{1,1},
\end{equation}
where $\hat{H}_1$ is the Hamiltonian for one atom in $\ket{1}$ coupled to $\ket{r}$ and $\hat{H}_{1,1}$ is the two-atom coupling, including interactions between Rydberg states. 
\begin{equation}
    \hat{H}_1 =
    -\hbar \Delta_\mathrm{eff} \ketbra{r}{r}
    + \frac{\hbar \Omega_\mathrm{eff}}{2}\left( \ketbra{r}{1}
    +\ketbra{1}{r} \right),
    \label{eq:OneAtomRydbergHamiltonian}
\end{equation}
where $\Omega_{\mathrm{eff}}$ and $\Delta_{\mathrm{eff}}$ are the effective Rabi frequency and detuning of the laser-atom coupling. Finally, the entangling two-atom Hamiltonian is
\begin{equation}
\begin{aligned}
    \hat{H}_{1,1} &
    = \ketbra{1}{1} \otimes \hat{H}_1 +
    \hat{H}_1 \otimes \ketbra{1}{1} +
    V\ketbra{r,r}{r,r}
    \\ &
    = -\hbar\Delta_\mathrm{eff}
    \left(\ketbra{b}{b} + \ketbra{d}{d}\right)
    \\ &
    + \left(V-2\hbar\Delta_\mathrm{eff}\right)
    \ketbra{r,r}{r,r}
    \\ &
    + \frac{\hbar}{2}\sqrt{2} \Omega_\mathrm{eff}
    \left(\ketbra{b}{1,1} +
    \ketbra{r, r}{b}+ +
    \text{h.c.} \right),
    \label{eq:TwoAtomRydbergHamiltonian}
\end{aligned}
\end{equation}
where $V$ is the atom-atom potential energy arising from interaction when both atoms are in $\ket{r}$,
and 
\begin{equation}
\begin{aligned}
\ket{b} & \equiv \frac{1}{\sqrt{2}} \left(\ket{1,r} + \ket{r,1}\right),
\\
\ket{d} &\equiv \frac{1}{\sqrt{2}} \left(\ket{1,r} - \ket{r,1}\right)
\end{aligned}
\end{equation}
are the bright and dark states, respectively, for symmetric coupling \cite{keating2015robust, mitra2020robust, robicheaux2021photon, mitra2023neutral}, which we introduced in Chap.~\ref{chap:NeutralAtoms}. When $|V|\gg \hbar \Omega_\mathrm{eff},\hbar| \Delta_\mathrm{eff}|$, excitation to the doubly excited Rydberg state is strongly blockaded  \cite{lukin2001dipole, saffman2010quantum, keating2015robust, mitra2020robust, mitra2023neutral}. In that case, we can reduce this Hamiltonian to a two-atom, two-level system
\begin{equation}
    \hat{H}_{1,1} \approx -\hbar\Delta_\mathrm{eff}\ketbra{b}{b}
    + \frac{\hbar}{2}\sqrt{2} \Omega_\mathrm{eff}
    \left(\ketbra{b}{1,1} +  \ketbra{1,1}{b}  \right).
    \label{eq:TwoAtomPerfectBlockade}
\end{equation}
The effect of the blockade is seen explicitly in the driving of $\ket{1,1}$ to the entangled bright state $\ket{b}$.

\begin{figure}
  \centering
    \includegraphics[width=0.9\textwidth]{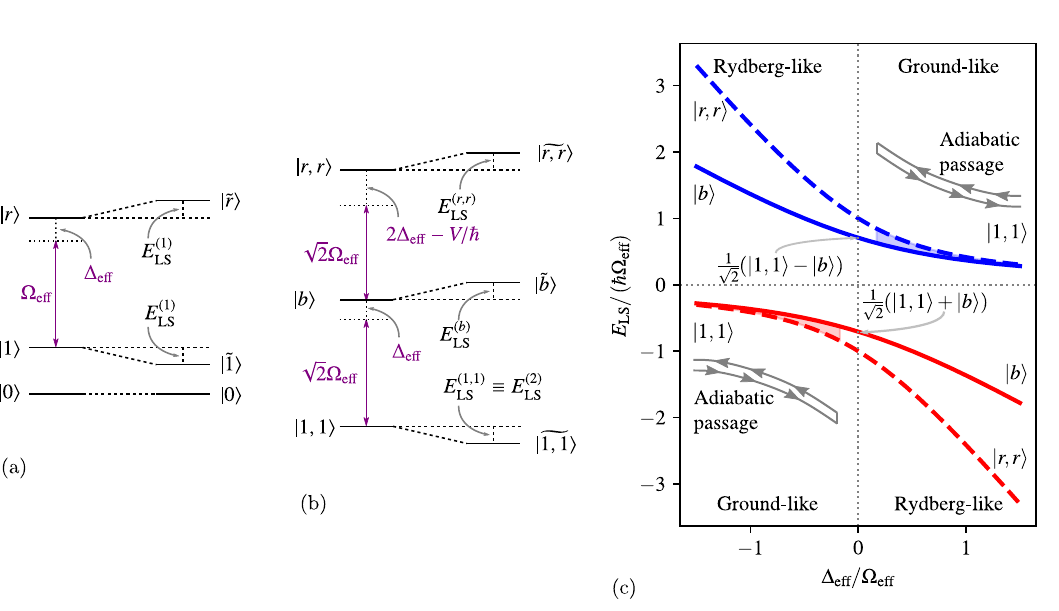}
    \caption{
    Energy levels for implementing adiabatic Rydberg dressing.
    (a) Energy levels and light shifts in one-atom dressing, where each atom is dressed independently.
    (b) Energy levels and light shifts in two-atom dressing, where both atoms are dressed together in the presence of interaction energy $V$.
    (c) Energy shifts of atomic states as a function of detuning, in the strong blockade ($\hbar\Omega_{1r} \ll |V|$, $\hbar|\Delta_{1r}| \ll |V|$) case, which play a role in the adiabatic passage between ground-like states and Rydberg-like states. The shaded region shows the entangling energy [Eq.~\eqref{eq:EntanglingEnergy}], which is used to accumulate the entangling phase.}
    \label{fig:RydbergDressingEnergyLevels}
\end{figure}

The eigenstates of the Hamiltonian in Eq.~\eqref{eq:Hamiltonian} are the dressed states. In particular, we denote the dressed clock states (computational basis states) $\{ \ket{0,0}, \ket*{0, \tilde{1}}, \ket*{\tilde{1},0}, \ket*{\widetilde{1,1}} \}$. The eigenvalues $E_{0,\tilde{1}} = E_{\tilde{1},0}$ and $E_{\widetilde{1,1}}$ contain contributions from light shifts, $E_{\mathrm{LS}}^{(1)}$ with one atom or $E_{\mathrm{LS}}^{(2)}$ with two atoms coupled to the Rydberg state. The entangling energy, denoted by  $\hbar \kappa$, is the energy difference between the interacting and noninteracting atoms,
\begin{equation}
   \kappa 
=
   \frac{1}{\hbar} \left(
   E_{\mathrm{LS}}^{(2)} - 2E_{\mathrm{LS}}^{(1)} \right)
   \approx \frac{\Delta_\mathrm{eff}}{2} \pm \frac{1}{2}\left(\sqrt{2 \Omega^2_\mathrm{eff} + \Delta^2_\mathrm{eff}}
   - 2\sqrt{ \Omega^2_\mathrm{eff} + \Delta^2 _\mathrm{eff}}\right),
   \label{eq:EntanglingEnergy}
\end{equation}
where the approximation holds only in the limit of a perfect blockade, with entangling Hamiltonian Eq.~\eqref{eq:TwoAtomPerfectBlockade}, and $\pm$ refers to the two branches of the dressed states in Fig.~\ref{fig:RydbergDressingEnergyLevels}. On resonance $\kappa = \pm \left(1 - 1/\sqrt{2}\right) \Omega_{\mathrm{eff}} \approx \pm 0.29 \Omega_{\mathrm{eff}}$, where $\Omega_{\mathrm{eff}}/2\pi$ can be as large as a few MHz. For weak dressing, $|\Delta_{\mathrm{eff}}|\gg \Omega_{\mathrm{eff}}$, $\kappa \approx - \Omega_{\mathrm{eff}}^4 / 8\Delta^3_{\mathrm{eff}}$, which will generally be smaller than the rate of photon scattering, which scales as $1/\Delta^2_{\mathrm{eff}}$ \cite{keating2015robust, mitra2020robust}. Therefore, weak dressing leads to a faster reduction in the entangling energy $\hbar \kappa$, than the photon scattering rate, implying that Rydberg states decay by scattering photons faster than entanglement is generated. In~Sec.~\ref{sec:DressingEntanglingEnergy}, we go beyond the perfect blockade.

The dressed energy levels provide an adiabatic passage from the one-atom ground state $\ket{1}$ to the one-atom Rydberg state $\ket{r}$ and from the two-atom ground state $\ket{1,1}$ to the two-atom entangled bright state $\ket{b}$, as shown in Fig.~\ref{fig:RydbergDressingEnergyLevels}(c). Assuming adiabatic evolution, we consider sweeping the detuning from $\ket{1,1}$ toward $\ket{b}$ and then back to $\ket{1,1}$, yielding an entangling phase given by $\varphi_2 = \int \kappa(t') \mathrm{d} t' $. 

%
\section{Entanglers with adiabatic Rydberg dressing}
\label{sec:EntanglinerAdiabaticRydbergDressing}
%
To understand the general class of gates enabled by the phases accumulated in adiabatic evolution and their sensitivity to errors, we consider the Hamiltonian in the dressed qubit (DQ) subspace spanned by $\{\ket{00}, \ket*{0\tilde{1}}, \ket*{\tilde{1}0}, \ket*{\widetilde{11}}\}$,  where $\ket*{\tilde{1}}$ is the one-atom dressed ground state that is a superposition of $\ket{1}$ and $\ket{r}$ with dressing angle given by $\tan \theta = \Omega_{\mathrm{eff}}/(-\Delta_{\mathrm{eff}})$. Let $\hat{\sigma}_z = \ketbra*{\tilde{1}}{\tilde{1}} - \ketbra{0}{0}$ be the adiabatic Pauli operator on one atom. In the dressed atomic basis, the Hamiltonian in the ground subspace can be written as
\begin{equation}
  \label{eq:RydbergDressedGroundHammy}
  \hat{H}_{\mathrm{DQ}} =
  - \left(E_\mathrm{LS}^{(1)} + \frac{\hbar\kappa}{2}\right) 
  \left(\frac{\hat{\sigma}_z}{2} \otimes \mathds{1} + \mathds{1} \otimes \frac{\hat{\sigma}_z}{2} \right)
  + \hbar\kappa \left( \frac{\hat{\sigma}_z}{2} \otimes \frac{\hat{\sigma}_z}{2} \right)
  .
\end{equation}
The linear terms with sums of one-spin Pauli operators generate $\mathbb{SU}(2)$ rotations on the collective spin, while the quadratic term with tensor product of Pauli operators on both spins ``twists" the collective spin and also generates two-qubit entanglement. Adiabatic evolution under this Hamiltonian generates unitaries of the form
\begin{equation}
\begin{aligned}
  \hat{U}_{\kappa}(\varphi_1, \varphi_2)
  & =
  \exp\left(- \mathrm{i} \varphi_2 \left(\frac{\hat{\sigma}_z}{2} \otimes \frac{\hat{\sigma}_z}{2}\right)\right)
  \\ & \times
  \exp\left(- \mathrm{i} \varphi_1 \left(\ident \otimes \frac{\hat{\sigma}_z}{2} + \frac{\hat{\sigma}_z}{2} \otimes \ident\right)\right),
  \label{eq:UnitaryAdiabaticPassage}
\end{aligned}
\end{equation}
where $\varphi_1 = -\int \left(E_\mathrm{LS}^{(1)} + \frac{\hbar \kappa}{2}\right) \mathrm{d} t $ is the rotation angle generated by the linear terms, and $\varphi_2 = \int \kappa \mathrm{d} t$ is the twist angle generated by the quadratic term in the Hamiltonian. An entangling gate is achieved through the dynamical phase accumulated from the entangling energy $\varphi_2$ \cite{keating2015robust, jau2016entangling, lee2017demonstration, mitra2020robust, martin2021molmer, schine2022long}, with the dynamical phase $\varphi_1$, accumulated from the local terms affect the local, separable components of the implemented gate. When the twist angle $\varphi_2 = \pm \pi$, these gates are perfect entanglers \cite{zhang2003geometric, zhang2004minimum, zhang2005generation, goerz2015optimizing}, which are gates that take a product state to a maximally entangled state. Examples of perfect entanglers of this kind are the Controlled Z (CZ) gate ($ \varphi_1 = \mp \pi/2, \varphi_2 = \pm \pi$) and the M{\o}lmer-S{\o}rensen (MS) gate ($\varphi_1 = 0, \varphi_2 = \pi$) In the latter, the axis is changed to $\mu \in\{x, y\}$. We review some properties of two-qubit entanglers in Appendix~\ref{app:TwoQubitGatesReview}. A CZ gate is achieved by adjusting the phases accumulated due to the independent one-atom light shifts, $E_{\mathrm{LS}}^{(1)}$, to the appropriate value of $\varphi_1 = \pm \pi/2$ \cite{keating2015robust}. In contrast, the MS gate is achieved by removing \textit{all} single qubit phases contributing to $\varphi_1$ \cite{mitra2020robust, martin2021molmer, mitra2023macrostates}. While this difference is theoretically trivial, the dominant source of gate infidelity is errors in $\varphi_1$, making the MS gate more robust than the CZ gate, as we will see below.

\begin{figure}
  \centering
  \includegraphics[width=0.6\textwidth]{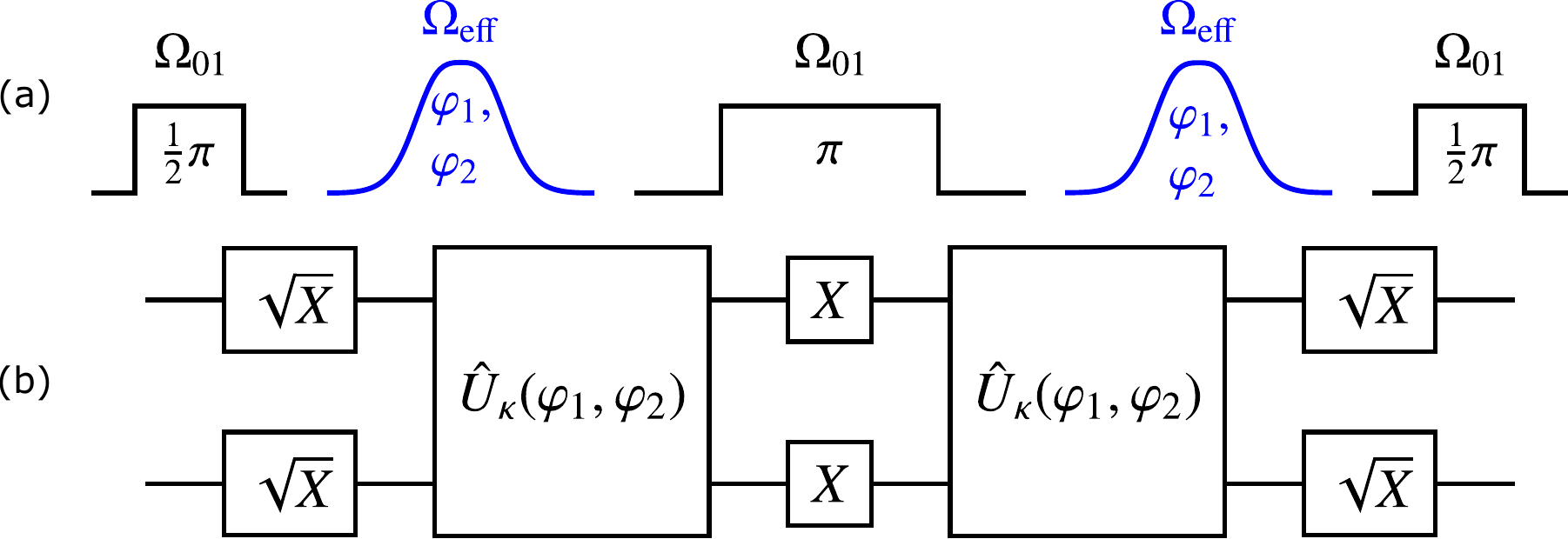}
  \caption{Adiabatic passages interleaved in a spin-echo sequence. (a) Pulse and ramp sequence. (b) Equivalent circuit diagram. When $\varphi_2 =\pi/2$ the result is the $\mathrm{MS}_{yy}$ gate [Eq.~\eqref{eq:MSyyGate}].}
  \label{fig:SpinEchoAdiabaticPassage}
\end{figure}

The MS gate is generated using a spin-echo sequence, as shown in Fig.~\ref{fig:SpinEchoAdiabaticPassage} \cite{mitra2020robust, martin2021molmer, schine2022long, mitra2023neutral}. The sequence consists of a $\pi/2$ pulse about the $x$-axis, followed by an adiabatic ramp accumulating non-local phase $\varphi_2 = \int \kappa(t') \mathrm{d} t'$, an echo $\pi$ pulse about the $x$-axis, followed by another adiabatic ramp accumulating nonlocal phase $\varphi_2 = \int \kappa(t') \mathrm{d} t'$ , and a final $\pi/2$ pulse about the $x$-axis, as shown in Fig.~\ref{fig:SpinEchoAdiabaticPassage}(a). An equivalent circuit diagram with the shorthand $\sqrt{X}$ representing a $\pi/2$ pulse about the $x$-axis, $X$ representing a $\pi$ pulse about the $x$-axis and $\hat{U}_{\kappa}(\varphi_1, \varphi_2)$ representing the unitary generated during each adiabatic passage, is shown in Fig.~\ref{fig:SpinEchoAdiabaticPassage}(b). Importantly, the spin-echo removes all phases, $\varphi_1$, arising for single atom-light shifts, including the dominant errors arising from atom thermal motion \cite{wilk2010entanglement, de2018analysis, graham2019rydberg}, and the resulting inhomogeneities \cite{mitra2020robust, martin2021molmer, schine2022long}. Designing the adiabatic ramps such that $\varphi_2 = \pi/2$ in each ramp, the resulting unitary transformation is a M{\o}lmer-S{\o}rensen YY-gate ($\mathrm{MS}_{yy}$),
\begin{equation}
  \hat{U}_{\mathrm{MS}_{yy}} =
  \exp(-\mathrm{i} \frac{\pi}{4}\hat{\sigma}_y \otimes \hat{\sigma}_y),
  \label{eq:MSyyGate}
\end{equation}
 which is a perfect entangler for the qubits. Details of deriving this form of the gate are discussed in Appendix~\ref{app:MSGateDetails}. Off-resonant coupling to the intermediate state leads to additional light shifts and potential noise due to intensity fluctuations. The spin echo removes this noise in its contribution to the single-atom light shift. There will still be some residual error that remains and cannot be canceled in the spin echo, but this is minimal and in practice can be reduced with further robust control techniques.

%
\subsection{Entangling gate fidelity and robustness}
\label{sec:EntanglingGateFidelity}
%

\begin{figure}[htbp]
  \centering
  \includegraphics[width=0.7\textwidth]{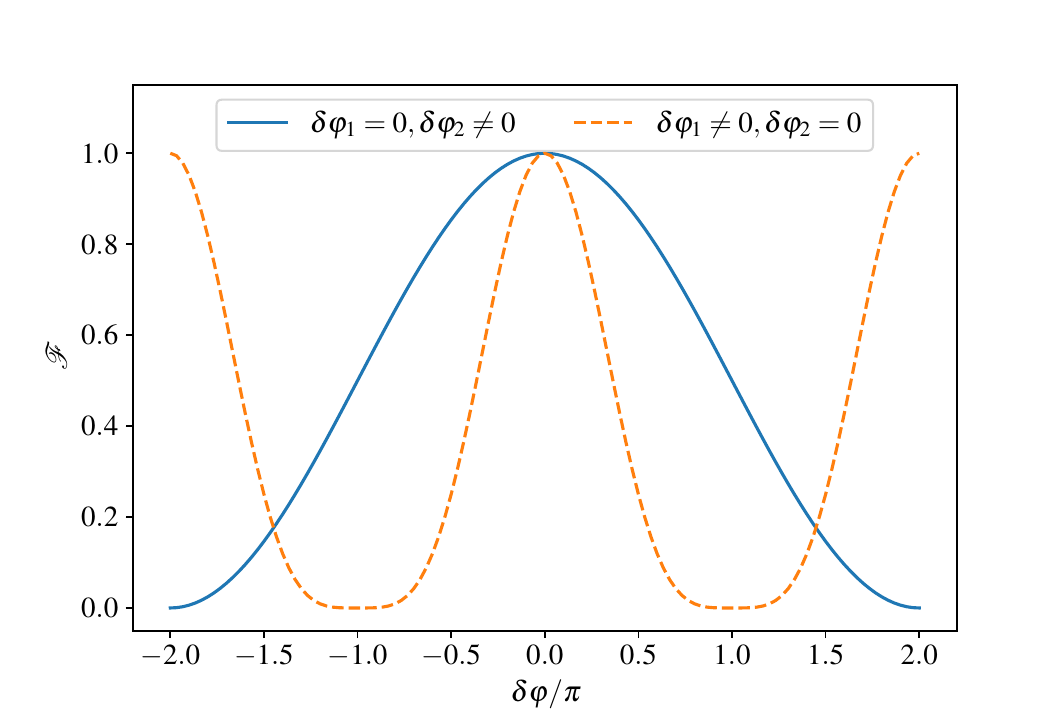}
  \caption{Fidelity between a target and implemented unitary transformation, each of the form given in (Eq.~\eqref{eq:UnitaryAdiabaticPassage}), as a function of the error in the two-qubit rotation angle, $\delta \varphi_2$, with zero one-qubit rotation error, $\delta \varphi_1 = 0$ (blue, solid line) and as a function of error in the one-qubit twist angle $\delta \varphi_1$, with zero two-qubit rotation error, $\delta \varphi_2 = 0$ (orange, dotted line). The implementation of the CZ gate is very sensitive to errors caused by inhomogeneities, as it is dominated by $\delta \varphi_1$, (orange, dotted line). In contrast, the MS gate is more robust, as it is prone to only errors in $\delta \varphi_2$ (blue, solid line).}
  \label{fig:UnitaryTransformations}
\end{figure}

We assess the performance of the gate by considering the fidelity between the implemented two-qubit gate $\hat{U}$ and the target ideal unitary transformation $\hat{V}$ defined using a normalized Hilbert Schmidt inner product between them 
\begin{equation}
    \mathcal{F} = \frac{1}{16} \left\lvert \mathrm{Tr}(\hat{V}^\dagger \hat{U}) \right\rvert^2
    ,
\end{equation}
which estimates how well any input basis is mapped to the corresponding target output basis, by the implemented unitary  \cite{pedersen2007fidelity}. Focusing on errors that can arise from inhomogeneities or coherent errors in the accumulated phases, the fidelity depends on the difference between twist angles $\delta \varphi_2$ and the difference between rotation angles $\delta \varphi_1$ of the implemented and target unitary maps as \cite{mitra2020robust}
\begin{equation}
\label{eq:FidelityOneAxistTwistRotate}
  \mathcal{F} (\delta \varphi_1, \delta \varphi_2)
  = \frac{1}{4}
  \left(1 + \cos^2(\delta \varphi_1)
  + 2 \cos(\delta \varphi_1)
  \cos(\frac{\delta \varphi_2}{2})
  \right).
\end{equation}
Importantly, the fidelity is much more sensitive to $\delta \varphi_1$ than it is to $\delta \varphi_2$. The twist angle $\varphi_2$ depends solely on the entangling energy $\kappa$. As this is the {\em difference} of two light shifts, it has some common mode cancellation of errors in the light shifts, while $\varphi_1$ has a contribution from independent single-atom light shifts with no such cancellation. This effect is seen in Fig.~\ref{fig:UnitaryTransformations} which shows the fidelity plotted as a function of $\delta \varphi_1$ when $\delta \varphi_2 = 0$, that is,
\begin{equation}
  \mathcal{F} (\delta \varphi_1=0, \delta \varphi_2)
  = \frac{1}{2} 
  \left(1 + \cos\left(\frac{\delta \varphi_2}{2}\right)\right),
\end{equation}
and as a function of $\delta \varphi_2$ when $\delta \varphi_1 = 0$, that is,
\begin{equation}
  \mathcal{F} (\delta \varphi_1, \delta \varphi_2=0)
  = \frac{1}{4} 
  (1 + \cos(\delta \varphi_1))^2.
\end{equation}
Details of deriving these expressions are discussed in Appendix~\ref{app:SymmetricTwoQubitGates}.

Note, implementation of a CZ gate requires knowledge of $E_{\text{LS}}^{(1)}$ to adjust the single-atom contribution to the phase $\varphi_1$ \cite{keating2015robust, mitra2020robust}, and errors will contribute substantially to infidelity through $\delta \varphi_1$. In contrast, the MS gate is substantially less sensitive to such errors, as $\delta \varphi_1$ can be made zero by using a spin echo \cite{martin2018cphase, mitra2020robust, martin2021molmer, schine2022long, mitra2023neutral}.

%

%
\section{Implementation of Adiabatic Rydberg dressing}
\label{sec:AdiabaticPassageImplementation}
%
In the adiabatic Rydberg dressing paradigm, entanglement is generated by the adiabatic dressing of the $\ket{1}$-state through a one- or two-photon transition to an excited Rydberg state $\ket{r}$ with high principle quantum number $n_r$ \cite{keating2015robust, mitra2020robust, mitra2023neutral}. This is most naturally implemented using a one-photon transition between a clock state and a high-lying Rydberg state \cite{keating2015robust, jau2016entangling, zeiher2016many, zeiher2017coherent, borish2020transverse, mitra2020robust}. Such an approach requires a high-power ultraviolet laser which is technically challenging and can lead to adverse effects, such as photoelectric charging of dielectrics and spurious electric fields. Adiabatic Rydberg dressing would be more simply achieved through a standard two-photon transition that is typically used for Rydberg excitation \cite{wilk2010entanglement, isenhower2010demonstration, levine2018high, de2018analysis, levine2019parallel}, but this may lead to other challenges due to additional decoherence and spurious light shifts from off-resonant excitation to the intermediate state \cite{zhang2012fidelity, de2018analysis}.

 For the one-photon ultraviolet excitation, there is direct excitation of the $\ket{1} \leftrightarrow \ket{r}$ transition, with $\Omega_\mathrm{eff} = \Omega_{1r}$, $\Delta_\mathrm{eff} = \Delta_{\mathrm{1r}}$ in Eq.~\eqref{eq:OneAtomRydbergHamiltonian}, Eq.~\eqref{eq:TwoAtomRydbergHamiltonian}, Eq.~\eqref{eq:TwoAtomPerfectBlockade} and Eq.~\eqref{eq:EntanglingEnergy}. In this case, 
\begin{equation}
    \ket{r} \equiv \ket{(n_r p),\, {}^2 \mathrm{P}_J}
    ,
\end{equation}
 for alkali atoms and 
 \begin{equation}
    \ket{r} \equiv \ket{(nsn_r s),\, {}^3 \mathrm{S}_1}
    ,
 \end{equation}
 for alkaline earth atoms.
 
\begin{figure}
  \centering
    \includegraphics[width=0.7\textwidth]{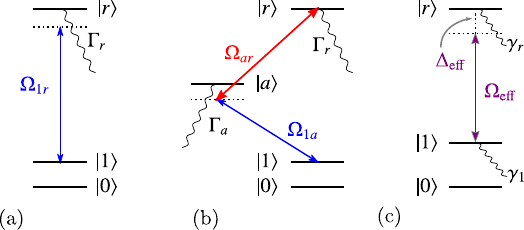}
    \caption{
    Schemes for ground-to-Rydberg excitation
    (a) One photon $\ket{1} \leftrightarrow \ket{r}$ transition, with Rabi frequency $\Omega_{1r}$ and Rydberg decay rate $\Gamma_r$.
    (b) Two photon $\ket{1} \leftrightarrow \ket{a} \leftrightarrow \ket{r}$ transition with Rabi frequencies $\Omega_{1a}$ and $\Omega_{ar}$ respectively, intermediate state decay rate $\Gamma_a$ and Rydberg decay rate $\Gamma_r$.
    (c) Effective three-level system in regime of adiabatically eliminating the intermediate state $\ket{a}$, with effective Rabi frequency $\Omega_{\text{eff}}$ and effective detuning $\Delta_{\text{eff}}$ due to the difference of light-shifts experienced by $\ket{a}$ and $\ket{r}$ and effective decay rate $\gamma_r$ from $\ket{r}$ and $\gamma_1$ from $\ket{1}$.}
    \label{fig:RydbergExcitationEnergyLevels}
\end{figure}

 In the two-photon case, there is indirect excitation to the Rydberg state using two laser fields addressing the $\ket{1} \leftrightarrow \ket{a}$ and $\ket{1} \leftrightarrow \ket{a}$ transitions. In this case, the states 
 \begin{equation}
     \ket{r} \equiv \ket{(n_r s),\, {}^2\mathrm{S}_{1/2}} 
     ,
 \end{equation}
 for alkali atoms, and 
\begin{equation}
    \ket{r} \equiv \ket{(ns n_r p),\, {}^3\mathrm{P}_J}
    ,
\end{equation}
 for alkaline earth atoms, with an intermediate auxiliary state 
 \begin{equation}
     \ket{a} \equiv \ket{(n_a p),\, {}^2 \mathrm{P}_J}
     ,
 \end{equation}
 or 
\begin{equation}
    \ket{a} \equiv \ket{(ns n_a s),\,{}^3\mathrm{S}_1}
    ,
\end{equation}
respectively. The Hamiltonian reads
\begin{equation}
\begin{aligned}
    \hat{H}_{\mathrm{2-photon}} 
    &
    = - \hbar \Delta_{1a} \ketbra{a}{a}
    - \hbar \left(\Delta_{1a} + \Delta_{ar} \right) \ketbra{r}{r}
    \\ &
    + \frac{\hbar\Omega_{1a}}{2} \left(\ketbra{a}{1} + \ketbra{1}{a}\right)
    + \frac{\hbar\Omega_{1a}}{2} \left(\ketbra{r}{a} + \ketbra{a}{r}\right)
    ,
\end{aligned}
\end{equation}
where we use the notation with the Rabi frequencies $\Omega_{\alpha \beta}$ and detunings $\Delta_{\alpha \beta}$ for each of the corresponding $\ket{\alpha} \leftrightarrow \ket{\beta}$ transitions as shown in Fig.~\ref{fig:RydbergExcitationEnergyLevels}. In this case, the generation of entanglement is fundamentally limited by decoherence due to the lifetime of $\ket{r}$ and $\ket{a}$, which depend on the choice of principal quantum numbers $n_r$ and $n_a$.

 For a two-photon excitation, we consider the regime $\Omega_{1a} \ll |\Delta_{1a}|$ so that the intermediate state can be adiabatically eliminated. This gives us,
 \begin{equation}
 \begin{aligned}
 \Omega_\mathrm{eff} & = \frac{\Omega_{1a}\Omega_{ar}}{2\Delta_{1a}}
 \\
 \Delta_\mathrm{eff} & = \Delta_{1a}+\Delta_{ar} + 
 \underbrace{
 \frac{\Omega^2_{1a}}{4\Delta_{1a}}}_{\delta_{1}}
 \underbrace{
 - \frac{\Omega^2_{ar}}{4\Delta_{ar}}
 }_{\delta_{r}}
 \end{aligned}
\end{equation}
 where $\delta_1$ and $\delta_r$ are the light shifts of levels $\ket{1}$ and $\ket{r}$ respectively due to their coupling to $\ket{a}$ in the regime $\Omega_{1a} \ll |\Delta_{1a}|$, where the auxilliary state $\ket{a}$ is adiabatically eliminated  \cite{de2018analysis, mitra2023neutral}, giving the effective Hamiltonians Eq.~\eqref{eq:OneAtomRydbergHamiltonian} and Eq.~\eqref{eq:TwoAtomRydbergHamiltonian}.

The fundamental source of decoherence is due to the decay of the Rydberg state at rate $\Gamma_r$ and the intermediate state at rate $\Gamma_a$. To good approximation, these decay processes will lead to leakage outside the qubit subspace. In that case, we can treat decoherence simply through a non-trace-preserving Schr\"{o}dinger evolution with a non-Hermitian Hamiltonian 
\begin{equation}
    \hat{H}_\mathrm{eff} = \hat{H} -\frac{\mathrm{i} \hbar}{2}\sum_\mu \hat{L}_\mu^\dag \hat{L}_\mu ,
\end{equation}
where $\{\hat{L}_\mu\}$ are the Lindblad jump operators. For one-photon excitation,
\begin{equation}
    \sum_\mu \hat{L}_\mu^\dag \hat{L}_\mu = \Gamma_r \ketbra{r}{r},
\end{equation}
for each atom, while for the two-photon excitation,
\begin{equation}
  \sum_\mu \hat{L}_\mu^\dag \hat{L}_\mu =
  \gamma_1 \ketbra{1}{1} +
  \gamma_r \ketbra{r}{r} +
  \gamma_{1r} \left(\ketbra{r}{1} + \ketbra{1}{r}\right),
\end{equation}
for each atom. Here levels $\ket{1}$ and $\ket{r}$ and their coherences decay due to off-resonant photon scattering with rates
\begin{equation}
  \gamma_1
  = \frac{\Omega_{1a}^2}{4\Delta_{1a}^2}\Gamma_a, \;
  \gamma_r
  = \frac{\Omega_{ar}^2}{4\Delta_{ar}^2}\Gamma_a+\Gamma_r, \;
  \gamma_{1r} = \frac{\Omega_{ra}\Omega_{1a}}{4\Delta_{1a}^2}\Gamma_a.
\end{equation}
High-fidelity gates for two-photon excitation require sufficiently long lifetimes of level $\ket{a}$.

As studied in \cite{mitra2020robust}, the highest fidelity gates are achieved for strong dressing, with Rydberg excitation resonance, $|\Delta_{\mathrm{eff}}| \ll \Omega_{\mathrm{eff}}$, and a large admixture of $\ket{b}$ in the dressed state $\ket*{\widetilde{1,1}}$. For a one-photon transition, we consider an adiabatic sweep involving a Gaussian laser intensity sweep and the linear detuning sweep, according to,
\begin{equation}
\begin{aligned}
  &
  \lvert\Delta_{1r}(t) \rvert
  =
  \begin{cases}
    \Delta_{\mathrm{max}} +
    \frac{\Delta_{\mathrm{max}}-\Delta_\mathrm{min}}{t_2-t_1} \times (t-t_1),
    & t_1 \leq t < t_2
    \\
    \Delta_{\mathrm{min}},
    & t_2 \leq t \leq t_3
    \\
    \Delta_{\mathrm{min}} +
    \frac{\Delta_{\mathrm{min}}-\Delta_\mathrm{max}}{t_4-t_3} \times (t-t_3),
    & t_3 < t \leq t_4
  \end{cases}
  \\ &
  \Omega_{1r}(t) =
  \begin{cases}
    \Omega_{\mathrm{min}} +
    \left(\Omega_\mathrm{max}-\Omega_{\mathrm{min}}\right)\exp\left(-\frac{(t - t_1)^2}{2t_w^2}\right),
    & t_1 \leq t < t_2
    \\
    \Omega_{\mathrm{max}},
    & t_2 \leq t \leq t_3
    \\
    \Omega_{\mathrm{min}} +
    \left(\Omega_\mathrm{max} - \Omega_{\mathrm{min}}\right)\exp\left(-\frac{(t - t_3)^2}{2t_w^2}\right),
    & t_3 < t \leq t_4
  \end{cases}.
  \label{eq:GaussianOmega_1rLinearDelta_1r}
\end{aligned}
\end{equation}
The resulting MS-gate was demonstrated in \cite{martin2021molmer, schine2022long}.

\begin{figure}
  \centering
  \includegraphics[width=0.99\textwidth]{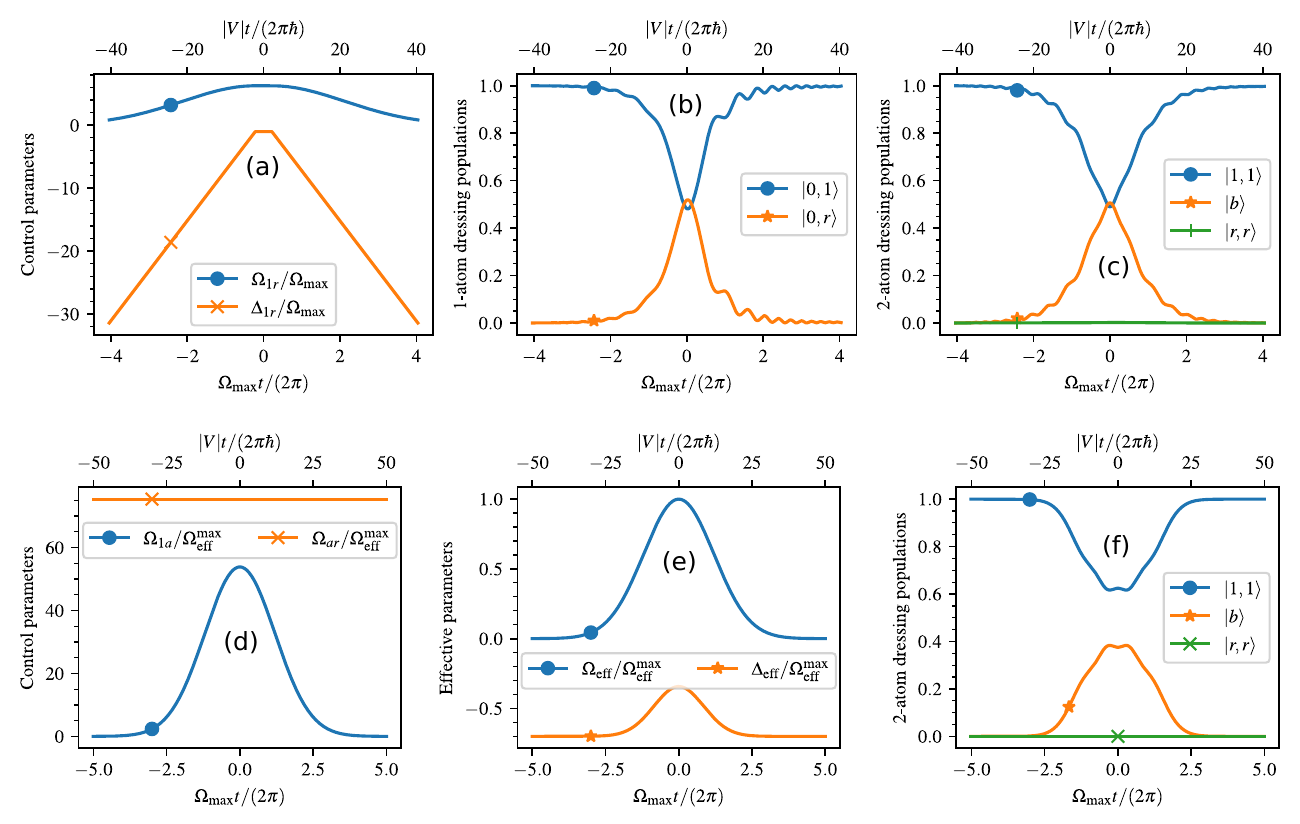}
  \caption{Adiabatic passages to implement $\hat{U}_{\kappa}(\varphi_1, \varphi_2)$ with $\varphi_2 = \pi/2$ [Eq.~\eqref{eq:UnitaryAdiabaticPassage}, Fig.~\ref{fig:SpinEchoAdiabaticPassage}]  in the strong blockade regime ($\hbar \Omega_{\mathrm{eff}} = 0.1 \vert V \vert$). (a) One-photon adiabatic passage Gaussian sweep of Rabi frequency and linear sweep of detuning as in \cite{mitra2020robust, martin2021molmer}. (b) 1-atom populations during a one-photon adiabatic passage  (c) 2-atom populations during a one-photon adiabatic passage. (d) Two-photon adiabatic passage using a Gaussian sweep of Rabi frequency $\Omega_{1a}$, with all other parameters fixed, which leads to an effective sweep of the two-photon Rabi frequency $\Omega_{\text{eff}}$ and two-photon detuning $\Delta_{\text{eff}}$ as shown in (e). (f) 2-atom populations during a two-photon adiabatic passage. The bottom axes show time measured in units of $2\pi/\Omega_{\max}$, and the top axes show time measured in units of $|V|t/(2 \pi \hbar)$. In the strong blockade, as expected, $|V|t/\hbar \gg \Omega_{\max}t$
}
\label{fig:StrongBlockadeAdiabaticPassage}
\end{figure}

For the two-photon case, the effect of the light shift arising from the intermediate detuning facilitates additional possibilities for coherent control \cite{mitra2023neutral}. We consider the case of exact two-photon resonance in the absence of the light shift, and a fixed Rabi frequency $\Omega_{ar}$ and detuning $\Delta_{ar}$ on the $\ket{a} \leftrightarrow \ket{r}$ transition. Adiabatic Rydberg dressing is achieved simply through a Gaussian ramp of the intensity of the laser driving the $\ket{1} \leftrightarrow \ket{a}$ according to the Rabi frequency,
\begin{equation}
  \Omega_{1a} =
  \begin{cases}
    \Omega_{1a}^{\mathrm{max}}, & -\lvert t_{\mathrm{stop}}\rvert \leq t \leq\lvert t_{\mathrm{stop}}\rvert\\
    \Omega_{1a}^{\mathrm{max}} \exp(-\frac{(t-\lvert t_{\mathrm{stop}\rvert})^2}{2 t_w^2}), & \text{  otherwise.}\\
  \end{cases}
  \label{eq:GaussianOmega_a}
\end{equation}
One can adjust $\vert t_{\mathrm{stop}}\vert$, the time after which the Rabi frequency remains constant, and $t_w$  the width of the Gaussian pulse, to obtain to the desired gate of interest \cite{mitra2023neutral}. Fig.~\ref{fig:StrongBlockadeAdiabaticPassage}(a, b) show examples of ramps for the one-photon and two-photon adiabatic passage as well the population as a function of time during the pulse sequence in the strong blockade regime with $V = 10 \hbar \Omega_{\max}$.

As discussed above, to implement the M{\o}lmer-S{\o}rensen gate we consider two adiabatic ramps intertwined by the spin echo sequence as shown in Fig.~\ref{fig:SpinEchoAdiabaticPassage}, similar to \cite{mitra2020robust, martin2021molmer, schine2022long}. The adiabatic ramps are obtained by numerically maximizing the fidelity defined using the Hilbert-Schmidt overlap,
\begin{equation}
    \mathcal{F}\left[\{c_{\mathrm{r}}\}\right]
    = \frac{1}{16}
    \left \lvert \mathrm{tr} \left(
    \hat{U}_{\mathrm{MS}_{yy}}^{\dagger}
    \hat{U}(\{c_{\mathrm{r}}\})
    \right) \right\rvert^2,
\end{equation}
with respect to ramp parameters $\{c_{\mathrm{r}}\}$ for both one photon and two photon cases; here $\hat{U}(\{c_{\mathrm{r}}\})$ is the unitary map implemented using the spin-echo sequence in Fig.~\ref{fig:SpinEchoAdiabaticPassage}. We elaborate on the details of the adiabatic passages in Appendix~\ref{app:RydbergAdiabaticPassage}. Replacing $\hat{H}$ with $\hat{H}_{\mathrm{eff}}$ gives an estimate of the fidelity including effects of finite lifetimes of the intermediate state $\ket{a}$ and the Rydberg state $\ket{r}$.
\begin{figure}
    \includegraphics[width=0.92\textwidth]{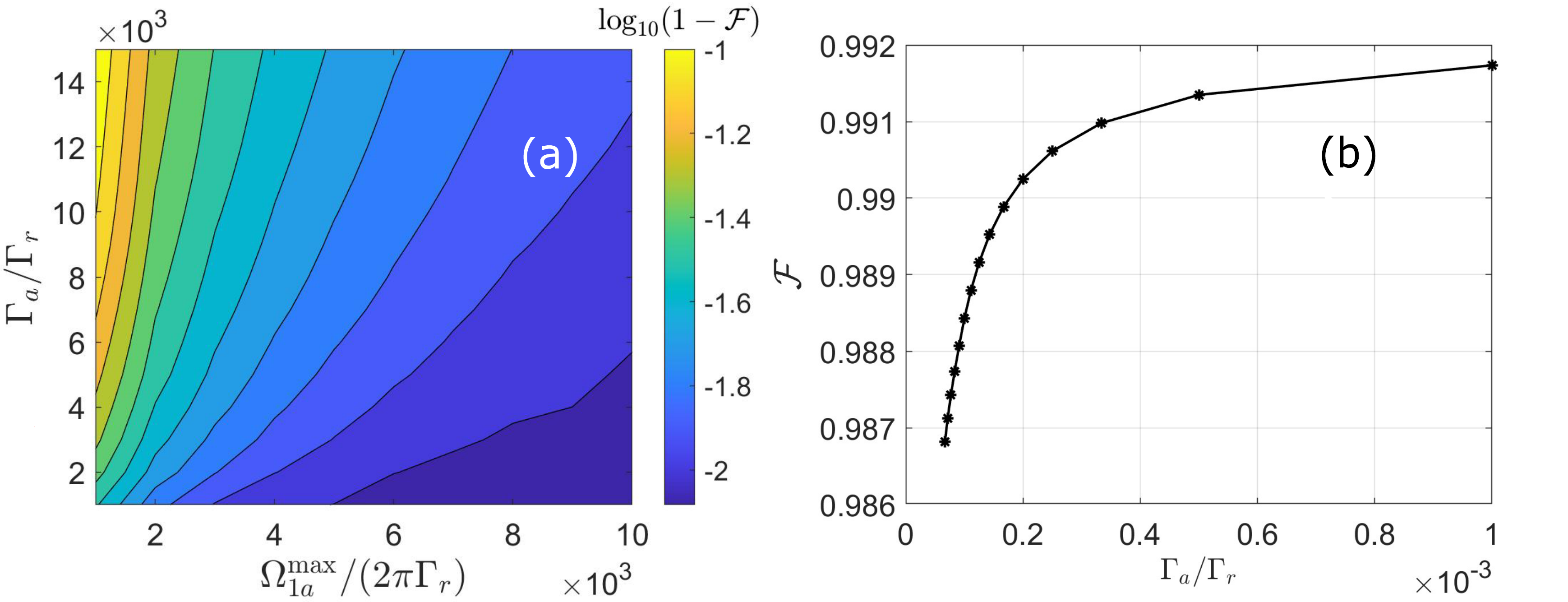}
    \caption{Dependence of the fidelity of the M{\o}lmer-S{\o}rensen gate on the intermediate state decay rate, $\Gamma_{a}$ and the Rabi frequency $\Omega_{1a}$, both measured in units of the Rydberg state decay rate $\Gamma_{r}$. Similar to the other two photon approaches the choice of an intermediate state with a smaller decay rate gives a better fidelity. Moreover, as expected a larger power gives better fidelity. However, this gives us the constraint that we need a larger $|V|$, thus posing some additional challenges. With reasonable experimental parameters, one could achieve an infidelity less than $10^{-2}$. The data are obtained by fixing the ratios $\Gamma_{a}/\Gamma_{r}$ and $\Omega_{1a}^{\max}/(2\pi\Gamma_{r})$ and optimizing over the choice of the detuning from the intermediate state $\Delta_{1a} = -\Delta_{ar}$.
    (a) Contour plot of the logarithm of infidelity, $\log_{10}(1-\mathcal{F})$ across different values of $\Gamma_{a}/\Gamma_{r}$ and $\Omega_{1a}^{\max}/(2\pi\Gamma_{r})$.
    (b) Fidelity, $\mathcal{F}$ as a function of the ratio $\Gamma_{a}/\Gamma_{r}$, for $\Omega_{1r} = 1.4\Omega_{1a}^{\mathrm{max}}$ and $\Omega_{1a}^{\mathrm{max}}/(2\pi\Gamma_{r})= 10^{4}$.}
    \label{fig:FidelityPlots}
\end{figure}

The short lifetime of the intermediate state $\ket{a}$ poses a challenge for implementing adiabatic passage using a two-photon scheme.
We explore the dependence of the achievable M{\o}lmer-S{\o}rensen gate fidelity on the intermediate state lifetime and the Rabi frequency in Fig.~\ref{fig:FidelityPlots}. We fix the Rydberg state decay rate $\Gamma_r$, vary the maximum Rabi frequency $\Omega_{1a}^{\max}$ and the intermediate state decay rate $\Gamma_a$, and then optimize over the intermediate state detuning $\Delta_{1a} = -\Delta_{ar}$ to maximize the fidelity. As in other two-photon approaches, the choice of an intermediate state with a larger lifetime gives a higher fidelity as this is the fundamental source of error in the model. Moreover, as expected a larger power gives higher fidelity, but in the perfect blockade regime, this is constrained by $\hbar \Omega_{\mathrm{eff}} \ll \vert V \vert$. With reasonable experimental parameters, one can achieve fidelity larger than $0.99$  as seen in Fig. \ref{fig:FidelityPlots}.

A key metric quantifying the temporal duration of the adiabatic Rydberg dressing passages is the time-integrated Rydberg population, summed over both atoms, $t_r$  \cite{saffman2010quantum, saffman2016quantum}. For the loss of fidelity due to Rydberg state decay to be small, we require $t_r \ll \tau_r$ where $\tau_r=1/\Gamma_r$ is the Rydberg state lifetime \cite{mitra2020robust, mitra2023macrostates}. For one photon adiabatic passages, we found $t_r \approx 0.89 \times 2\pi / \Omega_{\mathrm{eff}}^{\max}$, while for the two photon passage, we find $t_r \approx 0.95 \times 2\pi / \Omega_{\mathrm{eff}}^{\max}$, with initial state $\ket{1, 1}$. Initial states $\ket{0, 1}$ and $\ket{1, 0}$ lead to smaller time-integrated Rydberg population and initial $\ket{0, 0}$ does not lead to any Rydberg population \cite{mitra2020robust, mitra2023neutral}. In both one- and two-photon cases, since we are considering the strong blockade regime, the adiabatic passages, $t_r$ is significantly larger than $2\pi\hbar/|V|$, the time scale set by the interaction energy $V$. Nevertheless, using finely tuned parameters, adiabatic Rydberg dressing passages can be used to implement high-fidelity entangling gates.

%
\section{Robustness and error channels}
%
We consider the error channels and the intrinsic robustness of using adiabatic Rydberg dressing to implement the MS gate. Deleterious effects include thermal Doppler shifts and atomic motion in a spatially inhomogeneous exciting laser, imperfect blockade, and finite radiative lifetime of the Rydberg state. To see how these effects arise, let us revisit the dressed states, including the quantized motion. For generality, we include quantized atomic momenta $p_\alpha$ and $p_\beta$ of the two atoms in their Rydberg dressing interaction in addition to the electronic ground state and the bright and dark states. The bare states are the ground state
\begin{equation}
  \ket{G} = \ket{1, p_\alpha; 1, p_\beta},
\end{equation}
the bright state
\begin{equation}
  \ket{B} = \frac{1}{\sqrt{2}}
  \left( \ket{r, p_\alpha + \hbar k_{\mathrm{eff}}; 1, p_\beta} + \ket{1, p_\alpha; r, p_\beta + \hbar k_{\mathrm{eff}}}\right),
\end{equation}
and the dark state
\begin{equation}
  \ket{D} = \frac{1}{\sqrt{2}}
  \left(\ket{r, p_\alpha + \hbar k_{\mathrm{eff}}; 1, p_\beta} - \ket{1, p_\alpha; r, p_\beta + \hbar k_{\mathrm{eff}}}\right),
\end{equation}
where $k_{\mathrm{eff}}$ is the component effective wave vector of the Rydberg exciting laser(s) along the interatomic axis, $z$.
The two-atom Rydberg Hamiltonian now generalizes to \cite{keating2015robust, mitra2020robust}
\begin{equation}
\begin{aligned}
  \hat{H}_{\text{2atom}}(p_\alpha,p_\beta)
  &
  =-\left(\hbar\Delta_{\mathrm{eff}}-\frac{\hbar k_{\mathrm{eff}} P_{\mathrm{CM}}}{M}\right)
  \left(|{B}\rangle\langle{B}| + |{D}\rangle\langle{D}|\right)
  \\ &
  + \left(V - 2\left(\hbar \Delta_{\mathrm{eff}} - \frac{\hbar k_{\mathrm{eff}} P_{\mathrm{CM}}}{M}\right)\right)
  |{r,r}\rangle\langle{r,r}|
  \\ &
  + \frac{\hbar k_{\mathrm{eff}} p_{\mathrm{rel}}}{m}
  \left(|{B}\rangle\langle{D}| +
  \text{h.c.} \right)
  \\ &
  + \frac{\hbar}{2} \left(\frac{\Omega_\alpha + \Omega_\beta}{\sqrt{2}}\right)
  \left(|{B}\rangle\langle{G}| +
  |{r,r}\rangle\langle{B}| +
  \text{h.c.} \right)
  \\ &
  + \frac{\hbar}{2} \left(\frac{\Omega_\alpha - \Omega_\beta}{\sqrt{2}}\right)
  \left(|{D}\rangle\langle{G}| +
  |{r,r}\rangle\langle{D}| +
  \text{h.c.} \right)
\end{aligned}
\end{equation}
where $\Omega_\alpha = \Omega_{\mathrm{eff}}(z_\alpha)$ and $\Omega_\beta = \Omega_{\mathrm{eff}}(z_\beta)$ are the Rabi frequencies at the positions of atoms $\alpha$ and $\beta$; $P_{\mathrm{CM}} = p_\alpha + p_\beta$ and $p_{\mathrm{rel}} = (p_\alpha - p_\beta)/2$ are the center-of-mass and relative momenta of the atom $\alpha$ and $\beta$ respectively  \cite{keating2015robust, mitra2020robust}.

\begin{figure}[htbp]
    \centering
     \includegraphics[width=0.96\textwidth]{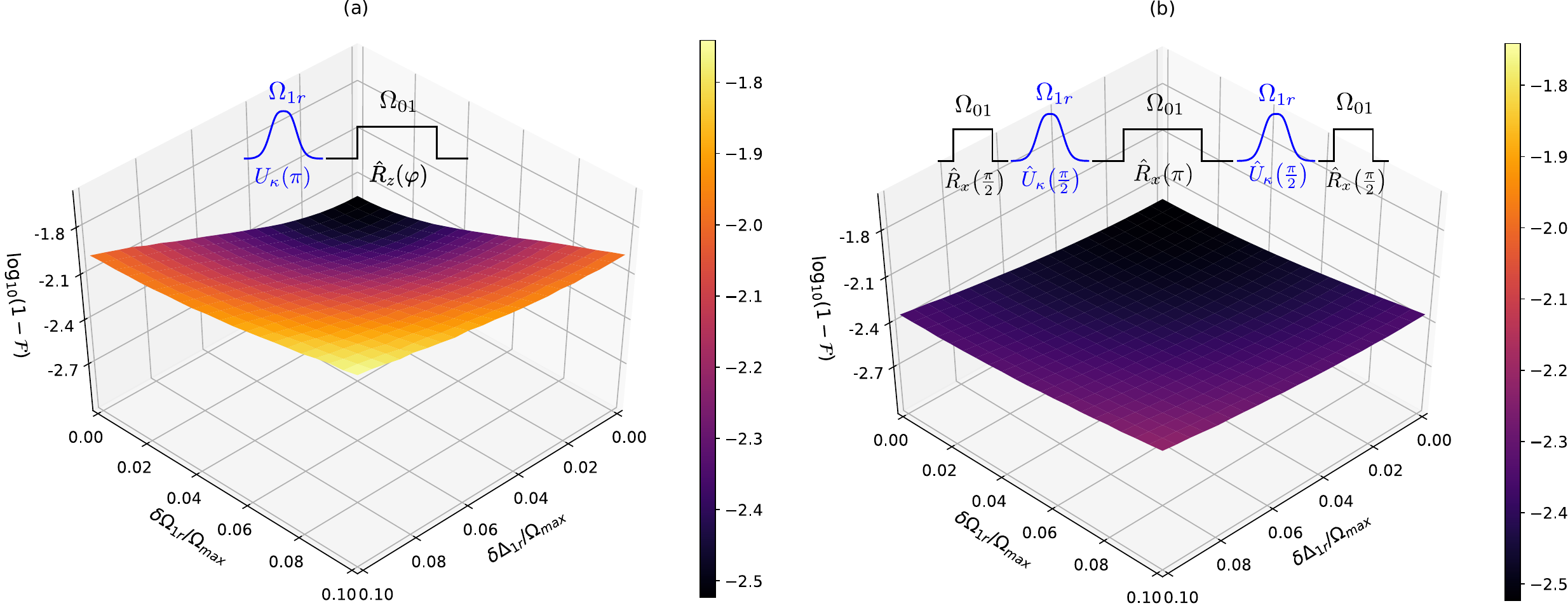}
     \caption{(a) Top: Implementing the CZ gate as proposed in  \cite{keating2015robust} using an adiabatic ramp, followed by removal of phases accumulated due to one-atom light shifts using a single qubit rotation $\hat{R}_z(\varphi)$, where $\varphi = \int \mathrm{d} t' E_{\text{LS}}^{(1)}(t')$.
    Bottom: Simulated infidelities of implementing the CZ gate with different levels of inhomogeneities in $\Delta_{1r}$ and $\Omega_{1r}$.
    (b) Top: Implementing the MS gate as done in  \cite{martin2018cphase} using two adiabatic ramps, with a spin echo in between.
    Bottom: Simulated infidelities of implementing the MS gate with different levels of inhomogeneities in $\Delta_{1r}$ and $\Omega_{1r}$.
    }
    \label{fig:Protocols}
\end{figure}

The standard protocol of Jaksch {\em et al.} \cite{jaksch2000fast} involves a pulse sequence on the control (c) and target (t) qubits, $\pi_{\mathrm{c}} - 2\pi_{\mathrm{t}} - \pi_{\mathrm{c}}$, ideally yielding a CZ gate. In the presence of thermal atomic velocity $v_{\mathrm{c}}$ for the control atom, the transformation on the logical states is
\begin{equation}
\begin{aligned}
  \ket{0, 0} \rightarrow & \ket{0, 0},
  \\
  \ket{0, 1} \rightarrow & -\ket{0, 1},
  \\
  \ket{1, 0} \rightarrow & -\mathrm{e}^{-\mathrm{i} k_{\mathrm{eff}} v_c\delta t}\ket{1, 0},
  \\
  \ket{1, 1} \rightarrow & -\mathrm{e}^{-\mathrm{i} k_{\mathrm{eff}} v_c \delta t}\ket{1, 1}
  .
\end{aligned}
\end{equation}
 Relative to the ideal CZ gate, there are additional phases due to the random Doppler shift acquired when the control atom stays in the Rydberg state for a time $\delta t$. For a thermal distribution of momenta, the random distribution of phases cannot be compensated, which causes gate errors \cite{wilk2010entanglement, keating2015robust, levine2018high, levine2019parallel, graham2019rydberg}. In contrast to the direct excitation to Rydberg states, for adiabatic Rydberg dressing, there are no random phases imparted to the qubits. Instead, the center-of-mass motion leads to a detuning error \cite{keating2015robust, mitra2020robust}, and the relative motion leads to coupling between bright and dark states \cite{keating2015robust, mitra2020robust}. However, while using an adiabatic ramp, this is suppressed due to the energy gap between the light-shifted bright state and the unshifted dark state. The residual off-resonance $\ket*{\tilde{B}} \leftrightarrow \ket{D}$ coupling leads to a small second-order perturbative shift on the dressed ground state \cite{keating2015robust}. Moreover, a nonuniform intensity in which atoms see different Rabi frequencies can introduce a coupling between the ground $\ket{1,1}$ and the dark state $\ket{D}$, which gives a small perturbative shift on the dressed ground state.

Finally, there is the effect of imperfect blockade. Whereas in the standard pulsed protocol, this can be a major source of error, gates based on adiabatic dressing are more resilient to this effect, and can successfully operate in the imperfect blockade as we will see in Sec.~\ref{sec:DressingEntanglingEnergy}. If we are close to the blockade radius, the gradient of dressed ground state energy as a function of separation between the atoms will be small, and there will be a force on the atoms due to the interactions. We return to this in Appendix~\ref{app:DressedAtomForces}. Of course, non-adiabatic effects such as resonant excitation to other doubly-excited Rydberg states can add additional errors, but these effects are not studied here. The width of the distribution of $\kappa$ due to a thermal Doppler width of the atomic momenta distribution can be estimated as
\begin{equation}
  \delta \kappa_{\text{th}}
  = k_{1r} \sqrt{\frac{k_{\mathrm{B}} T}{m}} 
  \sqrt{\left( \frac{\partial \kappa}{\partial \Delta_\alpha} \right)^2
    + \left( \frac{\partial \kappa}{\partial \Delta_\beta} \right)^2}
  ,
\end{equation}
where $\Delta_\alpha = \Delta_{1r} - k_{1r} p_{\alpha} / m$ and similarly for $\Delta_\beta$.

We model the experimental scenario by considering the detuning $\Delta_{1r}$ and Rabi frequency $\Omega_{1r}$ for each atom to be sampled from a normal distribution with mean equal to the fiducial value and standard deviation determined by the level of imperfections in the experiment. We simulate the implementation of the CZ gate using the protocol proposed earlier \cite{keating2015robust} and the implementation of the MS gate using two adiabatic ramps and a spin echo \cite{mitra2020robust, mitra2023neutral}, over a range of inhomogeneities $\delta \Delta_{1r}$ and $\delta \Omega_{1r}$. The gate fidelity including inhomogeneities, imperfect blockade, and Rydberg state decay (Rydberg lifetime $\tau_r \sim 10^3 (2\pi/\Omega_{\max})$) for $V = 10 \, \Omega_{\max}$, which is in the strong blockade regime, and the target gate is shown in Fig.~\ref{fig:Protocols} (a) for the CZ gate Fig.~\ref{fig:Protocols} (b) for the MS gate. As expected from Fig.~\ref{fig:UnitaryTransformations}, we see that implementing the MS gate using two adiabatic ramps and a spin echo is much more robust to inhomogeneities in $\Omega_{1r}$ and $\Delta_{1r}$ than implementing the CZ gate using an adiabatic ramp (Fig.~\ref{fig:Protocols}). For example, when we increase the level of imperfections from $0$ to about $10\%$ of the maximum Rabi frequency $\Omega_{\max}$ in the Rabi frequency and detuning, the MS gate fidelity falls from about $0.997$ to about $0.995$, while the CZ gate fidelity falls from about $0.997$ to about $0.986$. 

%
\section{Beyond the perfect blockade regime}
\label{sec:DressingEntanglingEnergy}

%
\begin{figure}
    \centering
    \includegraphics[width=0.48\textwidth]{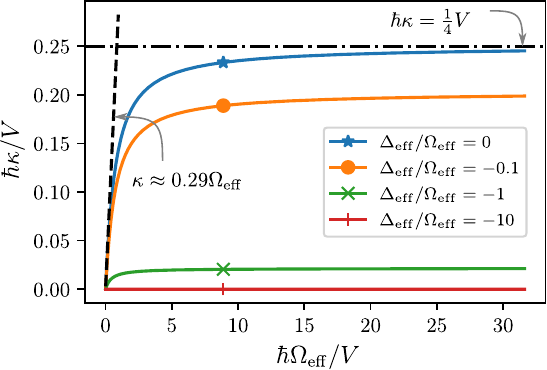}
    \caption{Entangling energy in units of the interaction energy as a function of the ground to Rydberg Rabi frequency in units of the interaction energy, for different detunings. For small detunings, in the strong blockade regime $\hbar\Omega_{\mathrm{eff}} \ll \vert V \vert$, the entangling energy scales linearly the Rabi frequency. and in the weak blockade regime $\hbar \Omega_{\mathrm{eff}} \gg \vert V \vert$, the entangling energy is independent of the Rabi frequency and scales linearly with the interaction energy. For large detunings, the entangling energy is negligible.}
    \label{fig:kappaOmegaVdd}
\end{figure}
\begin{figure}
    \centering
    \includegraphics[width=0.9\textwidth]{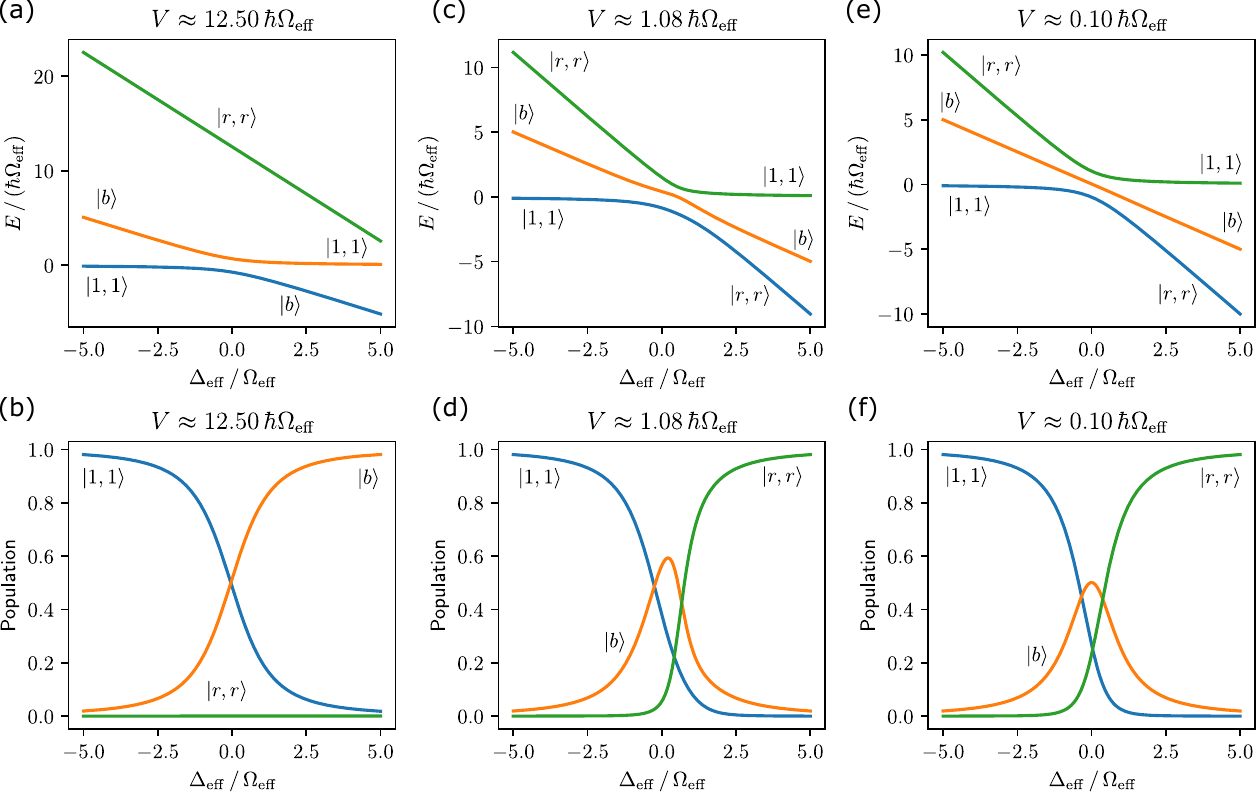}
    \caption{Dressed state energies and populations in the basis $\{\ket{1, 1}, \ket{b}, \ket{r, r}\}$ as a function of $\Delta_{\mathrm{eff}}/\Omega_{\mathrm{eff}}$ in different blockade regimes. (a, b) Strong blockade. (c, d) Intermediate blockade. (e, f) Weak blockade. (a, c, e) energy eigenvalues $V$, while (b, d, f) show populations of dressed states, when the initial detuning $\Delta_{\mathrm{eff}} < 0$.}
    \label{fig:BlockadeRegimesDressedStatePropertiess}
\end{figure}
\begin{figure}
    \centering
    \includegraphics[width=0.99\textwidth]{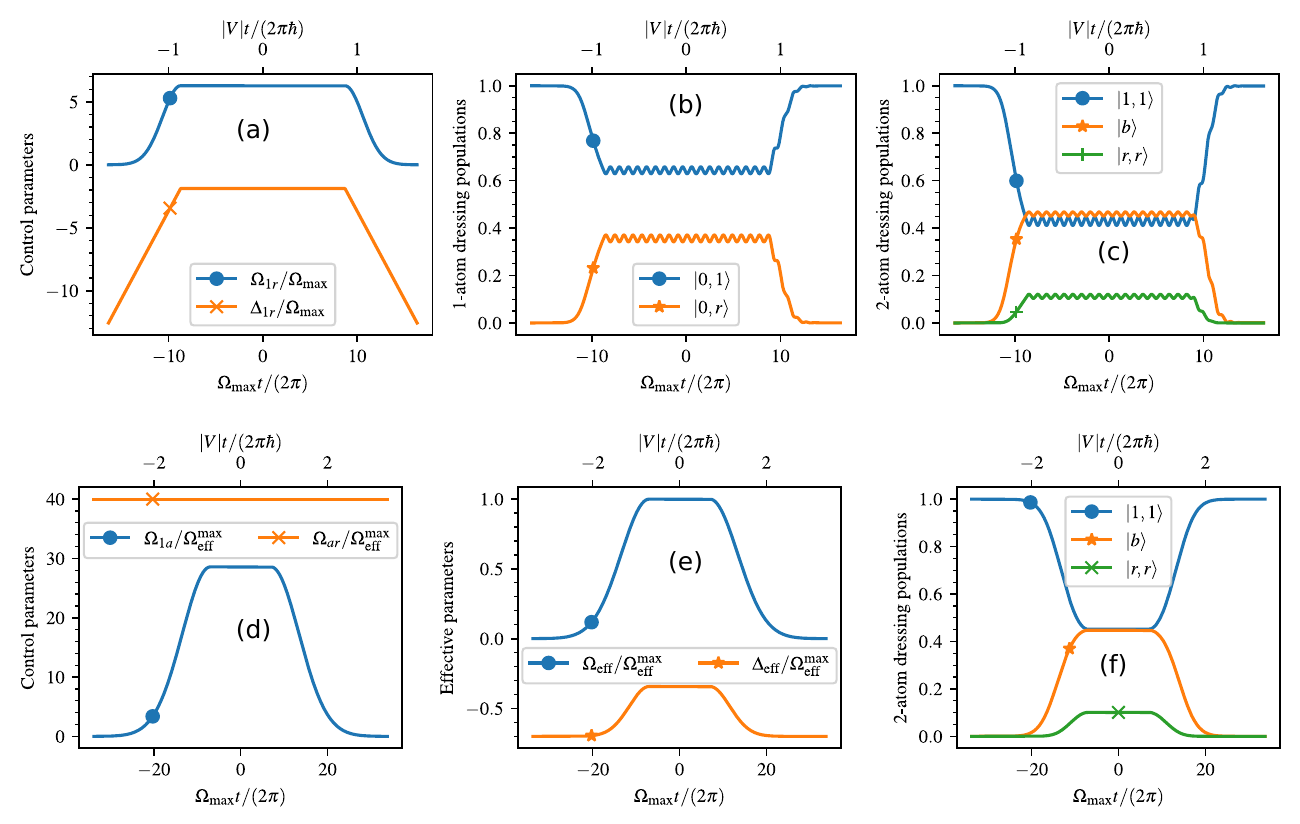}
    \caption{
    Adiabatic passages to implement $\hat{U}_{\kappa}(\varphi_1, \varphi_2)$ with $\varphi_2 = \pi/2$ [Eq.~\eqref{eq:UnitaryAdiabaticPassage}, Fig.~\ref{fig:SpinEchoAdiabaticPassage}] in the weak blockade regime ($\hbar \Omega_{\mathrm{eff}} = 0.1 \vert V \vert$). (a) One-photon adiabatic passage Gaussian sweep of Rabi frequency and linear sweep of detuning as in \cite{mitra2020robust, martin2021molmer}. (b) 1-atom populations during a one-photon adiabatic passage (c) 2-atom populations during a one-photon adiabatic passage. (d) Two-photon adiabatic passage using a Gaussian sweep of Rabi frequency $\Omega_{1a}$, with all other parameters fixed, which leads to an effective sweep of the two-photon Rabi frequency $\Omega_{\text{eff}}$ and two-photon detuning $\Delta_{\text{eff}}$ as shown in (e). (f) 2-atom populations during a two-photon adiabatic passage.
    The bottom axes show time measured in units of $2\pi/\Omega_{\max}$, and the top axes show time measured in units of $Vt/(2 \pi \hbar)$. In the weak blockade, as expected, $Vt/\hbar \ll \Omega_{\max}t$}

    \label{fig:WeakBlockadeAdiabaticPassage}
\end{figure}
\begin{figure}
    \centering
    \includegraphics[width=0.48\textwidth]{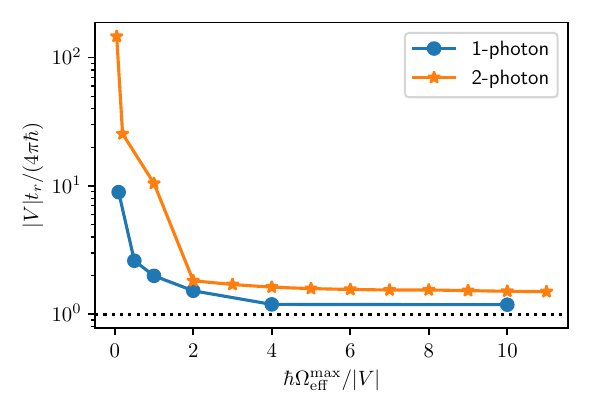}
    \caption{Time integrated Rydberg population $t_r$ as a function of the ratio $\hbar\Omega_{\mathrm{eff}}^{\max} / |V|$ for the one- and two- photon ramps. In both cases, the integrated Rydberg population becomes lower as we increase $\hbar \Omega_{\mathrm{eff}} / |V|$ for the two-photon adiabatic passage as in Eq.~\eqref{eq:GaussianOmega_a} and one-photon adiabatic passage as in Eq.~\eqref{eq:GaussianOmega_1rLinearDelta_1r}
    }
    \label{fig:IntegratedRydbergPopulation}
\end{figure}

In the previous section, we studied entangling gates in the case of a perfect Rydberg blockade, but this is not intrinsic to the adiabatic dressing protocol. Relaxing this assumption and studying protocols in the weak blockade regime is important to address the fundamental limits of Rydberg-atom quantum information processing, potentially improve the fidelity of our gates, and allow us to operate in new regimes. We note that in practice, quantum fluctuations in the atoms’ motional states always affect the fidelity of the implemented gate. How this uncertainty causes gate infidelity depends on the particular protocol. If atoms are released from a trap and are in free fall during the gate (as is commonly the case), the uncertainty in momentum can lead to Doppler shifts and the uncertainty in position can lead to fluctuations in the atom-atom coupling strength \cite{keating2015robust, mitra2020robust, robicheaux2021photon}. In principle, atoms can be cooled very close to the motional ground state of a sufficiently deep trap (a nearly pure state), and for some atomic species and specially chosen transitions, the gate can be done with the trap on \cite{madjarov2019strontium, madjarov2020high, wilson2022trapping}. If the motional state of the atom is not entangled with the internal state, there will be no error arising due to the position and momentum uncertainties. Loss of gate fidelity due to atomic motion, arising from uncertainties in the positions and momenta of the atoms have been considered in Refs.~\cite{keating2015robust, graham2019rydberg, mitra2020robust, robicheaux2021photon}.

 Besides limitations due to uncertainties in atomic motional states, no matter how cool the atoms are or how well we can remove these effects by special protocols, implementation of an entangling gate using Rydberg-meditated interactions is fundamentally limited by two energy-time scales -- the Rydberg state lifetime $\tau_r$ and the magnitude of the interatomic interaction energy $|V|$  \cite{saffman2010quantum, saffman2016quantum}. Wesenberg \textit{et al.} showed that the minimum time that the atoms need to spend in a Rydberg state to achieve a maximally entangling gate scales as $t_{r} \sim \hbar/|V| $ \cite{wesenberg2007scalable}. The standard protocols which employ a strong Rydberg blockade \cite{jaksch2000fast, levine2019parallel} cannot achieve this bound because the speed of the gates is set by $\Omega_{\rm{eff}}$, and since they require $\hbar \Omega_{\rm{eff}}\ll \vert V \vert$, we cannot make use of the full scale of the interaction energy \cite{saffman2010quantum}. Jo \textit{et. al.} implemented Rydberg-mediated entanglement outside the strong blockade regime using finely tuned two-atom Rabi oscillations \cite{jo2020rydberg}.

The minimum time scale for $t_r$ can be understood in a simple protocol using the limiting case of very large Rabi frequency $\hbar \Omega_{\mathrm{eff}}^{\max} / |V| \to \infty$. An entangling gate can be achieved using a collective $\pi$-pulse from $\ket{1}$ to $\ket{r}$ on both atoms, followed by an interaction for a time $|V|t_r/\hbar = \pi$ and a $\pi$-pulse from $\ket{r}$ to $\ket{1}$. In the limit of infinitesimally short $\pi$-pulses, the time spent in Rydberg states, or time-integrated Rydberg population, is $\pi\hbar/|V|$. All of the time spent in the Rydberg states is in the doubly excited Rydberg state $\ket{r,r}$, giving us the bound
\begin{equation}
    t_r \leq \frac{\pi\hbar}{|V|}
    \label{eq:TimeIntegratedRydbergPopulationBound}
    .
\end{equation}

While this simple protocol helps us understand the time scales, it is generally not practical for implementation. For small interatomic separations, the two-atom spectrum becomes a complex tangle of ``Rydberg spaghetti" \cite{keating2013adiabatic, jau2016entangling}. To achieve the fastest gates in this strongly interacting case, it is thus useful to avoid double Rydberg population which can lead to unexpected inelastic processes. In addition, the complex potential landscape at such small interatomic separations can lead to high sensitivity to atomic motion. In this section, we show that using adiabatic Rydberg dressing, we can get close to the minimum time scale $t_r$, while working in the weak blockade regime, $\hbar \Omega_{\mathrm{eff}}^{\max} \gg |V|$, without significant double Rydberg population. Moreover, for large interatomic separations protocols requiring a strong blockade would lead to exceedingly slow gates. The adiabatic dressing protocol considered here can achieve reasonably fast gates with high fidelity even for atoms separated beyond blockade radius.

To understand the different regimes of operation, we estimate how the interatomic interaction energy $V$ limits the entangling energy $\hbar \kappa$ in the strong blockade and weak blockade regimes. For simplicity, we consider the case in which the atoms see the same Rabi frequency, given in Eq.~\eqref{eq:TwoAtomRydbergHamiltonian}. It is useful to consider a pseudo-spin with $\ket{\uparrow_z} \equiv \ket{r}$ and $\ket{\downarrow_z} \equiv \ket{1}$. In this pseudo-spin picture, the two-atom Hamiltonian can be written as a sum of two terms
\begin{equation}
\begin{aligned}
    \hat{H}_{\text{int}}
    & = V \ketbra{r,r}{r,r} \equiv \frac{V}{2} \left(\hat{S}_z^2 + \hat{S}_z\right),
    \\
    \hat{H}_{\text{drive}}
    & \equiv -\hbar\Delta_{\mathrm{eff}} \, \ident - \hbar\Delta_{\mathrm{eff}} \, \hat{S}_z + \hbar\Omega_{\mathrm{eff}} \, \hat{S}_x
    \\ &
    \equiv -\hbar\Delta_{\mathrm{eff}} \, \ident + \hbar\sqrt{\Delta_{\mathrm{eff}}^2 + \Omega_{\mathrm{eff}}^2} \, \hat{S}_\theta,
\end{aligned}
\end{equation}
where $\hat{S}_{\mu}$ is the $\mu$-component of collective angular momentum operator $S_\mu = \ident \otimes \hat{\sigma}_\mu/2 + \hat{\sigma}_\mu/2 \otimes \ident$, $\hat{S}_\theta = \cos\theta \hat{S}_z + \sin\theta \hat{S}_x$ with $\tan \theta = \Omega_{\mathrm{eff}}/(-\Delta_{\mathrm{eff}})$. The collective symmetric spin-1 eigenstates of $S_z$ are the triplet of the pseudospins 
\begin{equation}
\begin{aligned}
    \ket{S=1, M_z=-1} 
    & \equiv \ket{1,1}
    , \\ 
    \ket{S=1, M_z=0} 
    & \equiv \frac{1}{\sqrt{2}} (\ket{1,r}+\ket{r,1}) 
    \equiv \ket{b}
    , \\ 
    \ket{S=1, M_z=+1} 
    & \equiv \ket{r,r}.
\end{aligned}
\end{equation}
The eigenvalues and eigenvectors of the driving Hamiltonian and the interaction Hamiltonian are in \cref{tab:EigenAtomLight} and \cref{tab:EigenAtomAtom} respectively.
\begin{table}[]
    \centering
    \begin{tabular}{ |c|c| }
 \hline
 $\text{Energy Eigenvalue}$ &$\text{Eigenvectors}$  \\
\hline
$-\hbar \Delta_{\mathrm{eff}} +\hbar \sqrt{\Omega_{\mathrm{eff}}^2 + \Delta_{\mathrm{eff}}^2}$ &
$\ket{\uparrow_\theta} \otimes \ket{\uparrow_\theta}$ \\
 \hline
 $-\hbar \Delta_{\mathrm{eff}} -\hbar \sqrt{\Omega_{\mathrm{eff}}^2 + \Delta_{\mathrm{eff}}^2}$ &
$\ket{\downarrow_\theta} \otimes \ket{\downarrow_\theta}$ \\
\hline
  $-\hbar \Delta_{\mathrm{eff}}$ &
  $\left(\ket{\uparrow_\theta} \otimes \ket{\downarrow_\theta} + \ket{\downarrow_\theta} \otimes \ket{\uparrow_\theta}\right)/\sqrt{2}$\\
\hline
\end{tabular}
\caption{Eigenvalues and eigenvectors of the atom-light Hamiltonian, $\hat{H}_{\mathrm{drive}}$. Here $\ket{\uparrow_\theta} \equiv \cos\left(\theta/2\right) \ket{\uparrow_z} + \sin\left(\theta/2\right) \ket{\downarrow_z}$, $\ket{\downarrow_\theta} \equiv \cos\left(\theta/2\right) \ket{\downarrow_z} - \sin\left(\theta/2\right) \ket{\uparrow_z}$ and $\tan\theta = \Omega_{\mathrm{eff}} / (-\Delta_{\mathrm{eff}})$. The first two rows represent the upper and lower branches of the single-atom dressed states.}
\label{tab:EigenAtomLight}
\end{table}
\begin{table}[]
    \centering
    \begin{tabular}{ |c|c|c| }
 \hline
 $\text{Energy Eigenvalue}$ &$\text{Eigenvectors}$  \\
\hline
 $V$  & $\ket{r,r}$ \\
\hline
     $0$  &
     $\ket{b},\;\ket{1,1}$ \\
\hline
\end{tabular}
\caption{Eigenvalues and eigenvectors of the atom-atom interaction Hamiltonian, $\hat{H}_{\mathrm{int}}$, in the symmetric subspace, spanned by $|1,1\rangle, |b\rangle, |r,r\rangle$.}
\label{tab:EigenAtomAtom}
\end{table}

First we consider the well-known strong blockade regime with $\vert V \vert \gg \hbar \Omega_{\mathrm{eff}}$, where the interaction term is the dominant Hamiltonian and the driving term is the perturbation. The zeroth-order eigenvectors are the states $\ket{S=1, M_z}$. The leading order correction is calculated using degenerate perturbation theory in the zero eigenvalue subspace spanned by $\ket{S=1, M_z=-1} \equiv \ket{1,1}$ and $\ket{S=1, M_z=0} \equiv \ket{b}$. Using $\mathscr{P}_{S, M_z}$ to denote the projector on the subspace of $S, M_z$,
\begin{equation}
\begin{aligned}
    \left(\mathscr{P}_{S=1, M_z=-1} + \mathscr{P}_{S=1, M_z=0}\right)
    \, & \hat{S}_{\theta} \,
    \left(\mathscr{P}_{S=1, M_z=-1} + \mathscr{P}_{S=1, M_z=0}\right)
    \\ &
    = - \cos(\theta) \ketbra{S=1, M_z=-1}{S=1, M_z=-1}
    \\ &
    + \frac{\sin(\theta)}{\sqrt{2}} \left(
    \ketbra{S=1, M_z=-1}{S=1, M_z=0}\right)
    \\ &
    + \frac{\sin(\theta)}{\sqrt{2}} \left(
    \ketbra{S=1, M_z=0}{S=1, M_z=-1}\right).
\end{aligned}
\end{equation}
The perturbative corrections to energy eigenvalues are the two-atom light shift experienced by the atoms together, in the presence of $V$.
The leading correction to the energy of the logical state $\ket{1, 1} \equiv \ket{S=1, M_z=-1}$ in perturbation theory, is the two-atom light shift under perfect blockade,
\begin{equation}
     E_{\text{LS}}^{(2)} = -\frac{\hbar \Delta_{\mathrm{eff}}}{2} \pm \frac{\hbar}{2} \sqrt{2\Omega_{\mathrm{eff}}^2 + \Delta_{\mathrm{eff}}^2}.
\end{equation}
Subtracting out the energy shifts in eigenstates of each atom to obtain the entangling energy $\kappa$ using Eq.~\eqref{eq:EntanglingEnergy},
\begin{equation}
\begin{aligned}
    &
    \lim_{\hbar\Omega_{\mathrm{eff}} / \vert V \vert \to 0}
    \kappa
    \\ &
    = - \frac{\Delta_{\mathrm{eff}}}{2}
    \pm
    \frac{1}{2}\left(\sqrt{\Delta_{\mathrm{eff}}^2 + 2\Omega_{\mathrm{eff}}^2}
    - 2 \sqrt{\Delta_{\mathrm{eff}}^2 + \Omega_{\mathrm{eff}}^2}\right).
\end{aligned}
\end{equation}
Note that here by design, $\hbar \vert \kappa \vert \ll \vert V \vert$, since we assumed $\hbar\Omega_{\mathrm{eff}} \ll \vert V \vert$. The maximum useful $\kappa$ scales with the Rabi frequency $\Omega_{\mathrm{eff}}$. Under a perfect Rydberg blockade regime $\vert V \vert \gg \hbar \Omega_{\mathrm{eff}}$, the state $\ket{r,r}$ is not populated. Thus, there is an adiabatic passage from the $\ket{1, 1}$ to $\ket{b}$ and back as shown in Fig.~\ref{fig:RydbergDressingEnergyLevels}(c).

\begin{table}[]
    \centering
    \begin{tabular}{ |c|c| }
 \hline
 $\text{Energy Eigenvalue}$ & $\text{Eigenvectors}$  \\
 \hline
     $-\frac{1}{2} \cos\theta + \frac{1}{2} \sqrt{\cos^2\theta + 2\sin^2\theta}$  &
     $\cos\frac{\Theta}{2} \ket{b} + \sin\frac{\Theta}{2} \ket{1,1}$
\\
\hline $-\frac{1}{2} \cos\theta - \frac{1}{2} \sqrt{\cos^2\theta + 2\sin^2\theta}$  &
     $\cos\frac{\Theta}{2} \ket{1,1} - \sin\frac{\Theta}{2} \ket{b}$
      \\
\hline
\end{tabular}
    \caption{Eigenvalues and eigenvectors of $\hat{S}_{\theta}$ in the zero-eigenvalue subspace of $\hat{H}_{\mathrm{int}}$. Here, $\tan \Theta = \sqrt{2}\Omega_\mathrm{eff} / (-\Delta_{\mathrm{eff}})$. The upper and lower rows represented the upper and lower branches of the two-atom dressed states in the perfect blockade regime, shown in Fig.~\ref{fig:RydbergDressingEnergyLevels}.}
\label{tab:EigenSthetaInProjAtomAtom}
\end{table}

Next, we consider the weak blockade regime where $\vert V \vert \ll \hbar \Omega_{\mathrm{eff}}$. In this case, the laser driving term is the dominant Hamiltonian and the interaction term is a perturbation. The eigenstates of the driving Hamiltonian are the one-atom dressed states, which are rotated spin-triplet states $\ket{S=1, M_\theta}$ given in \cref{tab:EigenAtomLight}. The energy eigenvalues correspond to the one-atom light shift. The entangling energy $\hbar\kappa$ can be estimated as the correction to the dress-ground state 
\begin{equation}
    \ket*{\widetilde{1,1}} 
    \equiv \ket{S=1, M_\theta=-1} 
    \equiv 
    \left( \cos\frac{\theta}{2} \ket{1} + \sin\frac{\theta}{2}\ket{r} \right)^{\otimes 2}.
\end{equation}
The unperturbed energies of the dominant Hamiltonian include the single-atom light shifts. Therefore the leading order correction to the non-interacting energy is the asymptotic value of $\hbar \kappa$,
\begin{equation}
\begin{aligned}
    &
    \lim_{ \vert V \vert / \hbar\Omega_{\mathrm{eff}} \to 0}
	\hbar \kappa
	= \left(\frac{1 \pm \cos \theta}{2}\right)^2 V,
	\label{eq:kappaAsymptoteWeakBlockade}
\end{aligned}
\end{equation}
where $\pm$ refers to the relative sign of the initial detuning and the detuning at peak dressing during an adiabatic passage, and the corresponding (unnormalized) dressed state, in leading-order perturbation theory, is
\begin{equation}
\begin{aligned}
    \ket*{\widetilde{1, 1}}
    &
    \equiv
    \left(\cos\frac{\theta}{2}\ket{1} +
    \sin\frac{\theta}{2}\ket{r}\right)^{\otimes 2}
    \\ &
    \pm
    \cos^2\left(\frac{\theta}{2}\right)\;
    \frac{V}{2\hbar \sqrt{\Omega_{\mathrm{eff}}^2 + \Delta_{\mathrm{eff}}^2}}
    \ket{r, r},
\end{aligned}
\end{equation}
now including the doubly excited Rydberg state.

We calculate the entangling energy $\hbar\kappa$ numerically beyond the strong and weak blockade regimes for different detunings as shown in Fig.~\ref{fig:kappaOmegaVdd}. We focus on entangling protocols that limit the population in the doubly-excited Rydberg state, $\ket{r, r}$, to avoid potentially deleterious decay and inelastic processes. To ensure this, we consider adiabatic ramps that are far from the anti-blockade condition, $V = 2\hbar\Delta_{\mathrm{eff}}$. In practice, this is done in the weak blockade case with a detuning at peak dressing (minimum $|\Delta_{\mathrm{eff}}|$) satisfying $\hbar |\Delta_{\mathrm{eff}}| \ll |V|$. As predicted from perturbation theory, we see that entangling energy scales with the Rabi frequency in the strong blockade regime and reaches $V/4$ at resonance, in the weak blockade regime.

Theoretically, all of the interaction energy $V$ is available as the Rydberg dressing entangling energy $\hbar \kappa$. However, this occurs when $\theta \in \left\{0, \pi\right\}$ or $\vert \Delta_{\mathrm{eff}} \vert / \Omega_{\mathrm{eff}} \to \infty$ when the dressed state is simply the bare atomic state $\ket{r, r}$. An adiabatic passage that starts far from ground-Rydberg resonance, goes close to resonance, and returns to far off-resonance, is most effective at limiting double Rydberg excitation \cite{mitra2020robust, mitra2023neutral}. In this weak blockade case, the adiabatic passage stays far from the anti-blockade condition, leading to a dressed state $\ket*{\widetilde{1, 1}}$ that is primarily an admixture of $\ket{1, 1}$ and the bright state $\ket{b}$, with a small $\ket{r, r}$ component.

In Fig.~\ref{fig:BlockadeRegimesDressedStatePropertiess} we consider examples of strong ($V\gg \hbar\Omega_{\mathrm{eff}}$), intermediate ($V\sim \hbar\Omega_{\mathrm{eff}}$), and weak ($V \ll \hbar\Omega_{\mathrm{eff}}$), showing the dressing energies and the populations of bare states $\ket{1, 1}, \ket{b}, \ket{r, r}$ in the dressed state $\ket*{\widetilde{1, 1}}$. Given the energy gaps, we see that the adiabatic dressing protocol allows for a gate as fast as a time scale of $\sim 2\pi\hbar/|V|$, and importantly, by sweeping the detuning close to resonance, while avoiding the anti-blockade condition, there is negligible excitation of the doubly excited Rydberg state $\ket{r, r}$. For example, we study $|V| = 0.1\hbar\Omega_{\mathrm{eff}}$ for both the one and two-photon excitation; the ramps are shown in Fig.~\ref{fig:WeakBlockadeAdiabaticPassage} using the same parameterization used for the strong blockade case, Eq.~\eqref{eq:GaussianOmega_1rLinearDelta_1r}, and Eq.~\eqref{eq:GaussianOmega_a}. Despite the weak blockade, we see that the time-integrated population in the state $\ket{r,r}$ is bounded, which overcomes one of the significant hurdles in going beyond perfect blockade.

Let us return to the question of the maximum possible achievable entangling gate fidelity, assuming the loss of fidelity due to uncertainties in atomic motion is negligible. When considering adiabatic Rydberg dressing, the entanglement is generated in the form of the dynamical phases from the entangling energy, $\varphi_2 \int \mathrm{d} t' \kappa(t')$  \cite{keating2015robust, mitra2020robust}. Fundamentally, the time spent in the Rydberg state is bounded by an energy scale proportional to the entangling energy $\hbar \kappa$. Using adiabatic Rydberg dressing in the strong-blockade regime leads to $t_{r}$ that scales inversely with the Rabi frequency as $\kappa \sim \Omega_{\max}$, and therefore is far from the minimum, $t_r \sim 2\pi/\Omega_{\max} \gg \pi \hbar/\vert V \vert$. In Fig.~\ref{fig:IntegratedRydbergPopulation}, we plot the time-integrated Rydberg population as a function of the ratio of Rabi frequency to the interatomic Rydberg interaction energy $\hbar \Omega_{\mathrm{eff}}^{\mathrm{max}}/ \vert V \vert$ for both the one-photon case using Eq.~\eqref{eq:GaussianOmega_1rLinearDelta_1r} and the two-photon ramps as given in Eq.~\eqref{eq:GaussianOmega_a}. The analysis indicates that the time-integrated population required to create the perfect entangler, while avoiding the anti-blockade condition, decreases as we increase the Rabi frequency $\hbar\Omega_{\max}$, compared to the interaction energy $|V|$ and it eventually saturates to slightly above $4\pi \hbar/|V|$.

This result is consistent with the bound found Eq.~\eqref{eq:TimeIntegratedRydbergPopulationBound}. Since the value of $\hbar |\kappa|$ reaches $|V|/4$, near resonance in the weak blockade regime [Eq.~\eqref{eq:kappaAsymptoteWeakBlockade}], the theoretically achievable maximum fidelity while limiting double Rydberg excitation is
\begin{equation}
     \mathcal{F} < 1 - \frac{4\pi\hbar}{\vert V \vert \tau_r},
\end{equation}
where $\tau_r$ is the Rydberg state lifetime. For contemporary experiments, with $\vert V \vert /(2\pi\hbar) = 40\;\mathrm{MHz}$ and $\tau_r = 150\;\mathrm{\mu s}$, the theoretical minimum infidelity is about $10^{-3}$. With cryogenically enhanced Rydberg lifetimes, around $\tau_r = 1 \; \mathrm{ms}$ and stronger interactions, $\vert V \vert/(2\pi\hbar) = 1$ GHz, the theoretical minimum infidelity would be $10^{-5}$. In practice achieving this would require working in the weak blockade regime, with large laser power such that $\hbar\Omega_{\text{eff}} \gg \vert V \vert$.

The ability to design gates with adiabatic dressing beyond the perfect blockade regime also loosens other constraints and potential sources of error. Maintaining atoms beyond the blockade radius reduces the requirement for transporting atoms, which leads to motional heating. Our results show that even for moderate Rydberg-Rydberg interaction, with $|V|/(2\pi\hbar)$ of a few $\mathrm{MHz}$, one can achieve fast gates with gates times of the order of a few $\mathrm{\mu s}$. Moreover, at moderate separations the shifted doubly excited states are well resolved and well defined, reducing spurious resonances. A potential downside to operation in this regime is the sensitivity of the entangling energy to atom separation and also the resulting forces on the atoms. We address this in Appendix~\ref{app:DressedAtomForces}.

%
\section{Conclusion}
\label{sec:RydbergDressingConclusion}
%
In this chapter, we explored how adiabatic Rydberg dressing provides a robust method for harnessing the interaction between Rydberg-excited atoms to generate entanglement between qubits encoded in atomic clock states. We introduced the neutral atom Mølmer-Sørensen gate, with interleaving of adiabatic Rydberg dressing and undressing with a spin echo on the qubit transition, which is robust to errors arising from experimental imperfections \cite{mitra2020robust}. During the adiabatic passages, entanglement is generated by the modification of ground state light shift introduced by the interaction energy between Rydberg atoms. We showed the implementation of these adiabatic Rydberg dressing passes via both one-photon and two-photon ground-to-Rydberg transitions. While for one-photon transitions, both laser frequency and amplitude need to be modulated, for two-photon transitions, adiabatic Rydberg dressing passages can be achieved by modulating only one laser amplitude as a function of time, with all laser frequencies fixed, allowing an easier experimental implementation and alleviating the need for a high power ultraviolet laser. This gate has been implemented in a first demonstration in Cesium \cite{martin2021molmer} and with high fidelity in Strontium \cite{schine2022long} atoms.

We also studied the fundamental limits of implementing an entangling gate using adiabatic Rydberg dressing of ground states, set by the finite Rydberg lifetime and the entangling energy obtained in the dressed states. In the well-known strong blockade regime, the entangling energy scale is limited by the ground-Rydberg Rabi frequency, that is, laser power, and in the weak blockade regime, the entangling energy is limited by the interaction energy between the atoms. Moreover, we showed proof-of-principle feasibility of rapid adiabatic passages without significant double-Rydberg population in strong, intermediate, and weak blockade regimes, thereby loosening the requirements of atoms being within a blockade radius for implementing entangling gates in a few $\mathrm{\mu s}$. A more precise model of the entangling energy using atomic species and Rydberg state-specific treatment, for example as in Ref.~\cite{de2018analysis}, can be used to design adiabatic passages for specific experiments.

In conclusion, adiabatic Rydberg dressing is a promising approach to implementing two-qubit entangling gates for neutral atoms. It can be implemented in several atomic species with one- or two-photon ground-to-Rydberg transitions and can be designed beyond the strong blockade regime to yield fast high-fidelity gates.

\chapter{Background: Ising models, quench dynamics and eigenstate thermalization}
\label{chap:SpinModels}

\section{Introduction}
A key question at the heart of many-body physics is how do large (of the order of $10^{24}$) degrees of freedom obeying the laws of quantum mechanics give rise to the macroscopic properties of matter \cite{akulin2014dynamics, cirac2021matrix, girvin2019modern, samajdar2022topological}. Initial attempts to answer this question have been based on models involving independent microscopic degrees of freedom, for example, Sommerfeld's model of metals using independent electrons \cite{sommerfeld1928elektronentheorie, sommerfeld1931statistical}. These primarily non-interacting or weakly interacting theories have been successful in explaining some properties of matter including some aspects of electrical and thermal conductivity in materials like metals, insulators, semiconductors, and superconductors. Nevertheless, many aspects of quantum matter cannot be explained through these theories and require a description using strongly interacting degrees of freedom. Thus, theories of quantum many-body systems have required descriptions using many-body wavefunctions with multi-partite entanglement, which is fundamentally different from the product wavefunctions that suffice in non-interacting and weakly interacting theories. Indeed, the structure of multi-partite entanglement, from short-range to long-range has added to the richness of the phenomenology of these strong-correlated systems.

Descriptions of macroscopic properties of matter are typically done with coarse-grained macro properties, that result from the collective behavior of several degrees of freedom, in the presence or absence of interactions \cite{kardar2007statisticalparticles, kardar2007statisticalfields, huang2008statistical, akulin2014dynamics, pathria2016statistical, girvin2019modern}. A quintessential domain of interest is to characterize phases of matter, from a description of its macroscopic contents. The Landau theory of phase transitions is based on the notion of symmetries of a system and whether states of the system respect the symmetry \cite{landau1937theory}. For concreteness, let us consider a quantum system with Hamiltonian $H$, which is invariant under the action of a symmetry group $\mathbb{G}$. Let $g \in \mathbb{G}$ be an element of a group. By the action of $g$ on the Hamiltonian, we mean the transformation of the Hamiltonian by a unitary representation of $g$, $U(g)$, as 
\begin{equation}
    H \to U^\dagger(g) H U(g)
    \label{eq:GroupActionHamiltonian}
    .
\end{equation}
The action of the group corresponds to transformations of the form Eq.~\eqref{eq:GroupActionHamiltonian} for every element $g \in \mathbb{G}$. Invariance of the Hamiltonian under a symmetry group $\mathbb{G}$ implies that the action of every element $g$ of the group $\mathbb{G}$ leaves the Hamiltonian invariant, $U^\dagger(g) H U(g) = H$ for all $g \in \mathbb{G}$. When the system is in a state that is not invariant under the action of at least one element $g' \in \mathbb{G}$, the system is in a symmetry-broken phase.

To distinguish between the symmetry-broken and symmetry-unbroken phases it is useful to consider an observable, $O$, that transforms nontrivially under $g'$, that is 
\begin{equation}
    U^\dagger(g') O U(g') \neq O
    \label{eq:GroupActionOrderParam}
    .
\end{equation}
This observable has a nonzero value in the symmetry-broken phase and is zero in the symmetry-unbroken phase and corresponds to an \textit{order parameter}. More precisely the order parameter is associated with its expectation value. For local order parameters, the symmetry-broken phase is characterized by long-range correlations of the order parameter.

In this chapter, we discuss a phase transition occurring in an interacting spin-1/2 model, the transverse field Ising model. This model will be used to study the role of many-body entanglement in quench dynamics probing phase transitions and critical phenomena at the phase transition. In the next two chapters, we will quantitatively study some critical properties near this phase transition. We introduce the model and discuss how its phases can be assessed through local order parameters. Next, we discuss the eigenstate thermalization hypothesis which forms a link between the equilibrium properties of a model and quantities observed during quenched dynamics of the model. Finally, we discuss concepts of microstates which play a role in studies of phenomena involving coarse-grained macro properties.

\section{Transverse field Ising model}
\label{sec:TFIMIntro}

A quintessential model of multiple interacting spin-1/2 degrees of freedom is the transverse field Ising model, which is sometimes called the quantum Ising model (TFIM), as it is considered a quantum version of the classical Ising model \cite{sachdev2011quantum}. This model serves as a ``fruit fly'' for many-body physics \cite{samajdar2022topological}. We consider a system of spin-1/2 degrees of freedom with a local Hilbert space spanned by the states $| \uparrow_z \rangle$ and $| \downarrow_z \rangle$, which are the $+1$ and $-1$ eigenvalue eigenvectors of the local $\sigma^z$ respectively. An Ising model has spin interacting with pairs of $\sigma^z$'s acting on spins. The TFIM has a transverse operator $\sigma^x$, due to a transverse magnetic field. This introduces a noncommuting term in the Hamiltonian, which reads
\begin{equation}
    H_{\mathrm{TFIM}}
    = - J_0 \sum_{l_1, l_2} \kappa_{l_1, l_2} \sigma^{z}_{l_1} \sigma^{z}_{l_2}
    - B \sum_{l} \sigma^{x}_{l}
    ,
    \label{eq:TFIMGeneric}
\end{equation}
where $B$ is the strength of the transverse magnetic field, $J_0$ characterizes the strength of the pairwise interactions between spins, and $\kappa_{l_1,l_2}$ is a dimensionless function of the pair of the spins, usually dependent only on the distance between spins $l_1$ and $l_2$, and normalized such that $\sum_{l_1,l_2} \kappa_{l_1,l_2}$ is $\mathcal{O}(n)$, where $n$ is the number of spins\footnote{From here onwards, we set $\hbar \equiv 1$.}.

For all values of the magnetic field energy $B$, interaction energy $J_0$ and interaction graph $\kappa_{l_1,l_2}$, this Hamiltonian is invariant under a symmetry transformation, $\Pi$, which flips the $z$-component of each spin: $\sigma^z_l \leftrightarrow - \sigma^z_l$, and does not affect the $x$-component of each spin $\sigma^x_l \leftrightarrow \sigma^x_l$. This symmetry operation can be interpreted as one where each spin is rotated by $\pi$ about the $x$-axis. Chosen with a global phase which makes $\Pi$ both unitary and Hermitian, $\Pi$ reads
\begin{equation}
  \Pi
  = \mathrm{i}^n 
  \bigotimes_{l=1}^{n} 
  \exp\left( - \mathrm{i} \pi \frac{\sigma^x_l}{2} \right)
  = \bigotimes_{l=1}^{n}\sigma^x_l
  ,
  \label{eq:RotateXPi}
\end{equation}
which is its own inverse, that is $\Pi^{-1} = \Pi^\dagger = \Pi$. Thus, the symmetry corresponds to a representation of the group $\mathbb{Z}_2$ with elements $\left\{ \mathds{1}, \Pi \right\}$, where $\mathds{1}$ is the identity operator acting on all spins and $\Pi^2 = \mathds{1}$.

Traditionally, studies in many-body physics are related to the equilibrium properties of a model. We begin with a discussion of the equilibrium properties of this model, specifically the ground state, which is also interpreted as the zero-temperature thermal state. It is useful to consider the two limiting cases, that of $B/J_0 \to 0$ and $J_0/B \to 0$.

First let us consider the limit $J_0/B \to 0$, where we can ignore the terms with $\sigma^z_{l_1} \sigma^z_{l_2}$ in Eq.~\eqref{eq:TFIMGeneric} and consider only the terms containing $\sigma^x_l$. The ground state corresponds to all spins being in the $+1$ eigenstate of $\sigma^x$, that is point along the $x$ axis, $| \uparrow_x \rangle^{\otimes n}$ as
\begin{equation}
    \left| \emptyset \left(\frac{J_0}{B} = 0\right) \right\rangle
    = | \uparrow_x \rangle^{\otimes n}
    ,
\end{equation}
where we use $| \emptyset \rangle$ to denote the ground state. The ground state in this regime is invariant under both elements $\mathds{1}$ and $\Pi$ of the group $\mathbb{Z}_2$. Moreover, in this regime, excitations above the ground state can be constructed by flipping some of the spins $| \uparrow_x \rangle \to | \downarrow_x \rangle$.

In the other extreme limit $B/J_0 \to 0$, we can ignore the terms containing $\sigma^x_l$ and consider only the terms with $\sigma^z_{l_1} \sigma^z_{l_2}$. The ground state is two-fold degenerate with either all spins being in the $+1$ eigenstate of $\sigma^z$ or all spins being the $-1$ eigenstate of $\sigma^z$. The ground subspace is spanned by the two degenerate states $| \uparrow_z \rangle^{\otimes n}$ and $| \downarrow_z \rangle^{\otimes n}$ as
\begin{equation}
    \left| \emptyset \left(\frac{B}{J_0} = 0\right) \right\rangle
    = \mathrm{Span} \left(| \uparrow_z \rangle^{\otimes n}, | \downarrow_z \rangle^{\otimes n} \right)
    .
    \label{eq:ParamagneticGroundStates}
\end{equation}
While all states in this degenerate subspace are trivially invariant under $\mathbb{1}$, only two states in the two-dimensional subspace are invariant under $\Pi$, that is the states
\begin{equation}
    \left| \emptyset \left(\frac{B}{J_0} = 0, \Pi = \pm 1 \right) \right\rangle
    = \frac{1}{\sqrt{2}} \left(| \uparrow_z \rangle^{\otimes n} \pm | \downarrow_z \rangle^{\otimes n} \right)
    ,
    \label{eq:FerromagneticGroundStates}
\end{equation}
with eigenvalues $\pm 1$ corresponding to the sign in the superposition.
Finding the TFIM in a ground state that is not invariant under the symmetry $\Pi$ is an example of the phenomenon of spontaneous symmetry breaking. Spontaneous symmetry breaking is the phenomenon of the existence of equilibrium states that do not respect the symmetry of the Hamiltonian which describes them \cite{beekman2019introduction}. In this regime, excitations above the ground state can be constructed by pairs of spins that are not aligned, which can be quantified using the correlation $\langle \sigma^z_{l_1} \sigma^z_{l_2} \rangle$ being $-1$.

%
\subsection{Quantum phase transition}
\label{sec:QuantumPhaseTransition}
%

In the limit of $B/J_0 = 0$, except for the two states in Eq.~\eqref{eq:FerromagneticGroundStates}, all other ground states are not invariant under the $\mathbb{Z}_2$-symmetry of the Hamiltonian. Therefore, we have reason to believe that at some critical value of the transverse field $B \equiv B_{\mathrm{c}}$, the nature of the ground state changes to being obligate symmetry respecting ($J_0/B = 0$) to a degenerate subspace where only two states are eigenstates of the symmetry operator ($B/J_0 = 0$) \cite{sachdev2011quantum}. This is an example of a \textit{quantum phase transition} (QPT) in the ground state, occurring solely due to the noncommutativity of different terms in the Hamiltonian, or equivalently the quantum fluctuations in the values of operators which have no definite value in the ground state dominated by other operators in the Hamiltonian \cite{sachdev2011quantum, chinni2022reliability, samajdar2022topological}. For equilibrium states, quantum phase transitions occur at zero temperature, or the ground state \cite{sachdev2011quantum, chinni2022reliability, samajdar2022topological}.

Corresponding to the element $\Pi$ in Eq.~\eqref{eq:RotateXPi}, the operator $\sigma^{z}_{l}$, representing the component of the $l$-th spin in the longitudinal direction, is associated with an order parameter. This operator transforms nontrivially under $\Pi$ as $\sigma^{z}_{l} \leftrightarrow - \sigma^{z}_{l}$. Symmetry-respecting or disordered states have zero expectation value $\langle \sigma^z_{l} \rangle = 0$, while symmetry-broken or ordered states have a nonzero expectation value  $\langle \sigma^z_l \rangle \neq 0$. Moreover, in the disordered phase, correlations in the expectation value at different sites are nonzero and expected to fall off exponentially with distance with distance as
\begin{equation}
    \langle \sigma^z_{l_1} \sigma^z_{l_2} \rangle 
    \propto 
    \exp\left( - \frac{|l_1-l_2|}{\xi} \right)
    ,
    \label{eq:ExponentialDecayCorrelations}
\end{equation}
where $\xi$ is the correlation length, parameterizing the exponential decay \cite{sachdev2011quantum, Kar2017}. This will play a role in our exploration of quench dynamics in Chap.~\ref{chap:QuenchChaos}. Near the critical point, the correlation length diverges $\xi \to \infty$, leading to a scale invariance in the properties of the system \cite{sachdev2011quantum}. The divergence of the correlation length is characterized by a dynamical critical exponent $\nu$ as
\begin{equation}
   \lim_{B \to B_{\mathrm{c}}} \xi \propto 
   \Lambda_{B}
   \left(\frac{B}{J_0} - \frac{B_{\mathrm{c}}}{J_0}\right)^{-\nu}
   \label{eq:CorrelationLengthDivergenceField}
   ,
\end{equation}
where $\Lambda_{B}$ is a wavelength scale associated with an effective momentum cutoff \cite{sachdev2011quantum}. 

The study of properties of thermal equilibrium states can be done through quantum quenches, where local observables equilibrate to thermal values at appropriate temperatures after short times \cite{d2016quantum, Kar2017, Tit2019}. In particular, the correlations and correlation length in Eq.~\eqref{eq:ExponentialDecayCorrelations} and their behavior near the critical point of the transition can be accessed through quantum quenches \cite{Kar2017, Tit2019, mitra2023macrostates}.

To make a connection with the macroscopic properties of a system with $n$ spin-1/2 degrees of freedom, it is useful to therefore use the collective magnetization $M_{z}$
\begin{equation}
  M_{z} = \frac{1}{n} \sum_{l} \langle \sigma^{z}_{l} \rangle
  \label{eq:MzDefinition}
  ,
\end{equation}
which distinguishes the ordered or \textit{paramagnetic} phase with $M_{z} = 0$ and the disordered or \textit{ferromagnetic} phase with $M_{z} \neq 0$. Associated with the change from $M_z$ from $1$ to $0$, which becomes discontinuous in the thermodynamic limit, $n \to \infty$, is the change behavior of the collective two-spin correlation
\begin{equation}
  M_{zz} = \frac{1}{n^2} 
  \sum_{l_1} \sum_{l_2}
  \langle \sigma^{z}_{l_1} \sigma^{z}_{l_2} \rangle
  \label{eq:MzzDefinition}
  ,
\end{equation}
whose behavior changes with the value of the transverse field $B$. At the critical transverse field, $B_{\mathrm{c}}$, $M_{zz}$ has a minima, which becomes singular in the thermodynamic limit.

%
\subsection{Thermal and dynamical phase transitions}
%

\begin{figure}
    \centering
    \includegraphics[width=0.9\textwidth]{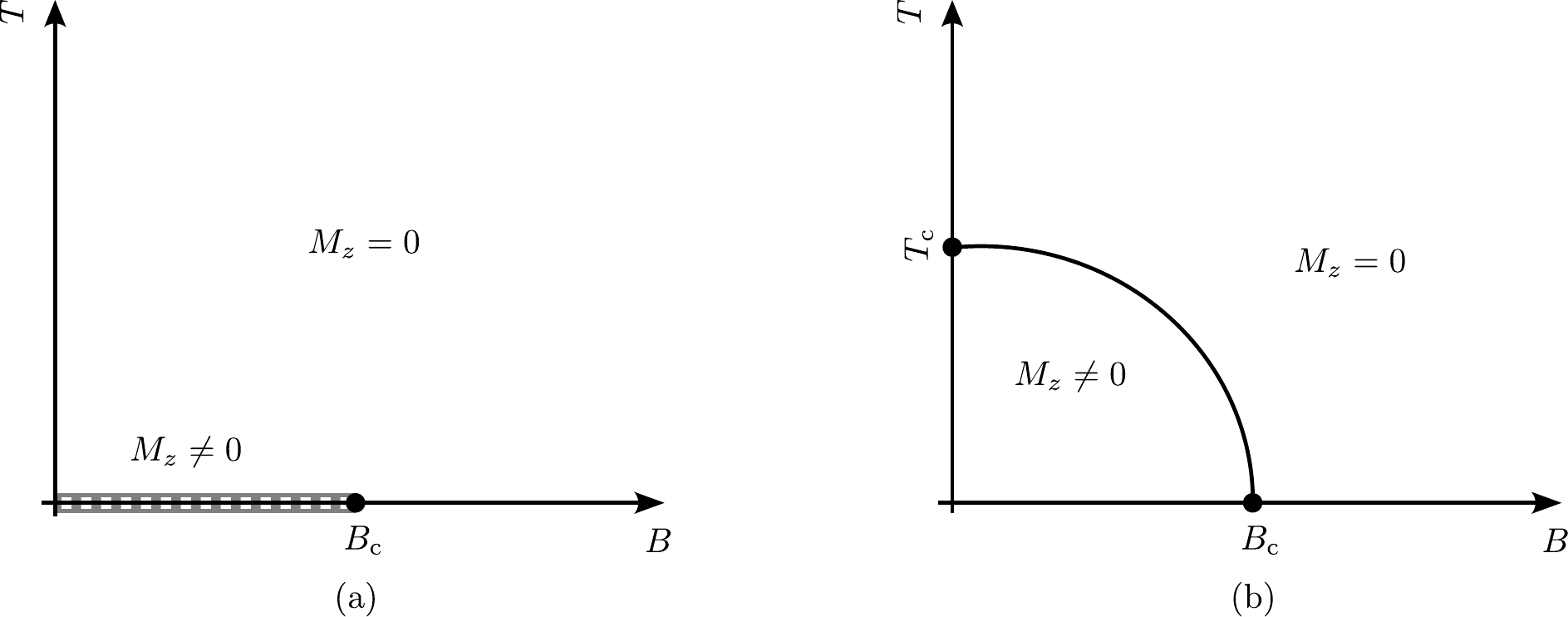}
    \caption{Phase diagram of transverse field Ising models (TFIMs). (a) For TFIMs with a small number of neighbors, in low spatial dimensions, there is the ferromagnetic phase ($M_{z} \neq 0$) only at zero temperature $T=0$, with a quantum critical point $B_{\mathrm{c}}$ separating it from a paramagnetic phase ($M_{z} = 0$). At zero temperature, there is a ferromagnetic phase ($M_z \neq 0$) for $B < B_{\mathrm{c}}$, shown by the shaded line. Everywhere else in the phase diagram, there is a paramagnetic phase ($M_z = 0$). (b) For TFIMs with a large number of neighbors, in high spatial dimensions, and/or with long-range interactions, a ferromagnetic phase ($M_z \neq 0$) exists at nonzero temperatures below a critical temperature $T < T_{\mathrm{c}}$. For transverse fields above the critical field $B > B_{\mathrm{c}}$ and temperature above the critical temperature, $T > T_{\mathrm{c}}$, there is only a paramagnetic phase ($M_z = 0$). The phase boundary is shown by a curve joining the critical temperature $T_{\mathrm{c}}$ and critical field $B_{\mathrm{c}}$.}
    \label{fig:TFIMPhaseDiagram}
\end{figure}

For models with a large number of interacting neighbors, in high spatial dimension and/or with long-range interactions, the nature of two different ground states survives for finite equilibrium temperature states for certain interaction graphs, specified by $\kappa_{l_1, l_2}$ in Eq.~\eqref{eq:TFIMGeneric}. For nonzero temperature $T>0$ below a critical temperature, $T < T_{\mathrm{c}}$, there exists a ferromagnetic phase with $M_{z} \neq 0$ and a thermal phase transition (TPT) to the paramagnetic phase ($M_{z} = 0$) \cite{sachdev2011quantum, chinni2022reliability, samajdar2022topological}. The TPT occurs due to thermal fluctuations in the terms contributing to the Hamiltonian -- thermal fluctuations are due to uncertainties arising from a thermal distribution of states. The ferromagnetic phase disappears for nonzero temperature $T > 0$ for low dimensional models with short-range interactions. In particular, the phase transition is absent in one-dimensional (1D) nearest neighbor models and present in one-dimension only if there are long-range interactions in the system \cite{dutta2001phase, Zun2018, piccitto2019dynamical}. This is because, for short-range interactions in lower spatial dimensional lattices, there is not enough influence of spins on each other to have a stable ferromagnetic phase at nonzero temperature \cite{chinni2022reliability}. The phase diagrams in the temperature magnetic field plane for short-range interactions in low spatial dimensions are shown in Fig.~\ref{fig:TFIMPhaseDiagram} (a) and for long-range interactions and/or high spatial dimensions are shown in Fig.~\ref{fig:TFIMPhaseDiagram} (b).

The order parameter $M_z$ in Eq.~\eqref{eq:MzDefinition} plays a role for the TPT analogous to its role in the QPTs. Moreover, the spatially averaged correlation $M_{zz}$ in Eq.~\eqref{eq:MzDefinition} also behaves similarly for the TPT between the ferromagnetic and paramagnetic phases as the QPT. Furthermore, near the critical temperature, there is a diverging correlation length $\xi$ of the spin correlations $\langle \sigma^z_{l_1} \sigma^z_{l_2} \rangle$ as
\begin{equation}
   \lim_{T \to T_{\mathrm{c}}} \xi \propto 
   \Lambda_{T}
   \left(\frac{k_{\mathrm{B}} T}{J_0} - \frac{k_{\mathrm{B}} T_{\mathrm{c}}}{J_0}\right)^{-\nu}
   \label{eq:CorrelationLengthDivergenceTemperature}
   ,
\end{equation}
where $k_{\mathrm{B}}$ is the Boltzmann's constant and $\Lambda_{T}$ is a wavelength scale associated with an effective momentum cutoff \cite{sachdev2011quantum}.

This finite temperature phase transition, which occurs due to the nature of the ground state being inherited by finite temperature states, also influences quench dynamics when states that are not eigenstates of the Hamiltonian Eq.~\eqref{eq:TFIMGeneric} are quenched to the Hamiltonian, in an example of a dynamical quantum phase transition (DQPT). This variety of DQPT will be the focus of Chap.~\ref{chap:QuenchDQPT}. While there are qualitative similarities between the TPT and the DQPT, the detailed relationship between the two is an area of ongoing work \cite{Zun2018, piccitto2019dynamical, halimeh2017dynamical, halimeh2019dynamical, mitra2023macrostates}.
%

%
\section{Quench dynamics}
\label{sec:QuenchDynamics}
%

A common paradigm of studying out-of-equilibrium many-body systems is that of quantum quenches \cite{mitra2018quantum}. The system is prepared in a stationary state, typically a ground state, of a Hamiltonian, and is then suddenly subjected to a different noncommuting Hamiltonian, usually through a sudden change of parameters. An example of a quantum quench would be an initial state corresponding to the ground state $B=0$ in Eq.~\eqref{eq:TFIMGeneric} being quenched to a nonzero value of $B$, which we will consider in Chap.~\ref{chap:QuenchDQPT}. In experimental practice, preparation of the initial state suffices without preparation of its parent Hamiltonian \cite{jurcevic2017direct, zhang2017observationmany}. The dynamics correspond to that of the time evolution of a nonstationary state under a time-independent Hamiltonian.

Consider an initial state $| \psi \rangle$ written as a super-position of eigenstates $| n \rangle$ of a quenched Hamiltonian $H$,
\begin{equation}
  | \psi \rangle = \sum_{n} c_{n} | n \rangle,
\end{equation}
or the density operator written as
\begin{equation}
  \rho
  = | \psi \rangle \langle \psi |
  = \left( \sum_{n_1} c_{n_1} | n_1 \rangle \right) \left( \sum_{n_2} c_{n_2}^* \langle n_2 | \right)
  =  \sum_{n_1, n_2} c_{n_1} c_{n_2}^* | n_1 \rangle \langle n_2 |.
\end{equation}
This representation makes the time-evolved state at time $t$ easy to write by including energy-dependent phases to each term in the superposition as
\begin{equation}
  | \psi (t) \rangle = \sum_{n} c_{n} \exp\left(-\mathrm{i} E_{n} t\right) | n \rangle,
  \label{eq:KetEvolveEigenbasis}
\end{equation}
which in turn gives the time-evolved density operator
\begin{equation}
  \begin{aligned}
    \rho(t)
    &
    = \sum_{n_1,n_2} c_{n_1} c_{n_2}^* \exp\left(-\mathrm{i} (E_{n_1} - E_{n_2}) t\right) | n_1 \rangle \langle n_2 |
    \\ &
    = \sum_{n} |c_{n}|^2 | n \rangle \langle n |
    + \sum_{n_1 \neq n_2} c_{n_1} c_{n_2}^* \exp\left(-\mathrm{i} (E_{n_1} - E_{n_2}) t\right) | n_1 \rangle \langle n_2 |.
    \label{eq:ETHDensityOperator}
  \end{aligned}
\end{equation}

From this the expectation value of an observable $O$ can be calculated as
\begin{equation}
\begin{aligned}
    \langle O \rangle
    &
    = \mathrm{Tr}\left(O \rho\right)
    \\ &
    = \sum_{n} |c_{n}|^2 O_{nn}
    + \sum_{n_1 \neq n_2} c_{n_1} c_{n_2}^* \exp\left(-\mathrm{i} (E_{n_1} - E_{n_2}) t\right) O_{n_1 n_2}
    \label{eq:ETHObservableExpectation}
    ,
\end{aligned}
\end{equation}
where the first sum corresponds to the contributions from the diagonal matrix element of $O$, and the second sum corresponds to the contributions from the off-diagonal matrix elements of $O$. In the long-time average, the off-diagonal terms average to zero, provided the degeneracies are either absent or subextensive \cite{d2016quantum}. While recurrences in these oscillations are possible, the frequencies being incommensurate make the time after a quench at which these recurrences occur, very large, often larger than any timescales accessible in physical systems \cite{d2016quantum, Mor2018}. Moreover, the expectation value stays close to the value described by the diagonal terms in Eq.~\eqref{eq:ETHObservableExpectation} for most times \cite{d2016quantum, Mor2018}.

%
\section{Equilibration and eigenstate thermalization}
\label{sec:ETHIntro}
%

Hitherto we have discussed equilibrium properties of the TFIM. While properties of equilibrium states forms a major part of the studies of statistical mechanics \cite{kardar2007statisticalfields, kardar2007statisticalparticles, huang2008statistical, pathria2016statistical}, the question of approach to equilibrium states from nonequilibrium states is of special interest for contemporary quantum experiments with multiple degrees of freedom \cite{d2016quantum, deutsch2018eigenstate, Mor2018}. A way to access equilibrium properties through quantum dynamics is to harness forms of equilibration which occur in nonequilibrium dynamics of ergodic systems and nearly ergodic systems, as specified by versions of the eigenstate thermalization hypothesis \cite{alba2015eigenstate, khodja2015relevance, d2016quantum, deutsch2018eigenstate, dymarsky2018subsystem, abanin2019colloquium, murthy2019bounds, noh2021eigenstate}. \textit{Equilibration} occurs for some quantities during quantum quenches, even though the state of the system keeps evolving under the Hamiltonian as in Eq.~\eqref{eq:KetEvolveEigenbasis}. \textit{Thermalization} may occur for quantities when these quantities equilibrate to values obtained from an appropriate thermal state. It is important to note that, even though the state keeps evolving, some quantities can experience equilibration and/or thermalization. For example, aspects of critical properties of equilibrium quantum phase transitions can be studied through quantum quenches that give quasi-steady state values for relevant order parameters \cite{mitra2018quantum, Kar2017, Zun2018, piccitto2019dynamical, Tit2019, zhang2017observationmany}. Moreover, dynamical properties involving transport and hydrodynamics can be accessed through quantum dynamics, albeit at exceedingly long times \cite{roberts2017chaos, leviatan2017quantum, abanin2019colloquium, rakovszky2022dissipation, vonkeyserlingk2022operator}.

Under reversible unitary dynamics of the time-dependent Schrodinger equation Eq.~\eqref{eq:IntroTimeDependentSchrodinger}, the irreversible phenomenon of thermalization, is far from evident. Indeed, there is no thermalization in the full quantum state of the many-body system during unitary dynamics. Von Neumann attempted to describe the process of thermalization using the thermalization of macroscopic observables \cite{neumann1929beweis}, instead of the wavefunction or density operator representing the state of the full system. Thermalization is described using the quantum ergodic theorem, which states that under unitary evolution expectation values of macroscopic observables evolve to a value that is close to the value obtained from the microcanonical ensemble with temperature set by the expectation value of the Hamiltonian in the initial state \cite{Gol2010}. \textit{Equilibration} occurs for an observable if its expectation value, during a quench, reaches a steady state value and does not deviate significantly from this value, for most times \cite{d2016quantum, Mor2018}. Moreover, \textit{thermalization} is also observable dependent and is said to occur if the expectation value of that observable reaches the value obtained from the microcanonical ensemble and does not deviate significantly from the latter for most times \cite{Gol2010, goldstein2015thermal, d2016quantum, Mor2018}.

It is important to note that we are considering thermalization of specific observables during unitary dynamics of the system, which is isolated and does not interact with external degrees of freedom. Thermalization can be interpreted as parts of the system acting as a thermal bath for the rest of the system, leading to effective thermal equilibrium between them, even under global unitary dynamics \cite{deutsch2018eigenstate, d2016quantum}.

While the quantum ergodic theorem was an important step towards understanding thermalization from unitary dynamics, it was far from complete as it did not make a distinction between integrable and nonintegrable systems and between the thermalization or lack thereof for specific observables in specific systems \cite{d2016quantum}.

Let us first consider the extreme limit of Hamiltonians being random matrices, sampled from random matrix ensembles. In this limit, random matrix theory (RMT) can be used to calculate these expectation values \cite{d2016quantum}. In this extreme case, the expectation value has no dependence on the energy of the system, which is formally equivalent to the infinite-temperature ensemble where there is no dependence of the populations $|c_n|^2$'s on the energies $E_n$ as
\begin{equation}
    \langle O \rangle 
    \approx 
    \mathrm{Tr}(O \rho_{\star})
    = \mathrm{Tr}(O)
    ,
\end{equation}
where $\rho_{\star}$ is the maximally mixed state proportional to the identity matrix $\mathds{1}$, which corresponds to an infinite-temperature ensemble. Nevertheless, for thermalization observed in physical systems, the expectation value of an observable depends on the energy of the system, the latter specifying the thermal ensembles. Sredenicki introduced an ansatz for the matrix elements of observables in the eigenbasis of Hamiltonian \cite{srednicki1994chaos, srednicki1996thermal, srednicki1999approach}
\begin{equation}
  O_{n_1 n_2} = O(\bar{E}) \delta_{n_1 n_2} 
  + \exp\left(-\frac{\mathcal{S}(\bar{E}}{2}\right) f_{O} (\bar{E}, \omega) R_{n_1 n_2}
  ,
  \label{eq:ETHMatrixElementsAnsatz}
\end{equation}
 where $\bar{E} \equiv (E_{n_1} + E_{n_2})/2$ is the mean energy of a pair of energy levels, $\omega \equiv E_{n_1} - E_{n_2}$ is the energy difference between the pair and $\mathcal{S}(\bar{E})$ is the entropy of the microcanonical ensemble at energy $\bar{E}$. Moreover, $O(\bar{E})$ and $f_O(\bar{E}, \omega)$ are smooth functions of their arguments, $O(\bar{E})$ is identical to the expectation value of the microcanonical ensemble at energy $\bar{E}$, and $R_{n_1 n_2}$ is a random real or complex variable with zero mean and unit variance ($\overline{R_{n_1 n_2}^2} = 1$ for $\overline{|R_{n_1 n_2}|^2} = 1$ respectively). Requiring the observable $O$ to be Hermitian, the function $f_O(\bar{E}, \omega)$ and the random numbers $R_{n_1 n_2}$ must satisfy the following relations:
\begin{equation}
\begin{aligned}
  R_{n_1 n_2} = R_{n_1 n_2},
  && f_O(\bar{E}, -\omega) = f_O(\bar{E}, \omega)
\end{aligned}
\end{equation}
for real matrix elements, and
\begin{equation}
\begin{aligned}
  R_{n_1 n_2}^* = R_{n_1 n_2},
  && f_O^*(\bar{E}, -\omega) = f_O(\bar{E}, \omega)
\end{aligned}
\end{equation}
for complex matrix elements.

This ansatz is known as the \textit{eigenstate thermalization hypothesis} (ETH). The ETH generalizes the predictions of RMT to have energy dependence and is expected to apply to observables in physical systems \cite{srednicki1994chaos, srednicki1996thermal, srednicki1999approach, d2016quantum}, Moreover, the applicability of ETH has been observed in systems with a moderate number of bosons \cite{rigol2008thermalization} and spins \cite{jensen1985statistical}. Indeed, when we zoom into a narrow energy window, which is small compared to the energy of the slowest time scales in the problem, the function $f_O(\bar{E}, \omega)$ can be treated as a constant, which leads to the ETH reducing to RMT. In other words, the structure of the eigenenergies and eigenvectors of a Hamiltonian is similar to that of a matrix described by RMT, for a sufficiently small energy window. The size of this energy window depends inversely on the square of the system size, $1/n^2$ and goes to zero in the thermodynamic limit \cite{d2016quantum}. Therefore, RMT has limited applicability in this limit.

Due to the properties of the off-diagonal elements in Eq.~\eqref{eq:ETHMatrixElementsAnsatz}, the phase differences in the off-diagonal terms in Eq.~\eqref{eq:ETHDensityOperator} quickly become zero and have small fluctuations. Therefore, there is a mechanism by which the expectation values of observables reach their thermal values and stay close to them. Both ETH and RMT posit that for observables $O$ whose eigenvalues do not scale with the size of the Hilbert space, which is the case for physical observables in finite Hilbert spaces, both the diagonal and off-diagonal matrix elements have fluctuations that are inversely proportional to the square root of the size of the Hilbert space \cite{d2016quantum}. Therefore, the off-diagonal terms in Eq.~\eqref{eq:ETHObservableExpectation} are quite small.

The ETH also conjectures that for a chaotic quantum system, eigenvectors of the Hamiltonian behave like statistical ensembles in several ways \cite{d2016quantum, deutsch2018eigenstate}. Chaos for quantum systems is usually understood in terms of level repulsion, that is whether energy eigenvalues tend to repel each other \cite{d2016quantum, serbyn2021quantum, moudgalya2022quantum}. Chaotic quantum systems have level repulsion, where the probability of finding eigenvectors with the same energy is small. In these eigenstates, for a small number of degrees of freedom of the system, the rest of the system acts as a thermal bath. This makes the eigenstates themselves appear thermal. In particular, the expectation values of local observables are well approximated by statistical ensembles. 

Let us consider the implication of this to quench dynamics. In Eq.~\eqref{eq:ETHDensityOperator} the diagonal elements (or populations) of the density operator are time-independent, while the off-diagonal elements (or coherences) have time-dependent phases determined by the difference in the energies of the two energy eigenvectors. If the $E_{n_1}$'s are incommensurate, the off-diagonal terms time-average to zero, giving the time-averaged state approximately equal to the diagonal ensemble
\begin{equation}
  \rho_{\mathrm{DE}} = \sum_{n} |c_{n}|^2 | n \rangle \langle n |.
  \label{eq:ETHDiagonalEnsemble}
\end{equation}
Moreover, expectation values of local observables $O$ obtained from Eq~\eqref{eq:ETHDiagonalEnsemble} are close to their microcanonical values \cite{d2016quantum, Mor2018}. Furthermore, the off-diagonal elements have small fluctuations about zero, and average out to zero due to energy level differences being incommensurate \cite{deutsch2018eigenstate, d2016quantum}. Indeed, the statistics of energy level difference are related to the chaoticity of a Hamiltonian and are used as a tool to assess the chaoticity of a Hamiltonian, as we will do in Chap.~\ref{chap:QuenchDQPT}. If the energy levels have signatures of level repulsion, the off-diagonal terms wash out rapidly, as is seen in chaotic systems.

While thermalization in classical systems is usually described using particle collisions and energy redistribution between different degrees of freedom, thermalization in quantum systems is seen to arise from chaos and ergodicity which are embedded in the energies and eigenvectors of Hamiltonians \cite{d2016quantum}. The approach of expectation values of observables to their equilibrium values is a result of the dephasing of their off-diagonal elements in the eigenbasis of the Hamiltonian \cite{d2016quantum, Mor2018}. Therefore, for quantum thermalization, the information about the final thermal state is hidden in the initial state and the spectrum of the Hamiltonian \cite{d2016quantum}. We will return to this in the context of quenches to Ising models in Chap.~\ref{chap:QuenchChaos}.

%
\subsection{Generalized eigenstate thermalization in integrable systems}
\label{sec:GeneralizedETH}
%
In integrable systems, there is no thermalization in the strict sense. In particular, the existence of conserved quantities constrains the processes of thermalization. The thermal reduced density operators under these constraints, constitute the generalized Gibbs ensemble (GGE) and can be written using a set of conserved quantities $\{Q_k\}$ as
\begin{equation}
   \rho_{\mathrm{GGE}}
   = \frac{\exp\left(- \sum_{k} \lambda_k Q_k \right)}
   {\mathrm{Tr}\left( \exp\left(- \sum_{k} \lambda_k Q_k \right) \right)}
   ,
\end{equation}
where $\lambda_k$'s are Lagrange multipliers determined by maximizing entropy under the condition of the expectation values of $\{Q_k\}$ being conserved \cite{d2016quantum, calabrese2012quantum2}.
 
In integrable systems, due to lack of thermalization, few-body observables do \textit{not} reach their thermal equilibrium values in general. They can, however, be described by the GGE. Nevertheless, for some special initial states and some special few body observables, the expectation value may reach close to the thermal expectation values in the steady state \cite{d2016quantum, Kar2017}. Thus, finding values of few-body observables close to thermal values does not imply thermalization has occurred, nor does it imply that the system is nonintegrable. Nevertheless, the equilibration of some observables and some initial states under quenches to integrable models can enable the study of equilibrium through quantum quenches, even for integrable models. We will see an example of such an integrable quench in Chap.~\ref{chap:QuenchChaos}.

%
\section{Macrostates and microstates}
\label{sec:MacroMicroIntro}
%

In the discussion of the TFIM in Sec.~\ref{sec:TFIMIntro} and thermalization in Sec.~\ref{sec:ETHIntro}, a central role was played by macroscopic observables, which are extensive sums of local observables. Since these macroscopic observables are related to macro properties of the system obtained through coarse-graining in the form of space and time averaging, it is helpful to relate this to notions of precise microstates and coarse-grained macrostates. Indeed, this coarse-graining of several microstates into an equivalence class of macrostates makes the study of quantum many-body systems more forgiving of both experimental imperfections and approximations \cite{preskill2012quantum, preskill2018quantum, deutsch2020harnessing, preskill2022physics, trivedi2022quantum}.

A microstate corresponds to a complete specification of the configuration of every component of the system. For the spin-1/2 TFIM considered in Sec.~\ref{sec:TFIMIntro}, a microstate involves a description of every spin in a system of $n$ spin-1/2 degree of freedom and the correlations between all spins, for example, the $n$-spin state $| \psi \rangle$. This state is a $2^n$-dimensional vector, which makes it intractable to keep track of for systems with large $n$.

The magnetization order parameter $M_{z}$ can be evaluated by summing the expectation values of $\sigma^{z}_{l}$ on each spin $l$ in the system. There can be several states that yield the same $M_{z}$. For simplicity consider the classical states corresponding to each spin being in one of the states $\{\uparrow_z, \downarrow_z\}$. The magnetization $M_{z}$ is determined by the number of spins pointing up, $n_{\uparrow}$ and the number of spins pointing down, $n_{\downarrow} = n - n_{\uparrow}$. Any permutation of $n_{\downarrow}$ spins pointing down and $n_{\uparrow}$ pointing up yields the same value of $M_{z} = n_{\uparrow} - n_{\downarrow}$. The number of such configurations is given by the binomial coefficient
\begin{equation}
  \binom{n}{n_{\uparrow}}
  = \binom{n}{n_{\downarrow}}
  = \frac{n!}{n_{\uparrow}! n_{\downarrow}!}
  .
\end{equation}
Thus the $2^n$ configurations are divided into $n+1$ values of $M_{z}$ in the range $-1 \leq M_z \leq +1$.

In the quantum case, where generally the state $|\psi\rangle$ is a superposition of all $2^{n}$ spin configurations, the magnetization is no longer quantized to $n+1$ values. Nevertheless, all states that yield the same value of $M_{z}$ are equivalent for using $M_{z}$ as an order parameter for the phase transition considered in Sec.~\ref{sec:TFIMIntro}. Therefore the set of all states with the same value of $M_{z}$ corresponds to a \textit{macrostate} of the system. As in statistical mechanics \cite{kardar2007statisticalparticles, kardar2007statisticalfields, huang2008statistical, pathria2016statistical}, the macrostate is determined by the values of macroscopic observables like temperature, pressure, here the macrostate is determined by the macroscopic observable $M_z$. However, to characterize the phases, we do not need a probability distribution over microstates $| \psi \rangle$ that belong to a macrostate specified by $M_{z}$.

In general, we will use \textit{macrostate} to denote a set of microstates of a system that yield a fixed value of an expectation value of an observable. For spin models like the TFIM in Sec.~\ref{sec:TFIMIntro}, we define macrostate as the set of all states that meet conditions involving the expectation value of the Hamiltonian, $\langle H \rangle$, the longitudinal magnetization consisting of an extensive sum of expectation values of $\sigma^z_l$ and the number of spins $n$ as
\begin{equation}
    \Omega(E, M_{z}, n)
    = \left\{
    | \psi \rangle \in \mathbb{C}^{2^n} :
    \left|\frac{1}{n} \sum_{l=1}^{n} \langle \sigma^z_l\rangle - M_z\right| < \epsilon_{M_z},
    |\langle H \rangle - E| < \epsilon_E
    \right\}
    ,
    \label{eq:MacrostateEnergyMz}
\end{equation}
where $\epsilon_{M_z}$ and $\epsilon_{E}$ are small tolerances in the macroscopic observables, which are determined by the context in which the macrostates are being considered. Similarly, we define a macrostate using the longitudinal spin-spin correlations as
\begin{equation}
    \Omega(E, M_{zz}, n)
    = \left\{
    | \psi \rangle \in \mathbb{C}^{2^n} :
    \left|\frac{1}{n^2} \sum_{l_1=1}^{n} \sum_{l_2=1}^{n} \langle \sigma^z_{l_1} \sigma^z_{l_2} \rangle - M_{zz}\right| < \epsilon_{M_{zz}},
    |\langle H \rangle - E| < \epsilon_E
    \right\}
    ,
    \label{eq:MacrostateEnergyMzz}
\end{equation}
where, again, $\epsilon_{M_{zz}}$ and $\epsilon_{E}$ are small tolerances in macroscopic observables.

When the observables of interest are local, these can be related to the local reduced states of the full microstates. Many pure states can yield the same local reduced states, on tracing out appropriate degrees of freedom. Each of these pure states would correspond to a \textit{microstate} while the set of all microstates, in Eq.~\eqref{eq:MacrostateEnergyMz} and/or Eq.~\eqref{eq:MacrostateEnergyMzz} in would be the \textit{macrostate}. We discuss the relationship between local observables and marginal states in Appendix~\ref{app:MarginalsExpectationValues}. This notion of macrostates will be useful in analyzing a DQPT in Chap.~\ref{chap:QuenchDQPT}.

%
\section{Conclusion}
%
In this chapter, we discussed the TFIM, and its symmetries. We argued that the phase transition in the TFIM can be studied using expectation values of local observables, without precise specification of the microstate. We also reviewed the eigenstate thermalization hypothesis, which posits a mechanism of a closed system reaching thermal equilibrium with parts of the system acting as thermal baths for the rest of the system.

With the background of the ideas reviewed in this chapter, we study dynamical quantum phase transitions in Ising models, and other critical properties accessible through quenches in the next two chapters. We will consider both dynamical and equilibrium phase transitions and aspects related to them.

%
\chapter{Simulating a dynamical quantum phase transition}
\label{chap:QuenchDQPT}
%

%
\section{Introduction}
\label{sec:QuenchDQPTIntro}
%

Quantum simulation of many-body physics has long been considered to be a potential application where noisy intermediate scale quantum (NISQ) processors \cite{preskill2018quantum} could exhibit a computational advantage over classical computers. While severely limited by control errors and decoherence, it is hoped that a NISQ device may still succeed when it probes ``universal properties'' of the model, characterized by local order parameters (typically one- or two-point correlation functions), that represent ``macrostates'' of the system rather than ``microstates'' that depend on the exact many-body state. Macro properties are more likely to be robust to small imperfections and decoherence and hence yield a good approximation to the exact solution. This robustness may also lend the estimation of these macro-properties to efficient classical methods. 

In this chapter, we study a dynamical quantum phase transition (DQPT) occurring in the quench dynamics of one-dimensional transverse field Ising models (TFIMs). This DQPT corresponds to a transition between the steady state value of order parameters obtained after a quantum quench to a TFIM, from an initial state that does not have the symmetry of the TFIM. We will focus on the estimation of order parameters to study the dynamical phases and the phase transition between them. We will also study the estimation of critical point and critical exponents of the phase transition from classical simulations with limited bond dimension matrix product states (MPS). While such 1D models are far from where one expects the greatest quantum advantage, we focus on them here because they are a standard paradigm for quench dynamics, and their simplicity will help us to reveal the robustness of the macro versus microstate in quantum simulation. We study classical simulations of critical properties near a $\mathbb{Z}_2$-symmetry breaking phase transitions in transverse field Ising models (TFIMs).

%
\section{Model, complexity and chaos}
\label{sec:Model}
%
We consider 1D TFIMs with variable-range interactions along the $z$-axis characterized by a power-law with exponent $\alpha \geq 0$ and a transverse field along the $x$-axis with magnitude $B$. For $n$ qubits the Hamiltonian is
\begin{equation} 
\begin{aligned}
  H = &
  - \frac{J_0}{\mathcal{N}} \sum_{\ell_1 < \ell_2}
  \frac{\sigma^z_{\ell_1} \sigma^z_{\ell_2}}
       {|\ell_1 - \ell_2|^{\alpha}}
  - B \sum_{\ell} \sigma^x_{\ell}
  ,
  \label{eq:HamiltonianPowerLawTFIM}
\end{aligned}
\end{equation}
where $\mathcal{N}$ is the Kac normalization factor \cite{halimeh2017prethermalization, halimeh2017dynamical, zhang2017observationmany, Zun2018, piccitto2019dynamical, lang2018dynamical}
\begin{equation}
  \mathcal{N}(\alpha, n) = \frac{1}{n-1}\sum_{\ell_1 < \ell_2} \vert \ell_1 -
  \ell_2 \vert^{-\alpha}
  = \frac{1}{n-1}\sum_{\ell=1}^{n-1} (n-\ell)\ell^{-\alpha}.
  \label{eq:KacNormalization}
\end{equation}
Such models have been implemented in atomic platforms, including trapped ions with $0 \leq \alpha \leq 3$ \cite{foss2013nonequilibrium, foss2013dynamical, bohnet2016quantum, jurcevic2017direct, zhang2017observationmany, zhang2017observationdiscrete, monroe2021programmable}. Extensions of this model with a longitudinal field have been implemented in Rydberg atom arrays with $\alpha = 6$ \cite{bernien2017probing,weimer2010rydberg, schauss2018quantum,browaeys2020many, scholl2021quantum,ebadi2021quantum, bluvstein2021controlling}. Experiments have studied quench and driven dynamics \cite{bohnet2016quantum, zhang2017observationmany,bernien2017probing, schauss2018quantum,bluvstein2021controlling, bluvstein2022quantum, choi2023preparing} as well as quasi-adiabatic passages \cite{bernien2017probing, ebadi2021quantum,scholl2021quantum, semeghini2021probing,ebadi2022quantum, kim2022rydberg, byun2022finding} to probe ground states in exotic quantum phases of matter \cite{bernien2017probing, ebadi2021quantum, scholl2021quantum, semeghini2021probing, ebadi2022quantum} and approximate solutions to graph problems encoded in Ising models \cite{ebadi2022quantum, kim2022rydberg, byun2022finding}.

For $\alpha = 0$ this is a fully connected model, and the Hamiltonian governs the integrable dynamics of the collective spin vector $\mathbf{S} = \sum_{\ell} \boldsymbol{\sigma}_{\ell}/2$ whose total angular momentum $\mathbf{S}^2$ is conserved. This is an example of a Lipkin-Meshkov-Glick (LMG) model \cite{Lip1965}. Here the thermodynamic limit is the mean-field limit, with all dynamics described by the motion of a unit vector on a spherical phase space \cite{munoz2020simulation,chinni2021effect, munoz2021nonlinear, chinni2022trotter, munoz2022floquet, munoz2023phase, munoz2020simulation}. When $\alpha \to \infty$ one obtains the integrable 1D nearest-neighbor TFIM \cite{lieb1961two, pfeuty1970one, pfeuty1971ising, derzhko1998numerical, sengupta2004quench, calabrese2011quantum, calabrese2012quantum1, calabrese2012quantum2}. Simulating dynamics of this model is classically tractable using the Onsager algebra of strings of local spin operators \cite{lychkovskiy2021closed} or using a transformation to Gaussian Fermionic Hamiltonians \cite{lieb1961two, pfeuty1970one, pfeuty1971ising, derzhko1998numerical, sengupta2004quench, calabrese2011quantum, calabrese2012quantum1, calabrese2012quantum2}. For all other values of $\alpha$ the system is nonintegrable.

We consider quench dynamics where all of the spins are initially polarized along some axis (a spin coherent state). The initial product state evolves according to the Hamiltonian in Eq.~\eqref{eq:HamiltonianPowerLawTFIM}, generally leading to volume-law entanglement, including for most integrable models \cite{calabrese2005evolution, schollwock2011density, zhang2017observationmany, bernien2017probing}, except in some special cases \cite{lerose2020origin,lerose2020bridging}. Small values of $\alpha \lesssim 0.5$ typically yield Hamiltonians close to the collective spin model that have slow growth of entanglement \cite{lerose2020origin, lerose2020bridging}, while large values of $\alpha \gtrsim 6$ lead to Hamiltonians close to the one-dimensional nearest-neighbor model that have fast growth of entanglement \cite{calabrese2005evolution}. 

We would like to assess the complexity of the simulation of the TFIMs described by Eq.~\eqref{eq:HamiltonianPowerLawTFIM}. Quantum chaos is one measure of complexity (not equivalent to integrability) that can affect the tractability of classical simulation of a given model. To assess the chaoticity of a model, we can consider the mean adjacent level spacing ratio, $\bar{r}$, which quantifies the level repulsion of the energy eigenvalues, defined as \cite{atas2013distribution}
\begin{equation}
\begin{aligned}
  \bar{r} = & \textsc{Mean}(r_{k}),\label{eq:MeanAdjacentLevelSpacingRatio}
  \\
  r_{k} = &
  \frac
      {\min\left(\left({E_{k} - E_{k-1}}\right),
        \left(E_{k+1} - E_{k}\right)\right)}
      {\max\left(\left({E_{k} - E_{k-1}}\right),
        \left(E_{k+1} - E_{k}\right)\right)},
\end{aligned}
\end{equation}
where $E_k$ is the $k$-th lowest energy eigenvalue. An integrable Hamiltonian would correspond to Poissonian level spacing statistics with $\bar{r} \approx 0.386$, while a chaotic Hamiltonian, in this case, would correspond to $\bar{r}$ of the Gaussian orthogonal ensemble (GOE) of random matrix theory, with $\bar{r} \approx 0.535$ \cite{atas2013distribution}. Another measure of the chaoticity of a model is the entanglement entropy of the Hamiltonian eigenvectors, which we use in Sec.~\ref{sec:RoleChaos}.

\begin{figure}
    \centering
    \includegraphics[width=0.48\textwidth]{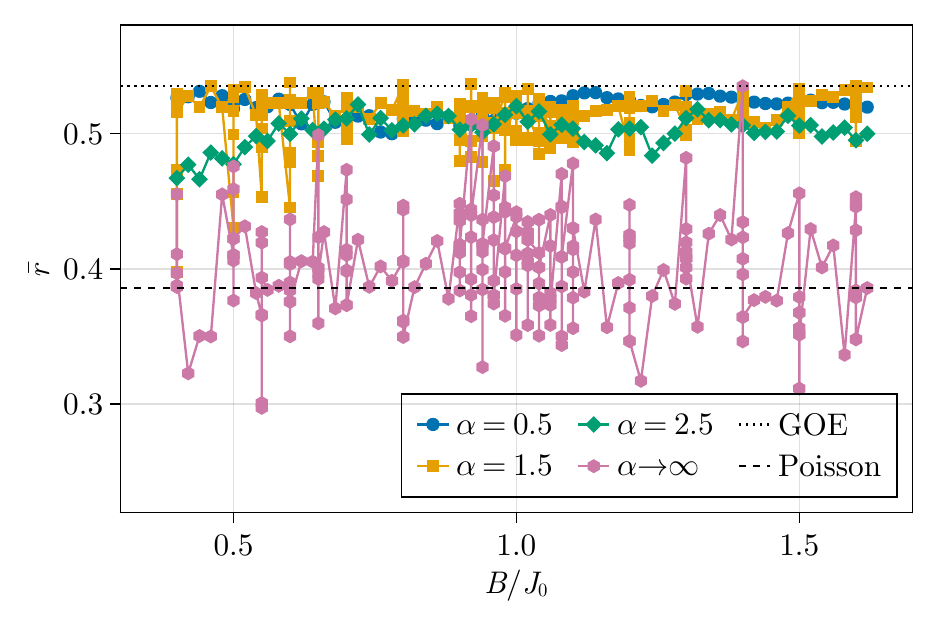}
    \caption{Mean adjacent level spacing ratio, $\bar{r}$, defined in Eq.~\eqref{eq:MeanAdjacentLevelSpacingRatio}, for the Hamiltonian in Eq.~\eqref{eq:HamiltonianPowerLawTFIM}, for a few values of $\alpha$, as functions of the magnetic field, $B$, for $n = 13$. The dotted line corresponds to $\bar{r} \approx 0.535$ for a typical chaotic Hamiltonian sampled from the Gaussian orthogonal ensemble and the dashed line corresponds to $\bar{r}\approx 0.386$ for a Hamiltonian with Poisson level spacing statistics.}
    \label{fig:MeanAdjacentLevelSpacingRatio}
\end{figure}

In Fig.~\ref{fig:MeanAdjacentLevelSpacingRatio} we plot $\bar{r}$ for TFIMs governed by Eq.~(\ref{eq:HamiltonianPowerLawTFIM}) for several values $\alpha$ and $n = 13$ spins. We see high chaoticity, approaching the expected value for the GOE, for $0.5\le\alpha\le 2$. For large $\alpha$ we approach level-repulsion associated with the Poisson statistics of regular systems. This has implications for equilibration associated with quench dynamics, which is typically understood using versions of the eigenstate thermalization hypothesis (ETH) \cite{alba2015eigenstate, khodja2015relevance, d2016quantum, deutsch2018eigenstate,dymarsky2018subsystem, murthy2019bounds, noh2021eigenstate}. 

The degree of chaos and ergodicity generated by dynamics determines the degree to which there is many-body ``scrambling" \cite{Furuya1998, Lakshminarayan2001, Hayden2007, Sekino2008, Liu2018}. Due to the growth of entanglement associated with scrambling, the ETH implies that the expectation values of local observables at sufficiently long times are well-approximated using thermal expectation values, with the temperature determined by the initial energy of the system. The exact time needed for equilibration depends on the Hamiltonian and its associated degree of chaos, the initial state, and other details of the model \cite{d2016quantum}. The most extreme equilibration occurs when the dynamics are maximally scrambling, and the many-body state has statistical properties of a typical state in Hilbert space \cite{Leb1993, Gol2006, Leb2007, Bar2009, Dub2012, San2012, Fac2015, d2016quantum, Mi2022} with near maximum entanglement \cite{Page1993, Foong1994, Sanchez1995, Sen1996}. In that case, the local reduced states are the infinite-temperature states or maximally mixed states of appropriate dimension. We expect that the most interesting circumstances for quantum advantage in quantum simulation would correspond to those cases where a sufficient amount of entanglement is generated to make classical simulation of the full many-body state intractable, but not enough to render the local reduced states close to the maximally mixed state.

We will focus on the problem of simulating critical properties of the TFIM. Models described by Eq.~(\ref{eq:HamiltonianPowerLawTFIM}) have a ground-state quantum phase transition around $B/J_0 \approx 1$ \cite{Kar2017, halimeh2017dynamical, halimeh2017prethermalization, Zun2018, piccitto2019dynamical,Tit2019}. For long-range interactions ($0\lesssim \alpha \lesssim 2$), the ground-state quantum phase transition survives as a thermal equilibrium phase transition at nonzero temperatures. Its properties can be accessed through long-time-averaged expectation values of local observables after a quench. The time and space averaged expectation values show a phase transition, which is considered a ``dynamical quantum phase transition" (DQPT) \cite{halimeh2017dynamical, halimeh2017prethermalization, zhang2017observationmany, Zun2018, piccitto2019dynamical}, which we will study in Sec~\ref{sec:DynamicalQuantumPhaseTransition}. For short-range interactions ($\alpha \gtrsim 2$) this is not the case; the thermal phase transition and dynamical quantum phase transition are absent. Nevertheless, signatures of the ground-state quantum phase transition persist, such as the critical behavior of the short-range spin-spin correlations near the critical point \cite{Kar2017, Tit2019}, as we will study in Sec.~\ref{sec:CorrelationLengthFromQuench}. These DQPTs are related to DQPTs occurring due to nonanalytic behavior of a Loschmidt echo amplitude  \cite{jurcevic2017direct, halimeh2017dynamical, halimeh2017prethermalization, Zun2018, piccitto2019dynamical}. Moreover, the DQPTs arising due to nonanalyticies in the Loschmidt amplitudes are due to both avoided crossing in the entanglement spectrum and qualitative changes in the overlap of the time evolved state with the initial state \cite{de2021entanglement}. In all cases, thermal and similar states can be effectively accessed through quench dynamics.

%
\subsection{Matrix product states: representations and dynamics}
\label{sec:MatrixProductStates}
%
Since we are considering one-dimensional systems, we use MPS to represent the quantum state \cite{perez2006matrix, schollwock2011density, vidal2003efficient,vidal2004efficient, daley2004time, white2004real, haegeman2011time, haegeman2016unifying,vanderstraeten2019tangent, zhou2020limits, cirac2021matrix}. We employ open boundary conditions, hence the MPS representation of the $n$ qubit system is written as
\begin{equation}
  \ket{\psi} = \sum_{z \in \left\{\uparrow, \downarrow \right\}^n} \sum_{j_1, \dots, j_{n}} (\mathcal{M}_1^{z_1})_{ j_1}(\mathcal{M}_2^{z_2})_{j_1 j_2}\dots(\mathcal{M}_n^{z_n})_{j_n} \ket{z} \ ,
  \label{eq:MatrixProductState}
\end{equation}
where $\left\{ \mathcal{M}_{\ell}^{z_\ell}\right\}$ are up to $\chi \times \chi$ matrices (the subscript $\ell$ labels the position, and the superscript $z_\ell$ labels the basis state). For open boundary conditions $\mathcal{M}_{1}^{z_1}$ and $\mathcal{M}_{n}^{z_n}$ are row and column vectors, respectively. The ``bond dimension" $\chi \le 2^{\lfloor n/2 \rfloor}$, and $\ket{z}$ is an $n$-qubit computational basis state \cite{perez2006matrix, schollwock2011density, vidal2003efficient,vidal2004efficient, daley2004time, white2004real, haegeman2011time, haegeman2016unifying, vanderstraeten2019tangent, cirac2021matrix}. While this representation can be used for an arbitrary quantum state with $\chi$ potentially scaling exponentially with $n$, it is useful for classical simulation when it can efficiently approximate the desired quantum state, meaning that the bond dimension$\chi$ scales polynomially with $n$ \cite{vidal2003efficient, vidal2004efficient}. The maximum entanglement content of a quantum state expressed in MPS form with bond dimension$\chi$ is related to $\chi$ via
\begin{equation}
    \mathcal{S}_0 = \ln(\chi),
    \label{eq:Renyi0EntropyBondDimension}
\end{equation}
where $\mathcal{S}_0$ is the R{\'e}nyi-0 (or Hartley) entanglement entropy of half of the spin array with the other half. All other R{\'e}nyi-$k$ entropies are bounded by the the R{\'e}nyi-0 entropy as $\mathcal{S}_k \leq \mathcal{S}_0$. Throughout this chapter and the next, we measure entropies in nats. The bond dimension thus acts as an entanglement measure for our classical simulation budget. Our goal is to study how well we can approximate local order parameters by truncating the bond dimension to some value maximum $\chi_{\max}$ where classical simulation is tractable. Such a truncation may lead to a poor approximation to the full many-body state (microstate) but an excellent approximation to the macrostate, as we will see.

To simulate dynamics for nearest-neighbor models we employ a 4th order time-evolving block decimation (TEBD) method \cite{vidal2003efficient,vidal2004efficient, daley2004time, white2004real}, and for long-range interacting models $\alpha \leq 3$, we use the time-dependent variational principle (TDVP) method \cite{haegeman2011time, haegeman2016unifying, vanderstraeten2019tangent,TimeEvoMPS}. In either case, we start with an initial spin coherent state in the MPS form, Eq.~\eqref{eq:MatrixProductState}. During each time step, we truncate the resultant MPS using a singular value decomposition (SVD) to bond dimension $\chi_{\max}$. In the following, we drop the subscript with the implicit assumption $\chi \equiv \chi_{\max}$. Both TEBD and TDVP have runtimes that scale polynomially in the system size $n$ and bond dimension$\chi$ as $\mathcal{O}(n \chi^3)$ \cite{vidal2004efficient, daley2004time, haegeman2011time, vanderstraeten2019tangent}. Therefore, being able to calculate quantities of interest to the desired precision using TEBD or TDVP with a polynomially scaling (with system size and time) $\chi$ implies an efficient classical simulation, regardless of the classical tractability of the full many-body wave function.

%
\subsection{Local observables and local reduced states}
\label{subsec:LocalReducedStates}
%
The expectation value of a local observable $\mathcal{A}$ is determined by the reduced density operator $\rho$ of the spins participating non-trivially in the observable~$\mathcal{A}$. Two simulations using different maximum bond dimension $\chi_1$ and $\chi_2$ typically give different density operators $\rho_1$ and $\rho_2$, and we can consider the squared difference in expectation value of the observable~$\mathcal{A}$ for the two simulations. This is upper-bounded by
\begin{eqnarray}
    \Tr\left(\left(\rho_1 - \rho_2\right) \mathcal{A}\right)^2 &\leq&  \Tr\left(\mathcal{A}^\dagger \mathcal{A}\right)  \Tr\left(\left(\rho_1 - \rho_2\right)^2\right) \nonumber \\
    &=& \Tr\left(\mathcal{A}^\dagger \mathcal{A}\right) \,  \mathcal{D}_{\mathrm{HS}}^2 \left(\rho_1, \rho_2\right),
    \label{eq:ExpectationValueBoundHSDistance}
\end{eqnarray}
where
\begin{equation}
    \mathcal{D}^2_{\mathrm{HS}}(\rho_1, \rho_2)  \equiv \Tr \left((\rho_1 - \rho_2)^\dagger (\rho_1 - \rho_2) \right)
    \label{eq:HSDistanceDefinition}
\end{equation}
is the squared Hilbert-Schmidt (HS) distance between the operators $\rho_1$ and $\rho_2$ (see Appendix~\ref{app:MarginalsExpectationValues} for additional details). Since multiple many-body pure states are consistent with the same reduced density operator, we consider the former to be microstates and the latter to be macrostates. 

%
\section{Dynamical quantum phase transition in transverse field Ising models}
\label{sec:DynamicalQuantumPhaseTransition}
%
We focus on a DQPT associated with long-range interacting one-dimensional TFIMs as in Eq.~\eqref{eq:HamiltonianPowerLawTFIM}. The phases can be characterized by dynamical order parameters that is the long-time average of expectation values of local observables. The order parameter characterizes two phases where the $\mathbb{Z}_2$ symmetry is broken and dynamically restored, respectively \cite{Yuz2006, Sci2010, zhang2017observationmany, Zun2018, piccitto2019dynamical, lang2018dynamical}. Phenomenology associated with this order parameter description has been studied theoretically in Refs. \cite{halimeh2017dynamical, halimeh2017prethermalization, zauner2017probing, lang2018dynamical, Zun2018, piccitto2019dynamical} and observed in a trapped ion quantum simulator with upto $53$ ions \cite{zhang2017observationmany}.

To observe the DQPT we initialize the state as $\vert \psi (0) \rangle = \vert\hspace{-0.1cm} \uparrow_z \rangle^{\otimes n}$, corresponding to one of the two degenerate ground states of the Hamiltonian with $B/J_0 = 0$ \cite{zhang2017observationmany, Zun2018, piccitto2019dynamical, lang2018dynamical}. This choice explicitly breaks the $\mathbb{Z}_2$ symmetry of the system associated with a global $\pi$-rotation about the $x$-axis. We then quench to different finite values of $B/J_0$ and let the system evolve accordingly. For $\alpha \lesssim 2$ there exists a critical value of the quench parameter $B/J_0$ that separates two different phases of the asymptotic long-time quantum state in the thermodynamic limit \cite{Zun2018, piccitto2019dynamical, lang2018dynamical, halimeh2017dynamical}, corresponding to the $\mathbb{Z}_2$ symmetry remaining broken ($B/J_0 < (B/J_0)_{\mathrm{c}}$) and being restored ($B/J_0 > (B/J_0)_{\mathrm{c}}$). A dynamical order parameter used to probe this phase transition is the time-averaged $z$-magnetization
\begin{eqnarray}
    M_{z}(n, t)&=& \frac{1}{t} \frac{1}{n} \sum_{\ell=1}^n \int_{0}^{t} \mathrm{d} t'  \langle \psi(t') \vert \sigma_{\ell}^{z} \vert \psi(t') \rangle \nonumber  \\
    &\equiv&    \frac{1}{t} \frac{1}{n}  \int_{0}^{t} \mathrm{d} t' \langle 2S_{z} \rangle (t') ,
  \label{eq:MzOrderParameter}
\end{eqnarray}
which, in the thermodynamic limit $n \to \infty$ and long-time limit $t \to \infty$, is zero in the symmetry-restored phase and nonzero in the symmetry-broken phase.

A semiclassical picture of this behavior is the following. The two ferromagnetic states can be thought of as degenerate local minima of the energy landscape, and they are separated by an energy barrier. For $B/J_0 < (B/J_0)_{\mathrm{c}}$ the time-evolved state remains confined to one side of the barrier and tunneling to the other side is suppressed in the thermodynamic limit, giving rise to a time-averaged non-zero magnetization. For $B/J_0 > (B/J_0)_{\mathrm{c}}$ the state is above the separatrix orbit and is no longer confined by the energy barrier to a single side. Oscillations occur between the two ferromagnetic states, resulting in a vanishing time-averaged magnetization in the long-time limit \cite{chinni2021effect, munoz2020simulation}.

A more convenient probe of the DQPT is the long-time averaged two-spin correlation,
\begin{eqnarray}
    M_{zz}(n, t)
    &
    =& 
    \frac{1}{t}
    \frac{1}{n^2}
    \sum_{\ell_1=1}^n \sum_{\ell_2=1}^n
    \int_{0}^{t} \mathrm{d} t'
    \langle \psi(t') \vert
    \sigma_{\ell_2}^{z} \sigma_{\ell_1}^{z}
    \vert \psi(t') \rangle \nonumber
    \\  &
    \equiv &
    \frac{1}{t}
    \frac{1}{n^2}
    \int_{0}^{t} \mathrm{d} t' \langle 4S_{z}^{2} \rangle (t')
    ,
  \label{eq:MzzOrderParameter}
\end{eqnarray}
which in the thermodynamic limit and long-time limit exhibits a minimum and discontinuity in the slope as a function of $B/J_0$ at the phase transition. At finite system size $n$, the discontinuity in the slope is smoothed out \cite{zhang2017observationmany}. The minimum tends to be less sensitive to taking the long-time limit than the phase transition in $M_z(n,t)$, hence it is a more convenient observable to use to identify the critical parameters. We elaborate further on the semiclassical picture for $M_{z}$ and $M_{zz}$ using the case of $\alpha = 0$ in Appendix~\ref{app:DQPTSimplePicture}.

We will focus on the case of $\alpha = 1.5$ because it is phenomenologically far from both the collective-spin regime ($\alpha = 0$) and the one-dimensional nearest neighbor model ($\alpha \to \infty$). We use MPS with TDVP as described in Sec.~\ref{sec:MatrixProductStates} to calculate the time evolution of the quantum state and the observables $M_z(n,t)$ (Eq.~\eqref{eq:MzOrderParameter}) and $M_{zz}(n,t)$ (Eq.~\eqref{eq:MzzOrderParameter}) to extract the critical properties in the thermodynamic limit and long-time limit. We will show that estimating these quantities is significantly easier to classically simulate in terms of the required maximum bond dimension of the MPS, compared to obtaining the full quantum state. We shall see that the robustness of the order parameters to bond dimension truncation arises because they involve spatial and temporal averaging and exhibit fast equilibration, which allows for accurate estimates at relatively short times.

%
\section{Microstates and macrostates}
\label{sec:QuenchDynamicsMicroMacro}
\begin{figure}[htbp]
   \centering
   (a){\includegraphics[width=0.44\textwidth]{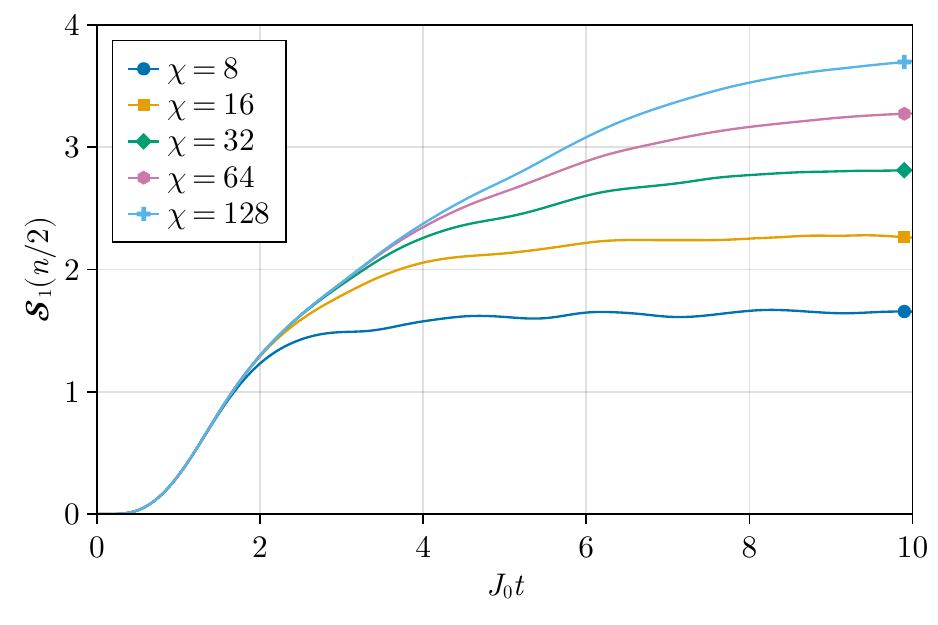}\label{fig:HSD}}
   (b){\includegraphics[width=0.44\textwidth]{figure_5_2a.pdf}\label{fig:EE}}
   \caption{Approximation to full many-body states and reduced states and entanglement in the many-body states. (a) The squared HS distance, Eq. (\ref{eq:HSDistanceDefinition}), for 2-site reduced density operator for $\chi_1 = 64$, and $\chi_2 = 128$. Other simulation parameters $n=50, \alpha=1.5, B/J_0 = 1, J_0 \Delta t = 10^{-2}$. The horizontal bar corresponds to the average-over-sites result for random MPS, sampled from the Gaussian tensor ensemble. The time- and site-averaged reduced density operator has almost 6 orders of magnitude smaller squared HS distance squared compared to the full state. (b) Half-system entanglement entropy for different bond dimension simulations. When a curve deviates from the rest gives an estimate of when the associated bond dimension is no longer sufficient to track the entanglement of the many-body state.}
   \label{fig:ApproximationsEntanglement}
\end{figure}

%
In the study of the dynamics of quantum quenches, we begin by comparing the tasks of approximating the full state (the microstate) and approximating space- and time-averaged reduced density matrices (the macrostates). To do this, we consider the squared overlap between the states $\ket{\psi_{\chi_1}(t)}$ and $\ket{\psi_{\chi_2}(t)}$ generated by our MPS simulations with maximum bond dimension$\chi_1$ and $\chi_2$ respectively,
\begin{equation}
    1 - \left| \braket{\psi_{\chi_1}(t)}{\psi_{\chi_2}(t)} \right|^2 
    = \frac{1}{2}\mathcal{D}^2_{\mathrm{HS}}\big(\rho_{\chi_1}(t),\rho_{\chi_2}(t)\big),
\end{equation}
where we have related the squared HS distance (Eq.~\eqref{eq:HSDistanceDefinition}) to the squared overlap of the full many-body pure state. We show this infidelity in Fig.~\ref{fig:ApproximationsEntanglement}(a) between bond dimensions $\chi = 64$ and $\chi = 128$ for a system of $n=50$ spins, where for $J_0 t \approx 5$ we see that the infidelity is above $10^{-1}$. Moreover, we see in Fig.~\ref{fig:ApproximationsEntanglement}(b) a difference in the half-system entanglement entropy $\mathcal{S}_1$ of the states evaluated with different bond dimensions, as $\mathcal{S}_1 \leq \mathcal{S}_0 = \ln(\chi)$, showing that the fixed bond dimension simulations for $\chi \leq 64$ are insufficient for capturing the full entanglement structure of the full quantum state for $J_0 t \gtrsim 5$.

We contrast this with the squared HS distance of the 2-site reduced density operators associated with the states $\ket{\psi_{\chi_1}(t)}$ and $\ket{\psi_{\chi_2}(t)}$ after the spatial and time averaging. The HS distance serves as an upper bound for the error of observables, as in Eq.~\eqref{eq:ExpectationValueBoundHSDistance} and further discussed in Sec.~\ref{app:MarginalsExpectationValues}. In addition to the squared HS distance between the full states with $\chi = 64$ and $\chi = 128$, we show in Fig.~\ref{fig:ApproximationsEntanglement}(a) the squared HS distance for two additional different cases:
\begin{enumerate}
    \item first averaging all instantaneous two-spin reduced density operators (Site-Averaged 2-RDM), then calculating the squared HS distance;
    \item first calculating the time-averaged two-spin reduced density operators, then averaging over sites  (Time- and Site-Averaged 2-RDM), and then calculating the squared HS distance.
\end{enumerate}

Spatial and temporal averaging significantly reduce the error, and at $J_0 t \approx 5$, the squared HS distance is closer to $10^{-7}$ for the 2-site averaged reduced density operator. This indicates that spatial and temporal averaging allows for more accurate estimates of order parameters for this system than the full-state infidelity might suggest. We note that randomly generated MPS with $n=50$ sites and with bond dimensions $64$ and $128$ have (on average) a space-averaged squared HS distance of $10^{-5}$.

\begin{figure}[ht] %
   \centering
   \includegraphics[width=0.48\textwidth]{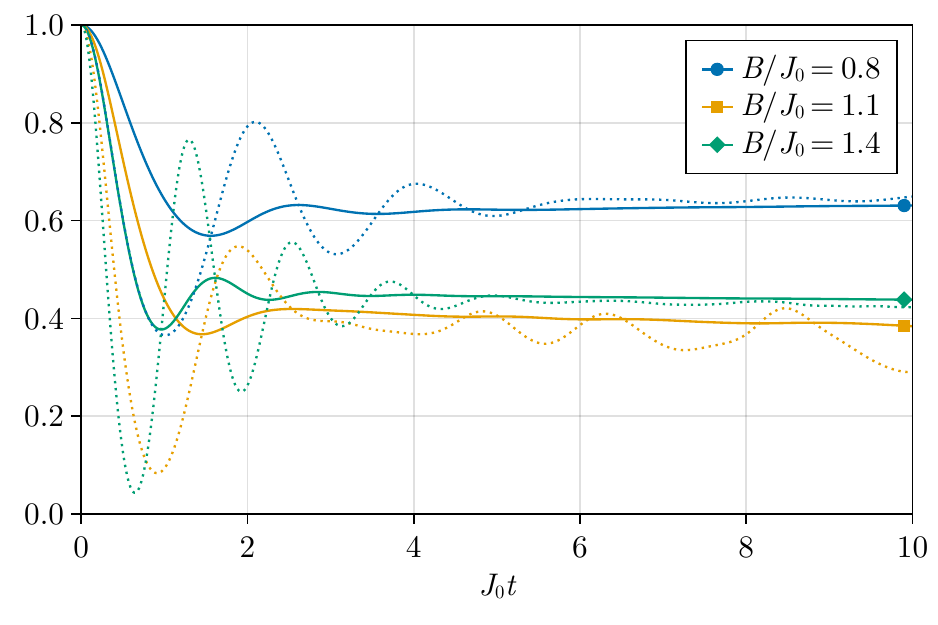}
   \caption{The evolution of the time- averaged magnetization squared $M_{zz}(n,t)$ (thick lines) and the instantaneous magnetization squared $\langle 4S_z^2 \rangle(t)/n^2$ (dotted lines). Simulation parameters are $n=50, \alpha = 1.5, J_0 \Delta t = 10^{-2}, \chi = 128$. The time-averaged observable is already close to its steady-state value by $J_0 t = 3$.}
   \label{fig:Mz}
\end{figure}

Another crucial ingredient for the classical tractability of the critical properties of the DQPT is the relatively rapid equilibration of $M_{zz}$ (we elaborate more on this property in Sec.~\ref{sec:RoleChaos}), allowing us to estimate long-time-averaged expectation values from relatively short time simulations. To illustrate this property, we visualize in Fig.~\ref{fig:Mz} the behavior of both $\langle 4 S_{z}^2 \rangle(t)$ and $M_{zz}(t)$ for different quench parameters $B/J_0$ starting from the initial state $\ket{\uparrow_z}$. After $J_0 t \approx 3$, the instantaneous observable oscillates close to a fixed value and the time-average approaches this value steadily. As we will demonstrate next, this allows for the extraction of critical properties of the DQPT using simulations with $3 \leq J_0 t \leq 10$ and with surprisingly low bond dimensions.
%
%
\section{Estimating the critical point of a DQPT using the minimum of $M_{zz}(t)$}
\label{sec:DQPTmin}
\begin{figure}[htbp] %
   \centering
   (a){\includegraphics[width=0.44\textwidth]{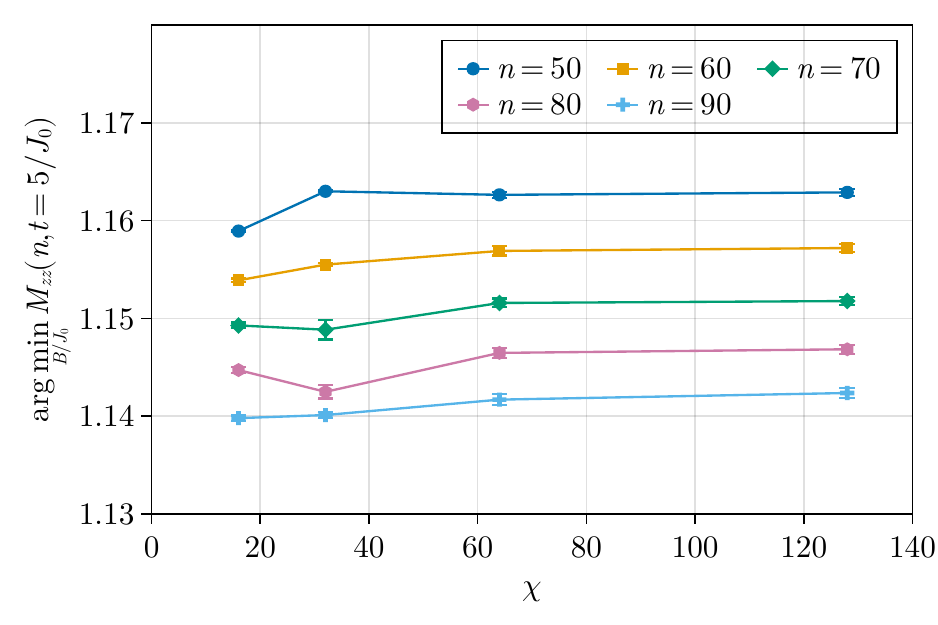}}
   (b){\includegraphics[width=0.44\textwidth]{figure_5_4a.pdf}}
   \caption{Estimating the critical point using the location of the minimum of $M_{zz}$ as a function of $B/J_0$.(a) $M_{zz}$ versus $B/J_0$ for $n=50$, $\alpha = 1.5$, $J_0 t = 5$, and different bond dimension$\chi$. (b) Fits of the location of the minimum of $M_{zz}$ with respect to $B/J_0$ for $\alpha = 1.5$, $J_0 t = 5$, different $n$ and bond dimension$\chi$; error bars are smaller than the marker size. These plots show the robustness to MPS truncation in the estimation of the minimum in the order parameter, even for a relatively large number of spins.}
   \label{fig:DQPTmin1}
\end{figure}

%
As a first demonstration of the robustness of estimating critical properties, we estimate the critical transverse field magnitude $(B/J_0)_{\mathrm{c}}$ by identifying the minimum in the curve of $M_{zz}$ as a function of $B/J_0$. We show the behavior of $M_{zz}(n,t)$ in Fig.~\ref{fig:DQPTmin1} at a final quench time of $J_0t = 5$ for $n=50$ for different bond dimensions. By fitting the points around the minimum with a parabola, we can extract the minima as a function of $n$ and $\chi$, an shown in Fig.~\ref{fig:DQPTmin1}. We observe that the predicted $(B/J_0)_{\mathrm{c}}$ has a weak dependence on the bond dimension for $\chi \geq 16$ for a fixed system size $n$. Additionally, we see an almost linear behavior as a function of system size for a fixed bond dimension, indicating that we are still seeing significant finite size effects for system sizes $n\leq 90$. We emphasize even though the overlap between the MPS with bond dimensions $\chi$ and $2 \chi$ is very small, they still give very similar estimates for the location of the minimum. As an example, for $n=90$ the overlap squared between states with bond dimension$\chi =16$ and $\chi = 32$ is $< 10^{-6}$ (not shown) but their estimates of the location of the minimum are within $10^{-2}$ of each other.

\begin{figure}[htbp] %
   \centering
   (a){\includegraphics[width=0.44\textwidth]{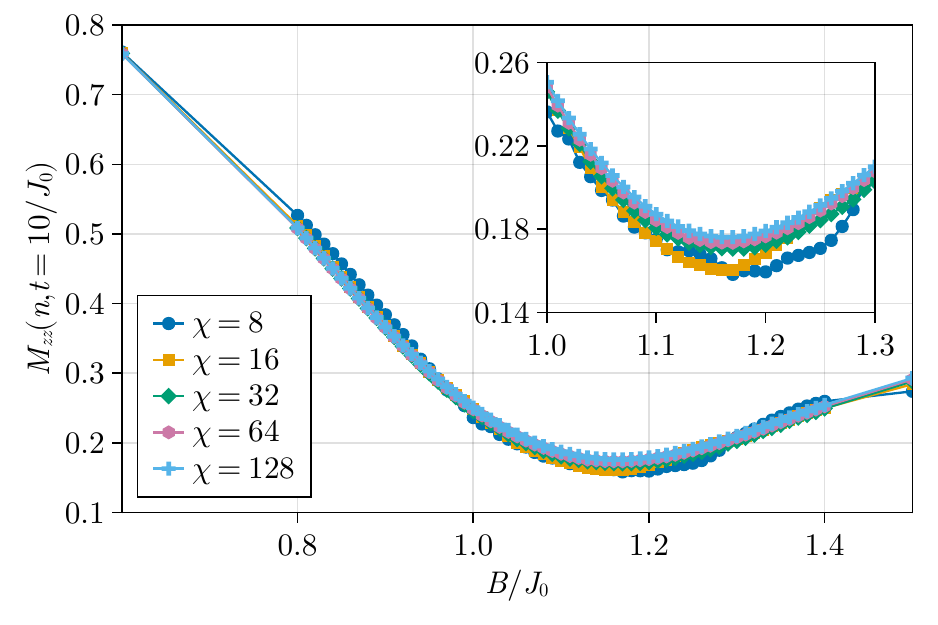}}
   (b){\includegraphics[width=0.44\textwidth]{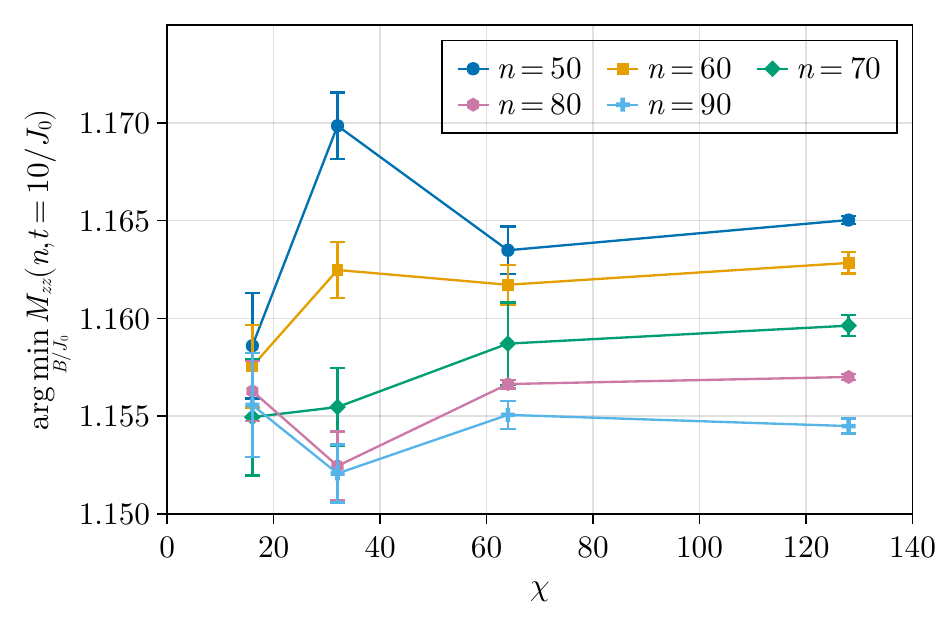}}
   \caption{ (a) $M_{zz}$ versus $B/J_0$ for $n=50$, $\alpha = 1.5$, $J_0 t = 10$ and different bond dimension$\chi$. (b) Fit of the location of the minimum of $M_{zz}$ with respect to $B/J_0$ for $\alpha = 1.5$, $J_0 t = 10$, shown for different system sizes $n$ and bond dimensions $\chi$. These plots show the robustness to MPS truncation in the estimation of the minimum in the order parameter, even for a relatively large number of spins and longer times.}
   \label{fig:DQPTmin2}
\end{figure}

Repeating the analysis for a later time $J_0 t = 10$, we observe that changing the final time $J_0 t$ changes the overall value of $M_{zz}(n,t)$ but does not significantly change the location of the minimum of $M_{zz}$ for bond dimensions $\chi \geq 16$. In Fig.~\ref{fig:DQPTmin2}, we show the behavior of $M_{zz}$ as a function of $B/J_0$ at this later time. The variation for the location of the minimum between different system sizes is smaller than observed for $J_0 t = 5$ in the main text, however, we observe similar behavior: there is a very small dependence on the bond dimension, and the location of the minimum steadily decreases with system size for all bond dimensions $\chi \geq 32$. Thus our results suggest that a bond dimension of $\chi = 16$ and a quench time of $J_0 t = 5$ should be sufficient for identifying the critical point via the minimum of $M_{zz}$ at arbitrary system sizes.

\begin{figure}[htbp] %
   \centering
   \includegraphics[width=0.46\textwidth]{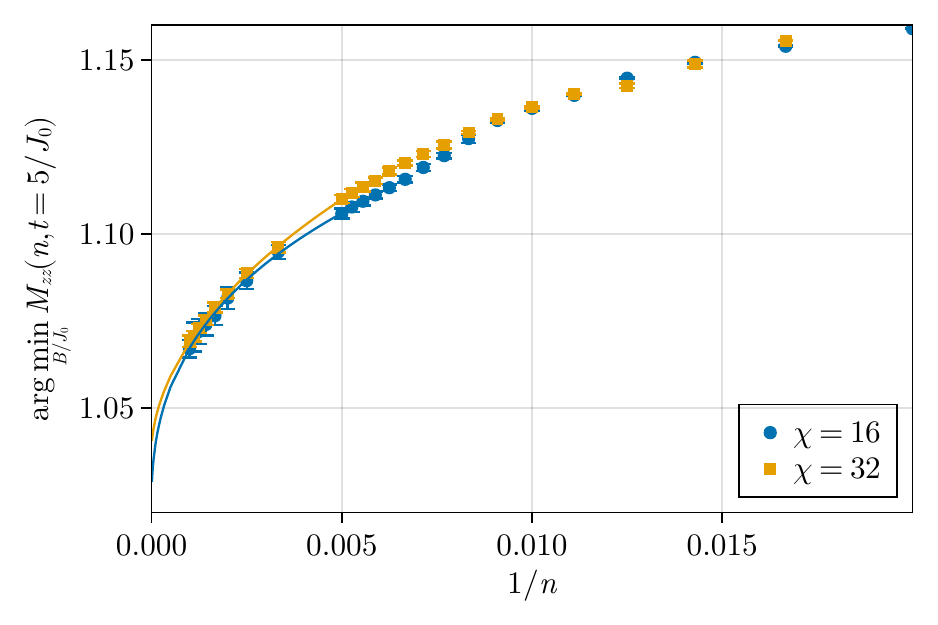}
   \caption{Fits of the the location of the minimum of $M_{zz}$ with respect to $B/J_0$ for $\alpha = 1.5$, $J_0 t = 5$, different $n$ and bond dimensions, $\chi =16$ and $32$. The solid lines correspond to a fit to the function $a + b n^{-c}\ln(n)$ using $n \geq 200$, with $ a = 1.02 \pm 0.04, b = 0.3 \pm 0.3, c = 0.5 \pm 0.3$ for $\chi = 16$ and $ a = 1.04 \pm 0.02, b = 0.5 \pm 0.4, c = 0.7 \pm 0.2$ for $\chi = 32$. The estimates for the location of the minimum in the thermodynamic limit are within the uncertainty of the fits.}
   \label{fig:DQPTminLargeFixedBonddim}
\end{figure}

Using the above hypothesis, in Fig.~\ref{fig:DQPTminLargeFixedBonddim} we show our results for the minimum of $M_{zz}$ up to $n=1000$ spins using $\chi = 16$ and $\chi = 32$, where the data at large sizes allows us to extrapolate the location of the critical point in the thermodynamic limit.

%

%
\section{Finite scaling analysis and universality}
\label{sec:FiniteSizeScaling}
%
To further study the critical behavior near the DQPT in the thermodynamic limit, we perform a finite-size scaling (FSS) analysis. To do so we use the data collapse method \cite{New1999,Bin2010} implemented in the \textit{pyfssa} scientific software package \cite{And2015,Mel2009} on the order parameters $M_z$ and $M_{zz}$, and we consider two different FSS ansatz.

The first is an ``equilibrium'' ansatz, where we assume that the behavior of $M_{z}$ and $M_{zz}$ near the DQPT for sufficiently long quench times $J_0 t$ only depends on the system size $n$ and not on the quench duration. While this seems reasonable given our observation from the previous section that $M_{zz}$ thermalizes rapidly and exhibits a weak dependence on the quench time $J_0 t$, the long-time limit of the FSS ansatz assumes global equilibration and not only local equilibration, which is unlikely to be satisfied for our relatively short time simulations of $J_0 t \ll n$. The second is a ``nonequilibrium'' FSS ansatz that reflects the fact that our simulations are relatively short-time simulations with $J_0 t \ll n$. Nonequilibrium data for FSS is known to be possible in a Monte-Carlo setting \cite{ozeki2007nonequilibrium}.

%
\subsection{Finite scaling analysis for $M_z$}
\label{sec:FiniteSizeScalingMz}
%

\begin{figure}[htbp] %
   \centering
   (a)
   {\includegraphics[width=0.44\textwidth]{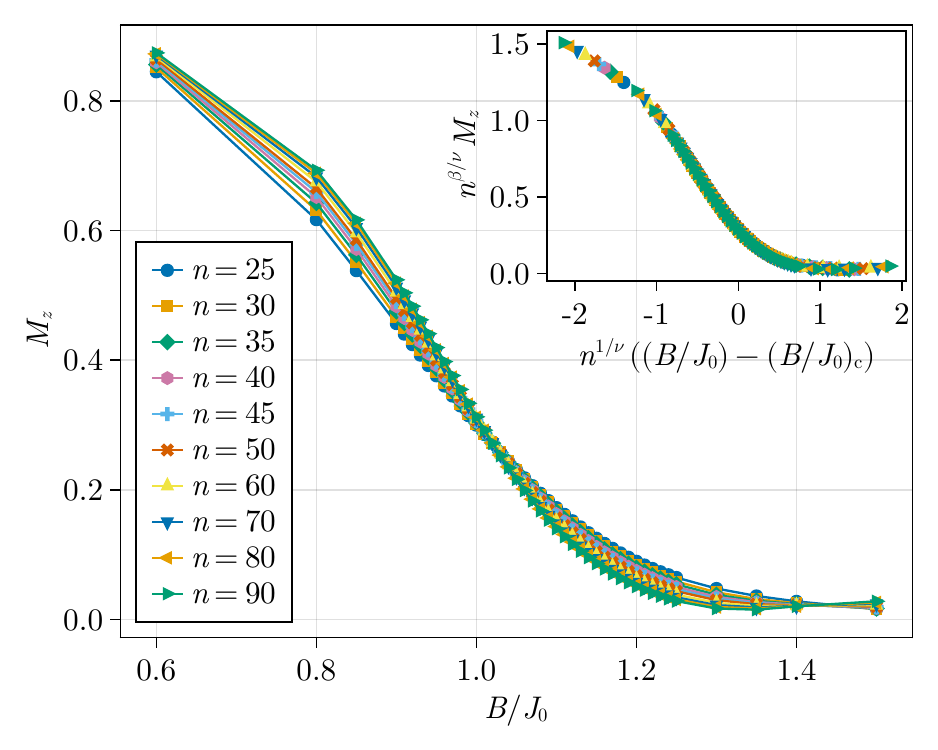} \label{fig:DataCollapseMzSpaceExample}}
   (b)
   {\includegraphics[width=0.44\textwidth]{figure_5_7a.pdf} \label{fig:DataCollapseSpaceMzBCritical}}
   (c)
   {\includegraphics[width=0.44\textwidth]{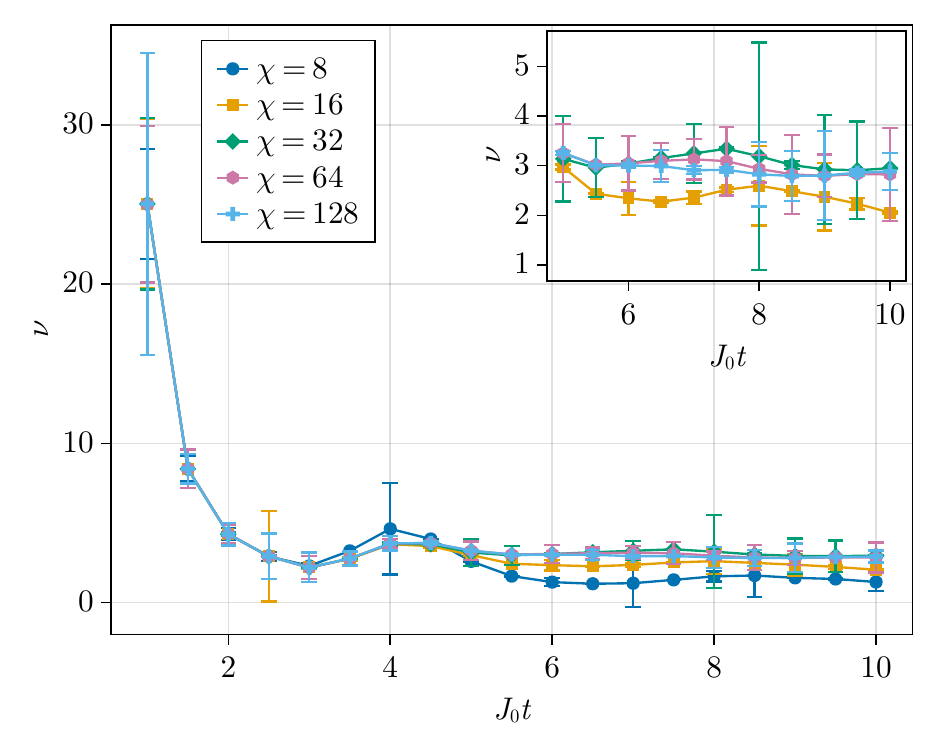} \label{fig:DataCollapseMzSpaceNu}}
   (d)
   {\includegraphics[width=0.44\textwidth]{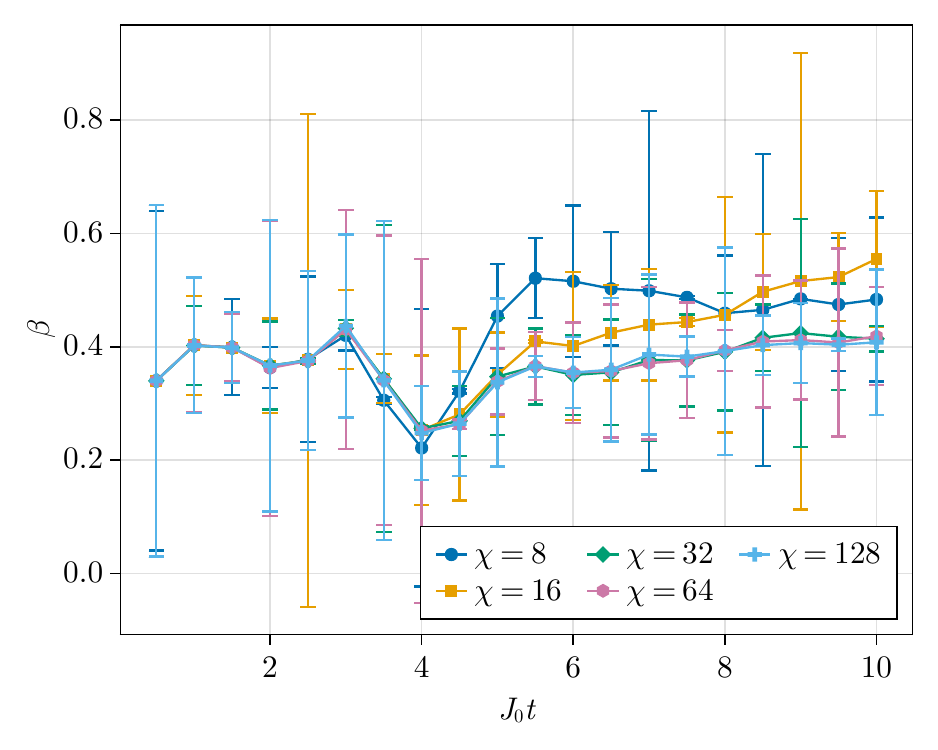} \label{fig:DataCollapseMzSpaceZeta}}
   \caption{Estimates of critical properties using the finite system size scaling ansatz [Eq.\eqref{eq:FiniteSizeScalingSpaceAnsatz}]. (a) Finite-size scaling analysis for the order parameter $M_{z}$ for simulations with parameters $\alpha = 1.5, J_0 t = 5.5, \chi = 64, J_0 \Delta t = 10^{-2}$. Using the data collapse method, the extracted critical parameters from this simulation for this total simulation time are $\left(B/J_0\right)_{\mathrm{c}} = 1.082 \pm 0.001, \nu = 3.008 \pm 0.008, \beta = 0.37 \pm 0.02$. We further study the dependence of the critical point (b) and critical exponents (c),(d),  on the total quench time for different bond dimensions $\chi \in \{8, 16, 32, 64, 128\}$. Insets show estimates for $J_0t \geq 5.0$, which are more consistent. Though the order parameter is defined in the asymptotic limit, it converges at relatively short times. The estimates of the critical point and critical exponents also converge at relatively small bond dimensions.}
   \label{fig:DataCollapseParametersMzSpace}
\end{figure}

We take as our equilibrium FSS ansatz
\begin{equation}
    M_{z}(n) = n^{-\beta/\nu} f\left(n^{1/\nu} \,
    \left(\frac{B}{J_0} - \left(\frac{B}{J_0}\right)_{\mathrm{c}}\right); J_0 t \right),
    \label{eq:FiniteSizeScalingSpaceAnsatz}
\end{equation}
where $f$ is a universal scaling function and the exponents $\beta$ and $\nu$ are the critical exponents that characterize the behavior of $M_{z}$ near the DQPT. Specifically, near the critical point and in the thermodynamic limit, the exponent $\nu$ characterizes the scaling of correlation lengths $\xi \sim | B/J_0 - (B/J_0)_{\mathrm{c}}|^{-\nu}$, and $\beta$ characterizes the scaling of $M_{z}(\infty)$ near the critical point:
\begin{equation}\label{eq:fssuniversal}
  f(x) \sim
  \begin{cases}
    |x|^{\beta} & \text{for }  x \to -\infty , \\
    \mathrm{constant} & \text{for } x \to +\infty .
 \end{cases}
\end{equation}
The data collapse method proceeds by identifying the critical parameters $\left\{ (B/J_0)_{\mathrm{c}}, \beta, \nu\right\}$ that give universal behavior for $M_{z}(n) n^{\beta/\nu}$ as a function of $n^{1/\nu} \left( B/J_0 - (B/J_0)_{\mathrm{c}} \right)$. We show an example of the results of the data collapse for a fixed $J_0 t$ in Fig.~\ref{fig:DataCollapseParametersMzSpace}(a).

We are able to perform the data collapse and identify the critical parameters for different fixed $J_0 t$ and bond dimension $\chi$, which we show in Fig.~\ref{fig:DataCollapseParametersMzSpace}(b)-(d). For short-duration quenches, $J_0t \lesssim 2$, the FSS results are inconsistent with those for longer quenches, implying that such time scales are too short to be used for FSS.

For longer duration quenches $J_0t \gtrsim 5$, the estimates of the critical parameters are more consistent with each other, exhibiting a negligible drift with $J_0 t$ and showing agreement within the error bars for bond dimensions $\chi \geq 16$ at a fixed $J_0 t$. We note that the critical point $\left(B/J_0 \right)_{\mathrm{c}}$ is already well estimated for simulation times of $J_0 t \gtrsim 3$. The estimate of $(B/J_0)_{\mathrm{c}} \approx 1.1$ is consistent with the analysis of the minimum of $M_{zz}$ as a function of $B/J_0$ performed in the previous section. This shows that relatively rapid equilibration allows us to estimate the critical point and critical exponents at short times $J_0 t \ll n$.

Furthermore, since the estimates of the critical parameters using $\chi = 16$ agree to within the error bars with the estimates using $\chi = 128$, it suggests that many-body entanglement beyond a R{\'e}nyi-0 entropy of $\mathcal{S}_0 = \ln(16)$ plays a negligible role in this phase transition.

\begin{figure}[htbp] %
   \centering
   (a)
   {\includegraphics[width=0.44\textwidth]{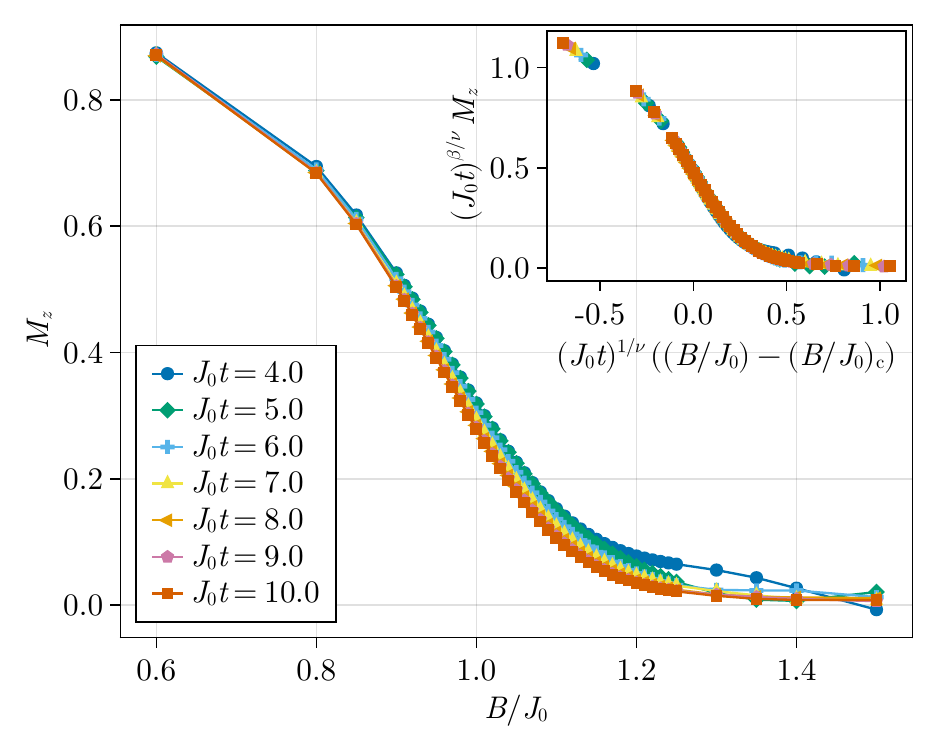} \label{fig:DataCollapseMzTimeExample}}
   (b)
   {\includegraphics[width=0.44\textwidth]{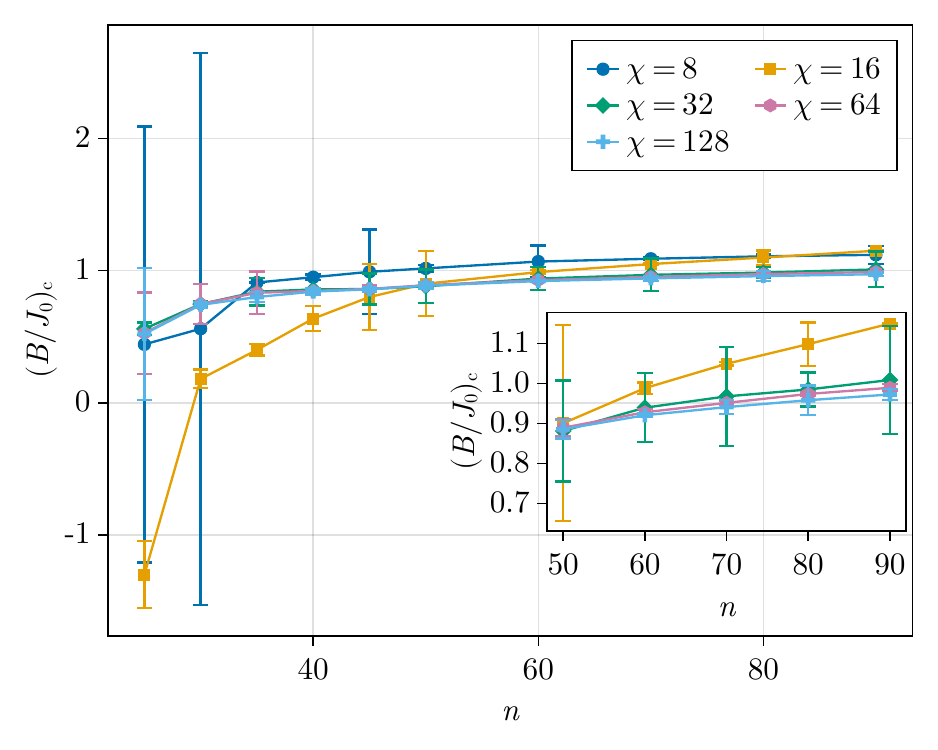} \label{fig:DataCollapseMzTimeBCritical}}
   (c)
   {\includegraphics[width=0.44\textwidth]{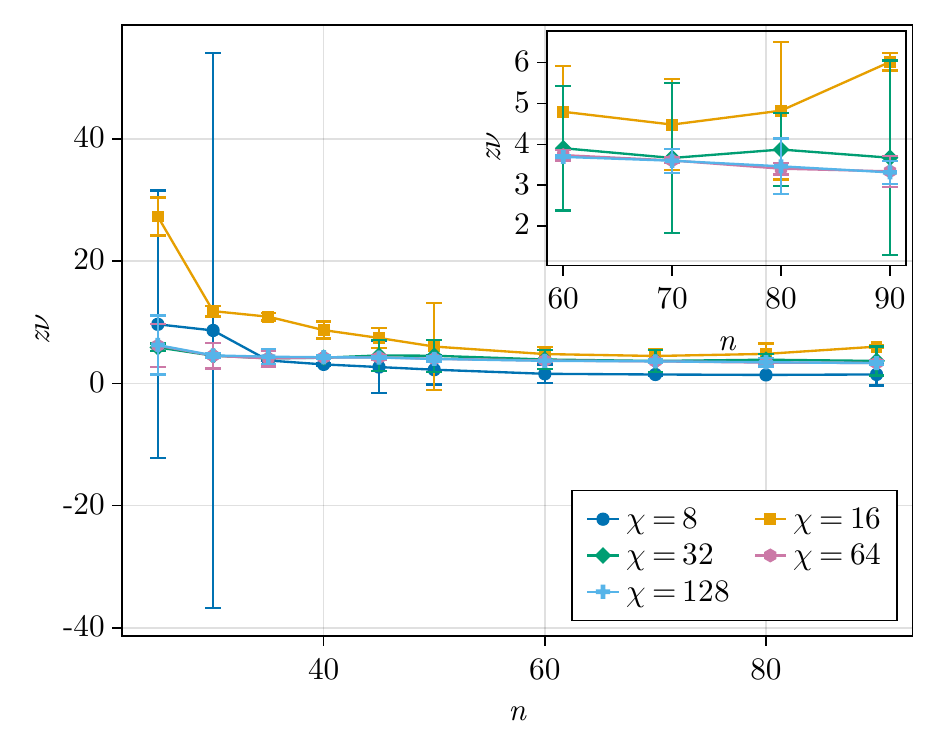} \label{fig:DataCollapseMzTimeNu}}
   (d)
   {\includegraphics[width=0.44\textwidth]{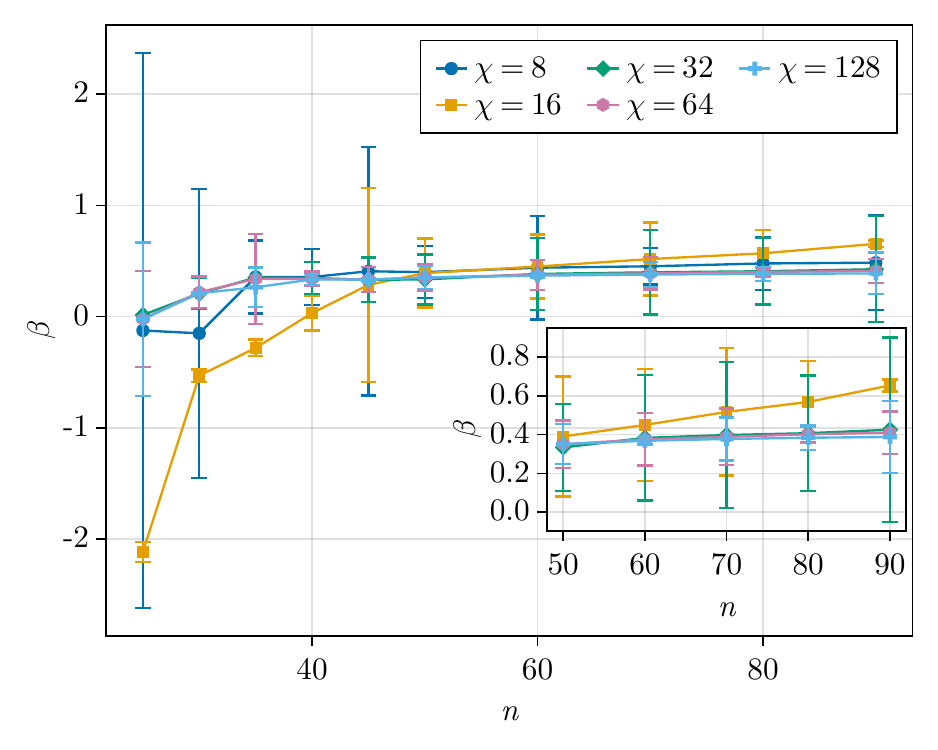} \label{fig:DataCollapseMzTimeZeta}}
    \caption{Estimates of critical properties using a finite quench duration scaling ansatz [Eq.\eqref{eq:FiniteSizeScalingTimeAnsatz}]. (a) Finite-size scaling analysis for order parameter $M_{z}$ for simulations using $\alpha =1.5, n = 80, \chi = 128, \Delta t = 10^{-2}$. Using the data collapse method, the extracted critical parameters for these simulations parameters from this simulation are $\left(B/J_0\right)_{\mathrm{c}} = 0.95 \pm 0.04, z\nu = 3.5 \pm 0.7, \beta = 0.38 \pm 0.6$. Using the data collapse method, the extracted critical point (b) and critical exponents (c), (d) as a function of system size $n$, for different bond dimensions $\chi \in \{8, 16, 32, 64, 128\}$. Insets show estimates for $n \geq 50$, which are more consistent. The estimates of the critical point and critical exponents are approximately the same across different bond dimensions.}
   \label{fig:DataCollapseParametersMzTime}
\end{figure}

We consider a non-equilibrium FSS ansatz with no system size dependence (we have now effectively taken the infinite size limit)
\begin{equation}
    M_{z}(t) = (J_0 t)^{-\beta/(z\nu)} \,
    g\left((J_0 t)^{1/(z\nu)} \left(\frac{B}{J_0} - \left(\frac{B}{J_0}\right)_{\mathrm{c}}\right);
	n \right),
    \label{eq:FiniteSizeScalingTimeAnsatz}
\end{equation}
where $g$ is a universal scaling function. Similar to our first FSS ansatz (Eq.~\eqref{eq:FiniteSizeScalingSpaceAnsatz}), near the critical point and in the infinite time limit we expect a diverging time scale $\tau \sim | B/J_0 - (B/J_0)_{\mathrm{c}}|^{-z \nu}$ with dynamical critical exponent $z \nu$. (We note that a critical slowing down is observed for the $\alpha = 0$ case as we approach the critical point.) We expect $g$ to behave in a similar manner to $f$, that is,
\begin{equation}
 g(x) \sim
    \begin{cases}
        |x|^{\beta} & \text{for }  x \to -\infty , \\
        \mathrm{constant} & \text{for } x \to +\infty .
    \end{cases}
\end{equation}
The data collapse method proceeds similar to before by identifying the critical parameters that give universal behavior for $M_{z}(t) (J_0 t)^{\beta/(z \nu)}$ as a function of $(J_0 t)^{1/(z\nu)} \left( B/J_0 - (B/J_0)_{\mathrm{c}} \right)$. We show an example of the results of the data collapse method in Fig.~\ref{fig:DataCollapseParametersMzTime}(a).

We perform the data collapse and identify the critical parameters for different system sizes $n$ and bond dimension $\chi$, which we show in Fig.~\ref{fig:DataCollapseParametersMzTime}(b)-(d). The data collapse is not as successful as for the equilibrium ansatz, suggesting that we are not as close to the steady state limit $J_0 t \to \infty$ as we are to the thermodynamic limit $n \to \infty$. For small system sizes $n < 50$ the FSS results are inconsistent with those for larger system sizes, suggesting that the assumption that $J_0 t \ll n$ is being violated for these smaller system sizes. For larger sizes $n \gtrsim 50$ the estimates of the critical parameters are more consistent with each other, exhibiting a negligible drift with $n$ and generally showing agreement within the error bars for bond dimensions $\chi \geq 32$ at a fixed $n$.

Our simulations show that the order parameters corresponding to the DQPT can be approximated classically using modest bond dimension MPS ($\chi \ll 2^{n/2}$), with a runtime that scales as $\mathcal{O}(n \chi^3)$. The simulations allow us to not only distinguish the symmetry-broken phase from the symmetry-restored phase but to also estimate the critical transverse field and critical exponents of the phase transition. The short duration of the quenches considered in our simulations affects our analysis of both order parameters, $M_{z}$ and $M_{zz}$. The drift in critical exponents for $M_{zz}$ as a function of $J_0 t$ is more pronounced, and the finite quench duration scaling for $M_{z}$ is not as successful as the finite size scaling. While a careful analysis of how these estimated values converge as a function of the computational cost is beyond the scope of this work, we believe that these methods can be used to quantitatively study the properties of the phenomenon without the overhead of an exponentially large state space. A detailed understanding of the critical behavior near the DQPT across different values of $\alpha$ remains an avenue of future work.

%

%
\subsection{Finite scaling analysis for $M_{zz}$}
\label{sec:FiniteSizeScalingMzz}
%

\begin{figure}[htbp] %
   \centering
   (a) {\includegraphics[width=0.44\textwidth]{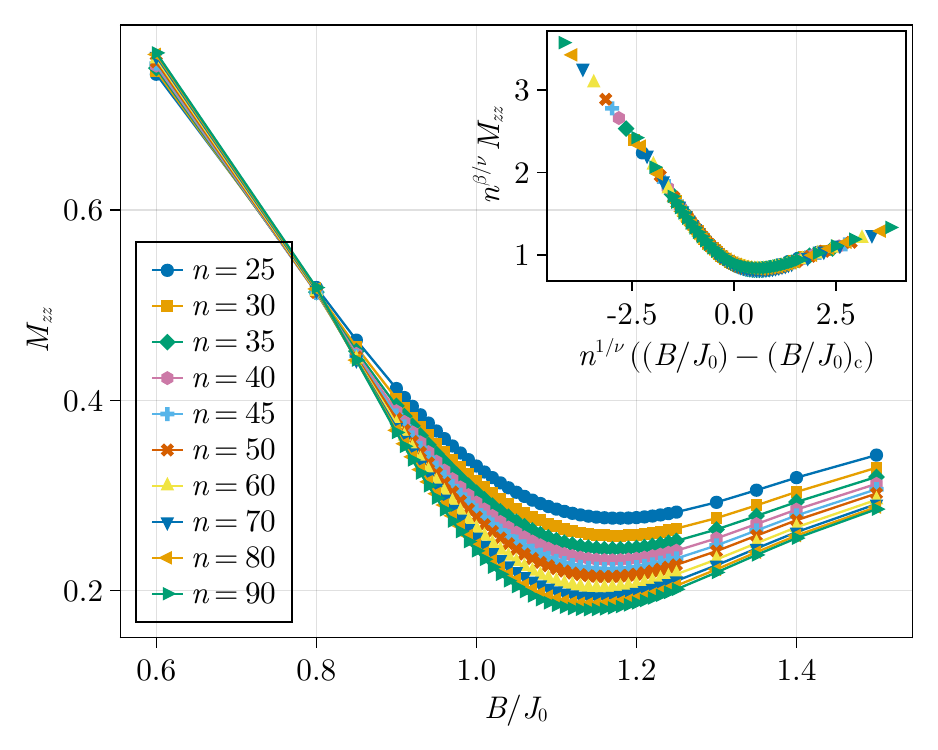} \label{fig:DataCollapseSpaceExample}}
   (b)   {\includegraphics[width=0.44\textwidth]{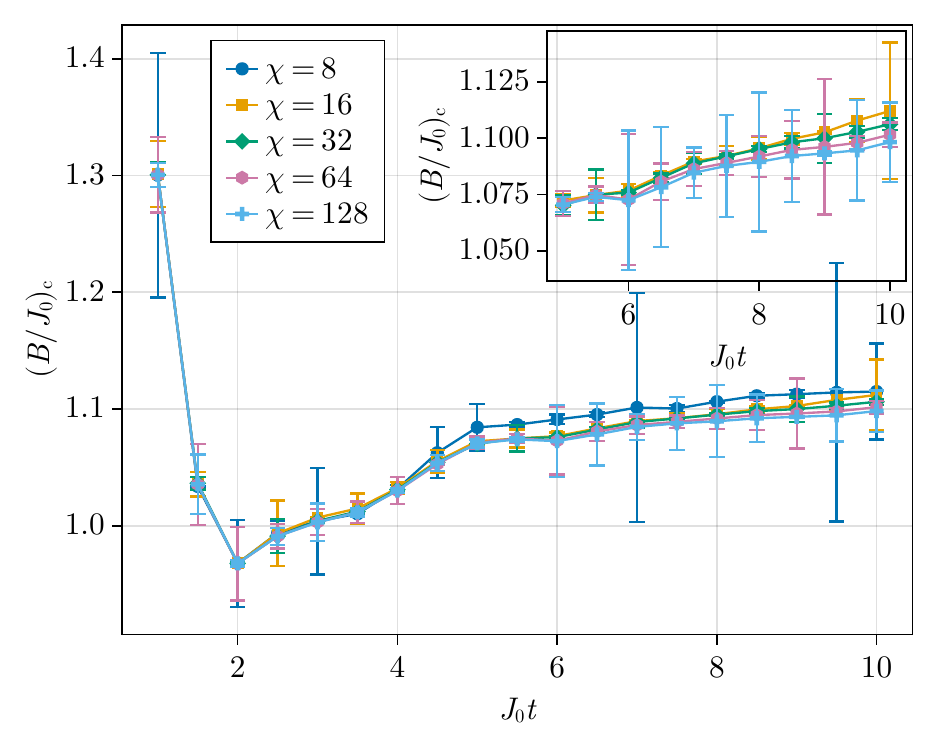} \label{fig:DataCollapseSpaceBCritical}}
   (c)   {\includegraphics[width=0.44\textwidth]{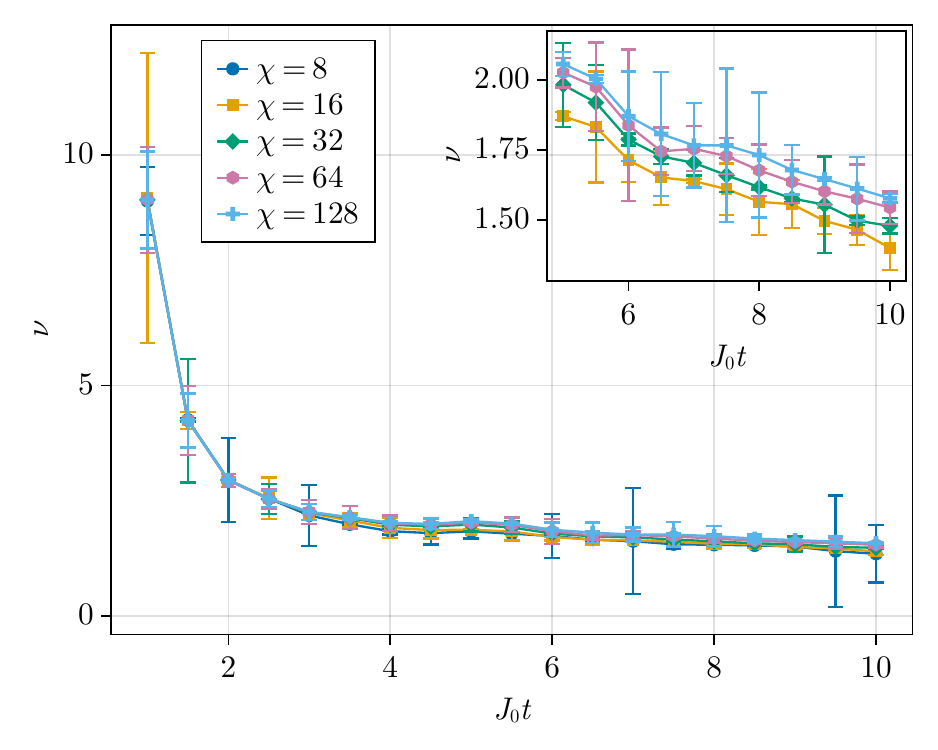} \label{fig:DataCollapseSpaceNu}}
   (d)   {\includegraphics[width=0.44\textwidth]{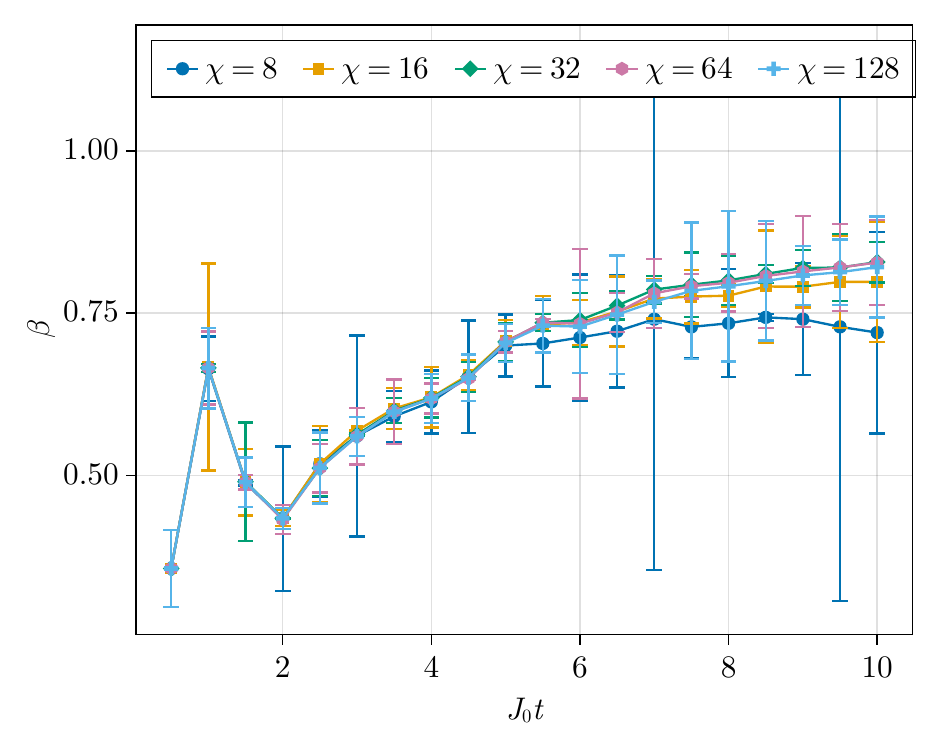} \label{fig:DataCollapseSpaceZeta}}
   \caption{ (a) Finite-size scaling analysis for order parameter $M_{zz}$ for simulations using $\alpha =1.5, J_0 t = 5, \chi = 128, J_0 \Delta t = 10^{-2}$. Using the data collapse method, the extracted critical parameters for these simulations parameters from this simulation are $\left(B/J_0\right)_{\mathrm{c}} = 1.071 \pm 0.003, \nu = 2.05 \pm 0.04, \beta = 0.70 \pm 0.03$. Using the data collapse method, the extracted critical point (b) and critical exponents (c), (d) as a function of dimensionless time, $J_0t$, for different bond dimensions $\chi \in \{8, 16, 32, 64, 128\}$. Insets show estimates for $J_0t \gtrsim 5.0$, which are more consistent. The estimates of the critical point and critical exponents are approximately the same across different bond dimensions.}
	\label{fig:DataCollapseParametersSpace}
\end{figure}

We next consider the finite-size scaling of the other order parameter, $M_{zz}$, to estimate the critical point and critical exponents of the DQPT we considered in Sec.~\ref{sec:DynamicalQuantumPhaseTransition}. As before, we use the data collapse method \cite{New1999, Bin2010}, assuming an equilibrium FSS ansatz
\begin{equation}
  M_{zz}(n) = n^{-\beta/\nu} F\left(n^{1/\nu}
    \left(\frac{B}{J_0} -
    \left(\frac{B}{J_0}\right)_{\mathrm{c}}\right); J_0 t \right),
    \label{eq:FiniteSizeScalingSpaceMzzAnsatz}
\end{equation}
with a universal scaling function $F$ (not necessarily the same function $f$ used for $M_{z}$ in the main text). We show an example of the results of the data collapse method in Fig.~\ref{fig:DataCollapseParametersSpace}(a). As with $M_{z}$, we obtain estimates of the critical parameters. In particular we show critical point estimates in Fig.~\ref{fig:DataCollapseParametersSpace}(b), and the estimates of critical exponent $\nu$ and $\beta$ in Fig.~\ref{fig:DataCollapseParametersSpace}(c) and Fig.~\ref{fig:DataCollapseParametersSpace}(d), respectively, for different times $t$ and bond dimension $\chi$. The function $F$ behaves as $F(x) \sim |x|^{\beta}$ as $|x| \to \infty$.

Similar to $M_{z}$ in the main text, for short duration quenches $J_0t \lesssim 2$ the FSS results for $M_{zz}$ are inconclusive, suggesting that local equilibration has not occurred. For longer quenches $J_0t \gtrsim 5$, the estimates are stable, but with a temporal drift. The rapid equilibration that we observed in Fig.~\ref{fig:Mz} enables us to observe characteristics of the phase transition at short times $J_0 t \ll n$.

We note that this data collapse and the one for $M_{z}$ in Sec.~\ref{sec:DynamicalQuantumPhaseTransition} find a finite $\nu$, which suggests a diverging correlation length at the critical point in the thermodynamic limit. However, Lieb-Robinson bounds for the $\alpha = 1.5$ power-law interacting system limit the rate of growth of correlations \cite{Tran2019, Chen2019}.
We expect the drift in the estimates of the critical exponents $\nu$ and $\beta$ with time to be a reflection of this growth.
An understanding of the nature of the evolving divergence and it relationship with Lieb-Robinson bounds and the thermal phase transition is an avenue for future work.

Even with the temporal drift, the estimates of the critical point $(B/J_0)_{\mathrm{c}}$ and critical exponents $\nu, \beta$ are approximately the same across different bond dimensions. Specifically estimates at $\chi = 16$ are quite close to the estimates at $\chi = 128$. Therefore, many-body entanglement beyond a R{\'e}nyi-0 entropy of $\mathcal{S}_0 = \ln(16)$ plays a negligible role in the order parameter $M_{zz}$ in this phase transition, just as they did with the order parameter $M_{z}$.

\begin{figure}[htbp] %
   \centering
   (a)   {\includegraphics[width=0.44\textwidth]{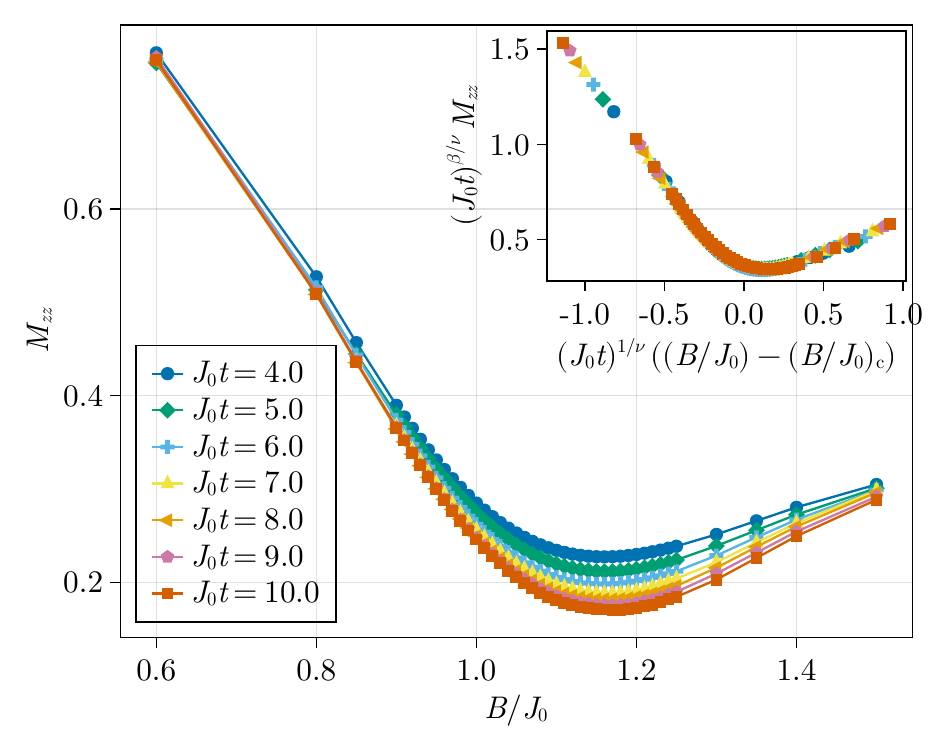} \label{fig:DataCollapseTimeExample}}
   (b)   {\includegraphics[width=0.44\textwidth]{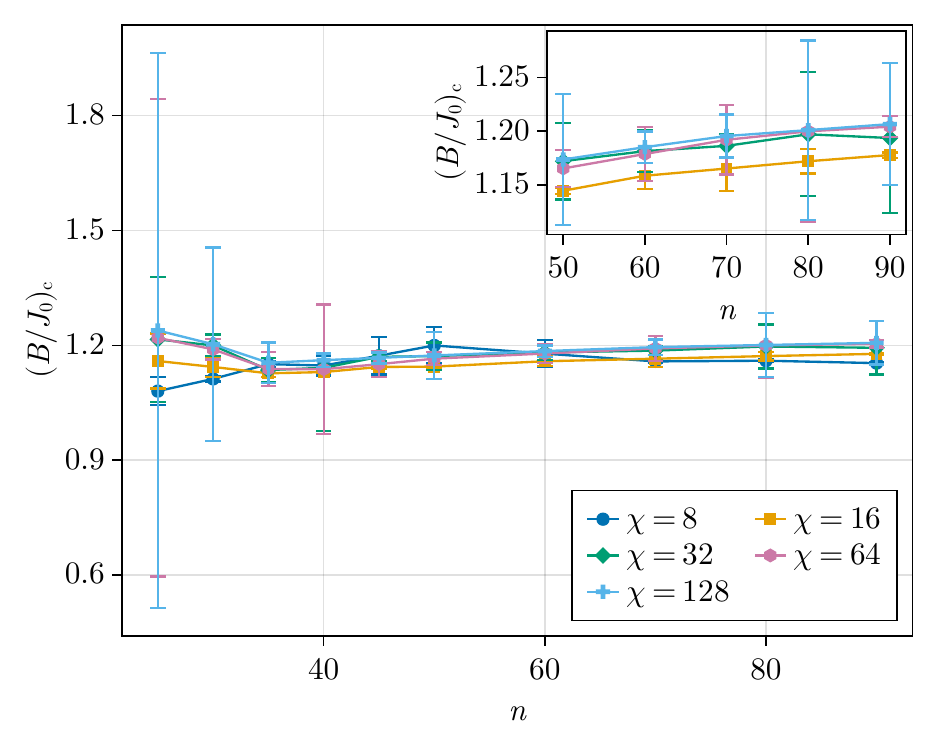} \label{fig:DataCollapseTimeBCritical}}
   (c)   {\includegraphics[width=0.44\textwidth]{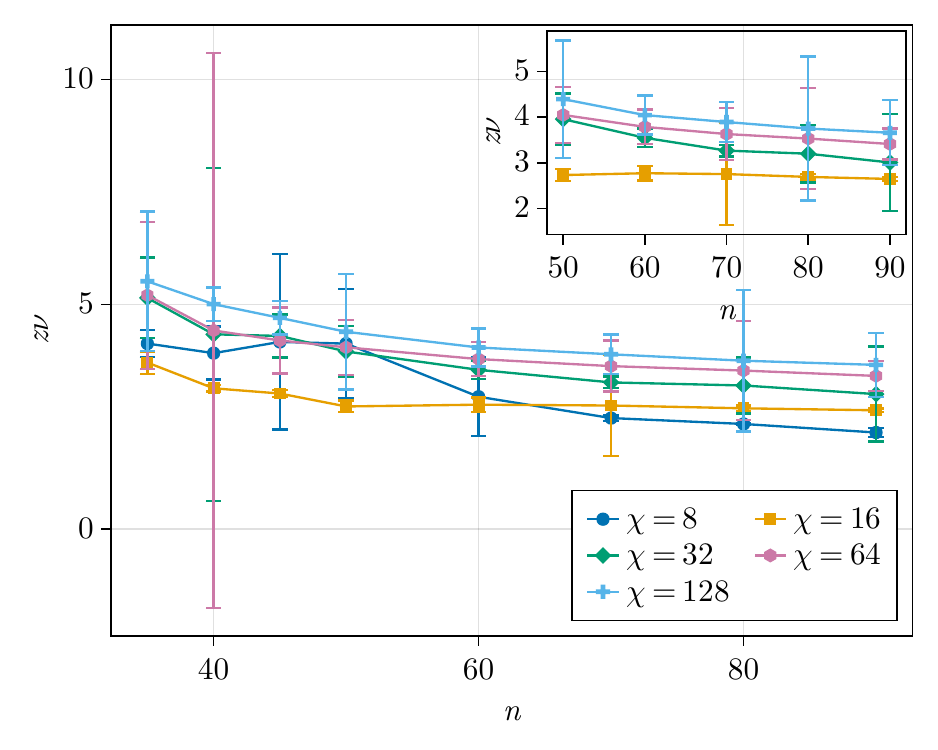} \label{fig:DataCollapseTimeNu}}
   (d)   {\includegraphics[width=0.44\textwidth]{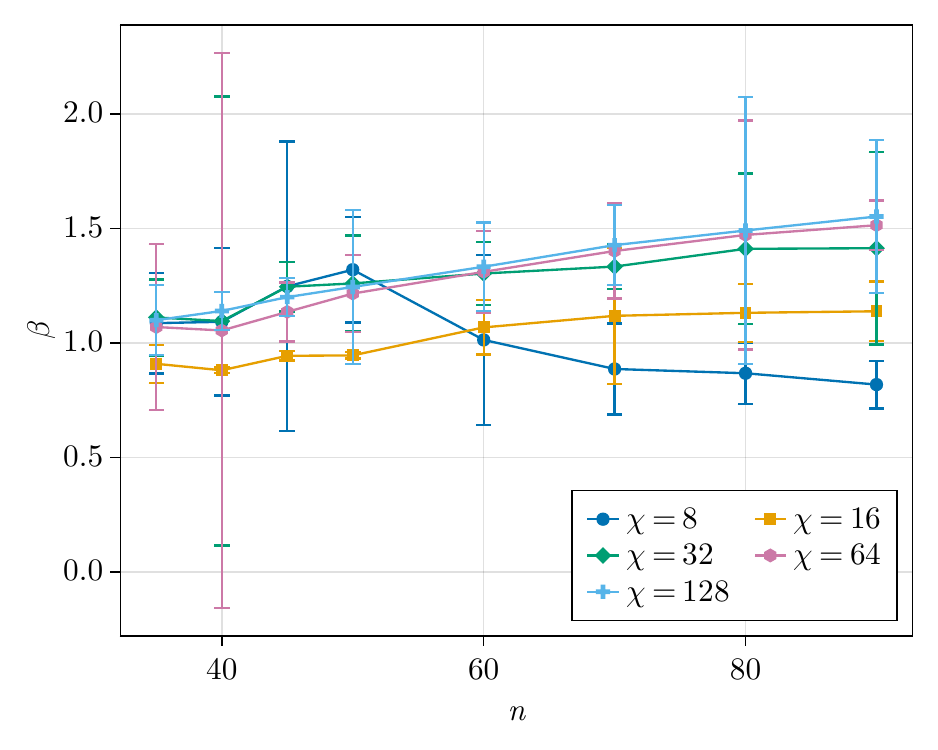} \label{fig:DataCollapseTimeZeta}}
   \caption{ (a) Finite-size scaling analysis for order parameter $M_{zz}$ for simulations using $\alpha =1.5, n = 50, \chi = 32, \Delta t = 10^{-2}$. Using the data collapse method, the extracted critical parameters for these simulations parameters from this simulation are $\left(B/J_0\right)_{\mathrm{c}} = 1.0979 \pm 0.0007, z\nu = 2.8 \pm 0.3, \beta = 0.85 \pm 0.09$. Using the data collapse method, the extracted critical point (b) and critical exponents (c), (d) as a function of system size $n$, for different bond dimensions $\chi \in \{8, 16, 32, 64, 128\}$. Insets show estimates for $n \geq 50$, which are more consistent. The estimates of the critical point and critical exponents are approximately the same across different bond dimensions.} \label{fig:DataCollapseParametersTime}
\end{figure}

We now consider a nonequilibrium FSS ansatz of the form
\begin{equation}
    M_{zz}(t) = (J_0 t)^{-\beta/z\nu} G\left((J_0 t)^{1/z\nu}
    \left(\frac{B}{J_0} -
    \left(\frac{B}{J_0}\right)_{\mathrm{c}}\right); n\right),
    \label{eq:FiniteSizeScalingTimeMzzAnsatz}
\end{equation}
for FSS analysis in the temporal direction. $G$ is a scaling function (not necessarily the same function $g$ used for $M_{z}$) and $z$ is a dynamical critical exponent. We give an example of the results of the data collapse method in Fig.~\ref{fig:DataCollapseParametersTime}(a). Performing this analysis for simulations for different system sizes $n$ and bond dimension $\chi$, we get different estimates of the critical parameters. In particular, we show the critical point estimates in Fig.~\ref{fig:DataCollapseParametersTime}(b), critical exponent $z \nu$ in Fig.~\ref{fig:DataCollapseParametersTime}(c), and critical exponent $\beta$ in Fig.~\ref{fig:DataCollapseParametersTime}(d). The function $G$ behaves as $G(x) \sim |x|^{\beta}$ as $|x| \to \infty$. As in our previous cases, the estimates of the critical properties are approximately the same across different bond dimensions.

%

%
\section{Conclusion}
\label{sec:QuenchDQPTConclusion}
%

In this chapter, we studied the extent to which we can classically simulate local expectation values in the context of quench dynamics of 1D TFIMs to study a DQPT. We showed that even a poor approximation to the exact many-body state may still yield an excellent approximation to the order parameters that characterize the dynamical phases associated with the DQPT. The Hamiltonian of the TFIM has a $\mathbb{Z}_2$ symmetry that is explicitly broken by the initial state, and one defines two dynamical phases in which the symmetry is either restored or remains broken by the long-time dynamics. These phases are characterized by local order parameters involving spatial and temporal averaging of expectation values of one-spin and two-spin observables. We find that the dynamical phases can be distinguished using low bond dimension MPS simulations. In particular, we find that even when the approximation of the full many-spin state is poor, the approximation of the local reduced density operators can be orders of magnitude better. We used our simulations to also estimate the critical point and critical exponents of the phase transition using finite-size scaling analyses, and we found these critical properties to be insensitive to the MPS bond dimension.

The microstate is specified by the full many-body quantum state, which can become intractable to simulate at long times and large enough system sizes when it becomes volume-law entangled. The macrostate on the other hand is specified by the order parameter, which can be consistent with a multitude of microstates. We therefore conclude that the macrostates associated with the DQPTs of 1D TFIM's can be well approximated with truncated MPS even if the microstates cannot.

%
\chapter{Many-body chaos and quench dynamics}
\label{chap:QuenchChaos}
%

As we discussed in Chap.~\ref{chap:QuenchDQPT}, a near-term goal of quantum computing is the study of many-body physics through the estimation of quantities of interest through experiments, referred to as quantum simulation. Continuing with the theme of simulation of quench dynamics using classical methods, we will consider a quench that is classically tractable due to special symmetries in the model, allowing us to benchmark the classical approximation methods. In the previous chapter, we considered a quantum quench to a nonintegrable model to probe a dynamical phase transition. The dynamical phases were characterized by order parameters that involved spatial and temporal averaging of expectation values of local observables. In this chapter, we consider problems associated with quench dynamics that do not involve spatial and temporal averaging of expectation values of local observables. The quantities of interest are derived from expectation values of specific local observables at specific times. This will show different tradeoffs in simulations of the full microstate and the role of integrability of the many-body dynamics.

%
\section{Estimation of correlation lengths from short-time quench dynamics}
\label{sec:CorrelationLengthFromQuench}
%
We consider 1D TFIMs with nearest neighbor along the $z$-axis characterized by a power-law with exponent $\alpha \geq 0$ and a transverse field along the $x$-axis with magnitude $B$. For $n$ qubits the Hamiltonian is
 \begin{equation} \label{eq:alphaInftyTFIM}
  H =
  - J_0 \sum_{\ell=1}^{n-1} \sigma_\ell^z \sigma_{\ell+1}^z
  - B \sum_{\ell=1}^n \sigma_\ell^x,
\end{equation}
which corresponds to taking $\alpha \rightarrow \infty$ in Eq.~\eqref{eq:HamiltonianPowerLawTFIM}. This model is integrable and many equilibrium properties and nonequilibrium dynamics can be calculated using a mapping to Gaussian Fermionic Hamiltonian~\cite{lieb1961two, pfeuty1970one, pfeuty1971ising, derzhko1998numerical, sengupta2004quench, calabrese2011quantum, calabrese2012quantum1, calabrese2012quantum2} or using the Onsager algebra of strings of local spin operators~\cite{lychkovskiy2021closed} or using a transformation to Gaussian Fermionic Hamiltonians. Moreover, this model does not have the DQPT that we considered in Chap.~\ref{chap:QuenchDQPT} \cite{Zun2018, piccitto2019dynamical, heyl2018dynamical, heyl2019dynamical}. Nevertheless, we can still probe critical properties of the finite-temperature quantum phase transition by looking at selected local expectation values after short-duration quenches~\cite{Kar2017, Tit2019}. We simulate quench dynamics to extract the long-time correlation length from correlations between spins with a given separation after specific quench durations. In particular, we consider the expectation value of the two-spin operator $\sigma^z_{b} \sigma^z_{b+\ell}$, acting between spins separated by a distance $\ell$ as
%
\begin{equation} 
  C_{zz}(\ell; t, n, b) 
  = \bra{\psi_n(t)}\sigma_b^z \sigma_{b + \ell}^z \ket{\psi_n(t)},
  \label{eq:Czz}
\end{equation}
where $| \psi(t) \rangle$ is the time-evolved state for a system with $n$-spins after a quench of duration $t$ and $b$ is reference spin, near the center of the chain for a finite system size with open boundary conditions. In our analysis, to avoid ambiguity, we consider chains of odd length. In the thermodynamic limit of an infinite chain, the dependence on $b$ disappears, therefore, we drop the argument for $C_{zz}$.

In the thermodynamic limit $n \rightarrow \infty$, and in the infinite-time limit, $J_0 t \rightarrow \infty$, when correlations have completely spread through the chain and equilibrated, when quenching from initial field $B_{\mathrm{i}}$ to final field $B_{\mathrm{f}}$ the correlation function takes the form~\cite{calabrese2012quantum1, calabrese2012quantum2}
\begin{equation}
    \lim_{n \to \infty} \lim_{J_0 t \to \infty}
    C_{zz}(\ell; t, n) \approx
    C_0(B_{\mathrm{f}},B_{\mathrm{i}}) 
    \exp\left(-\frac{\ell}{\xi(B_{\mathrm{f}},B_{\mathrm{i}})}\right) 
    \label{eq:CorrelationLength}
    ,
\end{equation}
where $C_0(B_{\mathrm{f}},B_{\mathrm{i}})$ is given by
\begin{equation}
  C_0(B_{\mathrm{f}},B_{\mathrm{i}})
  = \left[ \frac{(B_{\mathrm{i}}- B_{\mathrm{f}}) B_{\mathrm{f}}
      \sqrt{B_{\mathrm{i}}^2-1}}{(B_{\mathrm{i}}+ B_{\mathrm{f}})(B_{\mathrm{i}}B_{\mathrm{f}}-1)}
   \right]^{1/2} .
\end{equation}
Of particular interest is the correlation length $\xi$ at the critical point $B_{\mathrm{f}} = J_0$ (for $\alpha \to \infty$) of the ground state quantum phase transition. In the quench dynamics paradigm, this involves using ansatz of the form Eq.~\eqref{eq:CorrelationLength} and estimating the correlation length $\xi$ from the correlations $C_{zz}$, defined in Eq.~\eqref{eq:Czz}. We expect the dynamics of quenches to near the critical point $B/J_0 = 1$ to be the hardest to simulate as they lead to entanglement growing most strongly. Moreover, this point is of most interest as it can be related to the diverging correlation lengths near the thermal and quantum phase transition, and used to extract the correlation length exponents in Eq.\eqref{eq:CorrelationLengthDivergenceField} and Eq.\eqref{eq:CorrelationLengthDivergenceTemperature}. This would play a role in studies universality of the transition exhibited in the model~\cite{Kar2017, Tit2019}. Because this problem is integrable, long-studied known solutions~\cite{derzhko1998numerical, calabrese2011quantum, calabrese2012quantum1, calabrese2012quantum2} can be used to benchmark quantum simulators. 

While the exact correlation length is defined in the infinite-time limit as in Eq.~\eqref{eq:CorrelationLength}, Karl \textit{et al.} showed that, surprisingly, $\xi(B_{\mathrm{f}},B_{\mathrm{i}})$ can be extracted from \textit{short-time quenches} due to the different time scales for local and global equilibration~\cite{Kar2017}. This allows us to explore the robustness to truncation in the MPS simulation under consideration here. We saw in Chap.~\ref{chap:QuenchDQPT}, that for the $\mathbb{Z}_2$ symmetry breaking DQPT, expectation values of space and time averaged observables may be robust to MPS truncation errors, even up to times where the state deviates considerably from the true state. For the DQPT spatial and temporal averaging played an important role in the rapid convergence with bond dimension truncation. This is not the case for the correlation length which depends on the correlation function between specific spins at a specific time, as in Eq.~\eqref{eq:CorrelationLength}. This paradigm thus allows us to explore further distinctions between micro- and macrostates and the tractability of quantum simulation.

\begin{figure}[ht] 
   \centering
   {\includegraphics[width=0.48\textwidth]{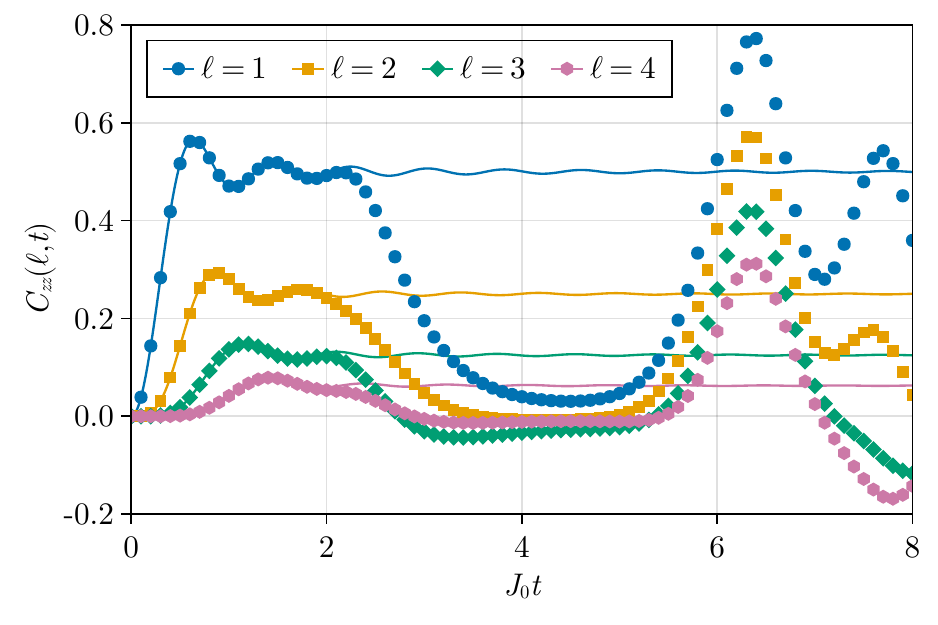}  } 
   \caption{Spatial correlation function in TFIM for $n=81$ spins when quenched to the critical point $B/J_0 = 1$, for different separations $\ell$ between two spins. The solid curves correspond to the exact results while the data points are for simulations with a bond dimension of 8. The step size used is $J_0 \Delta t = 10^{-2}$, but only data with a separation of 0.1 are shown. For early times ($J_0 t \leq 2$), the low bond dimension simulation tracks the exact results very well, but at later times it deviates dramatically from the exact values, calculated using the Gaussian Fermionic representation~\cite{derzhko1998numerical}.}
	\label{fig:CzzThermal}
\end{figure}

We consider here quenches from deep within the disordered phase ($B_{\mathrm{i}}/J_0 \to \infty$) to the critical point $B_{\mathrm{f}}/J_0 = 1$, where initially all spins are polarized along the $x$-direction, $\ket{\psi(0)} = \ket{\uparrow_x}^{\otimes n}$. Using 4th order TEBD to simulate the dynamics, Fig.~\ref{fig:CzzThermal} shows the behavior of $C_{zz}(t, \ell)$ for fixed distance $\ell$ as a function of the simulation time $J_0 t$. We observe that the equal-time correlation for each value of the site separation $\ell$ reaches a quasi-steady-state value, with smaller separations reaching it sooner than larger $\ell$ values. As we will see, we only need a fixed range of $\ell$ to accurately extract the correlation length, and thus we will not need to scale the simulation time with $\ell$. 

Fig.~\ref{fig:CzzThermal} additionally demonstrates how well a fixed bond dimension MPS simulation, in this case, $\chi = 8$, does in approximating the exact correlation function. While the MPS simulation reproduces the correlation function at short times when the fixed bond dimension is sufficient to capture the exact state, the approximation fails at later times. Nevertheless, the long-time correlation length can be reliably extracted from the short-time quench dynamics that we can accurately simulate, well before the full many-body state has thermalized.

\begin{figure}[ht] 
   \centering
   (a) {\includegraphics[width=0.44\textwidth]{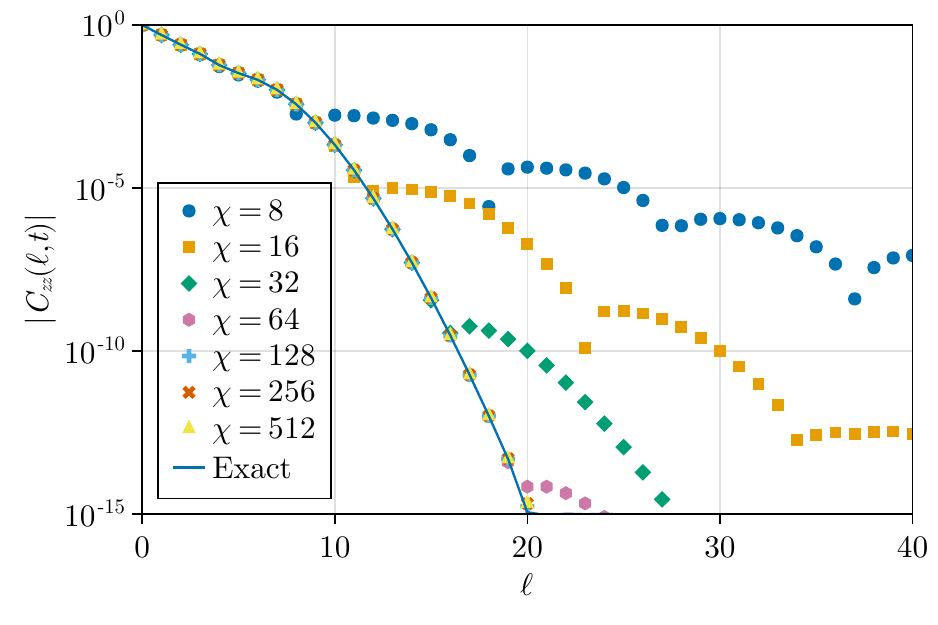}}
   (b) {\includegraphics[width=0.44\textwidth]{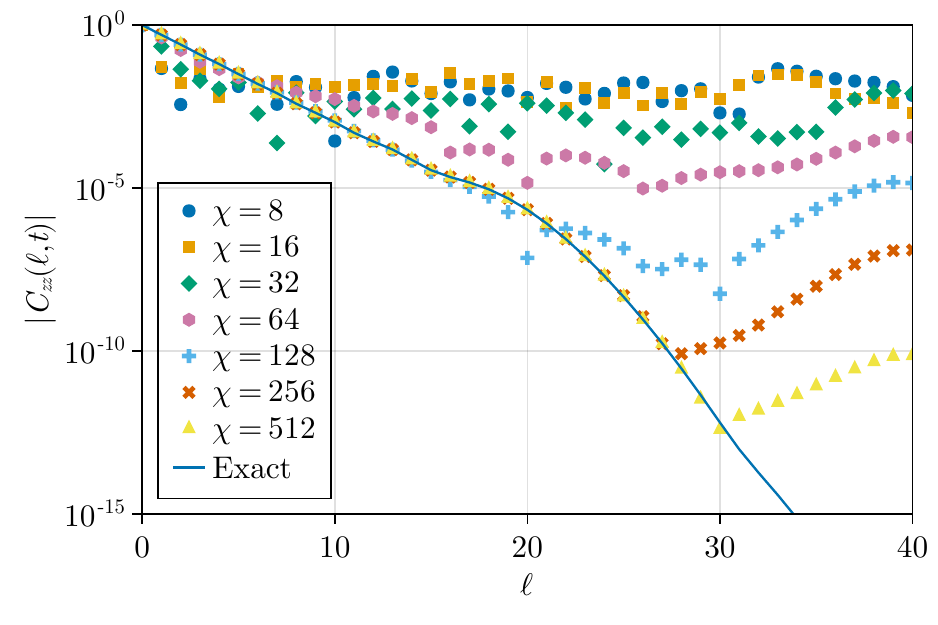}}
   \caption{Spatial correlation function in TFIM for $n=81$ spins when quenched to the critical point $B/J_0 = 1$, for different total quench times $J_0 t$. For small simulation times as in (a), low bond dimension simulations can accurately reproduce the spatial correlation function for an extended $\ell$ range. For example, $\chi = 16$ is sufficient to accurately predict the behavior up to $\ell =10$. However for larger simulation times as in (b), only significantly higher bond dimension simulations can reproduce the spatial correlation function for the same range. For example, $\chi = 128$ is needed to accurately predict the behavior up to $\ell =10$. Exact simulations are done using the Gaussian Fermionic Hamiltonian representation~\cite{derzhko1998numerical}.} 
   \label{fig:Czz}
\end{figure}

Fig.~\ref{fig:Czz} illustrates the behavior of $C_{zz}(t, \ell)$ for two different fixed values of $J_0 t$ as we vary the separation distance $\ell$. One clearly observes a transition between two different exponential curves, with characteristic decay lengths $\xi_1$ at short separation $\ell$ and a weaker decay $\xi_2$ at large separation, as observed in~\cite{Kar2017}. The latter results from ``prethermalization" before correlations have spread to the entire system. By increasing the simulation time $J_0 t$, the range of separation distances, $\ell$, over which the first decay occurs can be made larger, and importantly $\xi_1$ is independent of time. In the infinite time limit, $\xi_1$ defines pure exponential decay over the whole chain. Reproducing the full correlation function at large $\ell$, even at short times, requires a more accurate simulation using a larger bond dimension. However, the short-range behavior decay constant, which survives to
long times is well approximated by a highly truncated MPS.

\begin{figure}[h] 
   \centering
    \includegraphics[width=0.48\textwidth]{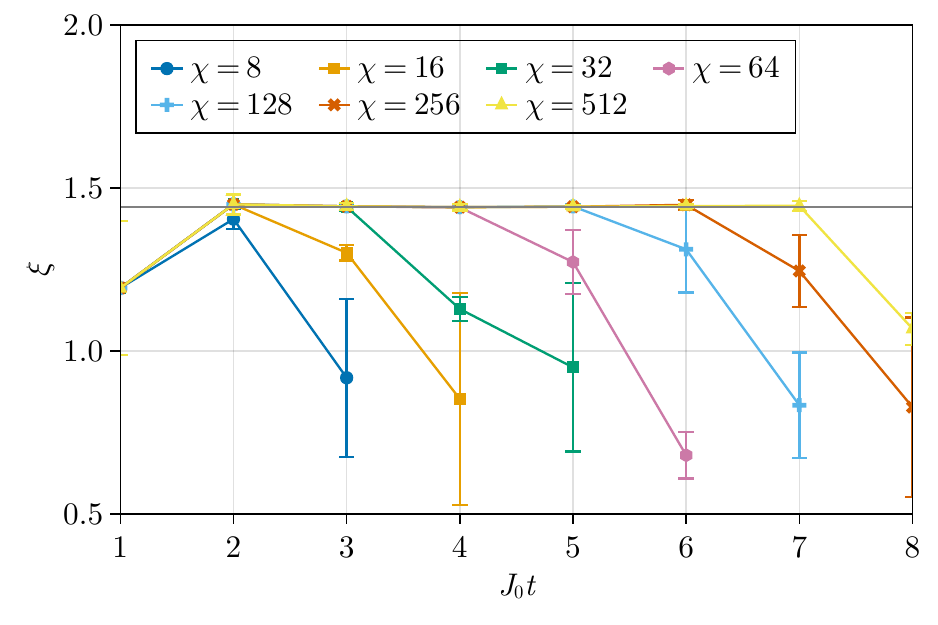}
    \caption{Results of fitting $|C_{zz}(t,\ell)|$ using the datapoints shown in Fig.~\ref{fig:Czz} at $\ell \leq 6$ to $\exp(-\ell/\xi)$. The thin black line corresponds to the long-time correlation length of $1/\ln(2)$. Error bars correspond to 95\% confidence interval. Moving from left to right after $J_0 t = 2$, the bond dimension must be doubled for every dimensionless unit of time increment to maintain an accurate estimate of the correlation length.}
   \label{fig:CorrelationLength}
\end{figure}

We show the correlation length $\xi_1$ extracted from the data in Fig.~\ref{fig:CorrelationLength}, which we obtain by fitting $|C_{zz}(t,\ell)|$ to $\exp(-\ell/\xi)$ using short-range correlations with $\ell \leq 6$ . If $J_0 t$ is too small $(J_0 t < 2)$, the extracted correlation length is not accurate even if the simulations are. However, already at $J_0 t=2$ we see that we can extract a very good estimate of the infinite-time correlation length, and we can do this using a relatively small bond dimension. Increasing $J_0 t$ does not practically improve our estimate of the infinite-time correlation length, but it does require simulations using a larger bond dimension to converge to the correct behavior. Specifically, increasing the dimensionless simulation time, $J_0 t$, by one unit requires us to increase the bond dimension by a factor of two to reproduce the correct correlation length. This can be seen in Fig.~\ref{fig:CorrelationLength}: at $J_0 t = 2$, bond dimensions  $\geq 16$ reproduce the correct correlation length; at $J_0 t = 3$ only bond dimensions $\geq 32$ reproduce the correct correlation length, etc. 

\begin{figure}[h] %
   \centering
   \includegraphics[width=0.48\textwidth]{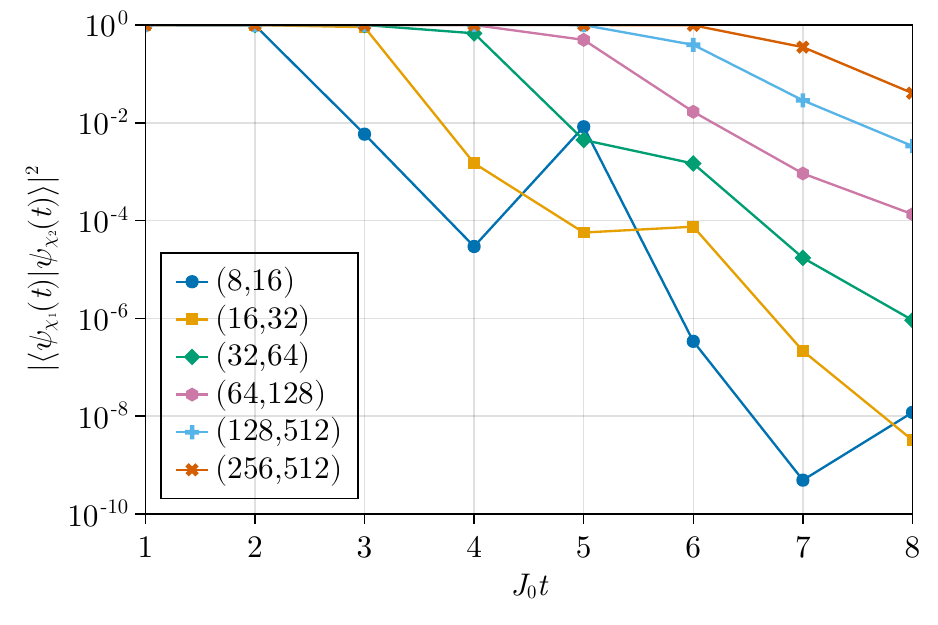} 
   \caption{Overlap squared of states simulated with different bond dimensions as a function of simulation time $T$ in the TFIM with $n=81$ spins quenched to the critical point $B/J_0 = 1$. The times at which each curve drops, which corresponds to when the two simulations no longer agree in terms of the many-body state with high fidelity, agrees with the times in Fig.~\ref{fig:CorrelationLength} when the estimate of the correlation length begins to fail.}
   \label{fig:CorrelationLengthOverlap}
\end{figure}

\begin{figure}[h] %
   \centering
   \includegraphics[width=0.48\textwidth]{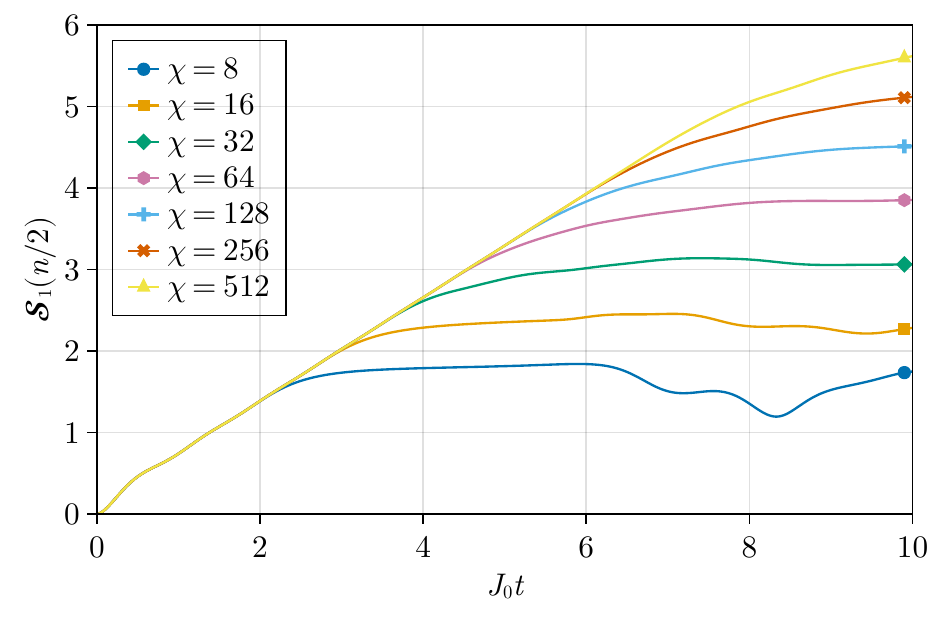} 
   \caption{Bipartite entanglement entropy between the two halves of the system for simulations	with different bond dimensions as a function of quench time $t$ in the TFIM for $n=81$ spins when quenched to the critical point $B/J_0 = 1$. The times at which each curve flattens, which corresponds to when the bond dimension is insufficient to capture the growth of entanglement, agrees with the times in Fig.~\ref{fig:CorrelationLength} when the estimate of the correlation length begins to fail.}
   \label{fig:CorrelationLengthEE}
\end{figure}

To further quantify the cost of simulation, in Fig.~\ref{fig:CorrelationLengthOverlap} we plot the squared overlap between states simulated with bond dimension $\chi$ and $2 \chi$. The time at which the extracted correlation length deviates from the true value for a fixed bond dimension $\chi$ in Fig.~\ref{fig:CorrelationLength} agrees with the time at which the overlap between the full state with bond dimension $\chi$ deviates from that with bond dimension $2 \chi$. For example, the overlap of the states with bond dimensions $(16, 32)$ begins to deviate at $J_0 t \geq 3$, indicating that the simulation with bond dimension $\chi = 16$ is no longer accurate. This is simultaneously the value of $J_0 t$ for which the extracted correlation length deviates from the true value for $\chi = 16$. Similarly, the overlap of the states with bond dimensions $(32, 64)$ begins to deviate at $J_0 t \geq 4$, which coincides with the time for which the extracted correlation length deviates from the true value for $\chi = 32$. Before these times, the states from the simulations have overlap squared almost equal to 1. We note that these times also coincide with the times when the bipartite entanglement entropy between the two halves of the system saturates, as shown in Fig.~\ref{fig:CorrelationLengthEE}.

From these simulations, we draw the following conclusions. In contrast to the DQPT, the correlation length does not involve spatial or time averaging. As such, it is more sensitive to the microstate. Thus one finds that accurate simulation of the correlation length requires accurate simulation of the full many-body state, which we see in the exponential scaling of the bond dimension with time, $\chi \sim \mathcal{O}(2^{J_0 t})$. Formally the correlation length is defined in the thermodynamic and infinite-time limits, and at the critical point the thermalized state exhibits volume-law entanglement, making classical simulation via MPS and TEBD intractable. For the 1D nearest-neighbor TFIM, one finds that the correlation length formally defined in the long-time limit is already available in the short-range short-time quench dynamics, making classical simulation tractable with a fixed bond dimension and runtimes that scale as $\mathcal{O}(n \chi^3)$. The exact microstate in the thermodynamic limit is thus not necessary to extract the order parameter of interest.

%

%
\section{Role of chaos}
\label{sec:RoleChaos}
%
As we discussed in Chap.~\ref{chap:QuenchDQPT}, the regime of interest for a quantum advantage in simulating local expectation values is where the full many-body state reaches volume-law entanglement, making classical simulation intractable, but where there is not complete scrambling to a typical quantum state~\cite{Leb1993, Gol2006, Leb2007, Bar2009, Dub2012, San2012, Fac2015, d2016quantum, Mi2022}, which would render the local state near maximally mixed. To better understand the findings related to the estimation of quantities in quantum quenches in  Sec.~\ref{sec:CorrelationLengthFromQuench} through the lens of quantum chaos, we revisit the framework of chaos and chaotic systems that we introduced in Chap.~\ref{chap:QuenchDQPT} and how it affects the growth of entanglement and equilibration of local reduced density operators. In what follows, we focus on 1D nearest neighbor models for simplicity.

In addition to the mean adjacent level spacing ratio that we considered in Chap.~\ref{chap:QuenchDQPT}, the entanglement entropy of a bipartition of the Hamiltonian eigenvectors is another measure of chaoticity of the Hamiltonian~\cite{Lakshminarayan2001, karthik2007entanglement, Oma2022}. For chaotic Hamiltonians, most of the eigenvectors with eigenvalues near the middle or bulk of the spectrum have volume-law scaling of entanglement entropy, with some states with area-law scaling entanglement entropy at the edges of the spectrum. For integrable Hamiltonians, there are eigenvectors with area-law scaling of entanglement entropy throughout the spectrum. We use this measure as it allows us to relate the chaoticity of the Hamiltonian to the entanglement entropies of the local reduced density operators in quantum quenches, which plays a role in determining the simulability of the local observables we have considered in Sec.~\ref{sec:DynamicalQuantumPhaseTransition} and ~\ref{sec:CorrelationLengthFromQuench}.

\begin{figure}[h!] %
    \centering
    (a) {\includegraphics[width=0.42\textwidth]{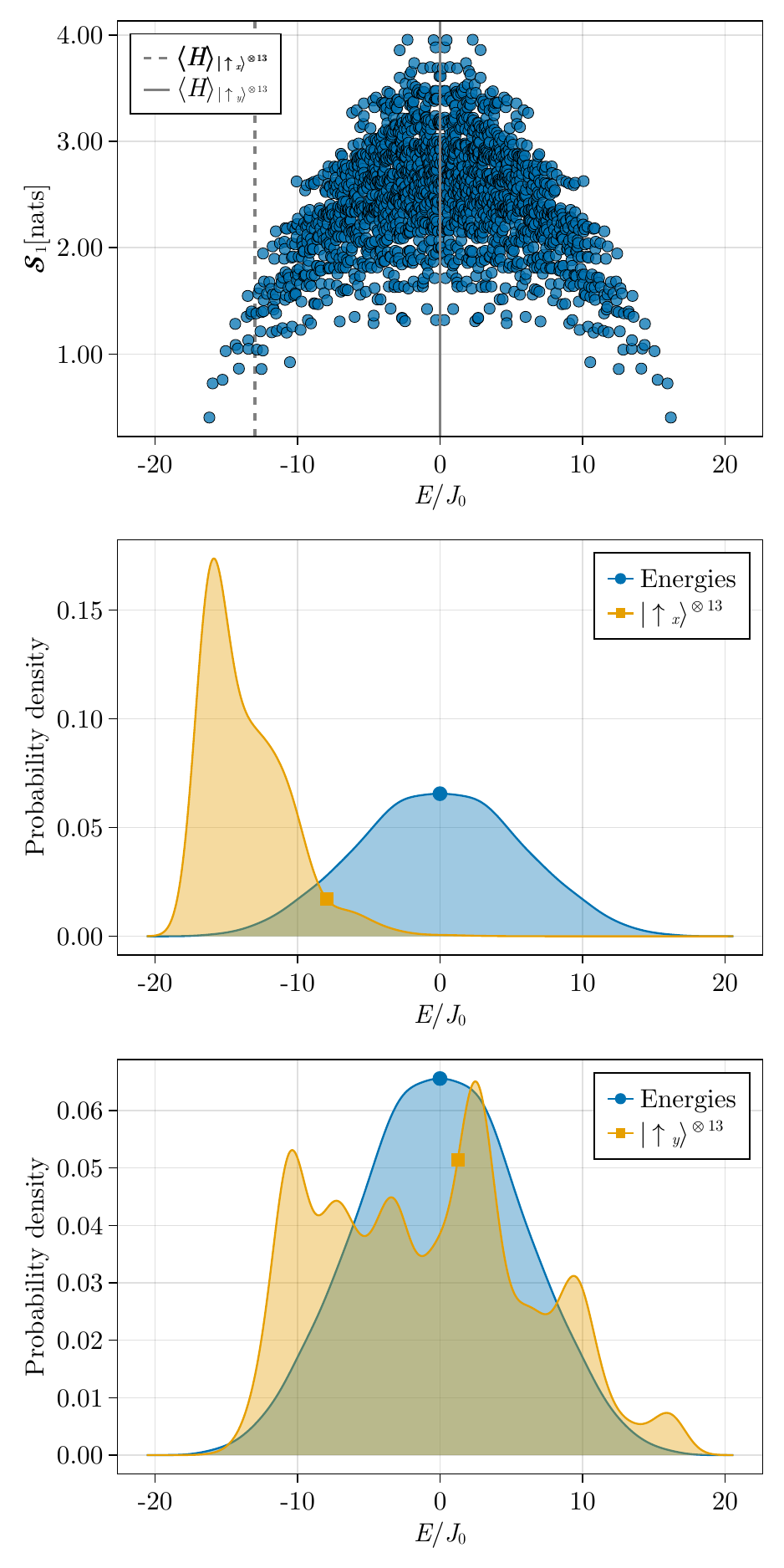} \label{fig:EigenvectorEntropiesPopulationTFIM}}
    (b) {\includegraphics[width=0.42\textwidth]{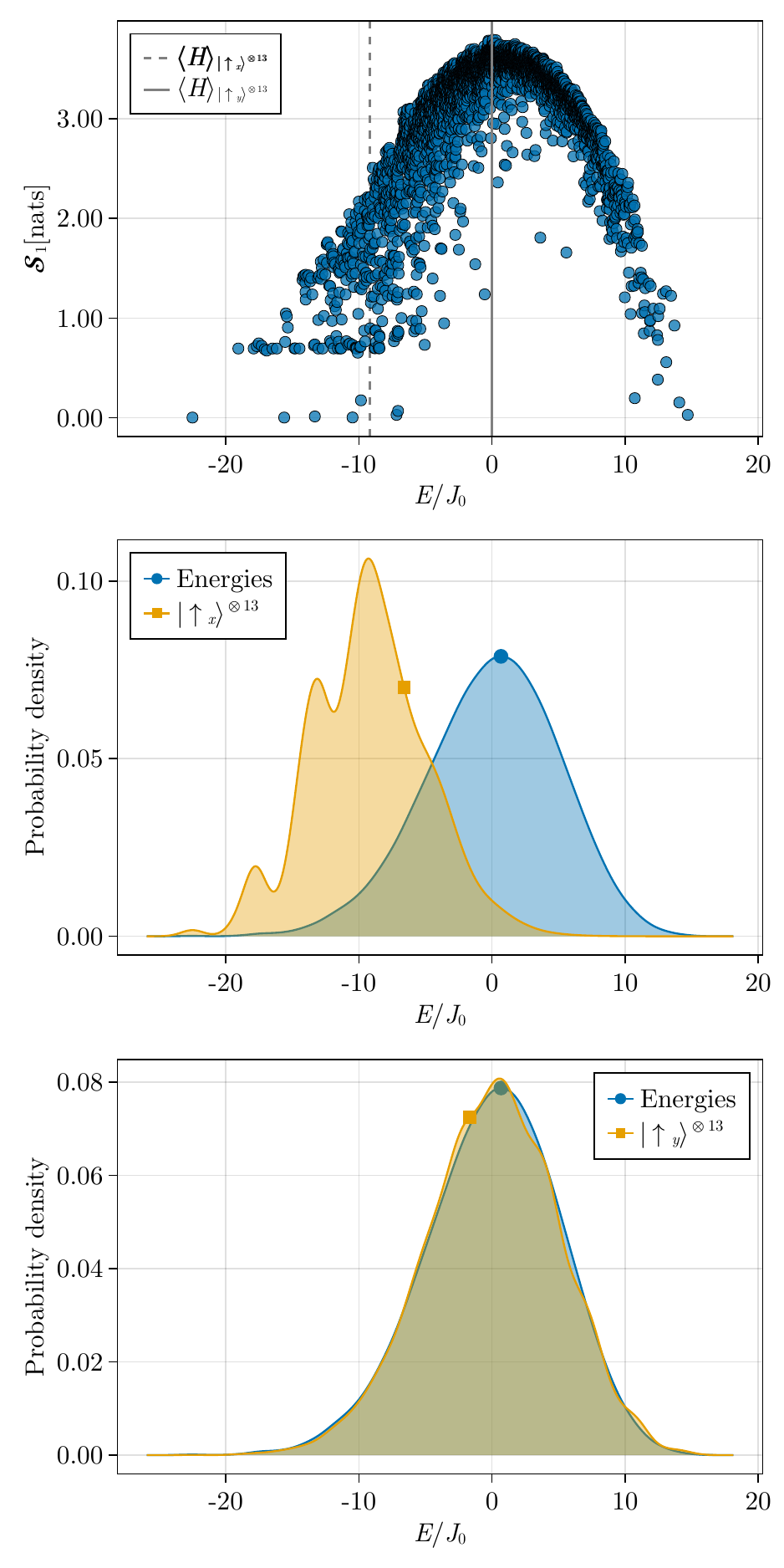}\label{fig:EigenvectorEntropiesPopulationSFIM}}
    \caption{Entanglement in TFIM and SFIM Hamiltonian eigenvectors and the population in energy eigenstates for different initial states. Top row: half array entanglement entropy, $\mathcal{S}_1$, in nats, of energy eigenvectors as a function of energy for (a) TFIM (b) SFIM with $\theta = \pi/4$. Vertical lines show the expectation value of the Hamiltonian in the states $|\uparrow_x\rangle^{\otimes 13}$ and $|\uparrow_y\rangle^{\otimes 13}$. Middle row: energy density of the initial state $| \uparrow_x \rangle^{\otimes 13}$ in the quenched eigenbasis and the Hamiltonian as a function of energy for (a) TFIM (b) SFIM with $\theta = \pi/4$. Bottom row: Energy density of the initial state $| \uparrow_y \rangle^{\otimes 13}$ in the quenched eigenbasis and the Hamiltonian as a function of energy for (a) TFIM (b) SFIM with $\theta = \pi/4$. The energy densities are calculated using kernel density estimation~\cite{sheather2004density, wkeglarczyk2018kernel} with Gaussian kernels whose widths are determined by Silverman's rule~\cite{silverman1981using, silverman1986density}.}
    \label{fig:EigenvectorEntropiesPopulation}
\end{figure}

The TFIM studied in Sec.~\ref{sec:CorrelationLengthFromQuench} is an integrable model, so to better understand the role of chaos we break the spin-flip symmetry by slanting the magnetic field in the $z-x$ plane to an angle $\theta$ with the $z$-axis. The Hamiltonian now reads
\begin{equation} 
\begin{aligned}
  H = &
  - J_0 \sum_{\ell} \sigma^z_{\ell} \sigma^z_{\ell + 1}
  - B \sum_{\ell} \left( \cos(\theta)
  \sigma^z_{\ell} + \sin(\theta) \sigma^x_{\ell} \right).
  \label{eq:HamiltonianPowerLawSFIM}
\end{aligned}
\end{equation}
We call this the slanted field Ising model (SFIM), as studied in~\cite{prosen2007efficiency, leviatan2017quantum, rakovszky2022dissipation, vonkeyserlingk2022operator, cotler2023emergent, choi2023preparing, Mil2022}. When $\theta =0$ or $\theta =\pi$ the model reduces to the classical Ising model, and for $\theta = \pi/2$ it is the TFIM from Eq.~\eqref{eq:alphaInftyTFIM}. To illustrate how different the properties of the spectra of these models are, we plot in the top row of Fig.~\ref{fig:EigenvectorEntropiesPopulation}(a), (b). the bipartite half-array von Neumann entanglement entropies of the eigenvectors of the Hamiltonian as a function of energy for the TFIM and SFIM with $B/J_0 = 1$. For the integrable TFIM, there are low entanglement eigenvectors throughout the spectrum, while for the chaotic SFIM most states near the bulk of the spectrum are highly entangled and states near the edges of the spectrum have low entanglement. Therefore, we generically expect larger entanglement entropy generation in the quench for the chaotic SFIM. By tuning $\theta$ we can probe the role of chaos and equilibration in the relative ease of approximating local expectation values. 

\begin{figure}[ht] 
    \centering
    \includegraphics[width=0.48\textwidth]{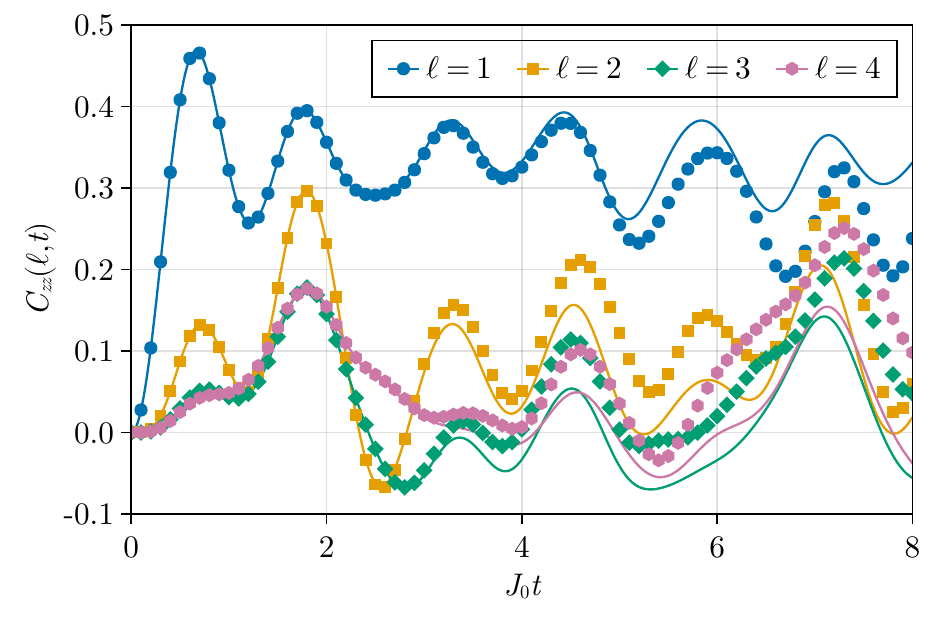} 
    \caption{Spatial correlation function in the SFIM with $\theta=\pi/4$, [Eq.~\eqref{eq:HamiltonianPowerLawSFIM}] with $n = 81$ spins when quenched to the critical point $B/J_0 = 1$, for different separations between two spins. The solid curves correspond to results using a maximum bond dimension of $512$ while the data points are for simulations with a bond dimension of 8. The step size used is $J_0 \Delta t = 10^{-2}$, but only data with a separation of 0.1 are shown. In contrast to Fig.~\ref{fig:CzzThermal}, the low bond dimension simulation approximates the correct behavior for longer times.}\label{fig:CzzHz}
\end{figure}

Consider the spin-spin correlation function for quenches from $|\uparrow_x\rangle^{\otimes n}$, as studied in Sec.~\ref{sec:CorrelationLengthFromQuench}, but now for the case of the SFIM with $\theta = \pi/4$. In Fig.~\ref{fig:CzzHz} we show the correlation functions simulated with bond dimensions $\chi = 8$ and $\chi = 512$. We see that the estimates of the correlation functions agree for a significantly longer time compared to the integrable case of $\theta = \pi/2$ studied in Sec.~\ref{sec:CorrelationLengthFromQuench}. Thus, with the introduction of a nonzero longitudinal field, the system becomes chaotic, and we find that lower bond dimension simulations are better able to approximate the results of higher bond dimension simulations.

\begin{figure}[ht]
  \centering
  \includegraphics[width=0.48\textwidth]{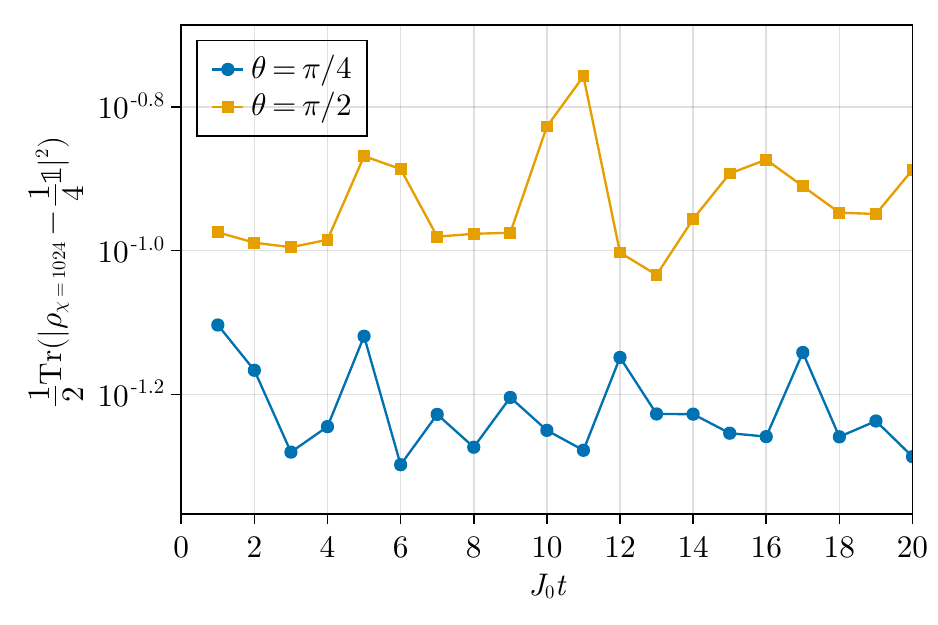}
  \caption{Comparison of the proximity of the two-spin reduced density operator to the maximally mixed state attained for the reduced states in the TFIM (circles) and SFIM with $\theta = \pi/4$ (squares).  Here we choose $n=20$ spins and an initial state, $|\uparrow_x\rangle^{\otimes 20}$, for a Hamiltoninan quenched to $B/J_0=1$. Plotted is the squared HS distance between the two-spin reduced density operator for nearest neighbor spins at the center of the chain evaluated with bond dimension $\chi=1024$ and the maximally mixed two-spin state as a function of time during quenches from $|\uparrow_x\rangle^{\otimes 20}$ to TFIM and $\theta = \pi/4$. The reduced density operator is closer to the maximally mixed case for the SFIM than the TFIM.}
  \label{fig:HSDistanceSFIMXPolarizedQuench}
\end{figure}

To understand the increased robustness of expectation values to MPS truncation, we study the correlation between this and the closeness of the reduced density operator to the maximally mixed state. To illustrate this we exactly simulate the TFIM and SFIM for $n = 20$, and in Fig.~\ref{fig:HSDistanceSFIMXPolarizedQuench} we plot the squared HS distance Eq.~\eqref{eq:HSDistanceDefinition} between the two-body reduced density operator for nearest neighbors at the center of the chain and the maximally mixed state. For the initial state $|\uparrow_x\rangle^{\otimes n}$, the two-spin reduced density operator is closer to the two-spin maximally mixed state for the $\theta = \pi/4$ SFIM than for the integrable model. These simulations therefore suggest an important relationship between the chaoticity of the quenched Hamiltonian and the efficiency with which we can simulate local expectation values via a truncated MPS representation. 

\begin{figure}[ht]
  \centering
  \includegraphics[width=0.48\textwidth]{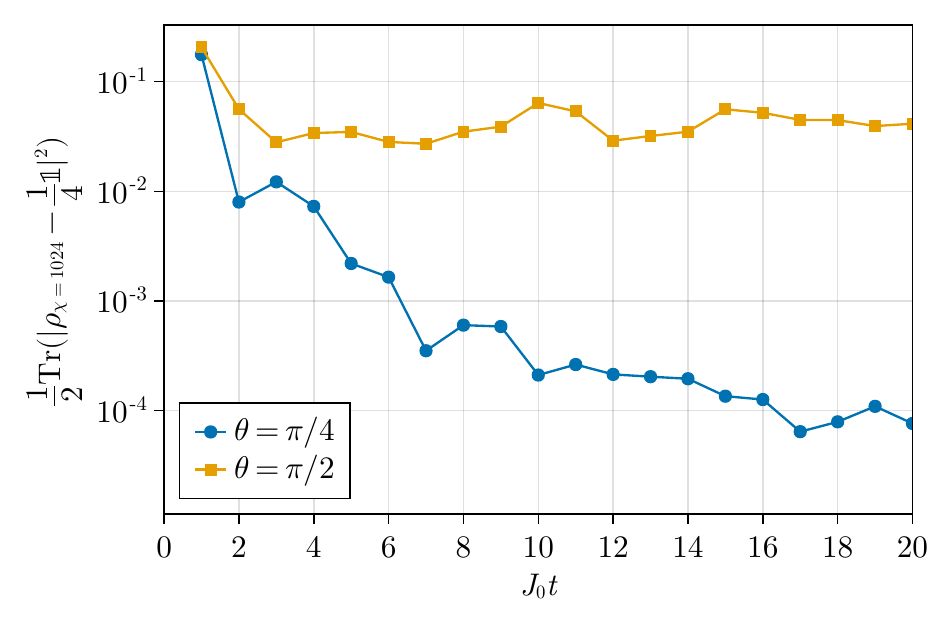}
  \caption{Same as Fig.~\ref{fig:HSDistanceSFIMXPolarizedQuench} but for an initial state $|\uparrow_y\rangle^{\otimes 20}$. In contrast to that figure, the reduced density operator of the SFIM with this initial condition is significantly closer to the maximally mixed state in comparison to the TFIM as this initial state thermalizes closer to a quantum typical state.}
  \label{fig:HSDistanceSFIMYPolarizedQuench}
\end{figure}

In addition to the Hamiltonian, the initial state plays an important role in the complexity of simulating quench dynamics. A chaotic Hamiltonian and an initial state with more support on energy eigenstates in the bulk of the spectrum, which are highly entangled, lead to higher effective temperature quenches with local reduced density operators being closer to the appropriate maximally mixed state. On the other hand, we expect nongeneric behavior for initial conditions with an average energy near the edges of the spectrum. For example, the $|\uparrow_x \rangle^{\otimes n}$ initial condition has an extensive average energy arising from the transverse field and is primarily supported in the low energy part of the spectrum for both the TFIM and $\theta = \pi/4$ SFIM (middle row of Fig.~\ref{fig:EigenvectorEntropiesPopulation}(a), (b)). In this case, we can expect a low effective temperature for the long-time evolved state. In contrast, the $|\uparrow_y \rangle^{\otimes n}$ initial condition has an average energy of zero and is supported throughout the spectrum of the TFIM and $\theta = \pi/4$ SFIM (bottom row of Fig.~\ref{fig:EigenvectorEntropiesPopulation}(a), (b). In this case, we can expect a significantly higher effective temperature for this initial state. This is borne out when we consider a quench with this initial state (Fig.~\ref{fig:HSDistanceSFIMYPolarizedQuench}), where the difference between the integrable TFIM and chaotic SFIM is clearer -- the two-spin reduced density operator is significantly closer to the maximally mixed state for the chaotic SFIM than for the TFIM. 

%

%
\section{Conclusion}
\label{sec:QuenchChaosConclusion}
%

In this chapter, we considered the quench dynamics of the 1D nearest neighbor TFIM to estimate the infinite-time correlation length after a quench to the critical point of the ground state quantum phase transition. In this case, the relevant correlation function is more sensitive to the microstate than it was in the study of a DQPT in Chap.~\ref{chap:QuenchDQPT} as it is defined by an expectation value of local observables of specific spins at a given time as a function of their separation. While the correlation length is defined by the infinite time and thermodynamic limit when correlations have spread throughout the long chain, in practice the desired correlation length can be extracted at surprisingly short times, as found in Ref.~\cite{Kar2017}. The dynamics are such that local observables equilibrate rapidly to a steady state of a generalized Gibbs ensemble~\cite{calabrese2012quantum2, Kar2017}, allowing one to extract the correlation length very quickly, well before volume-law entanglement for the quantum state is reached. This allowed us to extract the correlation length accurately with short-time MPS simulations using very small bond dimensions. Long-time simulations however required exponentially growing (with time) bond dimensions to keep a high accuracy estimate of the correlation length. In both cases, a faithful simulation of the full many-body state was required to accurately estimate the correlation length. Therefore, in contrast to the study of DQPT in Chap.~\ref{chap:QuenchDQPT}, we find that the local properties of this quench require an excellent approximation of the microstate when using MPS, which cannot be done accurately for long times.

We can understand the preceding results by the role of chaos and local equilibration according to the ETH in quench dynamics. The 1D TFIM with nearest-neighbor interactions is integrable, so equilibrated local density matrices retain certain structures that can depend sensitively on the full quantum state. This is in contrast to the fully chaotic models such as the SFIM that includes both transverse and longitudinal $B$-fields. The chaotic SFIM Hamiltonian we study leads to more entanglement generation during the quench than the TFIM and the equilibrated local state is the maximally mixed state, which has no useful information and leads to better approximations of local expectation values with truncated MPS. 

These findings reveal a somewhat counterintuitive relationship between quantum chaos and the efficiency of classical simulation of expectation values of local observables in quench dynamics. The more chaotic the Hamiltonian, as measured, for example, by level statistics, or entanglement entropy of its eigenvectors, the more resilient are these simulations to MPS bond dimension truncation. Local expectation values in quench dynamics are easier to approximate for highly chaotic systems because at short times there is little entanglement and at longer times the system thermalizes such that the marginal states are close to maximally mixed. For such chaotic systems, there is little information of interest in the marginal states. Our conclusion is opposite to what one might draw when considering other representations, such as in the Heisenberg picture in the evolution of operators, where an operator evolves into an admixture of exponentially many operators with time, in different operator bases~\cite{parker2019universal, patramanis2022probing, heveling2022numerically}. Indeed, the operator spreading seen in operator entanglement of the SFIM led Prosen and {\ifmmode \check{Z}\else \v{Z}\fi{}nidari\ifmmode \check{c}\else \v{c}\fi{}} to ask, ``Is the efficiency of classical simulations of quantum dynamics related to integrability?'' with the reverse conclusion~\cite{prosen2007efficiency}. The choice of representation is thus essential when drawing conclusions about classical tractability. 

On the other hand, in the context of foundations of statistical mechanics and the justifications for ensembles thereof, the fact that ``typical" pure quantum states are compatible with predictions of statistical mechanics has been pointed out in the form of ETH or typicality~\cite{pop06ent, Gol2006, rei07typ, sug13can, alba2015eigenstate, khodja2015relevance, d2016quantum, deutsch2018eigenstate,dymarsky2018subsystem, murthy2019bounds, noh2021eigenstate}. From this perspective, chaotic time evolution leading to more robust simulability may not appear as surprising, since such nonintegrability will more likely lead the state to become a random ``typical" state, which is arguably the very ingredient that enables a powerful effective theory such as statistical mechanics to emerge. 
%
\chapter{Summary and Outlook}
\label{chap:SummaryOutlook}
%

In this dissertation, we studied quantum entanglement in the dynamics of spin models, both for implementing of two-qubit entangling gates between neutral atoms and for its role in using classical approximation methods to simulate analog quantum experiments probing critical phenomena. We first studied the implementation of entangling gates for neutral atoms, an essential ingredient for implementing quantum information processing. We studied the adiabatic Rydberg dressing paradigm and showed its robustness to static imperfections and favorable scaling with the fundamental time and energy scales in Rydberg-mediated entanglement between neutral atoms. In the context of analog quantum experiments, we studied critical phenomena in quantum quenches to interacting spin models. We established that the quantities of interest which characterize these critical phenomena depend on macro properties, and can be well approximated using classical approximation methods that truncate the entanglement content in the quantum states.

We introduced the neutral atom Mølmer-Sørensen gate, as an entangling gate for neutral atoms using the adiabatic Rydberg dressing paradigm \cite{mitra2020robust, martin2021molmer}. This involved rapid adiabatic Rydberg dressing passages, modulating the amplitude and frequency of the Rydberg-exciting laser such that the Hamiltonian changes slowly compared to the timescales corresponding to the energy gaps in the instantaneous Hamiltonian, but fast compared to the timescale of the finite Rydberg state lifetime. During the passage, the ground qubit states accumulate dynamical phases during this adiabatic passage, and the adiabatic passage allows a delineation of local (one-qubit) and non-local (two-qubit entangling) phases. We showed that in implementing two-qubit entangling gates, the dominant source of error arises from inhomogeneities in the one-atom phases and that they could be corrected using a simple spin echo sequence. This gate has been implemented in a first demonstration in Cesium \cite{martin2021molmer} and with high fidelity in Strontium \cite{schine2022long} atoms.

While adiabatic rapid passages that achieve these dynamical phase-based entanglement schemes are easily implemented using one-photon ground-to-Rydberg transitions, we subsequently showed that we can use two-photon ground-to-Rydberg transitions to implement these adiabatic rapid passages. By adiabatically eliminating the intermediate state, we need to use only one of two laser amplitudes to modulate both Rabi frequency and detuning. We also estimated the fundamental limits to achievable fidelity using adiabatic Rydberg dressing and showed that we can asymptotically reach the fundamental limit set by the Rydberg state lifetime and the interaction energy by appropriately designed adiabatic passages, outside the perfect Rydberg blockade regime \cite{mitra2023neutral}. This opens avenues for entangling gates between neutral atoms that are spatially separate beyond the Rydberg blockade radius and would be useful when more laser power becomes available for implementing these gates. Avenues for future work include the synthesis of entangling gates between neutral atoms, and avoiding crosstalk between atoms in an array of atoms with minimal rearrangement. These studies would benefit from a detailed analysis of the atomic species and their Rydberg state structure as done  \cite{de2018accurate, de2018analysis}. Moreover, adiabatic Rydberg dressing using Rydberg macrodimer states \cite{sassmannshausen2016observation, hollerith2021realizing} is another avenue for future work.

The other aspect of this dissertation has been understanding the role of quantum entanglement as a challenge posed for classical computational methods in problems of interest for quantum simulation \cite{preskill2012quantum, preskill2018quantum}. Understanding the capabilities of classical descriptions and the regimes of applicability is an important aspect of designing quantum information processing systems. We considered a classical simulation of critical phenomena in a nonequilibrium situation involving quench dynamics of transverse field Ising models (TFIMs) using matrix product states (MPS). These are similar to the quantum quench experiments that one would consider in atomic platforms like neutral atoms and trapped ions. We focused on both a $\mathbb{Z}_2$-symmetry breaking/restoring dynamical quantum phase transition (DQPT) that is seen in the quench dynamics and on accessing critical properties through quantum quenches even when there is no DQPT. The order parameters for the DQPT involve expectation values of local observables with space and time averaging. We showed that not only the order parameters but also the critical point and critical exponents of the phase transition are insensitive to extreme truncation of the MPS bond dimension, even though our approximations to the microstate were quite bad. The quantities characterizing critical behavior near the phase transition are macro properties and are not very sensitive to entanglement content in the wavefunctions used to estimate them.

For a quantum quench to integrable model without a DQPT, we tried to estimate correlation lengths that depended on specific expectation values of local observables at a specific time, with no space or time averaging. We found that while this required much better approximation to the wavefunction, requiring a large amount of entanglement, rapid equilibration of local expectation values allowed us to estimate the correlation length from short duration quenches, as prescribed by Ref. \cite{Kar2017} when the wavefunction did not develop much entanglement. We also considered the role of chaos and equilibration in the estimation of macro properties from quench dynamics. Looking for the sweet spot between low-entanglement wavefunctions which are easily approximated, and high-entanglement wavefunctions where local observables have little information, we found evidence for an inverse relationship between the proximity of local marginal steady states to the maximally mixed state and the ease of estimation of macro properties from local observables from low bond dimension MPS \cite{mitra2023macrostates}.

As we suggested in Chap.~\ref{chap:QuenchChaos}, useful quantum simulation of local order parameters in quench dynamics will require well-chosen problems whereby there is sufficient scrambling to render the problem intractable by classical simulation but not enough scrambling to erase all information of interest in the local observables. In this dissertation, we considered problems involving estimating equal-time local correlation functions. An avenue of future work is to understand the robustness to MPS truncation for two-time correlations, as in the study of out-of-time-ordered correlation functions (OTOCs). 

Another area where quantum simulation of quench dynamics may show a quantum advantage is in the study of local transport of globally conserved quantities in a many-body system through hydrodynamic and diffusion models. However, in a similar spirit to our work, transport properties for certain models may be accessible through appropriately chosen representations, such as truncated MPS \cite{leviatan2017quantum} and the Heisenberg evolution of operators of interest \cite{vonkeyserlingk2022operator, rakovszky2022dissipation}, even when the description of the microscopic details of the dynamics is intractable. A key open question is the precise conditions under which these truncation schemes break down, and local correlation functions can not be efficiently simulated classically. 

In many of these quantum dynamics problems, including the estimation of critical points and critical exponents of the DQPT and the estimation of correlation lengths, we needed to take both thermodynamic and steady-state limits. While we showed these quantities are insensitive to low-entanglement MPS, the relationship between the two infinite limits in specific cases remains unclear. With the appropriate order of limits, it is expected that the estimates of critical properties will converge. This remains an avenue for future work in quantum simulation of long-time dynamics.

Our study of many-body quantum systems has been restricted to closed quantum systems. The presence of noise will naturally modify the probability distribution of measurements and affect the output of a noisy intermediate scale quantum (NISQ) device and the potential tractability of quantum simulation. There have been studies exploring how truncated descriptions of the microstate can reliably model the output to within the errors expected by noise. Zhou \textit{et al.} showed that for realistic errors, a truncated MPS could efficiently simulate the microstate of 1D random circuits in a manner such that the truncation leads to an infidelity that is comparable with the gate infidelity of the noisy two-qubit gates \cite{zhou2020limits}. More detailed realistic modeling of decoherence using open quantum systems matrix product operators (MPOs) by Noh \textit{et al.} \cite{noh2020efficient} and matrix product density operators (MPDOs) by Cheng \textit{et al.} \cite{cheng2021simulating} showed similar results and set a bar for when random circuit sampling in 1D can achieve quantum advantage. The MPDO approach can accurately simulate the true probability distribution of a noisy quantum circuit when the noise is sufficiently large. More recently, Aharonov \textit{et al.} have shown that there is a polynomial-time classical algorithm with which one can sample the output distribution of a noisy random quantum circuit for sufficiently deep circuits with a constant depolarizing error rate per gate \cite{aharonov2022polynomial}. Their method, similar to Ref.~\cite{vonkeyserlingk2022operator}, uses a truncation of ``Pauli paths,'' and a corresponding truncation of operator entanglement \cite{aharonov2022polynomial} when viewed in the Heisenberg picture. Approaches based on truncating Pauli paths has been successfully realized \cite{Fontana2023, Rudolph2023, Shao2023} to reproduce the results of a recent experiment of Floquet dynamics on a gate-based quantum device \cite{kim2023evidence}. 

The general question remains -- what is the relationship between the robustness of a quantum simulation problem to noise and the complexity of the simulation? Quantum simulation is seen as a potential near-term goal for NISQ devices without the need for full fault-tolerant error correction because of the robustness of macro properties and order parameters describing phases of matter. Our preliminary studies here show that such macrostates can have an efficient description using highly truncated MPS in closed quantum systems. Phenomena that require a more microscopic representation are likely to be more complex, but then they are likely to be more sensitive to noise and decoherence. A more complete study of the efficiency with which we can represent \textit{open quantum systems} will be essential to demarcate the quantum-to-classical transition and its implications for achieving a useful quantum advantage in quantum simulation.


\appendix
%
\chapter{Symmetric, one axis two-qubit gates}
\label{app:SymmetricTwoQubitGates}
%

In this appendix, we review some properties of symmetric one-axis two-qubit gates. Two types of gates are implemented by adiabatic Rydberg dressing passages -- the Controlled-Z type gate \cite{keating2015robust} and the Mølmer-Sørensen type gates \cite{mitra2020robust}. The Controlled-Z type gate is of the form
\begin{equation}
    \exp(-\mathrm{i} \vartheta_2 \left(\frac{\hat{\sigma}_z}{2} \otimes \frac{\hat{\sigma}_z}{2}\right)
    -\mathrm{i} \vartheta_1 \left(\frac{\hat{\sigma}_z}{2} \otimes \mathds{1} + \mathds{1} \otimes \frac{\hat{\sigma}_z}{2}\right))
    ,
\end{equation}
and the Mølmer-Sørensen type gate is of the form
\begin{equation}
    \exp(-\mathrm{i} \vartheta_2 \left(\frac{\hat{\sigma}_y}{2} \otimes \frac{\hat{\sigma}_y}{2}\right))
    .
\end{equation}
These are symmetric one-axis two-qubit unitary operators. We call them ``symmetric'' as they are symmetric in the exchange of two qubits and ``one-axis'' as they are about a single axis. A symmetric one-axis two-qubit gate can be written using the collective spin operators, for example, $\hat{S}_\mu$, along the axis $\mu$ given as
\begin{equation}
    \hat{S}_\mu = \mathds{1} \otimes \frac{1}{2} \hat{\sigma}_\mu
    + \frac{1}{2} \hat{\sigma}_\mu \otimes \mathds{1}
    \label{eq:TwoQubitCollectiveSpin}
    .
\end{equation}
The square of $\hat{S}_\mu$, that is $\hat{S}_\mu^2$ is therefore
\begin{equation}
\begin{aligned}
    \hat{S}_\mu^2
    &
    = \frac{1}{4}(
    (\hat{\sigma}_\mu \otimes \mathds{1})^2 +
    (\mathds{1} \otimes \hat{\sigma}_\mu)^2 +
    2 (\hat{\sigma}_\mu \otimes \hat{\sigma}_\mu) )
    = \frac{1}{4} (
    2 (\mathds{1} \otimes \mathds{1}) +
    2 (\hat{\sigma}_\mu \otimes \hat{\sigma}_\mu) )
    \\ &
    = \frac{1}{2} (
    (\mathds{1} \otimes \mathds{1}) +
    (\hat{\sigma}_\mu \otimes \hat{\sigma}_\mu) )
    .
\end{aligned}
\end{equation}
Thus the two-qubit term $\hat{\sigma}_\mu \otimes \hat{\sigma}_\mu$ can be written using $\hat{S}_\mu^2$, up to and additive multiple of identity. Therefore, using the collective spin notation, a symmetric one-axis two-qubit unitary about axis $\mu$ can be written as
\begin{align}
    \exp(-\mathrm{i} \vartheta_1 \hat{S}_\mu - \mathrm{i} \vartheta_2 \frac{1}{2} \hat{S}_\mu^2)
    ,
\end{align}
ignoring an overall phase. Notice this gate has a rotation component, which is generated by the linear term $\hat{S}_\mu$, and a twist component, which is generated by the quadratic term $\hat{S}_\mu^2$.

To avoid confusion with spin quantum numbers, we will call the qubit levels $\uparrow$ and $\downarrow$ in place of $0$ and $1$, with the spin operators defined as
\begin{align}
    \hat{\sigma}_z = \ketbra{\uparrow}{\uparrow} - \ketbra{\downarrow}{\downarrow}
    ,
    &&
    \hat{\sigma}_x = \ketbra{\uparrow}{\downarrow} + \ketbra{\downarrow}{\uparrow}
    &&
    ,
    \hat{\sigma}_y = -\mathrm{i} \ketbra{\uparrow}{\downarrow} + \mathrm{i} \ketbra{\downarrow}{\uparrow}
    .
\end{align}

We consider a symmetric one-axis two-qubit gate along the $z$ axis. In the eigenbasis of $S^2$ and $\hat{S}_z$, the rotation can be expressed as
\begin{align}
    \exp(-\mathrm{i} \vartheta_1 \hat{S}_z)
    = \sum_{s \in {0, 1}} \sum_{m_z=-s}^{s}
    \exp(-\mathrm{i} \vartheta_1 m_z)
    \ketbra{s, m_z}{s, m_z}
    ,
\end{align}
and the twist can be expressed as
\begin{align}
    \exp(-\mathrm{i} \vartheta_2 \frac{1}{2} \hat{S}_z^2)
    = \sum_{s \in {0, 1}} \sum_{m_z=-s}^{s}
    \exp(-\mathrm{i} \vartheta_2 \frac{m_z^2}{2})
    \ketbra{s, m_z}{s, m_z}
    .
\end{align}
Putting them together, we have
\begin{align}
    \exp(-\mathrm{i} \vartheta_1 \hat{S}_z - \mathrm{i} \vartheta_2 \frac{1}{2} \hat{S}_z^2)
    = \sum_{s \in {0, 1}} \sum_{m_z=-s}^{s}
    \exp(-\mathrm{i} \vartheta_1 m_z -\mathrm{i} \vartheta_2 \frac{m_z^2}{2})
    \ketbra{s, m_z}{s, m_z}
    .
\end{align}
Expanding the sum, we have the explicit form of the gate
\begin{equation}
\begin{aligned}
    \exp(-\mathrm{i} \vartheta_1 \hat{S}_z - \mathrm{i} \vartheta_2 \frac{1}{2} \hat{S}_z^2)
    &
    = \exp(-\mathrm{i} \vartheta_1 -\mathrm{i} \vartheta_2 \frac{1}{2})
    \ketbra{s=1, m_z=+1}{s=1, m_z=+1}
    \\ &
    + \exp(+\mathrm{i} \vartheta_1 -\mathrm{i} \vartheta_2 \frac{1}{2})
    \ketbra{s=1, m_z=-1}{s=1, m_z=-1}
    \\ &
    + \ketbra{s=1, m_z=0}{s=1, m_z=0}
    \\ &
    + \ketbra{s=0, m_z=0}{s=0, m_z=0}
    .
\end{aligned}
\end{equation}

Next, we write this in the computational basis
\begin{equation}
\begin{aligned}
    & 
    \exp(-\mathrm{i} \vartheta_1 \hat{S}_z - \mathrm{i} \vartheta_2 \frac{1}{2} \hat{S}_z^2)
    \\ &
    = \exp(-\mathrm{i} \vartheta_1 -\mathrm{i} \vartheta_2 \frac{1}{2})
    \ketbra{\uparrow\uparrow}{\uparrow\uparrow}
    + \exp(+\mathrm{i} \vartheta_1 -\mathrm{i} \vartheta_2 \frac{1}{2})
    \ketbra{\downarrow\downarrow}{\downarrow\downarrow}
    \\ &
    + \frac{1}{\sqrt{2}} (\ket{\uparrow\downarrow} + \ket{\downarrow\uparrow})
    \frac{1}{\sqrt{2}} (\bra{\uparrow\downarrow} + \bra{\downarrow\uparrow})
    + \frac{1}{\sqrt{2}} (\ket{\uparrow\downarrow} - \ket{\downarrow\uparrow})
    \frac{1}{\sqrt{2}} (\bra{\uparrow\downarrow} - \bra{\downarrow\uparrow})
    \\ &
    = \exp(-\mathrm{i} \vartheta_1 -\mathrm{i} \vartheta_2 \frac{1}{2})
    \ketbra{\uparrow\uparrow}{\uparrow\uparrow}
    + \exp(+\mathrm{i} \vartheta_1 -\mathrm{i} \vartheta_2 \frac{1}{2})
    \ketbra{\downarrow\downarrow}{\downarrow\downarrow}
    \\ &
    + \ketbra{\uparrow\downarrow}{\uparrow\downarrow}
    + \ketbra{\downarrow\uparrow}{\downarrow\uparrow}
    .
\end{aligned}
\end{equation}

Written as a matrix, this is
\begin{align}
    \exp(-\mathrm{i} \vartheta_1 \hat{S}_z - \mathrm{i} \vartheta_2 \frac{1}{2} \hat{S}_z^2)
    \doteq
    \begin{pmatrix}
    \mathrm{e}^{-\mathrm{i} \vartheta_1 -\mathrm{i} \vartheta_2 \frac{1}{2}} & 0 & 0 & 0 \\
    0 & 1 & 0 & 0 \\
    0 & 0 & 1 & 0 \\
    0 & 0 & 0 & \mathrm{e}^{+\mathrm{i} \vartheta_1 -\mathrm{i} \vartheta_2 \frac{1}{2}}
    \end{pmatrix}
    .
\end{align}

The Controlled-Z has $\mu=z$, $\vartheta_2 = \pm \pi$ and $\vartheta_1 = \mp \frac{1}{2} \pi$. A Mølmer-Sørensen gate $y$ about has $\mu=y$, $\vartheta_2 = \pm \pi$ and $\vartheta_1 = 0$.

Another family of gates that have been of interest is a controlled-phase gate, which is parameterized by the phase $\varphi$.
\begin{equation}
    \hat{U}_{\text{Cphase}}
    = \ketbra{\uparrow\uparrow}{\uparrow\uparrow}
    + \ketbra{\uparrow\downarrow}{\uparrow\downarrow}
    + \ketbra{\downarrow\uparrow}{\downarrow\uparrow}
    + \mathrm{e}^{\mathrm{i}\varphi} \ketbra{\downarrow\downarrow}{\downarrow\downarrow}
\end{equation}
The Controlled-Z gate itself is a controlled-phase gate with $\varphi = \pi$. Comparing this to the expression we had for symmetric one-axis two-qubit gates with rotation angle $\vartheta_1$ and twist angle $\vartheta_2$, we obtain
\begin{align}
    \frac{\vartheta_2}{2} - \vartheta_1 = 0
    \implies \vartheta_1 = \frac{\vartheta_2}{2}
    ; &&
    \frac{\vartheta_2}{2} + \vartheta_1 = \varphi
    \implies \frac{\vartheta_2}{2} = \varphi - \vartheta_1
    \\
    \implies \vartheta_1 = - \varphi - \vartheta_1
    &&
    \implies \vartheta_1 = -\frac{\varphi}{2}; \qquad \vartheta_2 = - \frac{\varphi}{2}
\end{align}
Thus a controlled-phase gate with a phase of $\varphi$ is a symmetric one-axis two-qubit gate with rotation angle $\vartheta_1 = -\frac{\varphi}{2}$ and twist angle $\vartheta_2 = -\frac{\varphi}{2}$.

%
\section{Gate fidelities}
\label{sec:SymmetricTwoQubitGateFidelities}
%

How well do we implement a symmetric one-axis two-qubit gate? A Mølmer-Sørensen type gate is a specific form of a Controlled-Z type gate with $\vartheta_1 = 0$. Therefore, we consider the fidelity between unitary $\hat{U}'$ and $\hat{U}$
\begin{align}
    \hat{U} = \exp (-\mathrm{i} \vartheta_1 \hat{S}_z - \mathrm{i} \vartheta_2 \frac{1}{2} \hat{S}_z^2)
    &&
    \hat{U}' = \exp (-\mathrm{i} \vartheta'_1 \hat{S}_z - \mathrm{i} \vartheta'_2 \frac{1}{2} \hat{S}_z^2)
\end{align}
using the trace overlap between them $\mathrm{Tr} (\hat{U}^\dagger \hat{U}')$. In particular, the fidelity is
\begin{align}
    \mathcal{F} = \frac{1}{4^2}
     |\mathrm{Tr} (\hat{U}^\dagger \hat{U}')|^2
    = \frac{1}{16}  |\mathrm{Tr} (\hat{U}^\dagger \hat{U}')|^2
\end{align}
The unitary $\hat{U}^\dagger$ is
\begin{equation}
\begin{aligned}
    \hat{U}^\dagger = \exp (+\mathrm{i} \vartheta_1 \hat{S}_z + \mathrm{i} \frac{\vartheta_2}{2} \hat{S}_z^2)
    &
    = \exp (+\mathrm{i} \vartheta_1 + \mathrm{i} \frac{\vartheta_2}{2})
    \ketbra{s=1, m_z=+1}{s=1, m_z=+1}
    \\ &
    + \exp (+\mathrm{i} \vartheta_1 + \mathrm{i} \frac{\vartheta_2}{2})
    \ketbra{s=1, m_z=-1}{s=1, m_z=-1}
    \\ &
    + \ketbra{s=1, m_z=0}{s=1, m_z=0}
    \\ &
    + \ketbra{s=0, m_z=0}{s=0, m_z=0}
    ,
\end{aligned}
\end{equation}
and the unitary $\hat{U}'$ is
\begin{equation}
\begin{aligned}
    \hat{U}' = \exp (-\mathrm{i} \vartheta'_1 \hat{S}_z - \mathrm{i} \frac{\vartheta'_2}{2} \hat{S}_z^2)
    &
    = \exp (-\mathrm{i} \vartheta'_1 -\mathrm{i} \frac{\vartheta'_2}{2})
    \ketbra{s=1, m_z=+1}{s=1, m_z=+1}
    \\ &
    + \exp (+\mathrm{i} \vartheta'_1 -\mathrm{i} \frac{\vartheta'_2}{2})
    \ketbra{s=1, m_z=-1}{s=1, m_z=-1}
    \\ &
    + \ketbra{s=1, m_z=0}{s=1, m_z=0}
    \\ &
    + \ketbra{s=0, m_z=0}{s=0, m_z=0}
    .
\end{aligned}
\end{equation}
The product is
\begin{equation}
\begin{aligned}
    \hat{U}' \hat{U}^\dagger 
    & 
    =
    \exp (-\mathrm{i}  (\vartheta'_1 - \vartheta_1) \hat{S}_z - \mathrm{i} \frac{ (\vartheta'_2 - \vartheta_2)}{2} \hat{S}_z^2)
    \\ &
    = \exp (-\mathrm{i}  (\vartheta'_1 - \vartheta_1) -\mathrm{i} \frac{ (\vartheta'_2 - \vartheta_2)}{2})
    \ketbra{s=1, m_z=+1}{s=1, m_z=+1}
    \\ &
    + \exp (+\mathrm{i}  (\vartheta'_1 - \vartheta_1) -\mathrm{i} \frac{ (\vartheta'_2 - \vartheta_2)}{2})
    \ketbra{s=1, m_z=-1}{s=1, m_z=-1}
    \\ &
    + \ketbra{s=1, m_z=0}{s=1, m_z=0}
    \\ &
    + \ketbra{s=0, m_z=0}{s=0, m_z=0}
    .
\end{aligned}
\end{equation}
The trace of this expression is simply
\begin{equation}
\begin{aligned}
    \mathrm{Tr} (\hat{U}' \hat{U}^\dagger) 
    & =
    \mathrm{Tr} (\exp (-\mathrm{i}  (\vartheta'_1 - \vartheta_1) \hat{S}_z - \mathrm{i} \frac{ (\vartheta'_2 - \vartheta_2)}{2} \hat{S}_z^2))
    \\ &
    = \exp (-\mathrm{i}  (\vartheta'_1 - \vartheta_1) -\mathrm{i} \frac{ (\vartheta'_2 - \vartheta_2)}{2})
    + \exp (+\mathrm{i}  (\vartheta'_1 - \vartheta_1) -\mathrm{i} \frac{ (\vartheta'_2 - \vartheta_2)}{2})
    + 1
    + 1
    \\ &
    = 2 + \exp (-\mathrm{i} \frac{ (\vartheta'_2 - \vartheta_2)}{2}) 2 \cos (\vartheta'_1 - \vartheta_1) 
    \\ &
    = 2  (1 + \exp (-\mathrm{i} \frac{ (\vartheta'_2 - \vartheta_2)}{2}) \cos (\vartheta'_1 - \vartheta_1))
    .
\end{aligned}
\end{equation}
Using $\delta \vartheta_2 = \vartheta'_2 - \vartheta_2$ and $\delta \vartheta_1 = \vartheta'_1 - \vartheta_1$, this simplifies to
\begin{equation}
\mathrm{Tr} (\hat{U}' \hat{U}^\dagger) =
    \mathrm{Tr} (\exp (-\mathrm{i}  (\vartheta'_1 - \vartheta_1) \hat{S}_z - \mathrm{i} \frac{ (\vartheta'_2 - \vartheta_2)}{2} \hat{S}_z^2))
    = 2  (1 + \exp (-\mathrm{i} \frac{\delta\vartheta_2}{2}) \cos (\delta\vartheta_1))    
    .
\end{equation}
Therefore the fidelity is
\begin{equation}
\begin{aligned}
    \mathcal{F} 
    & 
    = \frac{1}{4^2}
     | 2  (1 + \exp (-\mathrm{i} \frac{\delta\vartheta_2}{2}) \cos (\delta\vartheta_1)) |^2
    \\ &
    = \frac{1}{4} \left( 1 + \cos^2 (\delta \vartheta_1) + \cos (\delta \vartheta_1)
    \cos (\frac{\delta\vartheta_2}{2}) \right)
    .
\end{aligned}
\end{equation}

Therefore, the fidelity between Controlled-Z type gates $\hat{U}$ and $\hat{U}'$ is
\begin{align}
    \mathcal{F}_{\text{CZ}} = \frac{1}{4}
     \left(1 + \cos^2 (\delta \vartheta_1)
    + 2\cos (\delta \vartheta_1)
    \cos (\frac{\delta \vartheta_2}{2})\right)
    .
\end{align}

For Mølmer-Sørensen type gates, $\vartheta_1 = 0$ and $\vartheta'_1 = 0$, therefore $\delta \vartheta_1 = 0$. The fidelity between two Mølmer-Sørensen type gates $\hat{U}$ and $\hat{U}'$ is
\begin{align}
    \mathcal{F}_{\text{MS}} = \frac{1}{4}
     \left(1 + 1
    + 2 \cos (\frac{\delta \vartheta_2}{2})\right)
    = \frac{1}{2} \left(1 + \cos (\frac{\delta \vartheta_2}{2})\right)
    = \cos^2 \left(\frac{\delta \vartheta_2}{4}\right)
    .
\end{align}

From these expressions, we see that the fidelity of implementing Controlled-Z type gate has contributions from both the difference in the rotation angle $\delta \vartheta_1$ and the difference in the twist angle $\delta \vartheta_2$, while the fidelity of implementing a Mølmer-Sørensen gate has contributions only from the difference in the twist angle $\delta \vartheta_2$. Moreover, the fidelity of a Controlled-Z type gate is dominated by the difference in the rotation angle $\delta \vartheta_1$, which is absent in the fidelity of a Mølmer-Sørensen gate. Thus, the Mølmer-Sørensen type gate is more robust to errors in implementation, as long as we implement a Mølmer-Sørensen type gate.

Finally, we calculate the fidelity between controlled-phase gates $\hat{U}$, with phase $\varphi$ and $\hat{U}'$, with phase $\varphi'$. The product $\hat{U}' \hat{U}^\dagger$ is
\begin{equation}
\begin{aligned}
    \hat{U}' \hat{U}^\dagger
    & 
    = \ketbra{s=1, m_z=+1}{s=1, m_z=+1}
    \\ &
    + \ketbra{s=1, m_z=0}{s=1, m_z=0}
    \\ &
    + \ketbra{s=0, m_z=0}{s=0, m_z=0}
    \\ &
    + \mathrm{e}^{+\mathrm{i}  (\varphi - \varphi')}
    \ketbra{s=1, m_z=-1}{s=1, m_z=-1}
    .
\end{aligned}
\end{equation}
The trace of this expression is
\begin{equation}
    \mathrm{Tr} (\hat{U}' \hat{U}^\dagger)
    = 3 + \mathrm{e}^{+\mathrm{i}  (\varphi - \varphi')}
    .
\end{equation}
Thus the fidelity is
\begin{equation}
\begin{aligned}
    \mathcal{F}_{\text{Cphase}} 
    &
    = \frac{1}{4^2}
     |3 + \mathrm{e}^{+\mathrm{i}  (\varphi - \varphi')}|^2
    \\ &
    = \frac{1}{8}  (
    5 + 3 \cos (\varphi - \varphi') )
    .
\end{aligned}
\end{equation}
\chapter{Adiabatic passages for Rydberg dressing}
\label{app:RydbergAdiabaticPassage}

In this appendix, we discuss rapid adiabatic passages from ground states to Rydberg states and back used to implement entangling gates for neutral atoms. Since the gates are symmetry under the exchange of the two qubits, it is convenient to use the two-qubit collective spin operators as defined in Eq.~\eqref{eq:TwoQubitCollectiveSpin}.

%
\section{Controlled-Z}
%
The Controlled-Z gate using adiabatic Rydberg dressing involves a single adiabatic ramp implementing
\begin{equation}
\begin{aligned}
    \exp\left(-\mathrm{i} \varphi_1 \hat{S}_z - \mathrm{i} \varphi_2 \tfrac{1}{2} \hat{S}_z^2\right)
    & 
    = \exp\left(-\mathrm{i} \varphi_1 -\mathrm{i} \varphi_2 \tfrac{1}{2}\right)
    \ketbra{\uparrow\uparrow}{\uparrow\uparrow}
    + \exp\left(+\mathrm{i} \varphi_1 -\mathrm{i} \varphi_2 \tfrac{1}{2}\right)
    \ketbra{\downarrow\downarrow}{\downarrow\downarrow}
    \\ &
    + \ketbra{\uparrow\downarrow}{\uparrow\downarrow}
    + \ketbra{\downarrow\uparrow}{\downarrow\uparrow}
\end{aligned}
\end{equation}
Under the adiabatic approximation and assuming all population is returned to the qubit states $\uparrow, \downarrow$, the rotation angle $\varphi_1$ and the twist angle $\varphi_2$ are related to the dynamical phases accumulated from the one atom light shift $E_{\mathrm{LS}(1)}$ and the entangling energy $\hbar \kappa = E_{\mathrm{LS}(1)} - 2E_{\mathrm{LS}(2)}$ as
\begin{align}
    \varphi_1 = - \int_{\mathrm{ramp}} \mathrm{d} t' \left( \frac{1}{\hbar} E_{\mathrm{LS}(1)}(t') + \frac{1}{2} \kappa(t')\right)
    &&
    \varphi_2 = \int_{\mathrm{ramp}} \mathrm{d} t' \kappa(t')
\end{align}
A straightforward way to design a ramp is to choose one where the expression for $\varphi_2$ with the integral over time is $\pi$. In other words, the time integral of the entangling energy $\kappa(t)$ is $\pi$. This would given some value of $\varphi_1$, which can be calculated. Following this with a rotation about $z$ to make the cumulative rotation angle about $z$ be $\pi/2$ implements the controlled-Z gate.

%
\section{Mølmer-Sørensen gate}
%

The adiabatic passages interleaved in a spin echo sequence as in Fig.~\ref{fig:SpinEchoAdiabaticPassage} give a Mølmer-Sørensen type gate about the $y$ axis.
\begin{equation}
\begin{aligned}
    &
    \hat{R}_x\left(\tfrac{\pi}{2}\right)
    \exp\left(- \mathrm{i} \varphi_1 \hat{S}_z
    - \mathrm{i} \varphi_2 \tfrac{1}{2} \hat{S}_z^2\right)
    \hat{R}_x\left(\pi\right)
    \exp\left(- \mathrm{i} \varphi_1 \hat{S}_z
    - \mathrm{i} \varphi_2 \tfrac{1}{2} \hat{S}_z^2\right)
    \hat{R}_x\left(\tfrac{\pi}{2}\right)
    \\
    = &
    \exp\left(- \mathrm{i} 2 \varphi_2 \tfrac{1}{2} \hat{S}_y^2\right)
    .
\end{aligned}
\end{equation}
For this to be canonical Mølmer-Sørensen gate about $y$, which is a perfect entangler, $2\varphi_2 = \pi$. A way to design ramps is to look at the fidelity of the gate implemented by the spin echo adiabatic sequence in Fig.~\ref{fig:SpinEchoAdiabaticPassage} with the canonical Mølmer-Sørenson gate about $y$. This is the expression $\mathcal{F}_{\mathrm{ramp}} = \left|\mathcal{G}_{\mathrm{ramp}}\right|^2 / 4^2$, where $\mathcal{G}_{\mathrm{ramp}}$ is the expression
\begin{multline}
    \mathcal{G}_{\mathrm{ramp}}
    = \mathrm{Tr}\left(
    \hat{R}_x\left(\tfrac{1}{2} \pi\right) \hat{U}_{\kappa} \hat{R}_x\left(\pi\right) \hat{U}_{\kappa} \hat{R}_x\left(\tfrac{1}{2} \pi\right) \,
    \left[\exp\left(- \mathrm{i} \pi \tfrac{1}{2} \hat{S}_y^2\right)\right]^\dagger\right)
    \\
    = \mathrm{Tr}\left(
    \hat{R}_x\left(\tfrac{1}{2} \pi\right) \hat{U}_{\kappa} \hat{R}_x\left(\pi\right) \hat{U}_{\kappa} \hat{R}_x\left(\tfrac{1}{2} \pi\right) \,
    \exp\left(+ \mathrm{i} \pi \tfrac{1}{2} \hat{S}_y^2 \right)\right)
    ,
    \label{eq:GoalFunction}
\end{multline}
where $\hat{U}_\kappa$ is the unitary implemented during one of the adiabatic ramps in the spin echo adiabatic sequence in Fig.~\ref{fig:SpinEchoAdiabaticPassage}.

%
\section{Sweeping Rabi frequency and detuning}
%

%
\subsection{One photon excitation}
%

For implementations using a one photon ground-to-Rydberg excitation, we consider an adiabatic ramp which involves a Gaussian sweep of the ultraviolet intensity, proportional to $\Omega_{1, r}$ and a linear sweep of the ultraviolet frequency $\Delta_{1, r}(t)$ to and from resonance to maintain the adiabatic condition. During the dressing sweep the magnitude of the detuning is swept linearly from about $\Delta_{\mathrm{max}}$ to $\Delta_{\mathrm{min}}$ from $t_{\mathrm{begin}}$ to $t_{\mathrm{end}}$. Simultaneously, the Rabi frequency is changed as a half Gaussian from $\Omega_{\mathrm{min}}$ to $\Omega_{\mathrm{max}}$. 
\begin{equation}
\begin{split}
    \left|\Delta_{1, r}(t)\right|
    = \Delta_{\max}
    +\frac{ (\Delta_{\max} - \Delta_{\min}) }
       {(t_{\mathrm{end}} - t_{\mathrm{begin}})}
    \left(t - t_{\mathrm{begin}}\right)
    \\ 
    \Omega_{1, r}(t)
    = \Omega_{\mathrm{\min}}
    +  (\Omega_{\max} - \Omega_{\min})
    \exp\left( - \frac{(t - t_{\mathrm{begin}})^2}{2t_{\mathrm{width}}^2}\right)
\end{split}
\end{equation}
After dressing the Rabi frequency $\Omega_{1, r}$ and the detuning $\Delta_{1, r}$ are held constant for a duration $t_{\mathrm{constant}}$. The undressing sweep is the reverse of the dressing sweep where the magnitude of the detuning is swept linearly from about $\Delta_{\mathrm{min}}$ to $\Delta_{\mathrm{max}}$ and the Rabi frequency is changed as a half Gaussian from $\Omega_{\mathrm{max}}$ to $\Omega_{\mathrm{min}}$ from $t_{\mathrm{begin}}$ to $t_{\mathrm{end}}$. 
\begin{equation}
\begin{split}
    \left|\Delta_{1, r}(t) \right|
    = \Delta_{\min}
    +\frac{ (\Delta_{\min} - \Delta_{\max}) }
        {(t_{\mathrm{end}} - t_{\mathrm{begin}})}
    \left(t - t_{\mathrm{begin}}\right)
    \\ 
    \Omega_{1, r}(t)
    = \Omega_{\mathrm{\max}}
    +  (\Omega_{\min} - \Omega_{\max})
    \exp\left( - \frac{(t - t_{\mathrm{begin}})^2}{2t_{\mathrm{width}}^2}\right)
\end{split}
\end{equation}
The parameters of the ramp are detuning parameters, $\Delta_{\min}, \Delta_{\max}$ temporal duration parameters $t_{\mathrm{constant}}, t_{\mathrm{width}}$ and temporal instant parameters $t_{\mathrm{begin}}$ and $t_{\mathrm{end}}$. Thes are determined using the cost function $\mathcal{G}_{\mathrm{ramp}}$ in Eq.~\eqref{eq:GoalFunction}. The minimum Rabi frequency $\Omega_{\min} = 0$, other angular frequencies are measured in units of the maximum Rabi frequency $\Omega_{\max}$ and time is measured in units of $2\pi/\Omega_{\max}$. Since the undressing sweep is simply the reverse of the dressing sweep, $t_{\mathrm{begin}}, t_{\mathrm{end}}$ in the undressing sweep are determined by $t_{\mathrm{begin}}, t_{\mathrm{end}}$ in the dressing sweep.

%
\subsection{Two photon excitation}
%

For the two-photon case, the effect of the light shift arising from the intermediate state provides an additional control parameter for adiabatic passages \cite{mitra2023neutral}. Tuning to exact two-photon resonance in the absence of the light shift from the admixture of the intermediate state, and a fixed Rabi frequency $\Omega_{ar}$ and detuning $\Delta_{ar}$ on the $\ket{a} \leftrightarrow \ket{r}$ transition, adiabatic Rydberg dressing is achieved simply through a Gaussian ramp of the intensity of the laser driving the $\ket{1} \leftrightarrow \ket{a}$ according to the Rabi frequency,
\begin{equation}
  \Omega_{1a} =
  \begin{cases}
    \Omega_{1a}^{\mathrm{max}}, \mathrm{ }-\lvert t_{\mathrm{stop}}\rvert \leq t \leq\lvert t_{\mathrm{stop}}\rvert\\
    \Omega_{1a}^{\mathrm{max}} \exp(-\frac{(t-\lvert t_{\mathrm{stop}\rvert})^2}{2 t_w^2}), \mathrm{  otherwise.}\\
  \end{cases}
  \label{eq:GaussianOmega1aAppendix}
\end{equation}
We optimized parameters $t_{\mathrm{stop}}$, the time after which the Rabi frequency remains constant, and $t_w$  the width of the Gaussian pulse, to obtain the desired gate of interest \cite{mitra2023neutral} using function $\mathcal{G}_{\mathrm{ramp}}$ in Eq~\eqref{eq:GoalFunction}. 

%
\chapter{Implementation of the Mølmer-Sørensen gate}
\label{app:MSGateDetails}
%

In this appendix, we show that the sequence introduced in Chap.~\ref{chap:AdiabaticRydbergDressing} and shown in Fig.~\ref{fig:SpinEchoAdiabaticPassage} implements in a Mølmer-Sørensen gate. Let $\hat{R}_x(\Theta)$ denote a rotation by $\Theta$ about the $x$ axis and $\hat{U}_{\kappa}(\vartheta_1, \vartheta_2)$
denote the unitary implemented by one of the adiabatic dressing ramps. Specifically,
\begin{equation}
    \hat{U}_{\kappa}(\vartheta_1, \vartheta_2)
    = \exp(- \mathrm{i} \vartheta_1 \hat{S}_z
    - \mathrm{i} \vartheta_2 \tfrac{1}{2} \hat{S}_z^2)
\end{equation}
and
\begin{equation}
    \hat{R}_x(\Theta)
    = \exp(- \mathrm{i} \Theta \hat{S}_x).
\end{equation}

The sequence with two adiabatic passages interleaved in a spin echo, shown in Fig.~\ref{fig:SpinEchoAdiabaticPassage} is 
\begin{equation}
    \hat{R}_x(\tfrac{\pi}{2})
    \hat{U}_{\kappa}(\vartheta_1, \vartheta_2)
    \hat{R}_x(\pi)
    \hat{U}_{\kappa}(\vartheta_1, \vartheta_2)
    \hat{R}_x(\tfrac{\pi}{2}),
\end{equation}
which can be written as
\begin{multline}
    \hat{R}_x(\tfrac{1}{2} \pi)
    \hat{U}_\kappa(\vartheta_1, \vartheta_2)
    \hat{R}_x(\pi)
    \hat{U}_\kappa(\vartheta_1, \vartheta_2)
    \hat{R}_x(\tfrac{1}{2} \pi)
    =
    [
    \hat{R}_x(\tfrac{1}{2} \pi)
    \hat{U}_{\kappa}(\vartheta_1, \vartheta_2)
    \hat{R}_x(-\pi)
    \hat{U}_{\kappa}(\vartheta_1, \vartheta_2)
    \hat{R}_x(\tfrac{1}{2}\pi)
    ]
    \\
    = [
    \hat{R}_x(\tfrac{1}{2} \pi)
    \hat{U}_{\kappa}(\vartheta_1, \vartheta_2)
    \hat{R}_x(-\tfrac{1}{2} \pi)
    ]\,
    [
    \hat{R}_x(-\tfrac{1}{2} \pi)
    \hat{U}_{\kappa}(\vartheta_1, \vartheta_2)
    \hat{R}_x(\tfrac{1}{2}\pi)
    ]
\end{multline}
using $\hat{R}_x(\pi) = \hat{R}_x(-\pi)$. Using $\hat{R}_x(-\tfrac{1}{2}\pi) = \hat{R}_x^\dagger(\tfrac{1}{2}\pi)$
and that a $-\pi/2$ rotation about the $x$ axis converts the $z$ axis to the $y$ axis
\begin{equation}
    \hat{R}_x (+\tfrac{1}{2}\pi)
    \, \hat{S}_z \, 
    \hat{R}_x (-\tfrac{1}{2}\pi)
    = \hat{R}_x^\dagger (-\tfrac{1}{2} \pi)
    \, \hat{S}_z \,
    \hat{R}_x (-\tfrac{1}{2} \pi)
    = + \hat{S}_y,
\end{equation}
the sequence $\hat{R}_x (+\tfrac{1}{2}\pi)
    \, \hat{U}_{\kappa}(\vartheta_1, \vartheta_2) \, 
    \hat{R}_x (-\tfrac{1}{2}\pi)$ simplifies to
\begin{equation}
\begin{aligned}
    &
    \hat{R}_x (+\tfrac{1}{2}\pi)
    \, \hat{U}_{\kappa}(\vartheta_1, \vartheta_2) \, 
    \hat{R}_x (-\tfrac{1}{2}\pi)
    = \hat{R}_x^\dagger (-\tfrac{1}{2}\pi)
    \, \exp(- \mathrm{i} \vartheta_1 \hat{S}_z
    - \mathrm{i} \vartheta_2 \tfrac{1}{2} \hat{S}_z^2)
    \, \hat{R}_x(-\tfrac{1}{2}\pi)
    \\ &
    = \exp(- \mathrm{i} \vartheta_1 (+\hat{S}_y)
    - \mathrm{i} \vartheta_2 \tfrac{1}{2} (+\hat{S}_y)^2)
    \\ &
    = \exp(- \mathrm{i} \vartheta_1 \hat{S}_y
    - \mathrm{i} \vartheta_2 \tfrac{1}{2} \hat{S}_y^2).
\end{aligned}
\end{equation}

Moreover, using $\hat{R}_x(-\tfrac{1}{2}\pi) = \hat{R}_x^\dagger(\tfrac{1}{2}\pi)$
and that a $+\pi/2$ rotation about the $x$ axis converts the $z$ axis to the $-y$ axis
\begin{equation}
    \hat{R}_x (+\tfrac{1}{2}\pi)
    \, \hat{S}_z \, 
    \hat{R}_x (-\tfrac{1}{2}\pi)
    = \hat{R}_x^\dagger (+\tfrac{1}{2} \pi)
    \, \hat{S}_z \,
    \hat{R}_x(+\tfrac{1}{2} \pi)
    = - \hat{S}_y,
\end{equation}
the sequence $\hat{R}_x (-\tfrac{1}{2}\pi)
    \, \hat{U}_{\kappa}(\vartheta_1, \vartheta_2) \, 
    \hat{R}_x (\tfrac{1}{2}\pi)$ simplifies to
\begin{equation}
\begin{aligned}
    &
    \hat{R}_x (-\tfrac{1}{2}\pi)
    \, \hat{U}_{\kappa}(\vartheta_1, \vartheta_2) \, 
    \hat{R}_x (+\tfrac{1}{2}\pi)
    = \hat{R}_x^\dagger (+\tfrac{1}{2}\pi)
    \, \exp(- \mathrm{i} \vartheta_1 \hat{S}_z
    - \mathrm{i} \vartheta_2 \tfrac{1}{2} \hat{S}_z^2)
    \, \hat{R}_x(+\tfrac{1}{2}\pi)
    \\ &
    = \exp(- \mathrm{i} \vartheta_1 (-\hat{S}_y)
    - \mathrm{i} \vartheta_2 \tfrac{1}{2} (+\hat{S}_y)^2)
    \\ &
    = \exp(- \mathrm{i} \vartheta_1 \hat{S}_y
    - \mathrm{i} \vartheta_2 \tfrac{1}{2} \hat{S}_y^2).
\end{aligned}
\end{equation}

Putting these together, the sequence in Fig.~\ref{fig:SpinEchoAdiabaticPassage} implements
\begin{equation}
\begin{aligned}
    &
    \hat{R}_x(\tfrac{1}{2} \pi)
    \hat{U}_\kappa(\vartheta_1, \vartheta_2)
    \hat{R}_x(\pi)
    \hat{U}_\kappa(\vartheta_1, \vartheta_2)
    \hat{R}_x(\tfrac{1}{2} \pi)
    \\ &
    = \exp(- \mathrm{i} \vartheta_1 \hat{S}_y
    - \mathrm{i} \vartheta_2 \tfrac{1}{2} \hat{S}_y^2)
    \exp(+ \mathrm{i} \vartheta_1 \hat{S}_y
    - \mathrm{i} \vartheta_2 \tfrac{1}{2} \hat{S}_y^2)
    = \exp(- \mathrm{i} 2 \vartheta_2 \tfrac{1}{2} \hat{S}_y^2)
    \\ &
    = \exp(- \mathrm{i} \vartheta_2 \hat{S}_y^2),
\end{aligned}
\end{equation}
where $\vartheta_2$ is the entangling phase accumulated in one of the adiabatic ramps. To implement a perfect entangler, the $YY$ Mølmer-Sørensen gate, each adiabatic ramp needs to be designed with $\vartheta_2 = \pi/2$.

%
\chapter{Overview of two-qubit gates}
\label{app:TwoQubitGatesReview}
%

In this appendix, we review some properties of the Lie group $\mathbb{SU}(4)$, which is the group of unitary gates acting on two qubits. A two-qubit unitary gate may be a product unitary, acting on each qubit independently, or it may be an entangling unitary that generates entanglement from product states \cite{zhang2003geometric, zhang2004minimum, zhang2005generation, goerz2015optimizing}. The group Lie group $\mathbb{SU}(4)$ is generated by the Lie algebra $\mathfrak{su}(4)$, which is spanned by the tensor products of two-qubit Pauli operators. Let us use the shorthand $\hat{\sigma}^{\mu}_{a} \hat{\sigma}^{\nu}_{b}$ to represent $\hat{\sigma}^{\mu} \otimes \hat{\sigma}^{\nu}$ acting on qubits $a$ and $b$, with $\mu, \nu \in \{0, x, y, z\}$ and for ease of notation, $\sigma^{0} \equiv \mathds{1}_{2\times 2}$, the one-qubit identity operator. The Lie algebra can be expressed as
\begin{equation}
    \mathfrak{su}(4) = 
    \mathrm{i} \,
    \mathrm{Span}\left( 
    \left\{
    \frac{\hat{\sigma}^{\mu}_{a}}{\sqrt{2}} \frac{\hat{\sigma}^{\nu}_{b}}{\sqrt{2}} 
    \, \big\vert \, \mu, \nu \in \{0, x, y, z\}
    \right\}
    \setminus
    \left\{
    \frac{\hat{\sigma}^{0}_{a}}{\sqrt{2}} \frac{\hat{\sigma}^{0}_{b}}{\sqrt{2}} 
    \right\}
    \right),
\end{equation}
where $\setminus$ denotes the difference between sets, the tensor product of two one-qubit identity operators is excluded, and a factor of $\mathrm{i}$ is included to make the generators anti-Hermitian. We normalize each one-qubit Pauli operator with a factor of $1/\sqrt{2}$ so that they have a Hilbert-Schmidt norm of $1$ that is
\begin{equation}
    \mathrm{Tr}\left( \hat{W} \hat{W}^\dagger \right) = 1
    ,
\end{equation}
for all Pauli operators $\hat{W}$.

Product unitary gates, which act on each qubit independently belong to the subgroup $\mathbb{SU}(2) \otimes \mathbb{SU}(2)$, which is generated by one-qubit Pauli operators of the form $\sigma^{0}_{a} \sigma^{\mu}_{b}$ and $\sigma^{0}_{a} \sigma^{\mu}_{b}$, for $\mu \in \{x, y, z\}$. Moreover, entangling unitary gates, which generate entanglement between qubits are generated by elements of the Lie algebra $\mathfrak{su}(4)$ that have two-qubit Pauli operators of the form $\sigma^{\mu}_{a} \sigma^{\nu}_{b}$ with $\mu, \nu \in \{x, y, z\}$.

Formally, a Lie algebra $\mathfrak{g}$ has a Cartan decomposition
\begin{equation}
    \mathfrak{g} = \mathfrak{p} \oplus \mathfrak{k}
    \label{eq:CartanDecomposition}
    ,
\end{equation}
where $\mathfrak{k}$ is the subalgebra where commutators between two elements of the subalgebra are in $\mathfrak{k}$, $\mathfrak{p}$ is the subalgebra orthogonal to it, $\mathfrak{p} = \mathfrak{k}^{\perp}$, with commutators between two elements of $\mathfrak{p}$ being in $\mathfrak{k}$. Finally, commutators between an element in $\mathfrak{p}$ and an element in $\mathfrak{k}$ evaluate to an element in $\mathfrak{p}$ \cite{zhang2003geometric, zhang2004minimum}. This is summarized as \cite{zhang2003geometric, zhang2004minimum}
\begin{equation}
\begin{aligned}
  \left[\mathfrak{k}, \mathfrak{k}\right] 
  & \subset \mathfrak{k}
  , \\ 
  \left[\mathfrak{p}, \mathfrak{p}\right] 
  & \subset \mathfrak{k}
  , \\
  \left[\mathfrak{p}, \mathfrak{k}\right] 
  & \subset \mathfrak{p} .
\end{aligned}
\end{equation}
A maximal Abelian subalgebra $\mathfrak{a}$ contained in $\mathfrak{p}$ is a Cartan subalgebra for the decomposition Eq.~\eqref{eq:CartanDecomposition}. The subalgebra $\mathfrak{p}$ can written as the union of all subalgebras generated by all Cartan subalgebra and $\mathfrak{k}$ \cite{zhang2003geometric}.

In the case of the Lie algebra of interest for two-qubit gates, $\mathfrak{g} = \mathfrak{su}(4)$, we have,
\begin{equation}
    \mathfrak{k} = 
    \frac{\mathrm{i}}{2} \,
    \mathrm{Span}\left( 
    \left\{
    \hat{\sigma}^{x}_{a}, \hat{\sigma}^{y}_{a}, \hat{\sigma}^{z}_{a}, 
    \hat{\sigma}^{x}_{b}, \hat{\sigma}^{y}_{b}, \hat{\sigma}^{z}_{b}, 
    \right\}
    \right),
\end{equation}
where we have dropped the tensor product with the identity operator $\sigma^{0}$, for brevity, and
\begin{equation}
    \mathfrak{p} = 
    \frac{\mathrm{i}}{2} \,
    \mathrm{Span}\left( 
    \left\{
    \hat{\sigma}^{x}_{a} \hat{\sigma}^{x}_{b}, \hat{\sigma}^{y}_{a} \hat{\sigma}^{x}_{b}, \hat{\sigma}^{z}_{a} \hat{\sigma}^{x}_{b}, 
    \hat{\sigma}^{x}_{a} \hat{\sigma}^{y}_{b}, \hat{\sigma}^{y}_{a} \hat{\sigma}^{y}_{b}, \hat{\sigma}^{z}_{a} \hat{\sigma}^{y}_{b},
    \hat{\sigma}^{x}_{a} \hat{\sigma}^{z}_{b}, \hat{\sigma}^{y}_{a} \hat{\sigma}^{z}_{b}, \hat{\sigma}^{z}_{a} \hat{\sigma}^{z}_{b},
    \right\}
    \right).
\end{equation}
Moreover the Abelian subalgebra, $\mathfrak{a}$,
\begin{equation}
    \mathfrak{a} = 
    \frac{\mathrm{i}}{2} \,
    \mathrm{Span}\left( 
    \left\{
    \hat{\sigma}^{x}_{a} \hat{\sigma}^{x}_{b}, \hat{\sigma}^{y}_{a} \hat{\sigma}^{y}_{b}, \hat{\sigma}^{z}_{a} \hat{\sigma}^{z}_{b}
    \right\}
    \right).
\end{equation}
is contained in $\mathfrak{k}$ and is a maximal Abelian subalgebra.

Using the subalgebras $\mathfrak{k}$ and $\mathfrak{a}$, a convenient representation of a two-qubit gate is through a KAK decomposition \cite{zhang2003geometric, zhang2004minimum, goerz2015optimizing}, which we describe here. Any two-qubit unitary gate $\hat{G}$ can be expressed as
\begin{equation}
  \hat{G} = \hat{K}_1 \hat{A} \hat{K}_2
  \label{eq:TwoQubitGateKAKDecomposition}
  ,
\end{equation}
where $\hat{K}_1$ and $\hat{K}_2$ are product unitaries generated by the subalgebra $\mathfrak{k}$ and $\hat{A}$ is pure entangling unitary generated by the Abelian subalgebra $\mathfrak{a}$. The property of generating the entire subalgebra $\mathfrak{p}$ using composition of elements from $\mathfrak{k}$ and elements from $\mathfrak{a}$ facilitates this decomposition.

The two-qubit Pauli operators are the challenging parts of the generators, requiring interacting Hamiltonians to generate. Therefore, it is convenient to parameterize every two-qubit gate by parametrizing the pure entangling (with no product component) operator $\hat{A}$ in its KAK decomposition as
\begin{equation}
    \hat{A} = \exp \left(- \frac{\mathrm{i}}{2}\left(
    \theta_x \, \hat{\sigma}^x_{a} \hat{\sigma}^x_{b}  + 
    \theta_y \, \hat{\sigma}^y_{a} \hat{\sigma}^y_{b}  + 
    \theta_z \, \hat{\sigma}^z_{a} \hat{\sigma}^z_{b} 
    \right)
    \right)
    \label{eq:TwoQubitGateEntanglingParamerizing}
\end{equation}
where $\theta_x, \theta_y, \theta_z$ are angles determining the correlated rotation along $x, y, z$ axes generated by $\hat{A}$. The two qubit gate $\hat{G}$ is periodic in $\theta_x, \theta_y, \theta_z$, therefore the geometric structure is a $3$-torus \cite{zhang2003geometric, zhang2004minimum, zhang2005generation, goerz2015optimizing}. 

Removing the symmetries in these angles leads to geometric interpretation using the Weyl chamber construction \cite{zhang2003geometric, zhang2004minimum, goerz2015optimizing}, where each nonlocal two-qubit gate lies with a tetrahedron. While the details of this construction are beyond the scope of this Appendix, we use the simple fact that two-qubit gates that can be transformed into each other using only local gates are equivalent. For example, $\mathbb{SU}(2)$ transformations on one or both qubits can be used to change the axes of the generators. For example $x \leftrightarrow z$ can be achieved using the Hadamard gate \cite{nielsen2010quantum, goerz2015optimizing}
\begin{equation}
    \hat{U}_{\mathrm{Hadamard}}
    = 
    \mathrm{i}
    \exp\left(- \mathrm{i} \pi \frac{1}{2}
    \left(
    \frac{1}{\sqrt{2}} \hat{\sigma}^x + \frac{1}{\sqrt{2}} \hat{\sigma}^z
    \right) \right)
    \doteq
    \frac{1}{\sqrt{2}}
    \begin{pmatrix}
        1 & 1 \\ 1 & -1
    \end{pmatrix}
    ,
\end{equation}
which represents a $\pi$ rotation about an axis making angle $\pi/4$ from the $z$ axis in the $z-x$ plane.

Formally, two unitary gates $\hat{G}_1$ and $\hat{G}_{2}$ are called \textit{locally equivalent} if they can be transformed into each other only by local operators $\hat{G}_1 = \hat{L}_1 \hat{G}_2 \hat{L}_2$, or equivalently, $\hat{G}_1 = \hat{L}_1^\dagger \hat{G}_1 \hat{L}_2^{\dagger}$ where $\hat{L}_1, \hat{L}_2 \in \mathbb{SU}(2) \otimes \mathbb{SU}(2)$ are tensor products of one-qubit gates.

Next we review some well-known two-qubit gates. The Controlled-Z (CZ) gate, can be written as an exponential of its Hermitian generators as
\begin{equation}
    \hat{U}_{\mathrm{CZ}}
    = \exp\left( -\mathrm{i} \frac{\pi}{4} \right)
    \exp\left(
    \pm \mathrm{i} \frac{\pi}{2}
    \left(
    \frac{\hat{\sigma}^z_a}{2}
    + \frac{\hat{\sigma}^z_b}{2}
    \right)
    \mp \mathrm{i} \frac{\pi}{2}
    \frac{\hat{\sigma}^z_a \hat{\sigma}^z_b}{2}
    \right)
    \doteq
    \begin{pmatrix}
        1 & 0 & 0 & 0 \\ 
        0 & 1 & 0 & 0 \\
        0 & 0 & 1 & 0 \\
        0 & 0 & 0 & -1
    \end{pmatrix}
    ,
\end{equation}
which can be interpreted as having $\theta_z = \pi/2$. The Controlled-NOT (CNOT) gate, with $a$ as the control qubit and $b$ as the target qubit, written as the exponential of its Hermitian generators, reads
\begin{equation}
    \hat{U}_{\mathrm{CNOT}}
    = \exp\left( -\mathrm{i} \frac{\pi}{4} \right)
    \exp\left(
    \pm \mathrm{i} \frac{\pi}{2}
    \left(
    \frac{\hat{\sigma}^z_a}{2}
    + \frac{\hat{\sigma}^x_b}{2}
    \right)
    \mp \mathrm{i} \frac{\pi}{2}
    \frac{\hat{\sigma}^z_a \hat{\sigma}^x_b}{2}
    \right)
    \doteq
    \begin{pmatrix}
        1 & 0 & 0 & 0 \\ 
        0 & 1 & 0 & 0 \\
        0 & 0 & 0 & 1 \\
        0 & 0 & 1 & 0
    \end{pmatrix}
    ,
\end{equation}
which can be interpreted as having $\theta_x = \pi/2$ or $\theta_z = \pi/2$. The Mølmer-Sørensen (MS) gate about $x$, which is routinely implemented in trapped ion systems \cite{molmer1999multiparticle, sorensen1999quantum}, can be written as an exponential of its Hermitian generators as
\begin{equation}
    \hat{U}_{\mathrm{MS}_{xx}}
    = \exp\left(
    - \mathrm{i} \frac{\pi}{2}
    \frac{\hat{\sigma}^x_a \hat{\sigma}^x_b}{2}
    \right)
    \doteq
    \frac{1}{\sqrt{2}}
    \begin{pmatrix}
        1 & 0 & 0 & -\mathrm{i} \\
        0 & 1 & -\mathrm{i} & 0 \\
        0 & -\mathrm{i} & 1 & 0 \\
        -\mathrm{i} & 0 & 0 & 1         
    \end{pmatrix}
    ,
\end{equation}
which can be interpreted as having $\theta_x = \pi/2$. The Mølmer-Sørensen (MS) gate about $y$, which we described in Chap.~\ref{chap:AdiabaticRydbergDressing} for neutral atoms \cite{mitra2020robust, martin2021molmer, mitra2023neutral}, can be written as an exponential of its Hermitian generators as
\begin{equation}
    \hat{U}_{\mathrm{MS}_{yy}}
    = \exp\left(
    - \mathrm{i} \frac{\pi}{2}
    \frac{\hat{\sigma}^y_a \hat{\sigma}^y_b}{2}
    \right)
    \doteq
    \frac{1}{\sqrt{2}}
    \begin{pmatrix}
        1 & 0 & 0 & \mathrm{i} \\
        0 & 1 & -\mathrm{i} & 0 \\
        0 & -\mathrm{i} & 1 & 0 \\
        \mathrm{i} & 0 & 0 & 1         
    \end{pmatrix}
    ,
\end{equation}
which can be interpreted as having $\theta_y = \pi/2$. The CZ, CNOT, $\mathrm{MS}_{xx}$ and $\mathrm{MS}_{yy}$ gates have only one of angles $\theta_x, \theta_y, \theta_z$ nonzero and thus are locally equivalent to each other.

Other two-qubit gates which are nonlocal, but not locally equivalent to CZ, CNOT, $\mathrm{MS}_{xx}$ and $\mathrm{MS}_{yy}$ are iSWAP and SWAP. The iSWAP gate can be written as an exponential of its Hermitian generators as
\begin{equation}
    \hat{U}_{\mathrm{iSWAP}}
    = \exp\left(
    + \mathrm{i} \frac{\pi}{2}
    \left(
    \frac{\hat{\sigma}^x_a \hat{\sigma}^x_b}{2}
    +\frac{\hat{\sigma}^y_a \hat{\sigma}^y_b}{2}
    \right)
    \right)
\end{equation}
which can be interpreted as having $\theta_x = \theta_y = -\pi/2$. Finally, the SWAP gate can be written as an exponential of its Hermitian generators as
\begin{equation}
    \hat{U}_{\mathrm{SWAP}}
    = \exp\left( +\mathrm{i} \frac{\pi}{4} \right)
    \exp\left(
    - \mathrm{i} \frac{\pi}{2}
    \left(
    \frac{\hat{\sigma}^x_a \hat{\sigma}^x_b}{2}
    + \frac{\hat{\sigma}^y_a \hat{\sigma}^y_b}{2}
    + \frac{\hat{\sigma}^z_a \hat{\sigma}^z_b}{2}
    \right)
    \right)
\end{equation}
which can be interpreted as having $\theta_x = \theta_y = \theta_z = \pi/2$. The iSWAP and SWAP gates have two and three of the angles nonzero. These parameters for these gates are summarized in Table~\ref{tab:TwoQubitGates}
\begin{table}[]
    \centering
    \begin{tabular}{c|c|c|c}
       Gate  & $\theta_x$ & $\theta_y$ & $\theta_z$  \\
       \hline
       Controlled-NOT  & $\frac{\pi}{2}$ & 0 & 0  \\
        & 0 & 0 & $\frac{\pi}{2}$ \\
       Controlled-Z & 0 & 0 & $\frac{\pi}{2}$ \\
       Mølmer-Sørensen $XX$ & $\frac{\pi}{2}$ & 0 & 0 \\
       Mølmer-Sørensen $YY$ & 0 & $\frac{\pi}{2}$ & 0  \\
       iSWAP & $\frac{\pi}{2}$ & $\frac{\pi}{2}$ & 0 \\
       SWAP & $\frac{\pi}{2}$ & $\frac{\pi}{2}$ & $\frac{\pi}{2}$ \\
    \end{tabular}
    \caption{Angles paramterizing the pure entangling component, $\hat{A}$ in Eq.\eqref{eq:TwoQubitGateKAKDecomposition} of some well-known two-qubit gates.}
    \label{tab:TwoQubitGates}
\end{table}
%

%
\chapter{Quantifying force on Rydberg dressed atoms}
\label{app:DressedAtomForces}
%
In this appendix, we discuss the forces between Rydberg-dressed atoms. Outside the strong blockade regime, it is important to consider the interatomic forces that could potentially affect the motional state of the atoms. Two atoms directly excited into the Rydberg state will experience a large Van der Waals force from the electric dipole-dipole interaction. However, in adiabatic dressing, a force will arise from the spatial gradient of the light shift, i.e., the ``softcore" adiabatic potential force arising from the $\ket{r, r}$ component in the dressed state $\ket*{\widetilde{1,1}}$.

Consider, thus, the adiabatic interatomic potential experienced by atoms in instantaneous internal ``adiabatic state" $\ket{\psi(\vec{R})}$. We treat here the center of mass motion of the atoms classically, in which the interatomic force is given according to
\begin{equation}
  V_{\text{ad}}
  = \bra{\psi(\vec{R})} H(\vec{R}) \ket{\psi(\vec{R})}
  \implies \vec{F}
  = -\vec{\nabla} V_{\text{ad}},
\end{equation}
where $\ket{\psi(\vec{R})} = c_{11}(\vec{R}) \ket{1,1} + c_b(\vec{R}) \ket{b} + c_{rr} \ket{r,r}$. The coefficients depend on the interatomic distance $\vec{R}$. If the state is an eigenstate of $H$, for example, the adiabatic potential of the dressed ground state $\ket*{\widetilde{1,1}}$ is
\begin{equation}
  V_{\text{ad}} (\widetilde{1, 1})
  = E(\widetilde{1, 1})
  = \hbar\kappa(\vec{R}) + 2 E_{\text{LS}}^{(1)} + 2 E_{1},
\end{equation}
which gives a force
\begin{equation}
  \vec{F}(\widetilde{1, 1}) = -\hbar \vec{\nabla} \kappa(\vec{R}),
\end{equation}
as the one-atom light shift $E_{\text{LS}}^{(1)}$ and bare energy $E_1$ are independent of $\vec{R}$.

When the interatomic distance is well within the blockade radius, where we have a perfect blockade, $\kappa$ is
independent of $\vec{R}$. This leads to a ``soft-core'' potential which has been observed experimentally \cite{jau2016entangling, zeiher2016many, zeiher2017coherent, martin2021molmer}. This can also be analyzed in the basis of bare atomic orbitals. For simplicity consider the case of zero detuning, the adiabatic potential between dressed ground states is
\begin{equation}
\begin{aligned}
  V_{\text{ad}} (\widetilde{1, 1})  &
  = \sqrt{2} \Omega_{\mathrm{eff}} (
  \Re \left(c_{11}(\vec{R}) c_{b}^{*}(\vec{R})\right)
  \\ & +
  \Re \left(c_{11}(\vec{R}) c_{rr}^{*}(\vec{R})) \right)
  + \vert c_{rr}\vert^2 V(\vec{R}).
\end{aligned}
\end{equation}
Note, the force is not simply $\vert c_{rr}\vert^2 \vec{\nabla} V(\vec{R})$; the interference terms in the adiabatic potential reduce the otherwise large force.

\begin{figure}
    \centering
    \includegraphics[width=0.6\textwidth]{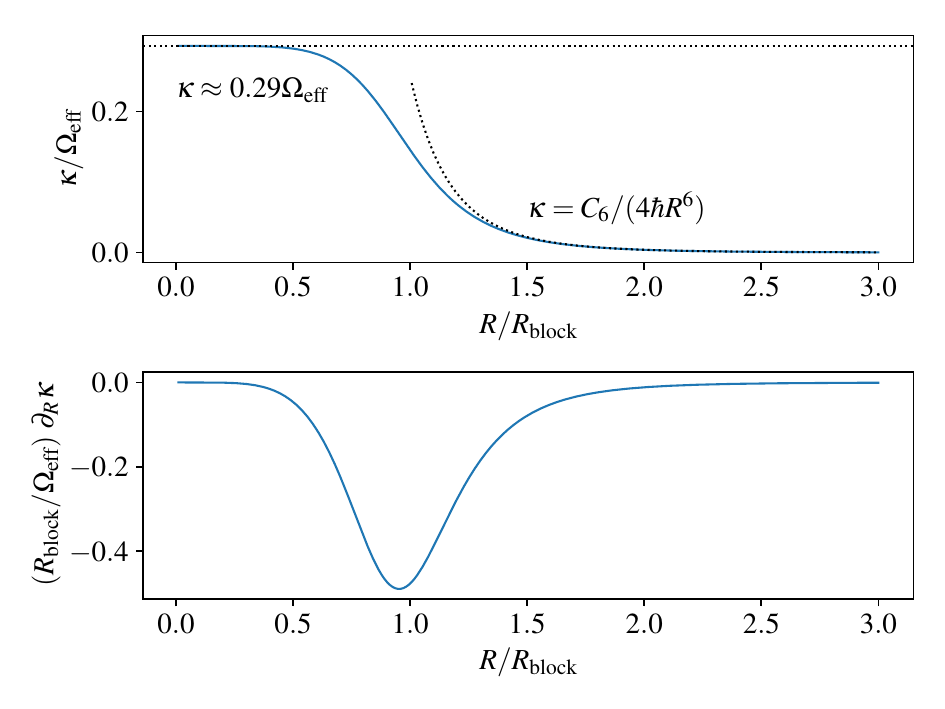}
    \caption{Interatomic potential energy and forces between atoms in the dressed state $\ket*{\widetilde{1,1}}$.
    (a) Entangling energy $\kappa$ in units of the Rabi frequency $\Omega_{\text{eff}}$ as a function of inter-atomic distance $R$, in units of the blockade radius $R_{\text{block}}$.
    (b) Gradient of the entangling energy along the inter-atomic direction $\partial_{R} \kappa$ in units of the ratio $\Omega_{\text{eff}}/R_{\text{block}}$.
    }
    \label{fig:InteratomicPotentialEnergy}
\end{figure}

For simplicity and generality, we calculate $\kappa$ as a function of distance using a Van der Waals potential, $V = C_6 |\vec{R}|^{-6}$,
and the interatomic force as a function of distance in Fig.~\ref{fig:InteratomicPotentialEnergy}. As is standard, we define the blockade radius where the energy of Rabi frequency of Rydberg excitation is equal to $V$, $\hbar \Omega_{\mathrm{eff}} = C_6 R_{\text{block}}^{-6}$. At short interatomic distances the adiabatic potential has a soft-core form and is the entangling energy $\hbar \kappa$, up to additive constants, as observed experimentally in  \cite{jau2016entangling, zeiher2016many, zeiher2017coherent, martin2021molmer}. At large distances, the interatomic potential asymptotes to a quarter of the vans Der Waals potential, $C_6 \left|\vec{R}\right|^{-6}/4$ for Van der Waals interactions. The transition occurs roughly between $\left|\vec{R}\right| /R_{\text{block}}\approx 1/2$ and $\left|\vec{R}\right| /R_{\text{block}}\approx 2$, where the potential energy has a nonzero gradient, giving rise to a non-trivial interatomic force (Fig.~\ref{fig:InteratomicPotentialEnergy}). From these results, we see that the operation of an adiabatic dressing gate outside the perfect blockade regime will lead to bounded perturbing forces on the atoms.
%

\chapter{Matrix product states}
\label{app:MatrixProductStates}

In this appendix, we review the matrix product state (MPS) representation of many-body wavefunctions representing quantum states of one-dimensional lattices. In the absence of fine-tuned parameters and special symmetries, exponentially large state spaces of quantum many-body systems make the exact simulation of quantum systems intractable except for small systems. Nevertheless, approximate simulations that capture quantities of interest have been used to study many-body physics, often with a limited entanglement budget, through appropriately chosen tensor networks. These tensor networks factor many-body wavefunctions into a graph of products of higher-order tensors, with entanglement content of the wavefunctions determining the size of the tensors \cite{orus2019tensor, biamonte2019lectures, cirac2021matrix}. Ground states of gapped one-dimensional (1D) systems, away from criticality, are believed to be tractable through matrix product states (MPS)~\cite{perez2006matrix, schollwock2011density, cirac2021matrix}. MPS \cite{perez2006matrix, schollwock2011density, vidal2003efficient, vidal2004efficient, daley2004time, white2004real, haegeman2011time, haegeman2016unifying, vanderstraeten2019tangent, cirac2021matrix}, are a class of one-dimensional tensor networks that factor a many-body wavefunction amplitude into a product of matrices, that have enjoyed tremendous success in representing quantum states of one-dimensional quantum systems \cite{vidal2003efficient, vidal2004efficient, schollwock2011density, halimeh2017prethermalization, Zun2018, paeckel2019timeevolution, zhou2020limits, cirac2021matrix}.

\begin{figure}
    \centering
    \includegraphics[width=0.9\textwidth]{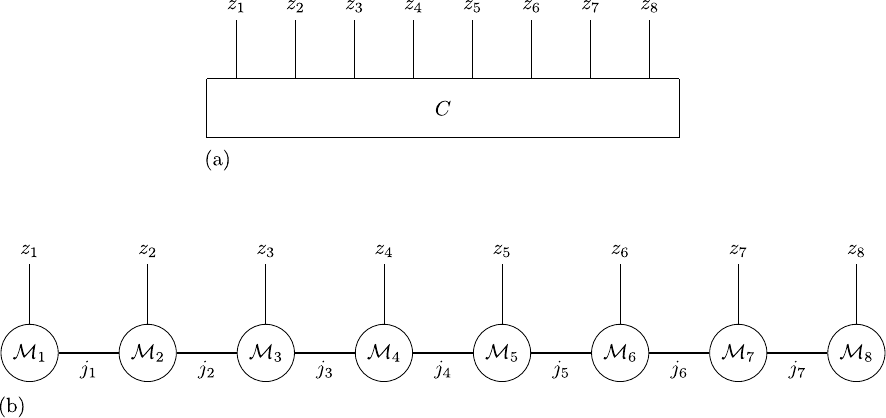}
    \caption{Matrix product state representation. (a) A wavefunction for $n=8$ degrees of freedom can be interpreted as an order-$n=8$ tensor. (b) MPS decomposition of the order-$n=8$ tensor into $n=8$ tensors, which are order-$3$ tensors in the bulk and order-$2$ at the boundary. Each site tensor inherits the physical index from the wavefunction and has virtual indices which represent correlations with its neighbors.}
    \label{fig:MatrixProductStateAppendix}
\end{figure}

\begin{figure}
    \centering
    \includegraphics[width=0.94\textwidth]{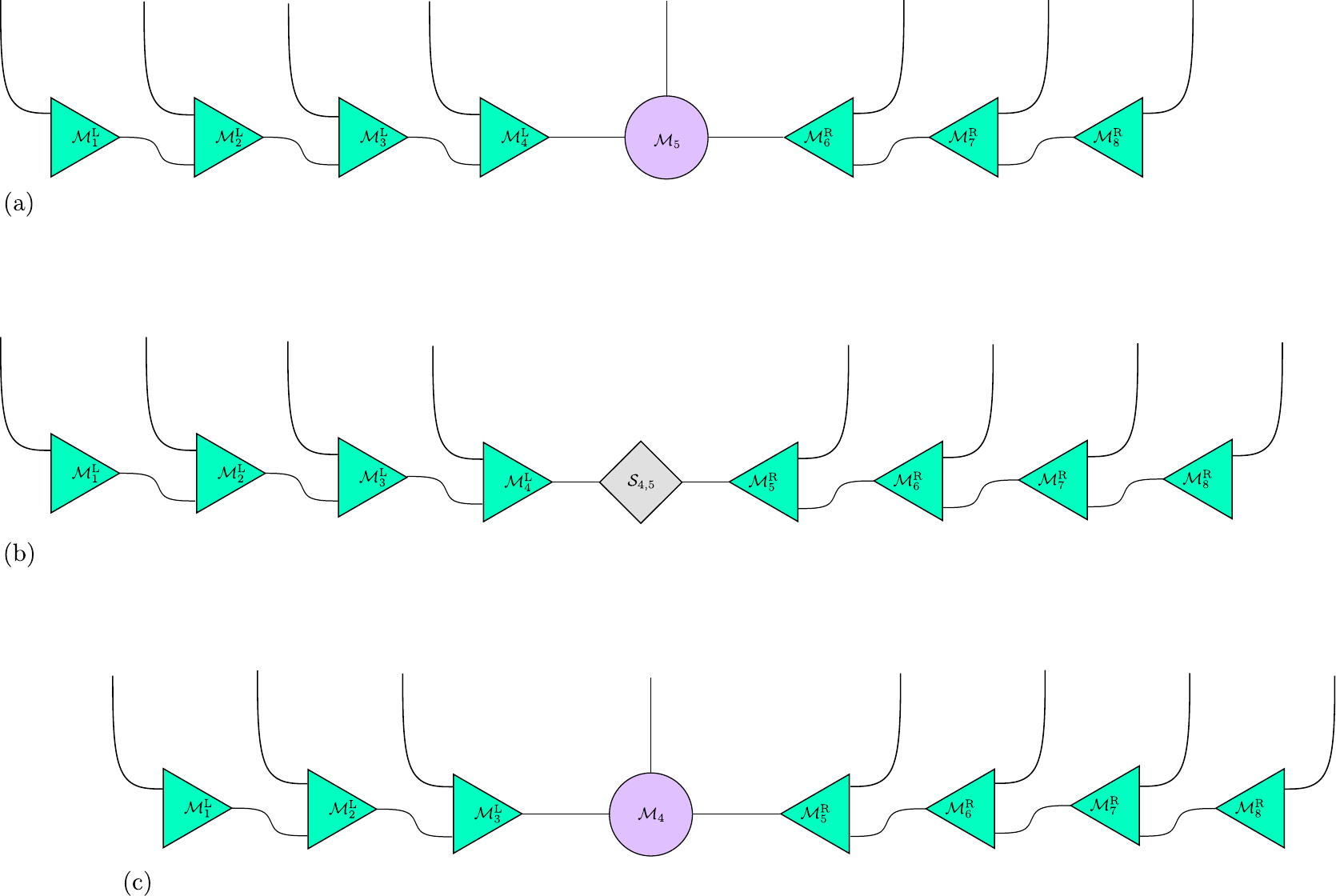}
    \caption{MPS canonical forms. (a) Mixed canonical form, with the canonical center at $5$, all tensors $\mathcal{M}_l$ for $l \leq 4$ are in left canonical form, and all tensors $\mathcal{M}_l$ for $l \geq 6$ are in right canonical form. (b) Canonical form with the singular values factored out of the site tensors, $\mathcal{M}$ in a bond tensor $\mathcal{S}_{4, 5}$. (c) Mixed canonical form, with the canonical center at $4$, all tensors $\mathcal{M}_l$ for $l \leq 3$ are in left canonical form, and all tensors $\mathcal{M}_l$ for $l \geq 5$ are in right canonical form.}
    \label{fig:MPSCanonicalForms}
\end{figure}

The Hilbert space for a quantum system with $n$ degrees of freedom each with $d$ levels is $d^n$, which is exponential in $n$. A pure state $|\psi\rangle$ describing the state of a system of $n$ degrees of freedom can be written as
\begin{equation}
    |\psi\rangle = 
    \sum_{z_1, z_2, \cdots, z_n}
    C_{z_1, z_2, \cdots, z_n}
    |z_1\rangle \otimes | z_2 \rangle \otimes \cdots \otimes | z_n\rangle
\end{equation}
where each $|z_l\rangle$ with $z_l \in [0, 1, \cdots, d-1]$ denotes the state of a single degree of freedom. The coefficient $C_{z_1, z_2, \cdots z_n}$ has $d^n$ entries and can be interpreted as a order-$n$ tensor as shown in Fig~\ref{fig:MatrixProductStateAppendix}(a). Therefore, representing an arbitrary pure state requires exponentially many parameters. A matrix product state (MPS) factorizes the amplitude tensor $C_{z_1, z_2, \cdots z_n}$ into a matrix product of order-$3$ tensors \cite{perez2006matrix, paeckel2019timeevolution, cirac2021matrix}
\begin{equation}
    C_{z_1, z_2, \cdots, z_n}
    = \sum_{j_1, \cdots, j_{n-1}}
    (\mathcal{M}_1)^{z_1}_{ j_1}
    (\mathcal{M}_2)^{z_2}_{j_1 j_2}
    \dots
    (\mathcal{M}_n)^{z_n}_{j_{n-1}}
    ,
\end{equation}
where $(\mathcal{M}_l)_{j_l, j_{l+1}}^{z_l}$ is an order-$3$ tensor, as shown in Fig.~\ref{fig:MatrixProductStateAppendix}(b). The $j_l$'s are virtual indices introduced to represent the correlations between the states of the individual degrees of freedom. The $z_l$'s are the physical indices that are inherited from the tensor $C$. In the mathematical physics literature, this is called a tensor-train decomposition of a tensor. Here we consider an open boundary MPS representation, where the tensors at the ends of the 1D array have only one virtual index. MPS with open boundary conditions are convenient for numerical calculations due to the existence of canonical forms, as shown in Fig.~\cref{fig:MPSCanonicalForms}, and efficient contraction algorithms \cite{schollwock2011density, paeckel2019timeevolution, cirac2021matrix}. This gives the MPS representation
\begin{equation}
  | \psi \rangle =
  \sum_{z \in \left\{\uparrow, \downarrow \right\}^n}
  \sum_{j_1, \dots, j_{n}}
  (\mathcal{M}_1)^{z_1}_{ j_1}
  (\mathcal{M}_2)^{z_2}_{j_1 j_2}
  \dots
  (\mathcal{M}_n)^{z_n}_{j_{n-1}} |z \rangle \ ,
  \label{eq:AppendixMatrixProductState}
\end{equation}
where we have used the abbreviated notation $|z \rangle \equiv |z_1\rangle \otimes |z_2 \rangle \otimes \cdots \otimes |z_n\rangle$.

The range of an index connecting these matrices is the bond dimension, $\chi$ \cite{perez2006matrix, schollwock2011density, vidal2003efficient, vidal2004efficient, daley2004time, white2004real, haegeman2011time, haegeman2016unifying, vanderstraeten2019tangent, cirac2021matrix}. While such an MPS representation can be used to represent an arbitrary quantum state, the representation is most convenient when it can efficiently approximate the desired quantum state, meaning with a bond dimension that does not scale exponentially with $n$, and be updated efficiently.

Since our focus is on dynamics, our primary aim is to represent the evolution of a quantum state according to the Hamiltonians as a series of transformations from one MPS representation into another, based on the time-dependent Schrödinger equation (TDSE),
\begin{equation}
    \frac{\partial}{\partial t} | \psi \rangle = -\mathrm{i} H |\psi\rangle
    ,
    \label{eq:TimeDependentSchrodingerEquationAppendix}
\end{equation}
where we have set $\hbar \equiv 1$. The solution to this can be formally written as
\begin{equation}
    |\psi(t) \rangle
    = \exp\left(- \mathrm{i} t H \right)
    |\psi(t) \rangle
    ,
    \label{eq:TimeEvolutionOperatorAppendix}
\end{equation}
where $\exp\left(- \mathrm{i} t H \right)$ is the time evolution operator generated by the Hamiltonian $H$. We use two methods two approximate Eq.~\eqref{eq:TimeEvolutionOperatorAppendix} as applied to the MPS representation. For one-dimensional nearest neighbor models considered in Chap.~\ref{chap:QuenchChaos}, we do this using the time-evolving block decimation (TEBD) method \cite{vidal2003efficient,vidal2004efficient, daley2004time, white2004real}, which involves a Trotterization of the time-evolution unitary as well as a truncation of the MPS. For long-range interacting models considered in Chap.~\ref{chap:QuenchDQPT}, we do this using the time-dependent variational principle \cite{haegeman2011time, haegeman2016unifying, vanderstraeten2019tangent} which involves solving a set of ordinary differential equations, for the tensors $\mathcal{M}$, which are obtained by projecting the wavefunction onto a manifold of bond dimension $\chi_{\max}$. We start with an initial state in the form \cref{eq:AppendixMatrixProductState} and compute the time evolution using TEBD or TDVP. After each time step, we truncate the resultant matrix product state using singular value decomposition to bond dimension $\chi_{\max}$, which parameterizes the simulation of time evolution. In the following, we will drop the subscript to mean $\chi \equiv \chi_{\max}$.

%

%
\section{Time evolving block decimation}
%

\begin{figure}
    \centering
    \includegraphics[width=0.96\textwidth]{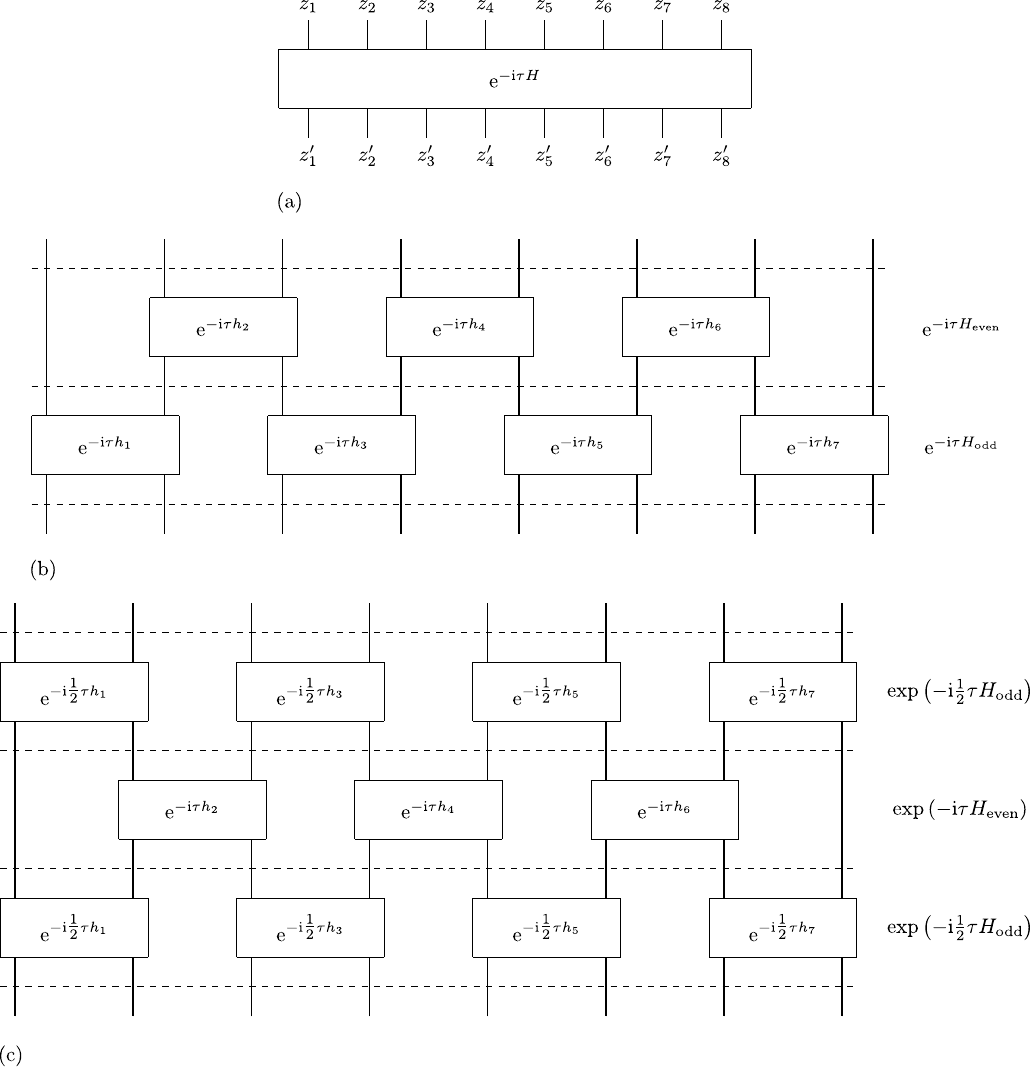}
    \caption{Trotter Suzuki decomposition of a time evolution operator. (a) The time evolution operator $\exp\left(-\mathrm{i} \tau H \right)$ as an order-$2n=16$ tensor with $n=8$ physical indices, $z_1', \cdots z_8'$, corresponding to the input, $n=8$ physical indices, $z_1, \cdots z_8$ corresponding to output. (b) First-order Trotter-Suzuki decomposition using Eq.~\eqref{eq:TrotterSuzuki1} and Eq.\eqref{eq:TwoSiteGatesEvenOdd}. (c) Second-order Trotter-Suzuki decomposition using Eq.~\eqref{eq:TrotterSuzuki2} and Eq.\eqref{eq:TwoSiteGatesEvenOdd}.}
    \label{fig:TEBDTrotterSuzuki}
\end{figure}

The time-evolving block decimation (TEBD) method is useful to approximate the time evolution operator generated by local Hamiltonians, with a few neighboring interactions, on a 1D lattice \cite{vidal2003efficient, vidal2004efficient, daley2004time, schollwock2011density, paeckel2019timeevolution}. The approximation involves, approximating the time evolution operator, $\exp\left(-\mathrm{i} t H \right)$, which is an order-$2n$ tensor, for $n$ degrees of freedom, as shown in Fig.~\ref{fig:TEBDTrotterSuzuki}(a). The approximation involves a Trotter-Suzuki decomposition of the time evolution operator into its local terms.

Consider a Hamiltonian with local and nearest-neighbor interaction terms. For example, the slanted field Ising model considered in Chap.~\ref{chap:QuenchChaos}, whose Hamiltonian reads
\begin{equation}
    H =
    - J \sum_{l} \sigma^z_{l} \sigma^z_{l+1}
    - B_z \sum_{l} \sigma^z_l
    - B_x \sum_{l} \sigma^x_l
    .
\end{equation}
This Hamiltonian can be written as a sum over even $l$ and a sum over odd $l$ as
\begin{equation}
    H = H_{\mathrm{even}} + H_{\mathrm{odd}}
    ,
\end{equation}
with
\begin{equation}
\begin{aligned}
    H_{\mathrm{even}}
    &
    = \sum_{l \in \mathrm{even}} h_{l}
    \\
    H_{\mathrm{odd}}
    &
    = \sum_{l \in \mathrm{odd}} h_{l}
    ,
\end{aligned}
\end{equation}
where $h_{l}$ is the two-local Hamiltonian acting on sites $l$ and $l+1$
\begin{equation}
    h_{l} =
    - J \sigma^z_{l} \sigma^z_{l+1}
    - \frac{1}{2} B_z \left(\sigma^z_l + \sigma^z_{l+1} \right)
    - \frac{1}{2} B_x \left(\sigma^x_l + \sigma^x_{l+1} \right)
    ,
    \label{eq:AppendixSFIMHamiltonian}
\end{equation}
where the factor of $1/2$ is introduced as the one-local terms, $\sigma^{z}_{l+1}$ and $\sigma^{x}_{l+1}$ are present in both $h_{l}$ and $h_{l+1}$. The overlap between $h_{l}$ and $h_{l+1}$ make them not commute, $[h_{l}, h_{l+1}] \neq 0$. Therefore, $H_{\mathrm{even}}$ does not commutes with $H_{\mathrm{odd}}$. However, every term in $H_{\mathrm{even}}$ commutes with every other term and every term in $H_{\mathrm{odd}}$ commutes with every other term.

Due to the noncommutativity of $H_{\mathrm{even}}$ and $H_{\mathrm{odd}}$, the propagator generated by them, during time evolution for a duration $t$, cannot be factorized as
\begin{equation}
    \mathrm{e}^{ -\mathrm{i} t \left(
      H_{\mathrm{even}} + H_{\mathrm{odd}}
    \right)}
    \neq
    \mathrm{e}^{-\mathrm{i} t
      H_{\mathrm{even}} }
    \mathrm{e}^{-\mathrm{i} t
      H_{\mathrm{odd}} }
    .
\end{equation}
Nevertheless, by for small time steps $\tau$, the propagator can be approximated by an order $p$ Trotter-Suzuki formula, $f_{p}$
\begin{equation}
    \mathrm{e}^{ -\mathrm{i} t \left(
      H_{\mathrm{even}} + H_{\mathrm{odd}}
    \right) }
    \approx
    f_p \left(
    \mathrm{e}^{ -\mathrm{i} t
      H_{\mathrm{even}} }
    ,
    \mathrm{e}^{ -\mathrm{i} t
      H_{\mathrm{odd}} }
    \right) 
    + \mathcal{O}(\tau^{p+1}).
\end{equation}
For the approximation to be useful, $\tau$ needs to be small compared to the inverse of the energy scales of the problem. Formally, we need $\tau \ll \Vert H \Vert^{-1}$ where $\Vert H \Vert$ is a norm of the Hamiltonian. While formally, this may be challenging due to a lack of knowledge of the spectrum of the Hamiltonian, which requires finding its eigenvalues, which is generally intractable, in practice choose a $\tau$ to be smaller than an inverse energy scale in the Hamiltonian, for example $J \tau \ll 1$ in Eq.~\eqref{eq:AppendixSFIMHamiltonian} suffices. The simplest Trotter-Suzuki formula is for $p=1$, which reads
\begin{equation}
    \mathrm{e}^{ -\mathrm{i} \tau \left(
      H_{\mathrm{even}} + H_{\mathrm{odd}}
    \right)}
    =
    \mathrm{e}^{-\mathrm{i} \tau
      H_{\mathrm{even}} }
    \mathrm{e}^{-\mathrm{i} \tau
      H_{\mathrm{odd}} }
    + \mathcal{O}(\tau^2)
    \equiv
    f_1 \left(
    \mathrm{e}^{-\mathrm{i} \tau
      H_{\mathrm{even}} } ,
    \mathrm{e}^{-\mathrm{i} \tau
      H_{\mathrm{odd}} }
    \right)
    + \mathcal{O}(\tau^2)
    ,
    \label{eq:TrotterSuzuki1}
\end{equation}
where the error introduced per step is second order in the time step \cite{schollwock2011density, paeckel2019timeevolution}. A better approximation is obtained using $p=2$,
\begin{equation}
\begin{aligned}
    \exp\left( -\mathrm{i} \tau \left(
      H_{\mathrm{even}} + H_{\mathrm{odd}}
    \right) \right)
    &
    =
    \exp\left(-\mathrm{i} \frac{\tau}{2}
      H_{\mathrm{even}} \right)
    \exp\left(-\mathrm{i} \tau
      H_{\mathrm{odd}} \right)
    \exp\left(-\mathrm{i} \frac{\tau}{2}
      H_{\mathrm{even}} \right)
    + \mathcal{O}(\tau^3)
    \\ &
    \equiv
    f_2 \left(
    \mathrm{e}^{-\mathrm{i} \tau
      H_{\mathrm{even}} } ,
    \mathrm{e}^{-\mathrm{i} \tau
      H_{\mathrm{odd}} }
    \right)
    + \mathcal{O}(\tau^3)
    ,
    \label{eq:TrotterSuzuki2}
\end{aligned}
\end{equation}
where the error introduced per step is third-order in the time step \cite{schollwock2011density, paeckel2019timeevolution}. Finally, a fourth-order approximation, $p=4$ uses the $p=2$ formula $f_2$ to reduce the error per step further to fifth-order in the time step as
\begin{equation}
\begin{aligned}
   \exp\left( -\mathrm{i} \tau \left(
      H_{\mathrm{even}} + H_{\mathrm{odd}}
    \right) \right)
    &
    = f_2(\tau_1)
    f_2(\tau_2)
    f_2(\tau_3)
    f_2(\tau_2)
    f_2(\tau_1)
    + \mathcal{O}(\tau^3)
    \\ &
    \equiv
    f_4 \left(
    \mathrm{e}^{-\mathrm{i} \tau
      H_{\mathrm{even}} } , 
    \mathrm{e}^{-\mathrm{i} \tau
      H_{\mathrm{odd}} }
    \right)
    + \mathcal{O}(\tau^5)
    ,
    \label{eq:TrotterSuzuki4}
\end{aligned}
\end{equation}
where
\begin{equation}
\begin{aligned}
    \tau_1 = \tau_2 = \frac{1}{4 - 4^{1/3}} \tau
    , &&
    \tau_3 = \tau - 2 \tau_1 - 2 \tau_2
    ,
\end{aligned}
\end{equation}
and for brevity skipped the dependence of $f_2$ on the Hamiltonians in the abbreviated notation
\begin{equation}
    f_2(\tau)
    \equiv
    f_2 \left(
    \mathrm{e}^{-\mathrm{i} \tau
      H_{\mathrm{even}} } ,
    \mathrm{e}^{-\mathrm{i} \tau
      H_{\mathrm{odd}} }
     \right)
     \equiv
     \exp\left(-\mathrm{i} \frac{\tau}{2}
      H_{\mathrm{even}} \right)
    \exp\left(-\mathrm{i} \tau
      H_{\mathrm{odd}} \right)
    \exp\left(-\mathrm{i} \frac{\tau}{2}
      H_{\mathrm{even}} \right)
      .
\end{equation}

Since every term in $H_{\mathrm{even}}$ commutes with every other term and every term in $H_{\mathrm{odd}}$ commutes with every other term, we can write
\begin{equation}
\begin{aligned}
  \mathrm{e}^{-\mathrm{i} \tau H_{\mathrm{even}}}
  =
  \prod_{l \in \mathrm{even}}
  \mathrm{e}^{-\mathrm{i} \tau h_l}
  , \\
  \mathrm{e}^{-\mathrm{i} \tau H_{\mathrm{odd}}}
  =
  \prod_{l \in \mathrm{odd}}
  \mathrm{e}^{-\mathrm{i} \tau h_l}
  \label{eq:TwoSiteGatesEvenOdd}
\end{aligned}
\end{equation}
as a product of two-qubit gates acting on sites $l$ and $l+1$.

\begin{figure}
    \centering
    \includegraphics[width=0.8\textwidth]{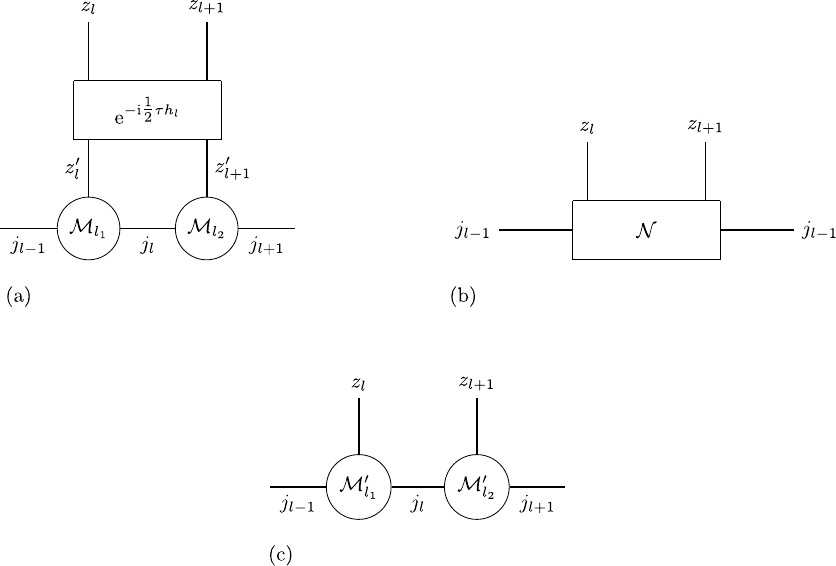}
    \caption{Application of a two-site time-evolution operator. (a) Contracting the two-site time evolution operator, $\exp\left(-\mathrm{i} \frac{1}{2} \tau H\right)$ to two adjacent MPS tensors, $\mathcal{M}_{l}$ and $\mathcal{M}_{l+1}$. (b) After the contract, a single two-site tensor $\mathcal{N}$ is obtained. (c) The two-site tensor $\mathcal{N}$ is split into two near single-site tensors $\mathcal{M}'_{l}$ and $\mathcal{M}'_{l+1}$, using SVD. Bond dimension is truncated by discarding the smallest few singular values as a part of the SVD.}
    \label{fig:TwoSiteGate}
\end{figure}

With these ingredients, we can describe the time evolution algorithms. Starting with an initial state expressed as MPS in Eq.~\eqref{eq:AppendixMatrixProductState}, we apply a series of two-site gates described by Eq.~\eqref{eq:TwoSiteGatesEvenOdd} with the appropriate Trotter-Suzuki formula (Eq.~\eqref{eq:TrotterSuzuki1}, Eq.~\eqref{eq:TrotterSuzuki2}, Eq.~\eqref{eq:TrotterSuzuki4}). After the application of every two-site gate, the MPS form needs to be restored by splitting the two-site tensor back to a matrix product of one-site tensors. This is done using a singular value decomposition (SVD) as shown in Fig.~\ref{fig:TwoSiteGate}. As a part of the SVD, the tensors are truncated to retain only $\chi$ largest singular values. A detailed description of the different ways of performing SVD and truncation can be found in original works in Refs.~\cite{vidal2003efficient, vidal2004efficient, daley2004time} and the reviews in Refs.~\cite{schollwock2011density, paeckel2019timeevolution}. This is where the approximation to the microstate, as described in Chap.~\ref{chap:QuenchDQPT} becomes worse as the entanglement content in the wavefunction is discarded. For the results presented in Chap.~\ref{chap:QuenchChaos}, we used $p=4$.

%

%
\section{Time dependent variational principle}
%

The time-dependent variational principle (TDVP) for matrix product states (MPS) is another way of approximating the dynamics of MPS by reducing the many-body time evolution to a series of local time evolution problems \cite{haegeman2011time, haegeman2016unifying, paeckel2019timeevolution}. TDVP involves constraining the time evolution to a specific manifold of MPS with a given bond dimension. TDVP applied to MPS is a convenient method for calculating the dynamics of long-range interacting models in 1D \cite{haegeman2016unifying, paeckel2019timeevolution}. It is also convenient for approximate time evolution which respects the conservation of quantities which are symmetries of the Hamiltonian.

In the one-site variant, 1TDVP, the action of the Hamiltonian is projected onto the manifold of MPS with the bond dimension specified by the initial bond dimension and the bond dimensional is not allowed to grow during the time evolution \cite{haegeman2011time}. This is achieved by solving differential equations for each MPS tensor, with a fixed bond dimension. In the two-site variant, 2TDVP, the bond dimension is allowed to grow, by considering the action of the Hamiltonian on a pair of neighboring tensors in the MPS \cite{haegeman2016unifying}. This gives a differential equation for every pair of neighbors. While the solution of these differential equations does not project each tensor onto a fixed bond dimension manifold, subsequent SVD and truncation do \cite{haegeman2016unifying}. The two-site variant, 2TDVP is more suitable for numerical calculations with the ability to dynamically adjust the bond dimension. Therefore, the calculations presented in Chap.~\ref{chap:QuenchDQPT} use 2TDVP. We discuss this method here.

In TDVP, the action of the Hamiltonian in the time-dependent Schr\"{o}dinger equation (TDSE) is used to derive an effective TDSE for tensors in the MPS. The effective Hamiltonian for the two-site tensors $\mathcal{M}_l$ and $\mathcal{M}_{l+1}$ is obtained by projecting the Hamiltonian onto the manifold where all other tensors $\mathcal{M}_{l'}$, with $l' \leq l-1$ and $l' \geq l+2$ are held fixed. The projection is obtained using the tangent projector $\mathcal{P}^{\mathrm{T}}_{\psi}$ \cite{haegeman2016unifying}
\begin{equation}
    \mathcal{P}^{\mathrm{T}}_{\psi}
    = \sum_{l=1}^{n-1}
    \mathcal{P}^{\mathrm{L}}_{1:l-1} \,
    \mathds{1}_{l} \, \mathds{1}_{l+1}
    \mathcal{P}^{\mathrm{R}}_{l+1:n} \,
    -
    \sum_{l=2}^{n-1}
    \mathcal{P}^{\mathrm{L}}_{1:l-1} \,
    \mathds{1}_{l} \,
    \mathcal{P}^{\mathrm{R}}_{l+1:n} \,
    ,
\end{equation}
where the left projector, $\mathcal{P}^{\mathrm{L}}_{1:l-1}$ and right projector $\mathcal{P}^{\mathrm{R}}_{l+2:n}$ are
\begin{equation}
\begin{aligned}
     \mathcal{P}^{\mathrm{L}}_{1:l-1}
     &
     = \mathcal{M}_{1}^{\mathrm{L}} \, \mathcal{M}_{2}^{\mathrm{L}} \, \cdots \,
     \mathcal{M}_{l-2}^{\mathrm{L}} \, \mathcal{M}_{l-1}^{\mathrm{L}}
     \\ 
     \mathcal{P}^{\mathrm{R}}_{l+2:n}
     &
     = \mathcal{M}_{l+2}^{\mathrm{R}} \, \mathcal{M}_{l+3}^{\mathrm{R}} \, \cdots \,
     \mathcal{M}_{n-1}^{\mathrm{R}} \, \mathcal{M}_{n}^{\mathrm{R}}
     ,
     \label{eq:MPSTangentSpaceProjectors}
\end{aligned}
\end{equation}
where we have suppressed the virtual and physical indices of the MPS tensors $\mathcal{M}_l$, for brevity. Moreover, we have used $\mathcal{M}^{\mathrm{L}}$ to represent a tensor in left canonical form and $\mathcal{M}^{\mathrm{R}}$ to represent a tensor in right canonical form.

Projecting the TDSE in Eq.~\eqref{eq:TimeDependentSchrodingerEquationAppendix} to the tangent manifold gives
\begin{equation}
    \frac{\partial}{\partial t} 
    \left( \mathcal{P}^{\mathrm{T}}_{\psi}
    | \psi \rangle 
    \right)
    = -\mathrm{i} 
    \left( \mathcal{P}^{\mathrm{T}}_{\psi}
    H |\psi\rangle
    \right)
    ,
    \label{eq:MPSTangentTDSE}
\end{equation}
which for 2TDVP can be written as
\begin{equation}
\begin{aligned}
    &
    \frac{\partial}{\partial t} 
    \left(\sum_{l=1}^{n-1}
    \mathcal{P}^{\mathrm{L}}_{1:l-1} \,
    \mathds{1}_{l} \, \mathds{1}_{l+1}
    \mathcal{P}^{\mathrm{R}}_{l+1:n} \,
    \right) 
    | \psi \rangle
    -
    \frac{\partial}{\partial t}
    \left(
    \sum_{l=2}^{n-1}
    \mathcal{P}^{\mathrm{L}}_{1:l-1} \,
    \mathds{1}_{l} \,
    \mathcal{P}^{\mathrm{R}}_{l+1:n} \,
    \right) 
    | \psi \rangle
    \\ &
    = -\mathrm{i}
    \left(\sum_{l=1}^{n-1}
    \mathcal{P}^{\mathrm{L}}_{1:l-1} \,
    \mathds{1}_{l} \, \mathds{1}_{l+1}
    \mathcal{P}^{\mathrm{R}}_{l+1:n} \,
    \right) 
    H | \psi \rangle
    + \mathrm{i}
    \left(
    \sum_{l=2}^{n-1}
    \mathcal{P}^{\mathrm{L}}_{1:l-1} \,
    \mathds{1}_{l} \,
    \mathcal{P}^{\mathrm{R}}_{l+1:n} \,
    \right) 
    H | \psi \rangle
    ,
    \label{eq:MPSTangentTDSEVerbose}
\end{aligned}
\end{equation}
for which an exact solution cannot be found due to the nonlinearity. The TDSE is nonlinear due to projector $\mathcal{P}^{\mathrm{T}}_{\psi}$ being dependent on the tensors in $|\psi\rangle$ \cite{haegeman2016unifying, paeckel2019timeevolution}. Nevertheless, Eq.~\eqref{eq:MPSTangentTDSEVerbose} can be approximately solved by solving each term individually and sequentially \cite{haegeman2016unifying, paeckel2019timeevolution}.

It is convenient to write a two-site tensor $\mathcal{N}_{l,l+1} \equiv \mathcal{M}_{l} \mathcal{M}_{l+1}$. The terms 
\begin{equation}
    \frac{\partial}{\partial t} 
    \left(
    \mathcal{P}^{\mathrm{L}}_{1:l-1} \,
    \mathds{1}_{l} \, \mathds{1}_{l+1}
    \mathcal{P}^{\mathrm{R}}_{l+1:n} \,
    \right) 
    | \psi \rangle
    = -\mathrm{i} \left(
    \mathcal{P}^{\mathrm{L}}_{1:l-1} \,
    \mathds{1}_{l} \, \mathds{1}_{l+1}
    \mathcal{P}^{\mathrm{R}}_{l+1:n} \,
    \right) 
    H | \psi \rangle
    ,
\end{equation}
give $n-1$ forward-evolving effective TDSE
\begin{equation}
     \frac{\partial}{\partial t} \mathcal{N}_{l+1}
     = - \mathrm{i} H^{\mathrm{eff}}_{l,l+1} \mathcal{N}_{l, l+1}
     ,
     \label{eq:TDVPSiteTensorTDSE}
\end{equation}
for the two-site tensors $\mathcal{N}_{l,l+1}$ $l \in [1, 2, \cdots, n-1]$. The terms
\begin{equation}
    \frac{\partial}{\partial t}
    \left(
    \mathcal{P}^{\mathrm{L}}_{1:l-1} \,
    \mathds{1}_{l} \,
    \mathcal{P}^{\mathrm{R}}_{l+1:n} \,
    \right) 
    | \psi \rangle
    = + \mathrm{i}
    \left(
    \sum_{l=2}^{n-1}
    \mathcal{P}^{\mathrm{L}}_{1:l-1} \,
    \mathds{1}_{l} \,
    \mathcal{P}^{\mathrm{R}}_{l+1:n} \,
    \right) 
    H | \psi \rangle
    ,
\end{equation}
give $n-2$ backward-evolving effective TDSE
\begin{equation}
     \frac{\partial}{\partial t} \mathcal{S}_{l, l+1}
     = + \mathrm{i} H^{\mathrm{eff}}_{l,l+1} \mathcal{S}_{l, l+1}
     ,
     \label{eq:TDVPBondTensorTDSE}
\end{equation}
for the bond-tensors $\mathcal{S}_{l, l+1}$ for $l \in [2, 3 \cdots, n-1]$.

The 2TDVP scheme involves integrating $n-1$ forward-evolving differential equations of the form of Eq.~\eqref{eq:TDVPSiteTensorTDSE} and $n-2$ backward-evolving differential equations of the form of Eq.~\eqref{eq:TDVPBondTensorTDSE}. After each two-site tensor and bond tensor is updated, it is split into two one-site tensors using SVD, similar to TEBD, as shown in Fig.~\ref{fig:TwoSiteGate}. The bond dimension is truncated as needed, as a part of the SVD. Discarding singular values leads to deterioration of the approximation to the microstate, as described in Chap.~\ref{chap:QuenchDQPT}, becoming worse with increasing entanglement in the MPS.
\chapter{Relating expectation values to states}
\label{app:MarginalsExpectationValues}

In this appendix, we show how to relate the difference in expectation values of observables estimated from two density matrices to the Hilbert-Schmidt (HS) distance between them. Let us consider an observable $\mathcal{A}$ whose expectation value is estimated from two density matrices $\rho_1$ and $\rho_2$ as
\begin{equation}
    \langle{\mathcal{A}}\rangle_{\rho_1} = \Tr\left(\rho_1 \mathcal{A}\right)
    ; \qquad
    \langle{\mathcal{A}}\rangle_{\rho_2} = \Tr\left(\rho_2 \mathcal{A}\right).
\end{equation}
For example, $\rho_1$ could be the reduced density operator calculated from an MPS with bond-dimension $\chi_1$, and $\rho_2$ could be the reduced density operator calculated from an MPS with bond-dimension $\chi_2$. The squared difference in the two estimates for the expectation value is given by
\begin{eqnarray}
    \big(\langle{\mathcal{A}}\rangle_{\rho_1} - \langle{\mathcal{A}}\rangle_{\rho_2}\big)^2
    &=& \big(\Tr\left(\rho_1 \mathcal{A}\right) - \Tr\left(\rho_2 \mathcal{A}\right)\big)^2 \nonumber \\
    &=& \big(\Tr\left( \left(\rho_1 - \rho_2\right) \mathcal{A} \big)\right)^2.
\end{eqnarray}
The last expression can be interpreted as the HS inner product or overlap between vectors $u \equiv \rho_1 - \rho_2$ and $v \equiv \mathcal{A}$
\begin{equation}
    \langle u, v \rangle
    = \Tr\left(u^\dagger v\right).
\end{equation}
Using the Cauchy-Schwarz inequality, 
\begin{equation}
    \left|\langle u, v \rangle\right|^2 \leq 
    \langle u, u \rangle
    \langle v, v \rangle,
\end{equation}
we have that
\begin{equation}
    \left|\Tr\left( u^\dagger v \right)\right|^2 \leq
    \Tr\left(u^\dagger u \right) \Tr\left( v^\dagger v \right).
\end{equation}

Since our operators are Hermitian, $u^\dagger = u$ and $v^\dagger = v$, we have
\begin{equation}
    \left|\Tr\left(u v \right) \right|^2 \leq
    \Tr\left(u^2 \right) \Tr\left(v^2 \right).
\end{equation}
Thus we have the inequality
\begin{eqnarray} \label{eq:ErrorBoundAppendix}
    \big(\Tr\left(\rho_1 - \rho_2 \right) \mathcal{A} \big)^2
    &\leq&
    \Tr\left(\mathcal{A}^2 \right) \Tr\left(\left(\rho_1 - \rho_2 \right)^2 \right)\nonumber \\
    &=& \Tr\left(\mathcal{A}^2 \right)
    \mathcal{D}_{\mathrm{HS}}^2 \left(\rho_1, \rho_2 \right),
\end{eqnarray}
where $\mathcal{D}_{\mathrm{HS}}^2 \left(\cdot, \cdot \right)$ is the squared HS distance between the operators:
\begin{equation} \label{eq:HSDistanceDefinitionAppendix}
    \mathcal{D}^2_{\mathrm{HS}}\left(\rho_1, \rho_2 \right) = \Tr\left(\left(\rho_1-\rho_2 \right)^\dagger \left(\rho_1-\rho_2 \right) \right).
\end{equation}
Thus we find that the squared difference in the two estimates of the observable is upper-bounded by the squared HS distance between their associated density operators.

We now aim to derive a similar relationship for the  time- and site- averaged expectation value of an observable. We assume that $\mathcal{A}^{(j,k)}$ is a two-spin observable acting non-trivially only on sites $j$ and $k$ and is identical on all pairs of sites, i.e. its operator representation is identical $\forall j, k$, which we denote by $\mathcal{A}^{(2)}$. The time- and site- averaged expectation value of the observable is given by
\begin{equation}
  \overline{
    \left\langle \frac{1}{n^2} \sum_{j,k} \mathcal{A}^{(j,k)} \right\rangle}
  = \frac{1}{T} \int_0^T dt
  \frac{1}{n^2} \sum_{j,k} \Tr \left( \mathcal{A}^{(j,k)} \rho(t) \right),
\end{equation}
where
\begin{equation}
  \overline{\bullet} = \frac{1}{T} \int_0^T dt \left( \bullet \right),
\end{equation}
denotes the time-average over a time interval from $0$ to $T$.
 
We first consider the average 2-site error of the reduced density operator. Letting $\rho^{(j,k)}$ denote the reduced density operator on sites $j$ and $k$, we find that
\begin{align} 
&\;\;\frac{1}{n^2} \sum_{j,k} \left| { \left \langle \mathcal{A}^{(j,k)}(t) \right \rangle_{\chi_1}}  -  { \left \langle  \mathcal{A}^{(j,k)}(t) \right \rangle_{\chi_2}}  \right|^2 \nonumber\\
&\;\;\;\;=\frac{1}{n^2}  \sum_{j,k} \left|  \mathrm{Tr} \left( \mathcal{A}^{(j,k)}  \left(\rho^{(j,k)}_{\chi_1}(t) - \rho^{(j,k)}_{\chi_2}(t) \right) \right) \right|^2 \nonumber \\
&\;\;\;\;=\frac{1}{n^2}  \sum_{j,k} \left|  \mathrm{Tr} \left( \mathcal{A}^{(2)}   \left(\rho^{(j,k)}_{\chi_1}(t) - \rho^{(j,k)}_{\chi_2}(t) \right) \right) \right|^2 \nonumber \\
	&\;\;\;\;\leq \frac{1}{n^2}   \sum_{j,k} \mathrm{Tr} \left( \mathcal{A}^{(2)\dagger} \mathcal{A}^{(2)} \right) \mathrm{Tr} \left( \left|  \rho^{(j,k)}_{\chi_1}(t) - \rho^{(j,k)}_{\chi_2}(t) \right|^2\right) \nonumber \\
&\;\;\;\;=\frac{4}{n^2}\sum_{j,k}\mathcal{D}^2_{\mathrm{HS}} \left( \rho^{(j,k)}_{\chi_1}(t), \rho^{(j,k)}_{\chi_2}(t) \right), \label{eq:Case1}
\end{align}
where we have used the Cauchy-Schwartz inequality. Similarly, the error of the 2-site averaged error is given by
\begin{align}
&\left| { \left \langle \frac{1}{n^2}\sum_{j,k} \mathcal{A}^{(j,k)}(t) \right \rangle_{\chi_1}}  -  { \left \langle \frac{1}{n^2}\sum_{j,k} \mathcal{A}^{(j,k)}(t) \right \rangle_{\chi_2}}  \right|^2 \nonumber\\
&\;\;= \left|  \mathrm{Tr} \left(  \mathcal{A}^{(2)}  \frac{1}{n^2}\sum_{j,k} \left(\rho^{(j,k)}_{\chi_1}(t) - \rho^{(j,k)}_{\chi_2}(t) \right) \right) \right|^2 \nonumber\\
&\;\;\leq  \mathrm{Tr} \left( \left|  \mathcal{A}^{(2)} \right|^2 \right) \mathrm{Tr} \left( \left| \frac{1}{n^2}\sum_{j,k} \left( \rho^{(j,k)}_{\chi_1}(t) - \rho^{(j,k)}_{\chi_2}(t) \right) \right|^2\right) \nonumber\\
&\;\;=4 \mathcal{D}^2_{\mathrm{HS}} \left( {\frac{1}{n^2}\sum_{j,k} \rho^{(j,k)}_{\chi_1} (t)},\; \frac{1}{n^2}{\sum_{j,k}\rho^{(j,k)}_{\chi_2} (t)} \right). \label{eq:Case2}
\end{align}
We can now easily consider the role of time-averaging by repeating the same calculation with the time-average of the expectation values:
\begin{align}
&\frac{1}{n^2} \sum_{j,k} \left| \overline{ \left\langle \mathcal{A}^{(j,k)} \right\rangle}_{\chi_1}  -  \overline{ \left\langle  \mathcal{A}^{(j,k)} \right\rangle}_{\chi_2}  \right|^2 \nonumber\\
&\;\;\;\;\;\;\leq \frac{4}{n^2}   \sum_{j,k}  \mathcal{D}^2_{\mathrm{HS}} \left( \overline{\rho^{(j,k)}_{\chi_1}(t)}, \overline{\rho^{(j,k)}_{\chi_2}(t)} \right). \label{eq:Case3}
\end{align} 
The above calculation corresponds to first calculating the time-averaged 2-site reduced density operator, then calculating the squared HS distance, and finally averaging over all sites. If we change the order to first averaging over sites and then over time before calculating the distance, we find that
\begin{align}
&\left| \overline{ \left\langle \sum_{j,k} \mathcal{A}^{(j,k)} \right\rangle}_{\chi_1}  -  \overline{ \left\langle \sum_{j,k} \mathcal{A}^{(j,k)} \right\rangle}_{\chi_2}  \right|^2 \nonumber\\
&\;\;\;\;\;\;\leq 4 \mathcal{D}^2_{\mathrm{HS}} \left( \overline{\frac{1}{n^2}\sum_{j,k} \rho^{(j,k)}_{\chi_2}}, \overline{\frac{1}{n^2}\sum_{j,k}\rho^{(j,k)}_{\chi_1}} \right) \label{eq:Case4}
\end{align}
The last term is the squared HS distance of the site- and time-averaged reduced density operator. Our expression provides an upper bound for the error of the expectation value of space- and time-averaged observables in terms of the squared HS distance between the site- and time-averaged reduced density operators.

The different inequalities derived in Eqs.~\eqref{eq:Case1}-\eqref{eq:Case4} correspond to whether we consider the site- and time-averaged reduced density operator. We use these different cases for the HS distance in Fig.~\ref{fig:ApproximationsEntanglement} of Chap.~\ref{chap:QuenchDQPT}.
%

\chapter{Mean-field description of the DQPT}
\label{app:DQPTSimplePicture}
%
In this appendix, we consider a mean-field description of the dynamical quantum phase transition (DQPT) considered in Chap.~\ref{chap:QuenchDQPT}.
A simple picture for DQPTs can be obtained by considering the $\alpha = 0$ case, which is a mean-field theory in the thermodynamic limit~\cite{zhang2017observationmany, Zun2018, lang2018dynamical, chinni2021effect, chinni2022trotter, munoz2020simulation, munoz2021nonlinear, munoz2022floquet, munoz2023phase}. This is also a more computationally tractable case. For $\alpha = 0$, the Hamiltonian in Eq.~\eqref{eq:HamiltonianPowerLawTFIM} can be expressed in terms of the total angular momentum operators, which simplify to powers of components of the total angular momentum operator $\vec{S} = \sum_{l} \vec{\sigma}/2$ as
\begin{equation}
H = - B (2 S_x) - J_0 \left( \left(2 S_z \right)^2 - \frac{1}{n} \ident \right),
\end{equation}
and it reduces to the LMG model \cite{Lip1965}. This Hamiltonian is invariant under permutation of the spins, and since the initial state $\ket{\uparrow_z}^{\otimes n}$ is permutation symmetric under spin permutation, the evolution under the Hamiltonian is restricted to the permutation symmetric subspace of dimension $n+1$. The polynomial scaling of the relevant space enables exact simulations for significantly larger sizes $n$ and larger simulation times $J_0 t$.

Another advantage of the case $\alpha = 0$ is that it lends itself to a semi-classical treatment. In the semiclassical limit, we can think of the Hamiltonian as giving rise to a potential energy density profile described by
\begin{equation}
    V(\theta, \phi) = -\frac{1}{2} B \sin(\theta) \cos(\phi) - \frac{1}{4} J_0 \cos^2(\theta) ,
\end{equation}
which gives rise to a double-well potential. The parameter $B/J_0$ controls the height of the barrier between the two minima of the double well. In the limit $B/J_0 \to \infty$, the barrier vanishes. For $B / J_0 = 0$, we can associate the two minima with the two ferromagnetic ground states, $\ket{\uparrow_z}^{\otimes n}$ and $\ket{\downarrow_z}^{\otimes n}$. In the DQPT quench, the initial state is one of these states (that is, in one of the ground states of the Hamiltonian with $B/J_0 =0$ with a high barrier), and we quench the Hamiltonian to $B/J_0 \neq 0$ with a lower barrier.

The initial state determines the energy of the state, which remains conserved during the evolution. Suppose we start with the $\ket{\uparrow_z}^{\otimes n}$ state, with a positive magnetization. If the energy of this state is below the height of the barrier of the quenched Hamiltonian, the semiclassical evolution is the system bouncing back and forth within the \emph{same} well it starts in. Thus the magnetization stays positive as a function of time, and thus $M_{z} > 0$.

If the energy of the state is above the height of the barrier, then the semiclassical evolution will be the system bouncing back and forth between \emph{both} wells. Thus we expect the magnetization to oscillate between positive and negative values, and hence a vanishing time-averaged magnetization $M_{z} = 0$. (At finite $n$ the oscillation has an amplitude that decays to zero.) Thus there is a critical value of $B/J_0$ that separates the two phases determined by when the energy of the initial state matches exactly the height of the barrier. This critical value turns out to be at $B/J_0 =1$ for $\alpha = 0$.

Let us understand how the minimum of $M_{zz}$ serves as a good indicator of the location of the phase transition. Suppose we approach the critical point from $B/J_0 <1$. While we expect $M_{z} > 0$ and the evolution to be periodic, the system spends a larger and larger fraction of the time in a single period near $\theta = \pi/2$ as $B/J_0 \to 1^{-}$. Thus, as we approach $B/J_0 \to 1^{-}$, not only does $M_{z}$ get smaller but so does $M_{zz}$. Similarly, if we approach the critical point from $B/J_0 > 1$, we have $M_{z} = 0$ because the evolution spends equal amounts of time with $\cos \theta >0$ as $\cos \theta < 0$. For $B/J_0 \gg 1$, the wells are almost indistinguishable, and the system spends almost equal time in all regions and gives $M_{zz} \approx 1/2$. However, as we approach $B/J_0 \to 1^{+}$, the fraction of time the evolution spends near $\cos \theta = 0$ grows and hence $M_{zz}$ gets smaller. Thus we find a global minimum in $M_{zz}$ at the critical point.

We can be more formal about this description. The semiclassical equations of motion are given by the equations of motion of a classical unit-norm vector in 3-dimensional space (see for example Ref.~\cite{Owerre2015}):
\begin{subequations}
\begin{align}
\frac{1}{2} \sin \theta(t) \frac{d}{dt} \theta(t) &= {B} \sin \theta(t) \sin \phi(t) \\
- \frac{1}{2} \sin \theta(t) \frac{d}{dt} \phi(t) &= \cos\theta(t) \left( 2 J_0 \sin \theta(t) - B \cos \phi(t) \right).
\end{align}
\end{subequations}
It then follows that $d V / dt = 0$. For our initial condition with $\theta(0) = 0$, we have $V/J_0 = -1$, so the energy conservation condition gives us $\sin \theta(t) = B/J_0 \cos \phi(t)$. Using this condition, we can simplify our expression for the dynamics of the angle $\theta(t)$:
\begin{equation}
  \frac{d}{dt} \theta(t) = \left\{ \begin{array}{lr}
  2 J \sqrt{ \left(\frac{B}{J_0} \right)^2 - \sin^2 \theta(t) } \ , & \sin \phi(t)  > 0 \\
  -2 J \sqrt{ \left(\frac{B}{J_0} \right)^2 - \sin^2 \theta(t) } \ , & \sin \phi(t)  < 0
  \end{array} \right.
\end{equation}
This gives us the turning points of the semiclassical evolution when the evolution is constrained to a single well: for $B/J_0 < 1$, the minimum and maximum angles are $0$ and $\sin^{-1} (B/J_0)$.

To calculate $M_{z}$ and $M_{zz}$, it suffices to consider the values these time-averaged quantities take during a single half-period of the evolution: for $B/J_0 < 1$, $\cos \theta(t)$ and $\cos^2 \theta(t)$ have the same functional form for the first half-period as the second half-period; for $B/J_0 > 1$, $\cos \theta(t)$ in the second half-period has the opposite sign to the first half-period, whereas $\cos^2 \theta(t)$ has the same functional form. Therefore, we can write:
\begin{eqnarray}
  M_{z} = \left\{
  \begin{array}{lr}
  \frac{1}{T_{1/2}} \int_0^{T_{1/2}} \cos \theta(t) dt = \frac{\int_0^{\theta_t} \frac{\cos \theta}{\sqrt{\left( \frac{B}{J_0} \right)^2 - \sin^2(\theta)}} d\theta}{\int_0^{\theta_t} \frac{1}{   \sqrt{\left( \frac{B}{J_0} \right)^2 - \sin^2(\theta)}} d\theta} \ , & B/J_0 < 1\\
  0 \ , & B/J_0 > 1
\end{array}\right.
\end{eqnarray}
\begin{eqnarray}
M_{zz} = \frac{1}{T_{1/2}} \int_0^{T_{1/2}} \cos^2 \theta(t) dt  = \left\{
\begin{array}{lr}
 \frac{\int_0^{\theta_t} \frac{\cos^2 \theta}{\sqrt{\left( \frac{B}{J_0} \right)^2 - \sin^2(\theta)}} d\theta}{\int_0^{\theta_t} \frac{1}{ \sqrt{\left( \frac{B}{J_0} \right)^2 - \sin^2(\theta)}} d\theta} \ , & B/J_0 < 1\\
 \frac{\int_0^{\pi/2} \frac{\cos^2 \theta}{ \sqrt{\left( \frac{B}{J_0} \right)^2 - \sin^2(\theta)}} d\theta}{\int_0^{\pi/2} \frac{1}{ \sqrt{\left( \frac{B}{J_0} \right)^2 - \sin^2(\theta)}} d\theta} \ , & B/J_0 > 1
\end{array}\right.
\end{eqnarray}
where $\theta_t = \sin^{-1}\left( \frac{B}{J_0} \right)$. We can now readily check that $M_{zz}$ attains a minimum as we take $B/J_0 \to 1$.

\clearpage

\addcontentsline{toc}{chapter}{Bibliography}
\bibliographystyle{alpha-letters}
\bibliography{references/literature}

\end{document}